\documentclass[a4paper,11pt,titlepage]{amsart}
\usepackage{amssymb}
\usepackage{xcolor}
\usepackage{graphicx}
\usepackage{tikz}
\usetikzlibrary{calc,decorations.markings,decorations.pathmorphing,fit,matrix,positioning,shapes.multipart}
\definecolor{linkblue}{rgb}{0.10,0.25,0.45}
\usepackage[
  colorlinks=true,
  linkcolor=linkblue,
  citecolor=linkblue,
  urlcolor=linkblue
]{hyperref}

\makeatletter
\def\@settitle{\thispagestyle{empty}\begin{center}%
  \Large\bfseries
  \@title\par
  \end{center}}
\def\@setauthors{\begin{center}%
  \large\authors\par
  \vspace{0.75em}%
  \small
  \def\author##1{}%
  \def\address##1##2{##2\par\smallskip}%
  \def\email##1##2{\href{mailto:##2}{\texttt{##2}}\par}%
  \addresses
  \end{center}%
  \global\let\addresses\@empty}
\def\@setthanks{\long\def\thanks##1{\par##1\@addpunct.}\thankses}
\def\section{\@startsection{section}{1}%
  \z@{.7\linespacing\@plus\linespacing}{.5\linespacing}%
  {\normalfont\bfseries\centering\large}}
\def\subsubsection{\@startsection{subsubsection}{3}%
  \z@{.5\linespacing\@plus.7\linespacing}{-.5em}%
  {\normalfont\sffamily}}
\renewcommand{\tocsubsection}[3]{%
  \indentlabel{\smaller{\@ifnotempty{#2}{\hskip2em\ignorespaces#1 #2.\quad}}#3}}
\renewcommand{\tocsubsubsection}[3]{%
  \indentlabel{\smaller\smaller{\@ifnotempty{#2}{\hskip4em\ignorespaces#1 #2.\quad}}#3}}
\makeatother

\renewcommand{\MR}[1]{\unskip\space
  \href{https://mathscinet.ams.org/mathscinet-getitem?mr=#1}{\textsc{mr}}}
\allowdisplaybreaks[1]
\newcommand\bra[1]{\left<#1\right|}
\newcommand\ket[1]{\left|#1\right>}
\newcommand\oh{{\frac{1}{2}}}
\newcommand\der{\partial}
\newcommand\psis{\psi^\star}
\newcommand\e[1]{\ifinner{e^{#1}}\else{e^{\textstyle#1}}\fi}
\newcommand\normord[1]{{\,:\!#1\!:\,}}
\newcommand\vcenterbox[1]{\vcenter{\hbox{#1}}}
\renewenvironment{quote}{\list{}{\rightmargin\leftmargin}\item[]\sl}{\endlist}
\newcommand{\codim}{\mathop{\hbox{codim}}\nolimits}
\newcommand{\mdeg}{\mathop{\hbox{mdeg}}\nolimits}
\newcommand{\rank}{\mathop{\hbox{rank}}\nolimits}
\newcommand{\dfn}{\em}
\newcommand{\wt}{\mathop{\hbox{wt}}\nolimits}
\renewcommand{\Im}{\mathop{\hbox{Im}}\nolimits}

\newcommand\integers{{\mathbb Z}}
\newcommand\iso{{\,\cong\,}}
\newcommand\Sym{{\rm Sym}}
\newsavebox\vcbox
\newsavebox\vclab
\newdimen\labshift
\newcommand\vcenterboxlabel[2]{\sbox{\vcbox}{#1}%
\sbox{\vclab}{#2}%
\labshift=\ht\vclab\advance\labshift by\dp\vclab%
\advance\labshift by\ht\vcbox\advance\labshift by\dp\vcbox%
\advance\labshift by 5pt%
\raise-0.5\labshift\hbox to 0pt{\hbox to \wd\vcbox{\hfill\usebox{\vclab}\hfill}\hss}%
\vcenter{\hbox{\box\vcbox}}%
}
\newdimen{\cellsize}
\newcommand\medboxes{\setlength{\cellsize}{14.22pt}\def\boxformat{}}
\newcommand\smallboxes{\setlength{\cellsize}{10pt}\def\boxformat{\scriptstyle}}

\medboxes
\tikzset{tableaubox/.style={draw=black,thin,sharp corners,solid,minimum size=\cellsize,inner sep=0pt}}
\tikzset{tableau/.style={matrix,name=tab,matrix anchor=tab-1-1.south west,inner sep=1pt,matrix of math nodes,cells={anchor=center,draw=black,thin,solid,arrows=-},nodes={tableaubox,execute at begin node=\boxformat},nodes in empty cells,row sep={\cellsize,between origins},column sep={\cellsize,between origins}}}
\newcommand\missingcell{|[draw=none]|}
\makeatletter
\newcommand\cellextra[1]{#1\expandafter\tikz@lib@matrix@start@cell}
\makeatother
\newcommand\hdotscell{\cellextra{\draw[dotted] (-0.5*\cellsize,0.5*\cellsize) --++(\cellsize,0);}\missingcell}
\newcommand\vdotscell{\cellextra{\draw[dotted] (-0.5*\cellsize,-0.5*\cellsize) --++(0,\cellsize);}\missingcell}
\newcommand\vhdotscell{\cellextra{\draw[dotted] (-0.5*\cellsize,-0.5*\cellsize) --++(0,\cellsize) --++(\cellsize,0);}\missingcell}
\def\activate#1{\begingroup
  \lccode`\~=`#1%
  \lowercase{\endgroup \let~#1}%
  \catcode`#1=13\relax}
\activate &
\newcommand\tableau[1]{\tikz[baseline=0]
\node[tableau]{#1};}

\makeatletter
\newcommand{\gettikzxy}[3]{
  \tikz@scan@one@point\pgfutil@firstofone#1\relax
\pgfmathsetmacro{#2}{\the\pgf@x/\linkpatternunit}
\pgfmathsetmacro{#3}{\the\pgf@y/\linkpatternunit}
}
\tikzset{label anchor/.code={%
    \let\tikz@auto@anchor=\pgfutil@empty
    \def\tikz@anchor{#1}
  },
  label anchor/.default=center
}
\makeatother
\tikzset{arrow/.style={postaction={decorate,thick,decoration={markings,mark = at position #1 with {\arrow{>}}}}},arrow/.default=0.5}
\tikzset{invarrow/.style={postaction={decorate,thick,decoration={markings,mark = at position #1 with {\arrow{<}}}}},invarrow/.default=0.5}
\newdimen\linkpatternunit%
\newdimen\linkpatternvertexradius
\newcount\linkpatternsize%
\newcount\lpsize
\newif\iflinkpatterninverted
\newif\iflinkpatterntikzstarted
\newif\iflinkpatternboxed
\newif\iflinkpatternaxis
\newif\iflinkpatternstraightlines
\newif\iflinkpatternnumbered
\newif\iflinkpatternalias
\newif\iflinkpatternnode
\newif\iflinkpatterncentered
\newcount\linkpatternfused
\pgfkeys{/linkpattern/.cd,centered/.is if=linkpatterncentered,inverted/.is if=linkpatterninverted,numbered/.is if=linkpatternnumbered,tikzstarted/.is if=linkpatterntikzstarted,straight lines/.is if=linkpatternstraightlines,boxed/.is if=linkpatternboxed,axis/.is if=linkpatternaxis,vertexcolor/.store in=\linkpatternvertexcolor,edgecolor/.store in=\linkpatternedgecolor,boxcolor/.code={\linkpatternboxedtrue\def\linkpatternboxcolor{#1}},tikzoptions/.style={every linkpattern/.append style={#1}},size/.code={\linkpatternsize=#1},numbering0/.code={\def\lpnumbering{#1}\def\linkpatternnumbering{#1}},numbering/.style={numbered,numbering0={#1}},unit/.code={\linkpatternunit=#1},height/.store in=\linkpatternheight,shape/.store in=\linkpatternshape,looseness/.store in=\linkpatternlooseness,squareness/.store in=\linkpatternsquareness,extra space/.store in=\linkpatternextraspace,width/.store in=\linkpatternwidth,alias/.is if=linkpatternalias,pos/.store in=\linkpatternpos,labeloptions/.style={labeloptionslist/.append style={#1}},labeloptionslist/.style={inner sep=2pt,font=\scriptsize,execute at begin node=$,execute at end node=$,label anchor={#1+180}},nodeon/.is if=linkpatternnode,node/.style={nodeon,nodeoptionslist/.append style={#1}},nodeoptionslist/.style={},
pipedream/.style={shape=pipedream,looseness=0,straight lines,numbering0=tangle},
tangle/.style={shape=tangle,numbering0=tangle},
every linkpattern/.style={x=\linkpatternunit,y=\linkpatternunit},
fused/.code={\linkpatternfused=#1},
endpoint/.style={circle,thin,solid,draw=black,fill=\linkpatternvertexcolor,minimum size={2\linkpatternvertexradius},inner sep=0pt,draw opacity=1,fill opacity=1,transform shape=false},
vertex/.style={/linkpattern/endpoint},
edge/.style={very thick,solid,draw=\linkpatternedgecolor,draw opacity=1}}
\linkpatterncenteredfalse%
\linkpatterninvertedfalse%
\linkpatternnumberedfalse%
\linkpatterntikzstartedfalse%
\linkpatternboxedfalse%
\linkpatternaxistrue%
\linkpatternaliastrue%
\linkpatternstraightlinesfalse%
\def\linkpatternlooseness{0.2}
\def\linkpatternsquareness{0.35}
\def\linkpatternvertexcolor{red}%
\def\linkpatternedgecolor{blue}%
\def\linkpatternboxcolor{none}%
\def\linkpatternheight{0}
\def\linkpatternwidth{0}
\def\linkpatternshape{default}
\def\linkpatternnumbering{default}
\def\linkpatternpos{(0,0)}
\def\linkpatternextraspace{0}
\linkpatternnodefalse
\def\firstchar#1#2\empty{#1}%
\def\linkpatterndo#1#2{
\edef\param{\csname linkpattern#2\endcsname}
\edef\firstcharparam{\expandafter\firstchar\param\empty}
\expandafter\ifcat\firstcharparam a
\expandafter\ifx\csname linkpattern#1\param\endcsname\relax
\csname linkpattern#1unknown\endcsname
\else
\csname linkpattern#1\csname linkpattern#2\endcsname\endcsname
\fi
\else
\csname linkpattern#1unknown\endcsname
\fi
}%
\def\linkpatterncoordtangle{\ifnum\x>\lphalfsize\pgfmathparse{\lpsize+1-\x}\xdef\lpcoordx{\pgfmathresult}\xdef\lpcoordy{\lpheight}\xdef\lpangle{270}\else\xdef\lpcoordx{\x}\xdef\lpcoordy{-\lpheight}\xdef\lpangle{90}\fi}
\def\linkpatterncoordpipedream{\ifnum\x>\lphalfsize\pgfmathparse{\lpsize+1-\x-0.5}\xdef\lpcoordx{\pgfmathresult}\xdef\lpcoordy{0}\xdef\lpangle{270}\else\pgfmathparse{0.5-\x}\xdef\lpcoordy{\pgfmathresult}\xdef\lpcoordx{0}\xdef\lpangle{0}\fi}
\def\linkpatterncoordrectangle{
\ifnum\x>\lptqsize
\pgfmathparse{\lpsize+1-\x-0.5}\xdef\lpcoordx{\pgfmathresult}\xdef\lpcoordy{0}\xdef\lpangle{270}
\else\ifnum\x>\lphalfsize
\pgfmathparse{\x-\lptqsize-0.5}\xdef\lpcoordy{\pgfmathresult}\xdef\lpcoordx{\linkpatternwidth}\xdef\lpangle{180}
\else\ifnum\x>\linkpatternheight
\pgfmathparse{\x-\linkpatternheight-0.5}\xdef\lpcoordx{\pgfmathresult}\xdef\lpcoordy{-\linkpatternheight}\xdef\lpangle{90}
\else
\pgfmathparse{0.5-\x}\xdef\lpcoordy{\pgfmathresult}\xdef\lpcoordx{0}\xdef\lpangle{0}
\fi\fi\fi
}%
\def\linkpatternsetsizeunknown{
\global\lpsize=\linkpatternsize
\if\linkpatternheight0
\xdef\maxsep{0}
\foreach \x/\xx in \mylist%
{%
\edef\tempx{\withoutprime{\x}}
\edef\tempxx{\withoutprime{\xx}}
\pgfmathparse{max(\maxsep,abs(\tempx-\tempxx))}
\xdef\maxsep{\pgfmathresult}
}%
\pgfmathparse{0.25+0.8*\linkpatternsquareness*\maxsep}
\xdef\lpheight{\pgfmathresult}
\else
\xdef\lpheight{\linkpatternheight}
\fi
}
\newcount\tempsize
\def\linkpatternrightmostunknown{
\global\lpsize=0
\global\tempsize=0
\foreach\x/\labx in \linkpatternnumbering
{
\edef\tempx{\withoutprime{\x}}
\ifnum\lpsize<\tempx\global\lpsize=\tempx\fi
\global\advance\tempsize by 1
}
\ifnum\tempsize>\lpsize\global\lpsize=\tempsize\fi
}%
\def\linkpatternrightmostdefault{
\global\lpsize=0
\global\tempsize=0
\foreach \x/\y in \mylist
{
\edef\tempx{\withoutprime{\x}}
\ifnum\lpsize<\tempx\global\lpsize=\tempx\fi
\ifx\x\y
\global\advance\tempsize by 1
\else
\edef\tempy{\withoutprime{\y}}
\ifnum\lpsize<\tempy\global\lpsize=\tempy\fi%
\global\advance\tempsize by 2
\fi
}
\ifnum\tempsize>\lpsize\global\lpsize=\tempsize\fi
}%
\def\linkpatternrightmosttangle{
\global\lpsize=0
\global\tempsize=0
\foreach \x/\y in \mylist
{
\edef\tempx{\withoutprime{\x}}
\ifnum\lpsize<\tempx\global\lpsize=\tempx\fi
\ifx\x\y
\global\advance\tempsize by 1
\else
\edef\tempy{\withoutprime{\y}}
\ifnum\lpsize<\tempy\global\lpsize=\tempy\fi%
\global\advance\tempsize by 2
\fi
}
\global\advance\lpsize by\lpsize
\ifnum\tempsize>\lpsize\global\lpsize=\tempsize\fi
}%

\newcommand\linkpattern[2][]{
{
\pgfkeys{/linkpattern/.cd,#1}
\edef\mylist{#2}
\def\primetest##1'{}%
\def\hasaprime##1{\expandafter\primetest##1''}
\def\internalwithoutprime##1'{##1}%
\def\withoutprime##1{\if\hasaprime##1 %
\expandafter\internalwithoutprime##1\else ##1\fi}%
\iflinkpatternnumbered%
\iflinkpatterninverted
\tikzset{/linkpattern/lbl/.style n args={3}{label={[/linkpattern/labeloptionslist=-##1,##3] ##1:##2}}}%
\else%
\tikzset{/linkpattern/lbl/.style n args={3}{label={[/linkpattern/labeloptionslist=##1,##3] ##1:##2}}}%
\fi%
\else%
\tikzset{/linkpattern/lbl/.style={}}%
\fi%
\tikzifinpicture{\linkpatterntikzstartedtrue%
\begin{scope}[shift/.expanded=\linkpatternpos,/linkpattern/every linkpattern]
}{%
\linkpatterntikzstartedfalse%
\iflinkpatterncentered
\begin{tikzpicture}[baseline=(current  bounding  box.center),/linkpattern/every linkpattern]%
\else
\begin{tikzpicture}[baseline=0,/linkpattern/every linkpattern]%
\fi
}%
\begin{scope}[local bounding box=link pattern box]
\iflinkpatterninverted%
\begin{scope}[yscale=-1]%
\fi%
\linkpatterndo{setsize}{shape}
\ifnum\lpsize=0
\linkpatterndo{rightmost}{numbering}
\fi
\pgfmathtruncatemacro{\lphalfsize}{\lpsize/2}
\linkpatterndo{numbering}{numbering}
\iflinkpatternboxed
\linkpatterndo{drawbox}{shape}
\else
\iflinkpatternaxis
\linkpatterndo{drawaxis}{shape}
\fi
\fi
\foreach\xx/\xlab/\opt in \lpnumbering
{
\ifx\xlab\opt\def\opt{}\fi
\if\hasaprime\xx %
\pgfmathtruncatemacro{\xx}{\lpsize+1-\withoutprime{\xx}}
\fi
\ifnum\linkpatternfused>1
\pgfmathsetmacro{\x}{0.4*(0.5+\linkpatternfused*(0.5+floor((\xx-1)/\linkpatternfused)))+0.6*\xx}
\else
\def\x{\xx}
\fi
\linkpatterndo{coord}{shape}
\iflinkpatternalias\def\xlabb{\xlab}\else\def\xlabb{\xx}\fi
\path (\lpcoordx,\lpcoordy) coordinate[/linkpattern/vertex,/linkpattern/lbl={\lpangle+180}{\xlab}{\opt},alias=v\xlabb] (v\xx) ++(\lpangle:\linkpatternunit) coordinate[alias=vv\xlabb] (vv\xx); 
}
\foreach \a/\b/\c in \mylist
{
\if\hasaprime\a %
\pgfmathtruncatemacro{\a}{\lpsize+1-\withoutprime{\a}}
\fi
\ifx\b\c\def\c{}\fi
\draw[/linkpattern/edge]
\ifx\a\b
(v\a)
\c
--
++(0,\lpheight);
\else
\pgfextra{
\if\hasaprime\b %
\pgfmathtruncatemacro{\b}{\lpsize+1-\withoutprime{\b}}
\fi
\gettikzxy{(v\a)}{\ax}{\ay}
\gettikzxy{(v\b)}{\bx}{\by}
\gettikzxy{(vv\a)}{\axx}{\ayy}
\gettikzxy{(vv\b)}{\bxx}{\byy}
\pgfmathsetmacro{\dist}{sqrt((\ax-\bx)*(\ax-\bx)+(\ay-\by)*(\ay-\by))}
\pgfmathsetmacro{\abx}{(\axx-\ax)*\dist*\linkpatternsquareness+(\bx-\ax)*\linkpatternlooseness)}
\pgfmathsetmacro{\aby}{(\ayy-\ay)*\dist*\linkpatternsquareness+(\by-\ay)*\linkpatternlooseness)}
\pgfmathsetmacro{\bax}{(\bxx-\bx)*\dist*\linkpatternsquareness+(\ax-\bx)*\linkpatternlooseness)}
\pgfmathsetmacro{\bay}{(\byy-\by)*\dist*\linkpatternsquareness+(\ay-\by)*\linkpatternlooseness)}
}
(v\a)
\c
\iflinkpatternstraightlines
\pgfextra{
\pgfmathsetmacro{\t}{((\ax-\bx)*\bay-(\ay-\by)*\bax)/(\aby*\bax-\abx*\bay)}
\pgfmathsetmacro{\abx}{\t*\abx}
\pgfmathsetmacro{\aby}{\t*\aby}
}
[rounded corners=0.2\linkpatternunit] -- ++(\abx,\aby) -- (v\b);
\else
.. controls ++(\abx,\aby) and ++(\bax,\bay) .. 
\fi
(v\b);
\fi
}
\end{scope}
\iflinkpatternnode
\node[fit=(link pattern box),/linkpattern/nodeoptionslist] {};
\fi
\iflinkpatterninverted
\end{scope}
\fi
\iflinkpatterntikzstarted
\end{scope}
\else%
\end{tikzpicture}%
\fi%
}}%
\newcommand\tanglelinkpattern[3][]{%
{
\pgfkeys{/linkpattern/.cd,#1}
\iflinkpatterninverted
\begin{tikzpicture}[/linkpattern/every linkpattern,baseline=\linkpatternunit]%
\else
\begin{tikzpicture}[/linkpattern/every linkpattern,baseline=-\linkpatternunit]%
\fi
\linkpattern[#1,tikzstarted,numbered=false]{#3}
\pgfmathtruncatemacro{\lptempsize}{2*\linkpatternsize}
\iflinkpatterninverted
\begin{scope}[yshift=0.5*\linkpatternunit]
\else
\begin{scope}[yshift=-0.5*\linkpatternunit]
\fi
\linkpattern[tangle,#1,tikzstarted,size=\lptempsize,
numbering=halftangle,
height=0.5]{#2}
\end{scope}
\end{tikzpicture}%
}}
\newcommand\schubdiag[4][]{%
\pgfkeys{/linkpattern/.cd,#1}
\iflinkpatterntikzstarted\else%
\begin{tikzpicture}[scale=0.5]
\fi%
\iflinkpatterninverted%
\begin{scope}[yscale=-1]%
\fi%
\draw (0,0) grid (#2,#3);
\edef\mylist{#4}
\foreach\y/\x/\z in \mylist
{
\ifx\x\z
\draw[decorate,decoration={zigzag,
amplitude=1pt,segment length=5pt}]
(\x-0.5,#3) -- (\x-0.5,\y-0.5) node[circle,fill=black,inner sep=2pt] {} -- (#2,\y-0.5);
\else
\node at (\x-0.5,\y-0.5) {$\z$};
\fi
}
\iflinkpatterninverted
\end{scope}
\fi
\iflinkpatterntikzstarted\else%
\end{tikzpicture}%
\fi%
}
\makeatletter
\tikzset{circle split part fill/.style  args={#1,#2}{%
 alias=tmp@name,
  postaction={%
    insert path={
     \pgfextra{%
     \pgfpointdiff{\pgfpointanchor{\pgf@node@name}{center}}%
                  {\pgfpointanchor{\pgf@node@name}{east}}%
     \pgfmathsetmacro\insiderad{\pgf@x}
      \fill[#1] (\pgf@node@name.base) ([xshift=-\pgflinewidth]\pgf@node@name.east) arc
                          (0:180:\insiderad-\pgflinewidth)--cycle;
      \fill[#2] (\pgf@node@name.base) ([xshift=\pgflinewidth]\pgf@node@name.west)  arc
                           (180:360:\insiderad-\pgflinewidth)--cycle;                    }}}}}  
 \makeatother
\tikzset{bdot/.style={circle,circle split,draw,circle split part fill={black,white},thin,inner sep=1pt}}%
\tikzset{wdot/.style={circle,circle split,draw,circle split part fill={white,black},thin,inner sep=1pt}}%
\newcommand\circlelinkpattern[2][]{
{
\pgfkeys{/linkpattern/.cd,#1}
\iflinkpatterntikzstarted\else%
\begin{tikzpicture}[/linkpattern/every linkpattern]%
\fi%
\iflinkpatterninverted%
\begin{scope}[yscale=-1]%
\fi%
\global\lpsize=\linkpatternsize
\edef\mylist{#2}
\foreach \x/\y in \mylist
{
\ifnum\x>\lpsize\global\lpsize=\x\fi
\ifnum\y>\lpsize\global\lpsize=\y\fi
}
\iflinkpatternaxis
\draw (0,0) circle (1);
\fi
\foreach\x in {1,...,\lpsize}
{
\pgfmathparse{(0.3*floor((\x-1)/\linkpatternfused)+0.7*((\x-0.5)/\linkpatternfused-0.5))*\linkpatternfused*360/\lpsize}
\coordinate[/linkpattern/vertex] (v\x) at (\pgfmathresult:1);
}
\foreach \x/\y/\z in \mylist
{
\ifx\y\z%
\draw[/linkpattern/edge] (v\x) .. controls ($0.5*(v\x)$) and  ($0.5*(v\y)$) .. (v\y);
\else
\draw[/linkpattern/edge] \z (v\x) .. controls ($0.5*(v\x)$) and  ($0.5*(v\y)$) .. (v\y);
\fi
}
\iflinkpatternnumbered%
\pgfmathparse{\lpsize/\linkpatternfused}
\global\lpsize=\pgfmathresult
\def\linkpatternnumbering{1,...,\lpsize}
\newdimen\angle
\foreach\x/\xx/\opt in \linkpatternnumbering
{
  \pgfmathsetmacro{\angle}{360/\lpsize*(\x-1)}
\ifx\xx\opt%
  \node[outer sep=1pt,anchor=180+\angle] at (\angle:1) {$\scriptstyle\xx$}; 
\else
  \node[outer sep=1pt,anchor=180+\angle,\opt] at (\angle:1) {$\scriptstyle\xx$}; 
\fi
}
\fi%
\iflinkpatterninverted%
\end{scope}
\fi%
\iflinkpatterntikzstarted\else%
\end{tikzpicture}%
\fi%
}}%
\newdimen{\loopcellsize}
\tikzset{bgplaq/.style={draw=black,fill=\linkpatternboxcolor}}
\def\plaqwest{}
\def\plaqeast{}
\def\plaqnorth{}
\def\plaqsouth{}
\def\plaqa{
\draw[/linkpattern/edge,\plaqwest,\plaqnorth] (0,0.5) .. controls (0,0.2) and (-0.2,0) .. (-0.5,0);
\draw[/linkpattern/edge,\plaqeast,\plaqsouth] (0,-0.5) .. controls (0,-0.2) and (0.2,0) .. (0.5,0);
}

\def\plaqc{
\draw[/linkpattern/edge,\plaqsouth,\plaqnorth] (0,0.5) -- (0,-0.5);
\draw[/linkpattern/edge,\plaqwest,\plaqeast] (0.5,0) -- (-0.5,0);
}
\def\plaqj{
\draw[/linkpattern/edge,\plaqwest,\plaqnorth] (0,0.5) .. controls (0,0.2) and (-0.2,0) .. (-0.5,0);
}

\pgfkeys{/linkpattern/.cd,west/.store in=\plaqwest,east/.store in=\plaqeast,north/.store in=\plaqnorth,south/.store in=\plaqsouth}
\def\plaqname{plaq}
\newcommand\plaq[2][]{
\node[bgplaq,rectangle,draw,minimum size=\loopcellsize,transform shape] (\plaqname) {};
\useasboundingbox;
\pgfkeys{/linkpattern/.cd,#1}
\ifx#2\empty\else
\begin{scope}[x=\loopcellsize,y=\loopcellsize]
\csname plaq#2\endcsname
\end{scope}\fi
}
\tikzset{loop/.code={\def\plaqname{loop-\the\pgfmatrixcurrentrow-\the\pgfmatrixcurrentcolumn}},loop/.append style={matrix,row sep={\loopcellsize,between origins},column sep={\loopcellsize,between origins}}}

\newcommand{\hdrdiracsea}[1]{%
  \begin{tikzpicture}[x=16pt,y=16pt,baseline=(current bounding box.center)]
    \useasboundingbox (0,0) rectangle (10,2);
    \draw[gray,densely dotted,line width=0.4pt] (5,0)--(5,2);
    \draw[line width=0.7pt] (0,1)--(10,1);
    \foreach \x in {0.5,1.5,...,4.5}
      \fill (\x,1) circle[radius=2.4pt];
    \foreach \x in {5.5,6.5,...,9.5}
      \filldraw[fill=white,draw=black,line width=0.6pt]
        (\x,1) circle[radius=2.4pt];
    \begin{scope}[overlay]
      \node[font=\tiny,anchor=base] at (5,-0.5) {$#1$};
    \end{scope}
  \end{tikzpicture}}

\newcommand{\hdrtransfermatrixlattice}{%
  \begin{tikzpicture}[x=0.6cm,y=0.6cm,baseline=(current bounding box.center)]
    \useasboundingbox (0,0) rectangle (5,2);
    \draw[line width=0.7pt] (0,1)--(5,1);
    \foreach \x in {1,...,4}\draw[line width=0.7pt] (\x,0)--(\x,2);
    \begin{scope}[overlay]
      \node[font=\tiny,anchor=base west] at (0,0.5) {$0$};
      \foreach \x in {1,...,4}
        \node[font=\tiny,anchor=base west] at ({\x+0.1},0) {$\x$};
    \end{scope}
  \end{tikzpicture}}

\newcommand{\hdrupperleftcorner}{%
  \tikz[baseline=0pt,x=10pt,y=10pt]{%
    \useasboundingbox (0,0) rectangle (0.7,1.2);
    \begin{scope}[overlay]
      \draw[line width=0.4pt] (-0.3,0)--(0.4,0)--(0.4,0.7);
      \node[anchor=base west,inner sep=0pt] at (-0.7,-0.1) {$\scriptstyle i$};
      \node[anchor=base west,inner sep=0pt] at (0.3,0.9) {$\scriptstyle j$};
    \end{scope}}}
\newcommand{\hdrlowerleftcorner}{%
  \tikz[baseline=0pt,x=10pt,y=10pt]{%
    \useasboundingbox (0,0) rectangle (0.7,1);
    \begin{scope}[overlay]
      \draw[line width=0.4pt] (-0.3,0.8)--(0.4,0.8)--(0.4,0.1);
      \node[anchor=base west,inner sep=0pt] at (-0.6,0.7) {$\scriptstyle i$};
      \node[anchor=base west,inner sep=0pt] at (0.3,-0.3) {$\scriptstyle j$};
    \end{scope}}}

\tikzset{
  hdr six arrow/.style={postaction={decorate,decoration={markings,
    mark=at position #1 with {\pgftransformxshift{2.10pt}\arrow[scale=2.15]{>}}}}},
  hdr six arrow/.default=0.5,
  hdr six invarrow/.style={postaction={decorate,decoration={markings,
    mark=at position #1 with {\pgftransformxshift{2.10pt}\arrow[scale=2.15]{<}}}}},
  hdr six invarrow/.default=0.5,
  hdr six small arrow/.style={postaction={decorate,decoration={markings,
    mark=at position #1 with {\pgftransformxshift{1.70pt}\arrow[scale=1.4]{>}}}}},
  hdr six small arrow/.default=0.5,
  hdr six small invarrow/.style={postaction={decorate,decoration={markings,
    mark=at position #1 with {\pgftransformxshift{1.70pt}\arrow[scale=1.4]{<}}}}},
  hdr six small invarrow/.default=0.5
}
\newcommand{\hdrsixhforward}[2]{%
  \draw[line width=0.65pt,hdr six arrow=0.5]
    (#1,#2) -- ++(1,0);}
\newcommand{\hdrsixhbackward}[2]{%
  \draw[line width=0.65pt,hdr six invarrow=0.5]
    (#1,#2) -- ++(1,0);}
\newcommand{\hdrsixvforward}[2]{%
  \draw[line width=0.65pt,hdr six arrow=0.5]
    (#1,#2) -- ++(0,1);}
\newcommand{\hdrsixvbackward}[2]{%
  \draw[line width=0.65pt,hdr six invarrow=0.5]
    (#1,#2) -- ++(0,1);}

\newcommand{\hdrsixhforwardshort}[2]{%
  \draw[line width=0.65pt,hdr six arrow=0.5]
    ({#1+0.34},#2) -- ({#1+0.66},#2);}
\newcommand{\hdrsixhbackwardshort}[2]{%
  \draw[line width=0.65pt,hdr six invarrow=0.5]
    ({#1+0.34},#2) -- ({#1+0.66},#2);}
\newcommand{\hdrsixvforwardshort}[2]{%
  \draw[line width=0.65pt,hdr six arrow=0.5]
    (#1,{#2+0.34}) -- (#1,{#2+0.66});}
\newcommand{\hdrsixvbackwardshort}[2]{%
  \draw[line width=0.65pt,hdr six invarrow=0.5]
    (#1,{#2+0.34}) -- (#1,{#2+0.66});}
\newcommand{\hdrsixhforwardsmall}[2]{%
  \draw[line width=0.65pt,hdr six small arrow=0.5]
    ({#1+0.34},#2) -- ({#1+0.66},#2);}
\newcommand{\hdrsixhbackwardsmall}[2]{%
  \draw[line width=0.65pt,hdr six small invarrow=0.5]
    ({#1+0.34},#2) -- ({#1+0.66},#2);}
\newcommand{\hdrsixvforwardsmall}[2]{%
  \draw[line width=0.65pt,hdr six small arrow=0.5]
    (#1,{#2+0.34}) -- (#1,{#2+0.66});}
\newcommand{\hdrsixvbackwardsmall}[2]{%
  \draw[line width=0.65pt,hdr six small invarrow=0.5]
    (#1,{#2+0.34}) -- (#1,{#2+0.66});}

\tikzset{
  hdr nilp arrow/.style={postaction={decorate,decoration={markings,
    mark=at position 0.5 with {\pgftransformxshift{0.95pt}\arrow[scale=1.8]{>}}}}},
  hdr nilp particle edge/.style={green!65!black,line width=0.5pt,
    dash pattern=on 2pt off 2pt},
  hdr nilp hole edge/.style={red,line width=0.5pt,
    dash pattern=on 2pt off 2pt}
}
\newcommand{\hdrnilpparticlegraph}[1][1cm]{%
  \begin{tikzpicture}[x=#1,y=#1,baseline=(current bounding box.center)]
    \useasboundingbox (-0.5,0) rectangle (6.5,4);
    \begin{scope}[hdr nilp particle edge]
      \foreach \x in {0,...,6}{
        \draw (\x,0) -- (\x,4);
        \foreach \y in {1,...,3}{
          \path[hdr nilp arrow] (\x,{\y-0.5}) -- (\x,{\y+0.5});
        }
      }
      \foreach \y in {0.5,1.5,2.5,3.5}{
        \draw (0,\y) -- (6,\y);
        \foreach \x in {0,...,5}{
          \path[hdr nilp arrow] (\x,\y) -- ({\x+1},\y);
        }
      }
    \end{scope}
    \foreach \x in {0,...,6}{
      \fill (\x,0) circle[radius=2.35pt];
      \fill (\x,4) circle[radius=2.35pt];
    }
  \end{tikzpicture}}

\newcommand{\hdrnilpholegraph}[1][1cm]{%
  \begin{tikzpicture}[x=#1,y=#1,baseline=(current bounding box.center)]
    \useasboundingbox (-0.5,0) rectangle (6.5,4);
    \begin{scope}[hdr nilp hole edge]
      \foreach \x in {0,...,6}{
        \draw (\x,0) -- (\x,4);
        \foreach \y in {0,...,3}{
          \path[hdr nilp arrow] (\x,\y) -- (\x,{\y+1});
        }
      }
      \foreach \x in {0,...,5}{
        \foreach \y in {0,...,3}{
          \draw (\x,{\y+1}) -- ({\x+1},\y);
          \path[hdr nilp arrow] ({\x+1},\y) -- (\x,{\y+1});
        }
      }
    \end{scope}
    \foreach \x in {0,...,6}{
      \filldraw[fill=white,draw=black,line width=0.7pt]
        (\x,0) circle[radius=2.35pt];
      \filldraw[fill=white,draw=black,line width=0.7pt]
        (\x,4) circle[radius=2.35pt];
    }
  \end{tikzpicture}}

\tikzset{
  hdr fermion boundary/.style={black,line width=0.52pt},
  hdr fermion separator/.style={gray,densely dotted,line width=0.52pt},
  hdr particle path/.style={green!65!black,line width=1.1pt,
    line cap=round,line join=round},
  hdr hole path/.style={red,line width=1.1pt,
    line cap=round,line join=round},
  hdr particle/.style={circle,draw=black,fill=black,line width=0.4pt,
    minimum size=4.7pt,inner sep=0pt},
  hdr hole/.style={circle,draw=black,fill=white,line width=0.7pt,
    minimum size=4.7pt,inner sep=0pt}
}
\newcommand{\hdrfermionparticles}[1]{%
  \foreach \p in {#1}{\node[hdr particle] at \p {};}}
\newcommand{\hdrfermionholes}[1]{%
  \foreach \p in {#1}{\node[hdr hole] at \p {};}}

\newcommand{\hdrlatticefermionfigure}{%
  \begin{tikzpicture}[x=0.90cm,y=0.90cm,baseline=(current bounding box.center)]
    \useasboundingbox (-0.65,-0.58) rectangle (13.15,4.58);
    \draw[hdr fermion boundary] (-0.5,0) -- (6.75,0);
    \draw[hdr fermion boundary] (-0.5,4) -- (6.75,4);
    \draw[hdr fermion separator] (2.5,-0.5) -- (2.5,4.5);

    \draw[hdr hole path] (1,0) -- (0,1) -- (0,4);
    \draw[hdr hole path] (2,4) -- (2,2) -- (4,0);
    \draw[hdr hole path] (3,4) -- (4,3) -- (4,1) -- (5,0);
    \draw[hdr hole path] (6,0) -- (6,2) -- (5,3) -- (5,4);
    \draw[hdr particle path] (6,4) -- (6,2.5) -- (5,2.5)
      -- (5,0.5) -- (3,0.5) -- (3,0);
    \draw[hdr particle path] (1,4) -- (1,0.5) -- (0,0.5) -- (0,0);
    \draw[hdr particle path] (2,0) -- (2,1.5) -- (3,1.5)
      -- (3,3.5) -- (4,3.5) -- (4,4);

    \hdrfermionparticles{(0,0),(1,1),(1,2),(1,3),(1,4),
      (2,0),(2,1),(3,0),(3,2),(3,3),(4,4),
      (5,1),(5,2),(6,3),(6,4)}
    \hdrfermionholes{(0,1),(0,2),(0,3),(0,4),(1,0),
      (2,2),(2,3),(2,4),(3,1),(3,4),
      (4,0),(4,1),(4,2),(4,3),(5,0),(5,3),(5,4),
      (6,0),(6,1),(6,2)}

    \node[inner sep=0pt] (hdrfermionyoungtop) at (8.25,4)
      {\begingroup\smallboxes\tableau{&&&\\&&\\\\}\endgroup};
    \node[anchor=west,inner sep=0pt] at (hdrfermionyoungtop.west |- 0,3)
      {\begingroup\smallboxes\tableau{&&&\\&\\\\}\endgroup};
    \node[anchor=west,inner sep=0pt] at (hdrfermionyoungtop.west |- 0,2)
      {\begingroup\smallboxes\tableau{&&\\&\\\\}\endgroup};
    \node[anchor=west,inner sep=0pt] at (hdrfermionyoungtop.west |- 0,1)
      {\begingroup\smallboxes\tableau{&&\\\\\\}\endgroup};
    \node[anchor=west,inner sep=0pt] at (hdrfermionyoungtop.west |- 0,0)
      {\begingroup\smallboxes\tableau{\\\\}\endgroup};
    \node[inner sep=0pt] at (11.35,2)
      {\begingroup\medboxes
        \tableau{\emptycell&1&1&3\\\emptycell&2&4\\1\\}%
       \endgroup};
  \end{tikzpicture}}

\newcommand{\hdrlatticefermionstandardfigure}{%
  \begin{tikzpicture}[x=0.90cm,y=0.90cm,baseline=(current bounding box.center)]
    \useasboundingbox (-0.65,-0.58) rectangle (11.55,4.58);
    \draw[hdr fermion boundary] (-0.5,0) -- (5.5,0);
    \draw[hdr fermion boundary] (-0.5,4) -- (5.5,4);
    \draw[hdr fermion separator] (2.5,-0.5) -- (2.5,4.5);

    \draw[hdr particle path] (0,0) -- (0,4);
    \draw[hdr particle path] (1,0) -- (1,1) -- (2,2)
      -- (2,3) -- (3,4);
    \draw[hdr particle path] (4,4) -- (4,3) -- (3,2)
      -- (3,1) -- (2,0);
    \draw[hdr hole path] (1,4) -- (1,2) -- (3,0);
    \draw[hdr hole path] (2,4) -- (4,2) -- (4,0);
    \draw[hdr hole path] (5,4) -- (5,0);

    \hdrfermionparticles{(0,0),(0,1),(0,2),(0,3),(0,4),
      (1,0),(1,1),(2,0),(2,2),(2,3),(3,1),(3,2),(3,4),
      (4,1),(4,3),(4,4)}
    \hdrfermionholes{(1,2),(1,3),(1,4),(2,1),(2,4),
      (3,0),(3,3),(4,0),(4,2),(5,0),(5,1),(5,2),(5,3),(5,4)}

    \node[inner sep=0pt] (hdrfermionstandardyoungtop) at (7.15,4)
      {\begingroup\smallboxes\tableau{&\\&\\}\endgroup};
    \node[anchor=west,inner sep=0pt] at (hdrfermionstandardyoungtop.west |- 0,3)
      {\begingroup\smallboxes\tableau{&\\\\}\endgroup};
    \node[anchor=west,inner sep=0pt] at (hdrfermionstandardyoungtop.west |- 0,2)
      {\begingroup\smallboxes\tableau{\\\\}\endgroup};
    \node[anchor=west,inner sep=0pt] at (hdrfermionstandardyoungtop.west |- 0,1)
      {\begingroup\smallboxes\tableau{\\}\endgroup};
    \node[anchor=west,inner sep=0pt] at (hdrfermionstandardyoungtop.west |- 0,0)
      {$\varnothing$};
    \node[inner sep=0pt] at (10.05,2)
      {\begingroup\medboxes\tableau{1&3\\2&4\\}\endgroup};
  \end{tikzpicture}}

\definecolor{hdrppgreen}{HTML}{D1FFC8}
\definecolor{hdrppblue}{HTML}{AEC0FF}
\definecolor{hdrpppink}{HTML}{FFCACA}
\definecolor{hdrpporange}{HTML}{E6842A}
\tikzset{
  hdr pp tile/.style={draw=black,line width=0.38pt,line join=round},
  hdr pp outline/.style={draw=black,line width=0.48pt,line join=round},
  hdr pp honeycomb/.style={draw=gray!78,dash pattern=on 1pt off 1pt,
    line width=0.43pt,line cap=butt,line join=miter},
  hdr pp dimer/.style={draw=black,line width=1.45pt,
    line cap=butt,line join=miter}
}
\newcommand{\hdrppface}[2]{\path[hdr pp tile,fill=#1] #2 -- cycle;}
\newcommand{\hdrpphex}[2]{%
  \draw[hdr pp honeycomb]
    ({#1-198.5},{#2-344.5}) -- ({#1+198.5},{#2-344.5})
    -- ({#1+397.5},{#2}) -- ({#1+198.5},{#2+344.5})
    -- ({#1-198.5},{#2+344.5}) -- ({#1-397.5},{#2}) -- cycle;}

\newcommand{\hdrplanepartitionfigure}{%
  \begin{tikzpicture}[x=0.00098cm,y=-0.00098cm,
      baseline=(current bounding box.center)]
    \useasboundingbox (1350,700) rectangle (12900,6000);

    \hdrppface{hdrppgreen}{(2100,1546)--(2700,1223)--(3300,1546)--(2700,1869)}
    \hdrppface{hdrppblue}{(2700,1869)--(2700,2515)--(3300,2192)--(3300,1546)}
    \hdrppface{hdrpppink}{(2100,1546)--(2100,2192)--(2700,2515)--(2700,1869)}
    \hdrppface{hdrpppink}{(3300,1546)--(3300,2192)--(3900,2515)--(3900,1869)}
    \hdrppface{hdrpppink}{(3900,1869)--(3900,2515)--(4500,2838)--(4500,2192)}
    \hdrppface{hdrpppink}{(4500,2192)--(4500,2838)--(5100,3162)--(5100,2515)}
    \hdrppface{hdrppblue}{(1500,1869)--(1500,2515)--(2100,2192)--(2100,1546)}
    \hdrppface{hdrppgreen}{(2700,2515)--(3300,2192)--(3900,2515)--(3300,2838)}
    \hdrppface{hdrppblue}{(1500,2515)--(1500,3162)--(2100,2838)--(2100,2192)}
    \hdrppface{hdrpppink}{(2100,2192)--(2100,2838)--(2700,3162)--(2700,2515)}
    \hdrppface{hdrppgreen}{(1500,3162)--(2100,2838)--(2700,3162)--(2100,3485)}
    \hdrppface{hdrppblue}{(2100,3485)--(2100,4131)--(2700,3808)--(2700,3162)}
    \hdrppface{hdrpppink}{(1500,3162)--(1500,3808)--(2100,4131)--(2100,3485)}
    \hdrppface{hdrpppink}{(2700,2515)--(2700,3162)--(3300,3485)--(3300,2838)}
    \hdrppface{hdrpppink}{(2700,3162)--(2700,3808)--(3300,4131)--(3300,3485)}
    \hdrppface{hdrppgreen}{(2100,4131)--(2700,3808)--(3300,4131)--(2700,4454)}
    \hdrppface{hdrppblue}{(3300,2838)--(3300,3485)--(3900,3162)--(3900,2515)}
    \hdrppface{hdrpppink}{(3900,2515)--(3900,3162)--(4500,3485)--(4500,2838)}
    \hdrppface{hdrpppink}{(4500,2838)--(4500,3485)--(5100,3808)--(5100,3162)}
    \hdrppface{hdrpppink}{(4500,3485)--(4500,4131)--(5100,4454)--(5100,3808)}
    \hdrppface{hdrppgreen}{(3900,4454)--(4500,4131)--(5100,4454)--(4500,4777)}
    \hdrppface{hdrppgreen}{(3300,4777)--(3900,4454)--(4500,4777)--(3900,5100)}
    \hdrppface{hdrppgreen}{(2700,4454)--(3300,4131)--(3900,4454)--(3300,4777)}
    \hdrppface{hdrppgreen}{(3300,3485)--(3900,3162)--(4500,3485)--(3900,3808)}
    \hdrppface{hdrpppink}{(3300,3485)--(3300,4131)--(3900,4454)--(3900,3808)}
    \hdrppface{hdrppblue}{(3900,3808)--(3900,4454)--(4500,4131)--(4500,3485)}
    \draw[hdr pp outline] (2700,1223)--(1500,1869)--(1500,3808)
      --(3900,5100)--(5100,4454)--(5100,2515)--cycle;

    \hdrppface{hdrpppink}{(6300,1869)--(6300,2515)--(6900,2838)--(6900,2192)}
    \hdrppface{hdrppgreen}{(6600,1223)--(7200,900)--(7800,1223)--(7200,1546)}
    \hdrppface{hdrppblue}{(7500,2192)--(7500,2838)--(8100,2515)--(8100,1869)}
    \draw[hdr pp dimer] (7020,1215)--(7425,1215);
    \draw[hdr pp dimer] (6515,2202)--(6710,2547);
    \draw[hdr pp dimer] (7697,2536)--(7907,2206);

    \foreach \x/\y in {9881.5/2159.5,11696.5/3165.5,
      11711.5/3855.5,12295.5/3496.5,12278.5/2810.5,
      10466.5/1815.5,10489.5/3194.5,11101.5/3518.5,
      9902.5/3549.5,10501.5/3882.5,11114.5/4211.5,
      11086.5/2836.5,11075.5/2144.5,11678.5/2477.5,
      11721.5/4545.5,12311.5/4191.5,9881.5/2849.5,
      10481.5/2504.5}{\hdrpphex{\x}{\y}}
    \foreach \xa/\ya/\xb/\yb in {
      9705/3195/10110/3195,12135/4530/12540/4530,
      9480/2160/9690/1830,9480/2835/9690/2505,
      11910/4185/12120/3855,11862/2123/12057/2468,
      12474/2476/12669/2821,12517/3840/12712/4185,
      10717/3517/10912/3862,10072/1815/10267/2160,
      11880/2805/12075/3150,10099/3878/10309/3548,
      10305/4221/10710/4221,10917/4557/11322/4557,
      11526/4887/11931/4887,11315/3858/11510/4203,
      11293/3156/11503/2826,10692/2844/10887/3189,
      11272/1800/11467/2145,10875/2484/11280/2484,
      10263/1464/10668/1464,12496/3154/12691/3499,
      11508/3510/11913/3510,9507/3552/9702/3897,
      10666/2140/10876/1802,10083/2507/10278/2852}
      {\draw[hdr pp dimer] (\xa,\ya)--(\xb,\yb);}

    \node[font=\normalsize] at (3084,5792) {(a)};
    \node[font=\normalsize] at (11240,5779) {(b)};
  \end{tikzpicture}}

\tikzset{
  hdr pp nilp/.style={hdr hole path}
}
\newcommand{\hdrplanepartitionnilptiling}{%
  \begin{tikzpicture}[x=0.00105cm,y=-0.00105cm,
      baseline=(current bounding box.center)]
    \useasboundingbox (3300,2350) rectangle (9250,8200);

    \foreach \p in {
      {(3938,3900)--(3600,4500)--(3938,5100)--(4275,4500)},
      {(4275,3300)--(3938,3900)--(4275,4500)--(4613,3900)},
      {(4613,3900)--(4275,4500)--(4613,5100)--(4950,4500)},
      {(4275,4500)--(3938,5100)--(4275,5700)--(4613,5100)},
      {(4950,4500)--(4613,5100)--(4950,5700)--(5288,5100)},
      {(4613,5100)--(4275,5700)--(4613,6300)--(4950,5700)},
      {(5288,5100)--(4950,5700)--(5288,6300)--(5625,5700)},
      {(4950,5700)--(4613,6300)--(4950,6900)--(5288,6300)},
      {(5288,2700)--(4950,3300)--(5288,3900)--(5625,3300)},
      {(5625,3300)--(5288,3900)--(5625,4500)--(5963,3900)},
      {(5963,3900)--(5625,4500)--(5963,5100)--(6300,4500)},
      {(7650,4500)--(7313,5100)--(7650,5700)--(7988,5100)},
      {(7650,5700)--(7313,6300)--(7650,6900)--(7988,6300)},
      {(6638,6300)--(6300,6900)--(6638,7500)--(6975,6900)},
      {(8663,5100)--(8325,5700)--(8663,6300)--(9000,5700)}}
      {\expandafter\hdrppface\expandafter{hdrppgreen}{\p}}

    \foreach \p in {
      {(4275,3300)--(4613,2700)--(5288,2700)--(4950,3300)},
      {(4950,6900)--(5288,6300)--(5963,6300)--(5625,6900)},
      {(5288,6300)--(5625,5700)--(6300,5700)--(5963,6300)},
      {(5963,6300)--(6300,5700)--(6975,5700)--(6638,6300)},
      {(5625,6900)--(5963,6300)--(6638,6300)--(6300,6900)},
      {(5963,5100)--(6300,4500)--(6975,4500)--(6638,5100)},
      {(6638,5100)--(6975,4500)--(7650,4500)--(7313,5100)},
      {(7650,5700)--(7988,5100)--(8663,5100)--(8325,5700)},
      {(7650,6900)--(7988,6300)--(8663,6300)--(8325,6900)},
      {(7313,7500)--(7650,6900)--(8325,6900)--(7988,7500)},
      {(6638,7500)--(6975,6900)--(7650,6900)--(7313,7500)},
      {(6638,6300)--(6975,5700)--(7650,5700)--(7313,6300)}}
      {\expandafter\hdrppface\expandafter{hdrpppink}{\p}}

    \foreach \p in {
      {(5625,6900)--(4950,6900)--(5288,7500)--(5963,7500)},
      {(6300,6900)--(5625,6900)--(5963,7500)--(6638,7500)},
      {(4950,3300)--(4275,3300)--(4613,3900)--(5288,3900)},
      {(5288,3900)--(4613,3900)--(4950,4500)--(5625,4500)},
      {(5625,4500)--(4950,4500)--(5288,5100)--(5963,5100)},
      {(5963,5100)--(5288,5100)--(5625,5700)--(6300,5700)},
      {(6638,5100)--(5963,5100)--(6300,5700)--(6975,5700)},
      {(7313,5100)--(6638,5100)--(6975,5700)--(7650,5700)},
      {(8325,4500)--(7650,4500)--(7988,5100)--(8663,5100)},
      {(5963,2700)--(5288,2700)--(5625,3300)--(6300,3300)},
      {(6300,3300)--(5625,3300)--(5963,3900)--(6638,3900)},
      {(6975,3300)--(6300,3300)--(6638,3900)--(7313,3900)},
      {(7650,3300)--(6975,3300)--(7313,3900)--(7988,3900)},
      {(6638,3900)--(5963,3900)--(6300,4500)--(6975,4500)},
      {(7313,3900)--(6638,3900)--(6975,4500)--(7650,4500)},
      {(7988,3900)--(7313,3900)--(7650,4500)--(8325,4500)},
      {(6638,2700)--(5963,2700)--(6300,3300)--(6975,3300)},
      {(7313,2700)--(6638,2700)--(6975,3300)--(7650,3300)},
      {(7313,6300)--(6638,6300)--(6975,6900)--(7650,6900)},
      {(8325,5700)--(7650,5700)--(7988,6300)--(8663,6300)}}
      {\expandafter\hdrppface\expandafter{hdrppblue}{\p}}

    \draw[hdr pp outline] (5288,7500)--(7988,7500)--(9000,5700)
      --(7313,2700)--(4613,2700)--(3600,4500)--cycle;

    \begin{scope}[hdr pp nilp]
      \draw (5625,7500)--(5288,6900)--(5963,5700)--(4613,3300)--(4950,2700);
      \draw (6300,7500)--(5963,6900)--(6638,5700)--(6300,5100)
        --(6638,4500)--(5625,2700);
      \draw (6975,7500)--(7313,6900)--(6975,6300)--(7313,5700)
        --(6975,5100)--(7313,4500)--(6300,2700);
      \draw (7650,7500)--(8325,6300)--(7988,5700)--(8325,5100)--(6975,2700);
    \end{scope}
    \foreach \x/\y in {4950/2700,5625/2700,6300/2700,6975/2700,
      5625/7500,6300/7500,6975/7500,7650/7500}
      {\filldraw[fill=white,draw=black,line width=0.7pt]
        (\x,\y) circle[radius=2.35pt];}
    \node at (3750,3450) {$a$};
    \node at (4050,6300) {$b$};
    \node at (6600,7950) {$c$};
  \end{tikzpicture}}

\newcommand{\hdrplanepartitionnilplattice}{%
  \begin{tikzpicture}[x=0.54cm,y=0.54cm,
      baseline=(current bounding box.center)]
    \useasboundingbox (-0.45,-0.45) rectangle (10.45,8.45);
    \begin{scope}[hdr pp nilp]
      \draw (6,0)--(5,1)--(5,3)--(1,7)--(1,8);
      \draw (7,0)--(6,1)--(6,3)--(5,4)--(5,5)--(2,8);
      \draw (8,0)--(8,1)--(7,2)--(7,3)--(6,4)--(6,5)--(3,8);
      \draw (9,0)--(9,2)--(8,3)--(8,4)--(4,8);
    \end{scope}
    \foreach \x in {0,...,10}
      \foreach \y in {0,...,8}
        {\node[hdr particle] at (\x,\y) {};}
    \hdrfermionholes{(6,0),(7,0),(8,0),(9,0),(10,0),
      (5,1),(6,1),(8,1),(9,1),(10,1),
      (5,2),(6,2),(7,2),(9,2),(10,2),
      (5,3),(6,3),(7,3),(8,3),(10,3),
      (4,4),(5,4),(6,4),(8,4),(10,4),
      (3,5),(5,5),(6,5),(7,5),(10,5),
      (2,6),(4,6),(5,6),(6,6),(10,6),
      (1,7),(3,7),(4,7),(5,7),(10,7),
      (1,8),(2,8),(3,8),(4,8),(10,8)}
  \end{tikzpicture}}

\newcommand{\hdrplanepartitionnilpfigure}{%
  \makebox[\textwidth][c]{\hdrplanepartitionnilptiling\hspace{0.65cm}%
    \hdrplanepartitionnilplattice}}

\newcommand{\hdrtsscppsector}[3]{%
  \foreach \p in {
    {(2668,4058)--(2446,4183)--(2446,4437)--(2668,4312)},
    {(2446,4437)--(2222,4565)--(2222,4818)--(2446,4690)},
    {(2222,4818)--(1999,4944)--(1999,5197)--(2222,5070)},
    {(1999,5197)--(1776,5325)--(1776,5578)--(1999,5452)},
    {(2446,5197)--(2222,5325)--(2222,5578)--(2446,5452)},
    {(2892,4944)--(2668,5070)--(2668,5325)--(2892,5197)},
    {(2668,4312)--(2446,4437)--(2446,4690)--(2668,4565)},
    {(2892,3931)--(2668,4058)--(2668,4312)--(2892,4183)},
    {(2892,3677)--(2668,3804)--(2668,4058)--(2892,3931)},
    {(3115,3297)--(2892,3423)--(2892,3677)--(3115,3551)},
    {(3337,4944)--(3115,5070)--(3115,5325)--(3337,5197)},
    {(3337,3931)--(3115,4058)--(3115,4312)--(3337,4183)},
    {(3337,3677)--(3115,3804)--(3115,4058)--(3337,3931)},
    {(3337,5452)--(3115,5578)--(3115,5832)--(3337,5705)},
    {(3337,4437)--(3115,4565)--(3115,4818)--(3337,4690)},
    {(3560,4312)--(3337,4437)--(3337,4690)--(3560,4565)},
    {(3560,5325)--(3337,5452)--(3337,5705)--(3560,5578)},
    {(3782,4944)--(3560,5070)--(3560,5325)--(3782,5197)},
    {(4230,5197)--(4007,5325)--(4007,5578)--(4230,5452)},
    {(4007,4818)--(3782,4944)--(3782,5197)--(4007,5070)},
    {(3782,5197)--(3560,5325)--(3560,5578)--(3782,5452)}}
    {\path[hdr pp tile,fill=#1] \p -- cycle;}
  \foreach \p in {
    {(1999,5452)--(2222,5578)--(1999,5705)--(1776,5578)},
    {(2222,5578)--(2446,5705)--(2222,5832)--(1999,5705)},
    {(2446,5705)--(2668,5832)--(2446,5959)--(2222,5832)},
    {(2668,5832)--(2892,5959)--(2668,6087)--(2446,5959)},
    {(2892,5705)--(3115,5832)--(2892,5959)--(2668,5832)},
    {(2668,5578)--(2892,5705)--(2668,5832)--(2446,5705)},
    {(2446,5452)--(2668,5578)--(2446,5705)--(2222,5578)},
    {(2222,5070)--(2446,5197)--(2222,5325)--(1999,5197)},
    {(2892,5197)--(3115,5325)--(2892,5452)--(2668,5325)},
    {(2446,4690)--(2668,4818)--(2446,4944)--(2222,4818)},
    {(2668,4818)--(2892,4944)--(2668,5070)--(2446,4944)},
    {(2668,4565)--(2892,4690)--(2668,4818)--(2446,4690)},
    {(2892,4690)--(3115,4818)--(2892,4944)--(2668,4818)},
    {(2892,4183)--(3115,4312)--(2892,4437)--(2668,4312)},
    {(2892,5959)--(3115,6087)--(2892,6213)--(2668,6087)},
    {(4230,5452)--(4007,5578)--(4230,5705)--(4453,5578)},
    {(4007,5578)--(3782,5705)--(4007,5832)--(4230,5705)},
    {(3782,5705)--(3560,5832)--(3782,5959)--(4007,5832)},
    {(3560,5832)--(3337,5959)--(3560,6087)--(3782,5959)},
    {(3337,5705)--(3115,5832)--(3337,5959)--(3560,5832)},
    {(3560,5578)--(3337,5705)--(3560,5832)--(3782,5705)},
    {(4007,5070)--(3782,5197)--(4007,5325)--(4230,5197)},
    {(3337,5197)--(3115,5325)--(3337,5452)--(3560,5325)},
    {(3115,5325)--(2892,5452)--(3115,5578)--(3337,5452)},
    {(3782,4690)--(3560,4818)--(3782,4944)--(4007,4818)},
    {(3560,4818)--(3337,4944)--(3560,5070)--(3782,4944)},
    {(3560,4565)--(3337,4690)--(3560,4818)--(3782,4690)},
    {(3337,4690)--(3115,4818)--(3337,4944)--(3560,4818)},
    {(3115,4818)--(2892,4944)--(3115,5070)--(3337,4944)},
    {(3337,4183)--(3115,4312)--(3337,4437)--(3560,4312)},
    {(3115,4312)--(2892,4437)--(3115,4565)--(3337,4437)},
    {(3337,5959)--(3115,6087)--(3337,6213)--(3560,6087)},
    {(3115,5832)--(2892,5959)--(3115,6087)--(3337,5959)},
    {(3115,6087)--(2892,6213)--(3115,6339)--(3337,6213)},
    {(3115,3551)--(2892,3677)--(3115,3804)--(3337,3677)},
    {(3782,5452)--(3560,5578)--(3782,5705)--(4007,5578)}}
    {\path[hdr pp tile,fill=#2] \p -- cycle;}
  \foreach \p in {
    {(1999,5197)--(2222,5325)--(2222,5578)--(1999,5452)},
    {(2446,5197)--(2668,5325)--(2668,5578)--(2446,5452)},
    {(2668,5325)--(2892,5452)--(2892,5705)--(2668,5578)},
    {(2892,5452)--(3115,5578)--(3115,5832)--(2892,5705)},
    {(2222,4818)--(2446,4944)--(2446,5197)--(2222,5070)},
    {(2446,4944)--(2668,5070)--(2668,5325)--(2446,5197)},
    {(2892,4944)--(3115,5070)--(3115,5325)--(2892,5197)},
    {(2668,4312)--(2892,4437)--(2892,4690)--(2668,4565)},
    {(2892,4437)--(3115,4565)--(3115,4818)--(2892,4690)},
    {(2892,3931)--(3115,4058)--(3115,4312)--(2892,4183)},
    {(2892,3677)--(3115,3804)--(3115,4058)--(2892,3931)},
    {(4230,5197)--(4453,5325)--(4453,5578)--(4230,5452)},
    {(4007,4818)--(4230,4944)--(4230,5197)--(4007,5070)},
    {(3115,3297)--(3337,3423)--(3337,3677)--(3115,3551)},
    {(3337,3677)--(3560,3804)--(3560,4058)--(3337,3931)},
    {(3560,4058)--(3782,4183)--(3782,4437)--(3560,4312)},
    {(3337,3931)--(3560,4058)--(3560,4312)--(3337,4183)},
    {(3782,4437)--(4007,4565)--(4007,4818)--(3782,4690)},
    {(3337,4944)--(3560,5070)--(3560,5325)--(3337,5197)},
    {(3560,4312)--(3782,4437)--(3782,4690)--(3560,4565)},
    {(3782,5197)--(4007,5325)--(4007,5578)--(3782,5452)}}
    {\path[hdr pp tile,fill=#3] \p -- cycle;}}

\tikzset{
  hdr tsscpp boundary/.style={draw=black,line width=1.45pt,
    line cap=round,line join=round},
  hdr tsscpp path/.style={draw=black,line width=1.1pt,
    line cap=round,line join=round},
  hdr tsscpp graph/.style={draw=gray!70,line width=0.5pt,
    dash pattern=on 2pt off 2pt,line cap=round,line join=round},
  hdr tsscpp point/.style={circle,draw=hdrpporange!65!black,
    fill=hdrpporange,line width=0.6pt,minimum size=4.7pt,inner sep=0pt},
  hdr tsscpp open point/.style={circle,draw=hdrpporange!75!black,
    fill=white,line width=0.7pt,minimum size=4.7pt,inner sep=0pt}
}

\newcommand{\hdrtsscppfull}{%
  \begin{tikzpicture}[x=0.00105cm,y=-0.00105cm,
      baseline=(current bounding box.center)]
    \useasboundingbox (300,190) rectangle (5910,6410);
    \foreach \ang/\ca/\cb/\cc in {
      0/hdrpppink/hdrppgreen/hdrppblue,
      60/hdrppgreen/hdrppblue/hdrpppink,
      120/hdrppblue/hdrpppink/hdrppgreen,
      180/hdrpppink/hdrppgreen/hdrppblue,
      240/hdrppgreen/hdrppblue/hdrpppink,
      300/hdrppblue/hdrpppink/hdrppgreen}{%
      \begin{scope}[rotate around={\ang:(3116,3295)}]
        \hdrtsscppsector{\ca}{\cb}{\cc}
      \end{scope}}
    \draw[hdr tsscpp boundary]
      (3115,3551)--(2892,3677)--(2892,3931)--(2668,4058)
      --(2668,4312)--(2446,4437)--(2446,4690)--(2222,4818)
      --(2222,5070)--(1999,5197)--(1999,5452)--(1776,5578)--(3115,6339);
    \draw[hdr tsscpp boundary]
      (3115,3551)--(3337,3677)--(3115,3804)--(3115,4312)
      --(3337,4437)--(3115,4565)--(3115,4818)--(3337,4944)
      --(3115,5070)--(3115,5325)--(3337,5452)--(3115,5578)
      --(3115,5832)--(3337,5959)--(3115,6087)--(3337,6213)--(3115,6339);
  \end{tikzpicture}}

\newcommand{\hdrtsscppfundamentaldomain}{%
  \begin{tikzpicture}[x=0.00105cm,y=-0.00105cm,
      baseline=(current bounding box.center)]
    \useasboundingbox (7270,1320) rectangle (9920,5840);
    \foreach \p in {
      {(8430,3987)--(8083,4185)--(8083,4579)--(8430,4382)},
      {(9124,3594)--(8777,3790)--(8777,4185)--(9124,3987)},
      {(8777,2610)--(8430,2806)--(8430,3198)--(8777,3000)},
      {(9124,2016)--(8777,2213)--(8777,2610)--(9124,2410)}}
      {\path[hdr pp tile,fill=hdrpppink] \p -- cycle;}
    \foreach \p in {
      {(7736,4382)--(8083,4579)--(7736,4776)--(7389,4579)},
      {(8083,4579)--(8430,4776)--(8083,4972)--(7736,4776)},
      {(8430,4776)--(8777,4972)--(8430,5172)--(8083,4972)},
      {(8777,4972)--(9124,5172)--(8777,5369)--(8430,5172)},
      {(9124,4776)--(9471,4972)--(9124,5172)--(8777,4972)},
      {(8777,4579)--(9124,4776)--(8777,4972)--(8430,4776)},
      {(8430,4382)--(8777,4579)--(8430,4776)--(8083,4579)},
      {(8083,3790)--(8430,3987)--(8083,4185)--(7736,3987)},
      {(9124,3987)--(9471,4185)--(9124,4382)--(8777,4185)},
      {(9471,4185)--(9818,4382)--(9471,4579)--(9124,4382)},
      {(8430,3198)--(8777,3394)--(8430,3594)--(8083,3394)},
      {(8777,3394)--(9124,3594)--(8777,3790)--(8430,3594)},
      {(8777,3000)--(9124,3198)--(8777,3394)--(8430,3198)},
      {(9124,3198)--(9471,3394)--(9124,3594)--(8777,3394)},
      {(9471,3394)--(9818,3594)--(9471,3790)--(9124,3594)},
      {(9124,2410)--(9471,2610)--(9124,2806)--(8777,2610)},
      {(9471,2610)--(9818,2806)--(9471,3000)--(9124,2806)},
      {(9124,5172)--(9471,5369)--(9124,5566)--(8777,5369)},
      {(9471,4972)--(9818,5172)--(9471,5369)--(9124,5172)},
      {(9471,5369)--(9818,5566)--(9471,5762)--(9124,5566)},
      {(9471,1424)--(9818,1620)--(9471,1818)--(9124,1620)}}
      {\path[hdr pp tile,fill=hdrppgreen] \p -- cycle;}
    \foreach \p in {
      {(7736,3987)--(8083,4185)--(8083,4579)--(7736,4382)},
      {(8430,3987)--(8777,4185)--(8777,4579)--(8430,4382)},
      {(8777,4185)--(9124,4382)--(9124,4776)--(8777,4579)},
      {(9124,4382)--(9471,4579)--(9471,4972)--(9124,4776)},
      {(8083,3394)--(8430,3594)--(8430,3987)--(8083,3790)},
      {(8430,3594)--(8777,3790)--(8777,4185)--(8430,3987)},
      {(9124,3594)--(9471,3790)--(9471,4185)--(9124,3987)},
      {(8777,2610)--(9124,2806)--(9124,3198)--(8777,3000)},
      {(9124,2806)--(9471,3000)--(9471,3394)--(9124,3198)},
      {(9124,2016)--(9471,2213)--(9471,2610)--(9124,2410)},
      {(9124,1620)--(9471,1818)--(9471,2213)--(9124,2016)}}
      {\path[hdr pp tile,fill=hdrppblue] \p -- cycle;}

    \begin{scope}[hdr tsscpp path]
      \draw (7736,4185)--(8083,4382)--(8430,4185)--(9471,4776);
      \draw (8083,3594)--(8777,3987)--(9124,3790)--(9471,3987);
      \draw (8430,3000)--(8777,2806)--(9471,3198);
      \draw (8777,2410)--(9124,2213)--(9471,2410);
      \draw (9108,1779)--(9455,1976);
    \end{scope}
    \foreach \x/\y in {7740/4211,8085/3610,8437/2999,8775/2395,
      9120/1845,9467/4764,9465/3970,9465/3196,9465/2407,9465/1994}
      {\node[hdr tsscpp point] at (\x,\y) {};}
  \end{tikzpicture}}

\newcommand{\hdrtsscppnilp}{%
  \begin{tikzpicture}[x=0.00126cm,y=-0.00126cm,
      baseline=(current bounding box.center)]
    \useasboundingbox (11220,1690) rectangle (12730,5450);
    \begin{scope}[hdr tsscpp graph]
      \draw (11525,4388)--(12401,5177)--(12618,4977);
      \draw (11471,4388)--(12562,3402);
      \draw (11908,3205)--(12562,2615);
      \draw (12180,2565)--(12618,2170);
      \draw (12401,1973)--(12618,1776);
      \draw (11963,3205)--(12618,3796)--(11743,4585);
      \draw (11743,3796)--(12618,4585)--(12180,4977);
      \draw (12562,4190)--(11963,4731);
      \draw (12618,4190)--(11963,3599);
      \draw (12618,4977)--(11743,4190);
      \draw (12126,2565)--(12618,3007)--(11743,3796);
      \draw (12401,5177)--(12618,5374);
    \end{scope}
    \begin{scope}[hdr tsscpp path]
      \draw (11505,4382)--(11725,4579)--(11942,4382)--(12597,4974);
      \draw (11725,3790)--(12163,4186)--(12380,3986)--(12597,4186);
      \draw (11942,3199)--(12163,3001)--(12597,3397);
      \draw (12163,2608)--(12380,2408)--(12597,2608);
      \draw (12406,1991)--(12625,2188);
    \end{scope}
    \foreach \x/\y in {11495/4374,11729/3801,11947/3228,12167/2599,
      12418/1987,12588/4965,12588/4182,12588/3401,12588/2599,12588/2198}
      {\node[hdr tsscpp point] at (\x,\y) {};}
    \foreach \x/\y in {12588/5366,12588/4564,12588/3801,12588/2999,12588/1815}
      {\node[hdr tsscpp open point] at (\x,\y) {};}
  \end{tikzpicture}}

\newcommand{\hdrtsscppfigure}{%
  \makebox[\textwidth][c]{\hdrtsscppfull\hspace{0.75cm}%
    \hdrtsscppfundamentaldomain\hspace{0.4cm}\hdrtsscppnilp}}

\newcommand{\hdrsmalltsscpp}[4]{%
  \begin{tikzpicture}[x=#1cm,y=#1cm,baseline=(current bounding box.center)]
    \begin{scope}[rotate=30]
      \begin{scope}[yscale=-0.8660254]
        \foreach \x/\y in {#2}{%
          \path[hdr pp tile,fill=hdrpppink]
            (\x,{\y-600})--({\x-300},\y)--(\x,{\y+600})--({\x+300},\y)--cycle;}
        \foreach \x/\y in {#3}{%
          \path[hdr pp tile,fill=hdrppblue]
            ({\x-450},{\y+300})--({\x-150},{\y-300})
            --({\x+450},{\y-300})--({\x+150},{\y+300})--cycle;}
        \foreach \x/\y in {#4}{%
          \path[hdr pp tile,fill=hdrppgreen]
            ({\x+150},{\y-300})--({\x-450},{\y-300})
            --({\x-150},{\y+300})--({\x+450},{\y+300})--cycle;}
      \end{scope}
    \end{scope}
  \end{tikzpicture}}

\newcommand{\hdrtsscppsizeone}{%
  \hdrsmalltsscpp{0.00077}%
    {0/-600,-900/0,900/0,0/600}%
    {-450/-900,450/-300,-450/300,450/900}%
    {450/-900,-450/-300,450/300,-450/900}}

\newcommand{\hdrtsscppsizetwoa}{%
  \hdrsmalltsscpp{0.000385}%
    {0/-1800,-300/-1200,300/-1200,-1800/-600,0/-600,1800/-600,
      -2100/0,-1500/0,1500/0,2100/0,-1800/600,0/600,1800/600,
      -300/1200,300/1200,0/1800}%
    {-1050/-2100,-450/-2100,-1350/-1500,-750/-1500,750/-900,1350/-900,
      450/-300,1050/-300,-1050/300,-450/300,-1350/900,-750/900,
      750/1500,1350/1500,450/2100,1050/2100}%
    {450/-2100,1050/-2100,750/-1500,1350/-1500,-1350/-900,-750/-900,
      -1050/-300,-450/-300,450/300,1050/300,750/900,1350/900,
      -1350/1500,-750/1500,-1050/2100,-450/2100}}

\newcommand{\hdrtsscppsizetwob}{%
  \hdrsmalltsscpp{0.000385}%
    {0/-1800,-900/-1200,900/-1200,-1800/-600,0/-600,1800/-600,
      -2100/0,-900/0,900/0,2100/0,-1800/600,0/600,1800/600,
      -900/1200,900/1200,0/1800}%
    {-1050/-2100,-450/-2100,-1350/-1500,450/-1500,-450/-900,1350/-900,
      -1350/-300,450/-300,-450/300,1350/300,-1350/900,450/900,
      -450/1500,1350/1500,450/2100,1050/2100}%
    {450/-2100,1050/-2100,-450/-1500,1350/-1500,-1350/-900,450/-900,
      -450/-300,1350/-300,-1350/300,450/300,-450/900,1350/900,
      -1350/1500,450/1500,-1050/2100,-450/2100}}

\newcommand{\hdrtsscppsizethreea}{%
  \hdrsmalltsscpp{0.0002567}%
    {0/-3000,-300/-2400,300/-2400,-600/-1800,0/-1800,600/-1800,
      -2700/-1200,-300/-1200,300/-1200,2700/-1200,
      -3000/-600,-2400/-600,0/-600,2400/-600,3000/-600,
      -3300/0,-2700/0,-2100/0,2100/0,2700/0,3300/0,
      -3000/600,-2400/600,0/600,2400/600,3000/600,
      -2700/1200,-300/1200,300/1200,2700/1200,
      -600/1800,0/1800,600/1800,-300/2400,300/2400,0/3000}%
    {-1650/-3300,-1050/-3300,-450/-3300,-1950/-2700,-1350/-2700,-750/-2700,
      -2250/-2100,-1650/-2100,-1050/-2100,1050/-1500,1650/-1500,2250/-1500,
      750/-900,1350/-900,1950/-900,450/-300,1050/-300,1650/-300,
      -1650/300,-1050/300,-450/300,-1950/900,-1350/900,-750/900,
      -2250/1500,-1650/1500,-1050/1500,1050/2100,1650/2100,2250/2100,
      750/2700,1350/2700,1950/2700,450/3300,1050/3300,1650/3300}%
    {450/-3300,1050/-3300,1650/-3300,750/-2700,1350/-2700,1950/-2700,
      1050/-2100,1650/-2100,2250/-2100,-2250/-1500,-1650/-1500,-1050/-1500,
      -1950/-900,-1350/-900,-750/-900,-1650/-300,-1050/-300,-450/-300,
      450/300,1050/300,1650/300,750/900,1350/900,1950/900,
      1050/1500,1650/1500,2250/1500,-2250/2100,-1650/2100,-1050/2100,
      -1950/2700,-1350/2700,-750/2700,-1650/3300,-1050/3300,-450/3300}}

\newcommand{\hdrtsscppsizethreeb}{%
  \hdrsmalltsscpp{0.0002567}%
    {0/-3000,-300/-2400,300/-2400,-1200/-1800,0/-1800,1200/-1800,
      -2700/-1200,-300/-1200,300/-1200,2700/-1200,
      -3000/-600,-2400/-600,0/-600,2400/-600,3000/-600,
      -3300/0,-2700/0,-1500/0,1500/0,2700/0,3300/0,
      -3000/600,-2400/600,0/600,2400/600,3000/600,
      -2700/1200,-300/1200,300/1200,2700/1200,
      -1200/1800,0/1800,1200/1800,-300/2400,300/2400,0/3000}%
    {-1650/-3300,-1050/-3300,-450/-3300,-1950/-2700,-1350/-2700,-750/-2700,
      -2250/-2100,-1650/-2100,750/-2100,-750/-1500,1650/-1500,2250/-1500,
      750/-900,1350/-900,1950/-900,-1950/-300,450/-300,1050/-300,
      -1050/300,-450/300,1950/300,-1950/900,-1350/900,-750/900,
      -2250/1500,-1650/1500,750/1500,-750/2100,1650/2100,2250/2100,
      750/2700,1350/2700,1950/2700,450/3300,1050/3300,1650/3300}%
    {450/-3300,1050/-3300,1650/-3300,750/-2700,1350/-2700,1950/-2700,
      -750/-2100,1650/-2100,2250/-2100,-2250/-1500,-1650/-1500,750/-1500,
      -1950/-900,-1350/-900,-750/-900,-1050/-300,-450/-300,1950/-300,
      -1950/300,450/300,1050/300,750/900,1350/900,1950/900,
      -750/1500,1650/1500,2250/1500,-2250/2100,-1650/2100,750/2100,
      -1950/2700,-1350/2700,-750/2700,-1650/3300,-1050/3300,-450/3300}}

\newcommand{\hdrtsscppsizethreec}{%
  \hdrsmalltsscpp{0.0002567}%
    {0/-3000,-300/-2400,300/-2400,-1200/-1800,0/-1800,1200/-1800,
      -2700/-1200,-900/-1200,900/-1200,2700/-1200,
      -3000/-600,-2400/-600,0/-600,2400/-600,3000/-600,
      -3300/0,-2700/0,-900/0,900/0,2700/0,3300/0,
      -3000/600,-2400/600,0/600,2400/600,3000/600,
      -2700/1200,-900/1200,900/1200,2700/1200,
      -1200/1800,0/1800,1200/1800,-300/2400,300/2400,0/3000}%
    {-1650/-3300,-1050/-3300,-450/-3300,-1950/-2700,-1350/-2700,-750/-2700,
      -2250/-2100,-1650/-2100,750/-2100,450/-1500,1650/-1500,2250/-1500,
      -450/-900,1350/-900,1950/-900,-1950/-300,-1350/-300,450/-300,
      -450/300,1350/300,1950/300,-1950/900,-1350/900,450/900,
      -2250/1500,-1650/1500,-450/1500,-750/2100,1650/2100,2250/2100,
      750/2700,1350/2700,1950/2700,450/3300,1050/3300,1650/3300}%
    {450/-3300,1050/-3300,1650/-3300,750/-2700,1350/-2700,1950/-2700,
      -750/-2100,1650/-2100,2250/-2100,-2250/-1500,-1650/-1500,-450/-1500,
      -1950/-900,-1350/-900,450/-900,-450/-300,1350/-300,1950/-300,
      -1950/300,-1350/300,450/300,-450/900,1350/900,1950/900,
      450/1500,1650/1500,2250/1500,-2250/2100,-1650/2100,750/2100,
      -1950/2700,-1350/2700,-750/2700,-1650/3300,-1050/3300,-450/3300}}

\newcommand{\hdrtsscppsizethreed}{%
  \hdrsmalltsscpp{0.0002567}%
    {0/-3000,-900/-2400,900/-2400,-1800/-1800,0/-1800,1800/-1800,
      -2700/-1200,-300/-1200,300/-1200,2700/-1200,
      -3000/-600,-1800/-600,0/-600,1800/-600,3000/-600,
      -3300/0,-2700/0,-1500/0,1500/0,2700/0,3300/0,
      -3000/600,-1800/600,0/600,1800/600,3000/600,
      -2700/1200,-300/1200,300/1200,2700/1200,
      -1800/1800,0/1800,1800/1800,-900/2400,900/2400,0/3000}%
    {-1650/-3300,-1050/-3300,-450/-3300,-1950/-2700,-1350/-2700,450/-2700,
      -2250/-2100,-450/-2100,1350/-2100,-1350/-1500,-750/-1500,2250/-1500,
      -2250/-900,750/-900,1350/-900,450/-300,1050/-300,2250/-300,
      -2250/300,-1050/300,-450/300,-1350/900,-750/900,2250/900,
      -2250/1500,750/1500,1350/1500,-1350/2100,450/2100,2250/2100,
      -450/2700,1350/2700,1950/2700,450/3300,1050/3300,1650/3300}%
    {450/-3300,1050/-3300,1650/-3300,-450/-2700,1350/-2700,1950/-2700,
      -1350/-2100,450/-2100,2250/-2100,-2250/-1500,750/-1500,1350/-1500,
      -1350/-900,-750/-900,2250/-900,-2250/-300,-1050/-300,-450/-300,
      450/300,1050/300,2250/300,-2250/900,750/900,1350/900,
      -1350/1500,-750/1500,2250/1500,-2250/2100,-450/2100,1350/2100,
      -1950/2700,-1350/2700,450/2700,-1650/3300,-1050/3300,-450/3300}}

\newcommand{\hdrtsscppsizethreee}{%
  \hdrsmalltsscpp{0.0002567}%
    {0/-3000,-900/-2400,900/-2400,-1800/-1800,0/-1800,1800/-1800,
      -2700/-1200,-900/-1200,900/-1200,2700/-1200,
      -3000/-600,-1800/-600,0/-600,1800/-600,3000/-600,
      -3300/0,-2700/0,-900/0,900/0,2700/0,3300/0,
      -3000/600,-1800/600,0/600,1800/600,3000/600,
      -2700/1200,-900/1200,900/1200,2700/1200,
      -1800/1800,0/1800,1800/1800,-900/2400,900/2400,0/3000}%
    {-1650/-3300,-1050/-3300,-450/-3300,-1950/-2700,-1350/-2700,450/-2700,
      -2250/-2100,-450/-2100,1350/-2100,-1350/-1500,450/-1500,2250/-1500,
      -2250/-900,-450/-900,1350/-900,-1350/-300,450/-300,2250/-300,
      -2250/300,-450/300,1350/300,-1350/900,450/900,2250/900,
      -2250/1500,-450/1500,1350/1500,-1350/2100,450/2100,2250/2100,
      -450/2700,1350/2700,1950/2700,450/3300,1050/3300,1650/3300}%
    {450/-3300,1050/-3300,1650/-3300,-450/-2700,1350/-2700,1950/-2700,
      -1350/-2100,450/-2100,2250/-2100,-2250/-1500,-450/-1500,1350/-1500,
      -1350/-900,450/-900,2250/-900,-2250/-300,-450/-300,1350/-300,
      -1350/300,450/300,2250/300,-2250/900,-450/900,1350/900,
      -1350/1500,450/1500,2250/1500,-2250/2100,-450/2100,1350/2100,
      -1950/2700,-1350/2700,450/2700,-1650/3300,-1050/3300,-450/3300}}

\newcommand{\hdrtsscppsizethreef}{%
  \hdrsmalltsscpp{0.0002567}%
    {0/-3000,-1500/-2400,1500/-2400,-1800/-1800,0/-1800,1800/-1800,
      -2700/-1200,-300/-1200,300/-1200,2700/-1200,
      -3000/-600,-1800/-600,0/-600,1800/-600,3000/-600,
      -3300/0,-2100/0,-1500/0,1500/0,2100/0,3300/0,
      -3000/600,-1800/600,0/600,1800/600,3000/600,
      -2700/1200,-300/1200,300/1200,2700/1200,
      -1800/1800,0/1800,1800/1800,-1500/2400,1500/2400,0/3000}%
    {-1650/-3300,-1050/-3300,-450/-3300,-1950/-2700,450/-2700,1050/-2700,
      -2250/-2100,-1050/-2100,-450/-2100,-1350/-1500,-750/-1500,2250/-1500,
      -2250/-900,750/-900,1350/-900,-2550/-300,450/-300,1050/-300,
      -1050/300,-450/300,2550/300,-1350/900,-750/900,2250/900,
      -2250/1500,750/1500,1350/1500,450/2100,1050/2100,2250/2100,
      -1050/2700,-450/2700,1950/2700,450/3300,1050/3300,1650/3300}%
    {450/-3300,1050/-3300,1650/-3300,-1050/-2700,-450/-2700,1950/-2700,
      450/-2100,1050/-2100,2250/-2100,-2250/-1500,750/-1500,1350/-1500,
      -1350/-900,-750/-900,2250/-900,-1050/-300,-450/-300,2550/-300,
      -2550/300,450/300,1050/300,-2250/900,750/900,1350/900,
      -1350/1500,-750/1500,2250/1500,-2250/2100,-1050/2100,-450/2100,
      -1950/2700,450/2700,1050/2700,-1650/3300,-1050/3300,-450/3300}}

\newcommand{\hdrtsscppsizethreeg}{%
  \hdrsmalltsscpp{0.0002567}%
    {0/-3000,-1500/-2400,1500/-2400,-1800/-1800,0/-1800,1800/-1800,
      -2700/-1200,-900/-1200,900/-1200,2700/-1200,
      -3000/-600,-1800/-600,0/-600,1800/-600,3000/-600,
      -3300/0,-2100/0,-900/0,900/0,2100/0,3300/0,
      -3000/600,-1800/600,0/600,1800/600,3000/600,
      -2700/1200,-900/1200,900/1200,2700/1200,
      -1800/1800,0/1800,1800/1800,-1500/2400,1500/2400,0/3000}%
    {-1650/-3300,-1050/-3300,-450/-3300,-1950/-2700,450/-2700,1050/-2700,
      -2250/-2100,-1050/-2100,-450/-2100,-1350/-1500,450/-1500,2250/-1500,
      -2250/-900,-450/-900,1350/-900,-2550/-300,-1350/-300,450/-300,
      -450/300,1350/300,2550/300,-1350/900,450/900,2250/900,
      -2250/1500,-450/1500,1350/1500,450/2100,1050/2100,2250/2100,
      -1050/2700,-450/2700,1950/2700,450/3300,1050/3300,1650/3300}%
    {450/-3300,1050/-3300,1650/-3300,-1050/-2700,-450/-2700,1950/-2700,
      450/-2100,1050/-2100,2250/-2100,-2250/-1500,-450/-1500,1350/-1500,
      -1350/-900,450/-900,2250/-900,-450/-300,1350/-300,2550/-300,
      -2550/300,-1350/300,450/300,-2250/900,-450/900,1350/900,
      -1350/1500,450/1500,2250/1500,-2250/2100,-1050/2100,-450/2100,
      -1950/2700,450/2700,1050/2700,-1650/3300,-1050/3300,-450/3300}}

\newcommand{\hdrtsscppcatalogue}{%
  \makebox[\textwidth][c]{\hdrtsscppsizeone\hspace{2.6cm}%
    \hdrtsscppsizetwoa\hspace{0.35cm}\hdrtsscppsizetwob}\par\vspace{0.3cm}%
  \makebox[\textwidth][c]{\hdrtsscppsizethreea\hspace{0.22cm}%
    \hdrtsscppsizethreeb\hspace{0.22cm}\hdrtsscppsizethreec\hspace{0.22cm}%
    \hdrtsscppsizethreed\hspace{0.22cm}\hdrtsscppsizethreee\hspace{0.22cm}%
    \hdrtsscppsizethreef\hspace{0.22cm}\hdrtsscppsizethreeg}}

\newcommand{\hdrsixvertexconfiguration}[1][1.125cm]{%
  \begin{tikzpicture}[x=#1,y=#1,baseline=(current bounding box.center)]
    \useasboundingbox (-1,-1) rectangle (3,3);
    \foreach \x in {-1,2}{\hdrsixhforward{\x}{2}}
    \foreach \x in {0,1}{\hdrsixhbackward{\x}{2}}
    \foreach \x in {-1,0,1,2}{\hdrsixhforward{\x}{1}}
    \foreach \x in {0,1}{\hdrsixhforward{\x}{0}}
    \foreach \x in {-1,2}{\hdrsixhbackward{\x}{0}}
    \foreach \y in {-1,2}{\hdrsixvforward{0}{\y}}
    \foreach \y in {0,1}{\hdrsixvbackward{0}{\y}}
    \foreach \y in {-1,0,1,2}{\hdrsixvbackward{1}{\y}}
    \foreach \y in {0,1}{\hdrsixvforward{2}{\y}}
    \foreach \y in {-1,2}{\hdrsixvbackward{2}{\y}}
  \end{tikzpicture}}

\newcommand{\hdrsixvertexdwbc}[1][0.82cm]{%
  \begin{tikzpicture}[x=#1,y=#1,baseline=(current bounding box.center)]
    \useasboundingbox (-1.08,-1.08) rectangle (4.08,4.08);
    \foreach \x in {-1,0}{\hdrsixhforward{\x}{3}}
    \foreach \x in {1,2,3}{\hdrsixhbackward{\x}{3}}
    \foreach \x in {-1,0,1}{\hdrsixhforward{\x}{2}}
    \foreach \x in {2,3}{\hdrsixhbackward{\x}{2}}
    \foreach \x in {-1,2}{\hdrsixhforward{\x}{1}}
    \foreach \x in {0,1,3}{\hdrsixhbackward{\x}{1}}
    \foreach \x in {-1,0,1}{\hdrsixhforward{\x}{0}}
    \foreach \x in {2,3}{\hdrsixhbackward{\x}{0}}
    \foreach \y in {1,2,3}{\hdrsixvforward{0}{\y}}
    \foreach \y in {-1,0}{\hdrsixvbackward{0}{\y}}
    \hdrsixvforward{1}{3}
    \foreach \y in {-1,0,1,2}{\hdrsixvbackward{1}{\y}}
    \foreach \y in {0,2,3}{\hdrsixvforward{2}{\y}}
    \foreach \y in {-1,1}{\hdrsixvbackward{2}{\y}}
    \foreach \y in {1,2,3}{\hdrsixvforward{3}{\y}}
    \foreach \y in {-1,0}{\hdrsixvbackward{3}{\y}}
  \end{tikzpicture}}

\newcommand{\hdrsixvertexrecurrence}[1][0.82cm]{%
  \begin{tikzpicture}[x=#1,y=#1,baseline=(current bounding box.center)]
    \useasboundingbox (-1.52,-1.42) rectangle (4.08,4.08);
    \begin{scope}[line width=0.65pt]
      \foreach \x in {0,1,2,3} \draw (\x,-1) -- (\x,4);
      \foreach \y in {0,1,2,3} \draw (-1,\y) -- (4,\y);
    \end{scope}
    \foreach \y in {0,1,2,3}{%
      \hdrsixhforwardshort{-1}{\y}%
      \hdrsixhbackwardshort{3}{\y}}
    \foreach \x in {0,1,2,3}{%
      \hdrsixvbackwardshort{\x}{-1}%
      \hdrsixvforwardshort{\x}{3}}
    \foreach \x in {0,1,2}{\hdrsixhbackwardshort{\x}{0}}
    \foreach \y in {0,1,2}{\hdrsixvforwardshort{0}{\y}}
    \foreach \y in {1,2,3}{\hdrsixhforwardshort{0}{\y}}
    \foreach \x in {1,2,3}{\hdrsixvbackwardshort{\x}{0}}
    \node[anchor=east] at (-1.12,0) {$x_1$};
    \node[anchor=north] at (0,-1.10) {$y_1$};
  \end{tikzpicture}}

\newcommand{\hdrpermutationdiagram}[4][0.56cm]{%
  \begin{tikzpicture}[x=#1,y=#1,baseline=(current bounding box.center)]
    \useasboundingbox (-1.08,-1.58) rectangle (4.08,4.08);
    \begin{scope}[line width=0.65pt]
      \foreach \y in {0,1,2,3} \draw (-1,\y) -- (4,\y);
      \draw (0,-1) -- (0,4);
      \draw (3,-1) -- (3,4);
      \ifcase#2\relax
        \draw (1,-1) -- (1,4);
        \draw (2,-1) -- (2,4);
      \or
        \draw[hdr six small arrow=0.97] (1,-1) -- (1,3.22)
          .. controls (1,3.45) and (2,3.55) .. (2,3.78) -- (2,4);
        \draw[hdr six small arrow=0.97] (2,-1) -- (2,3.22)
          .. controls (2,3.45) and (1,3.55) .. (1,3.78) -- (1,4);
      \or
        \draw (1,-1) -- (1,2.22)
          .. controls (1,2.45) and (2,2.55) .. (2,2.78) -- (2,4);
        \draw (2,-1) -- (2,2.22)
          .. controls (2,2.45) and (1,2.55) .. (1,2.78) -- (1,4);
      \or
        \draw[hdr six small arrow=0.90] (1,0) -- (1,-0.22)
          .. controls (1,-0.45) and (2,-0.55) .. (2,-0.78) -- (2,-1);
        \draw[hdr six small arrow=0.90] (2,0) -- (2,-0.22)
          .. controls (2,-0.45) and (1,-0.55) .. (1,-0.78) -- (1,-1);
        \draw (1,0) -- (1,4);
        \draw (2,0) -- (2,4);
      \fi
    \end{scope}
    \foreach \y in {0,1,2,3}{%
      \hdrsixhforwardsmall{-1}{\y}%
      \hdrsixhbackwardsmall{3}{\y}}
    \foreach \x in {0,3}{%
      \hdrsixvbackwardsmall{\x}{-1}%
      \hdrsixvforwardsmall{\x}{3}}
    \ifnum#2=1\relax
      \foreach \x in {1,2}{\hdrsixvbackwardsmall{\x}{-1}}%
    \else\ifnum#2=3\relax
      \foreach \x in {1,2}{\hdrsixvforwardsmall{\x}{3}}%
    \else
      \foreach \x in {1,2}{%
        \hdrsixvbackwardsmall{\x}{-1}%
        \hdrsixvforwardsmall{\x}{3}}%
    \fi\fi
    \node[anchor=north] at (1,-1.05) {#3};
    \node[anchor=north] at (2,-1.05) {#4};
  \end{tikzpicture}}

\newcommand{\hdrsixvertexice}[1][1.125cm]{%
  \begin{tikzpicture}[x=#1,y=#1,baseline=(current bounding box.center)]
    \useasboundingbox (-1,-1) rectangle (3,3);
    \foreach \x in {0,1,2} \draw[line width=0.65pt] (\x,-1) -- (\x,3);
    \foreach \y in {0,1,2} \draw[line width=0.65pt] (-1,\y) -- (3,\y);
    \foreach \x/\y in {
      0/2.667,-0.333/2,1/2.333,1/1.333,2.667/2,2/2.333,
      -0.333/1,-0.667/0,0/-0.333,1/-0.667,2/-0.667,2.333/0,
      1.667/0,2/0.667,1/0.333,0.667/0,0/0.333,2.667/1,
      0.333/2,0/1.333,0.667/1,1.667/1,2/1.667,1.333/2}
      \fill (\x,\y) circle[radius=0.10];
    \foreach \x in {0,1,2}
      \foreach \y in {0,1,2}
        \filldraw[fill=white,line width=0.4pt] (\x,\y) circle[radius=0.155];
  \end{tikzpicture}}

\newcommand{\hdrsixvertexpaths}[1][1.125cm]{%
  \begin{tikzpicture}[x=#1,y=#1,baseline=(current bounding box.center)]
    \useasboundingbox (-1,-1) rectangle (3,3);
    \foreach \x in {0,1,2}
      \draw[thin,gray,densely dotted] (\x,-1) -- (\x,3);
    \foreach \y in {0,1,2}
      \draw[thin,gray,densely dotted] (-1,\y) -- (3,\y);
    \begin{scope}[line width=2.2pt,line cap=butt,line join=miter]
      \draw (-1,2) -- (0,2) -- (0,3);
      \draw (3,2) -- (2,2) -- (2,1.34)
        .. controls (2,1.13) and (1.88,1) .. (1.67,1) -- (-1,1);
      \draw (3,1) -- (2.33,1)
        .. controls (2.12,1) and (2,0.87) .. (2,0.66)
        -- (2,0) -- (0,0) -- (0,-1);
    \end{scope}
  \end{tikzpicture}}

\newcommand{\hdrloopturnwn}[2]{%
  \draw ({#1-0.5},#2) -- (#1,#2) -- (#1,{#2+0.5});}
\newcommand{\hdrloopturnws}[2]{%
  \draw ({#1-0.5},#2) -- (#1,#2) -- (#1,{#2-0.5});}
\newcommand{\hdrloopturnne}[2]{%
  \draw (#1,{#2+0.5}) -- (#1,#2) -- ({#1+0.5},#2);}
\newcommand{\hdrloopturnse}[2]{%
  \draw (#1,{#2-0.5}) -- (#1,#2) -- ({#1+0.5},#2);}
\newcommand{\hdrloopconfigurationpaths}{%
  \begin{scope}[line width=0.65pt,rounded corners=8pt]
    \foreach \x in {0,1,2}{%
      \draw (\x,2.5) -- (\x,3);
      \draw (\x,-0.5) -- (\x,-1);}
    \foreach \y in {0,1,2}{%
      \draw (-1,\y) -- (-0.5,\y);
      \draw (2.5,\y) -- (3,\y);}
    \hdrloopturnwn{0}{2}\hdrloopturnse{0}{2}
    \hdrloopturnws{1}{2}\hdrloopturnne{1}{2}
    \hdrloopturnws{2}{2}\hdrloopturnne{2}{2}
    \hdrloopturnwn{0}{1}\hdrloopturnse{0}{1}
    \hdrloopturnwn{1}{1}\hdrloopturnse{1}{1}
    \hdrloopturnws{2}{1}\hdrloopturnne{2}{1}
    \hdrloopturnws{0}{0}\hdrloopturnne{0}{0}
    \hdrloopturnwn{1}{0}\hdrloopturnse{1}{0}
    \hdrloopturnwn{2}{0}\hdrloopturnse{2}{0}
  \end{scope}}

\newcommand{\hdrunorientedloopconfiguration}[1][1cm]{%
  \begin{tikzpicture}[x=#1,y=#1,baseline=(current bounding box.center)]
    \useasboundingbox (-1,-1) rectangle (3,3);
    \hdrloopconfigurationpaths
  \end{tikzpicture}}

\newcommand{\hdrorientedloopconfiguration}[1][1cm]{%
  \begin{tikzpicture}[x=#1,y=#1,baseline=(current bounding box.center)]
    \useasboundingbox (-1,-1) rectangle (3,3);
    \hdrloopconfigurationpaths
    \foreach \x in {-1,2}{\hdrsixhforwardshort{\x}{2}}
    \foreach \x in {0,1}{\hdrsixhbackwardshort{\x}{2}}
    \foreach \x in {-1,0,1,2}{\hdrsixhforwardshort{\x}{1}}
    \foreach \x in {0,1}{\hdrsixhforwardshort{\x}{0}}
    \foreach \x in {-1,2}{\hdrsixhbackwardshort{\x}{0}}
    \foreach \y in {-1,2}{\hdrsixvforwardshort{0}{\y}}
    \foreach \y in {0,1}{\hdrsixvbackwardshort{0}{\y}}
    \foreach \y in {-1,0,1,2}{\hdrsixvbackwardshort{1}{\y}}
    \foreach \y in {0,1}{\hdrsixvforwardshort{2}{\y}}
    \foreach \y in {-1,2}{\hdrsixvbackwardshort{2}{\y}}
  \end{tikzpicture}}

\newcommand{\hdrsixvertexstate}[6]{%
  \begin{scope}[shift={(#1,0)}]
    \draw[line width=0.65pt,#2=0.5] (-0.95,0) -- (0,0);
    \draw[line width=0.65pt,#3=0.5] (0,0) -- (0.95,0);
    \draw[line width=0.65pt,#4=0.5] (0,-0.95) -- (0,0);
    \draw[line width=0.65pt,#5=0.5] (0,0) -- (0,0.95);
    \node at (0,-1.38) {$#6$};
  \end{scope}}

\newcommand{\hdrsixvertexweights}{%
  \begin{tikzpicture}[x=1cm,y=1cm,baseline=(current bounding box.center)]
    \hdrsixvertexstate{0}{hdr six arrow}{hdr six arrow}{hdr six arrow}{hdr six arrow}{a_1}
    \hdrsixvertexstate{2.54}{hdr six invarrow}{hdr six invarrow}{hdr six invarrow}{hdr six invarrow}{a_2}
    \hdrsixvertexstate{5.08}{hdr six invarrow}{hdr six invarrow}{hdr six arrow}{hdr six arrow}{b_1}
    \hdrsixvertexstate{7.62}{hdr six arrow}{hdr six arrow}{hdr six invarrow}{hdr six invarrow}{b_2}
    \hdrsixvertexstate{10.16}{hdr six invarrow}{hdr six arrow}{hdr six arrow}{hdr six invarrow}{c_1}
    \hdrsixvertexstate{12.70}{hdr six arrow}{hdr six invarrow}{hdr six invarrow}{hdr six arrow}{c_2}
  \end{tikzpicture}}

\newcommand{\hdrsixvertexasm}{%
  \begin{tikzpicture}[x=1cm,y=1cm,baseline=(current bounding box.center)]
    \hdrsixvertexstate{0}{hdr six arrow}{hdr six arrow}{hdr six arrow}{hdr six arrow}{0}
    \hdrsixvertexstate{2.54}{hdr six invarrow}{hdr six invarrow}{hdr six invarrow}{hdr six invarrow}{0}
    \hdrsixvertexstate{5.08}{hdr six invarrow}{hdr six invarrow}{hdr six arrow}{hdr six arrow}{0}
    \hdrsixvertexstate{7.62}{hdr six arrow}{hdr six arrow}{hdr six invarrow}{hdr six invarrow}{0}
    \hdrsixvertexstate{10.16}{hdr six invarrow}{hdr six arrow}{hdr six arrow}{hdr six invarrow}{-1}
    \hdrsixvertexstate{12.70}{hdr six arrow}{hdr six invarrow}{hdr six invarrow}{hdr six arrow}{1}
  \end{tikzpicture}}

\newcommand{\hdrfpllocalstate}[3]{%
  \begin{scope}[shift={(#1,#2)},line cap=butt,line join=miter]
    \ifcase#3\relax
    \or
      \draw[line width=1.35pt] (0,0.48) -- (0,0) -- (0.48,0);
      \draw[thin,densely dotted] (0,-0.48) -- (0,0) -- (-0.48,0);
    \or
      \draw[line width=1.35pt] (-0.48,0) -- (0,0) -- (0,-0.48);
      \draw[thin,densely dotted] (0,0.48) -- (0,0) -- (0.48,0);
    \or
      \draw[line width=1.35pt] (-0.48,0) -- (0,0) -- (0,0.48);
      \draw[thin,densely dotted] (0,-0.48) -- (0,0) -- (0.48,0);
    \or
      \draw[line width=1.35pt] (0.48,0) -- (0,0) -- (0,-0.48);
      \draw[thin,densely dotted] (-0.48,0) -- (0,0) -- (0,0.48);
    \or
      \draw[line width=1.35pt] (-0.48,0) -- (0.48,0);
      \draw[thin,densely dotted] (0,-0.48) -- (0,0.48);
    \or
      \draw[line width=1.35pt] (0,-0.48) -- (0,0.48);
      \draw[thin,densely dotted] (-0.48,0) -- (0.48,0);
    \fi
  \end{scope}}

\newcommand{\hdrcpllocalstate}[2]{%
  \begin{scope}[shift={(#1,0)},line width=1.35pt,line cap=butt]
    \ifcase#2\relax
    \or
      \draw (-0.50,0) -- (-0.23,0) .. controls (-0.10,0) and (0,0.10) .. (0,0.23) -- (0,0.50);
      \draw (0,-0.50) -- (0,-0.23) .. controls (0,-0.10) and (0.10,0) .. (0.23,0) -- (0.50,0);
    \or
      \draw (0,0.50) -- (0,0.23) .. controls (0,0.10) and (0.10,0) .. (0.23,0) -- (0.50,0);
      \draw (-0.50,0) -- (-0.23,0) .. controls (-0.10,0) and (0,-0.10) .. (0,-0.23) -- (0,-0.50);
    \fi
  \end{scope}}
\newcommand{\hdrloopvertices}{%
  \begin{tikzpicture}[x=1cm,y=1cm,baseline=(current bounding box.center)]
    \useasboundingbox (-0.65,-1.10) rectangle (13.15,0.58);
    \hdrcpllocalstate{0}{1}
    \hdrcpllocalstate{1.70}{2}
    \node at (0.85,-0.90) {(a)};
    \foreach \x/\state in {3.75/1,5.45/2,7.15/3,8.85/4,10.55/5,12.25/6}
      {\hdrfpllocalstate{\x}{0}{\state}}
    \node at (8.00,-0.90) {(b)};
  \end{tikzpicture}}

\tikzset{
  hdr fpl four path/.style={blue,line width=0.65pt,line cap=butt,line join=miter},
  hdr fpl four grid/.style={draw=black!35,line width=0.22pt,line cap=round,
    dash pattern=on 0pt off 1.2pt}
}
\newcommand{\hdrfploccupiedvertex}[1]{%
  \begin{scope}[hdr fpl four path]
    \ifcase#1\relax
    \or\draw (0,0.5) -- (0,0) -- (0.5,0);
    \or\draw (-0.5,0) -- (0,0) -- (0,-0.5);
    \or\draw (-0.5,0) -- (0,0) -- (0,0.5);
    \or\draw (0.5,0) -- (0,0) -- (0,-0.5);
    \or\draw (-0.5,0) -- (0.5,0);
    \or\draw (0,-0.5) -- (0,0.5);
    \fi
  \end{scope}}
\newcommand{\hdrfplfourconfig}[1]{%
  \begin{tikzpicture}[x=0.316cm,y=0.316cm,baseline=(current bounding box.center)]
    \useasboundingbox (-1,-1) rectangle (4,4);
    \foreach \k in {0,...,3}{%
      \draw[hdr fpl four grid] (-1,\k) -- (4,\k);
      \draw[hdr fpl four grid] (\k,-1) -- (\k,4);}
    \foreach \state [count=\hdrindex from 0] in {#1}{%
      \pgfmathtruncatemacro{\hdrx}{mod(\hdrindex,4)}%
      \pgfmathtruncatemacro{\hdry}{3-int(\hdrindex/4)}%
      \begin{scope}[shift={(\hdrx,\hdry)}]
        \hdrfploccupiedvertex{\state}
      \end{scope}}
    \begin{scope}[hdr fpl four path]
      \draw (-1,3) -- (-0.5,3) (-1,1) -- (-0.5,1);
      \draw (3.5,2) -- (4,2) (3.5,0) -- (4,0);
      \draw (1,3.5) -- (1,4) (3,3.5) -- (3,4);
      \draw (0,-1) -- (0,-0.5) (2,-1) -- (2,-0.5);
    \end{scope}
  \end{tikzpicture}}
\newcommand{\hdrfplfourgroup}[1]{%
  \ifcase#1\relax
  \or
    \hdrfplfourconfig{2,6,4,3,6,6,6,4,3,6,6,6,4,3,6,1}%
  \or
    \hdrfplfourconfig{2,1,2,6,6,4,3,1,3,6,4,2,4,3,6,1}\hskip0.63cm%
    \hdrfplfourconfig{2,6,4,3,6,6,6,4,3,6,1,3,4,3,4,5}\hskip0.63cm%
    \hdrfplfourconfig{2,6,4,3,6,6,1,5,3,6,4,2,4,3,6,1}%
  \or
    \hdrfplfourconfig{2,1,2,6,6,4,3,1,3,1,5,2,4,5,2,1}\hskip0.63cm%
    \hdrfplfourconfig{2,1,2,6,1,2,6,1,5,3,1,2,4,5,2,1}\hskip0.63cm%
    \hdrfplfourconfig{2,6,4,3,6,6,1,5,3,1,5,2,4,5,2,1}%
  \or
    \hdrfplfourconfig{2,1,2,6,1,2,6,1,2,6,1,2,6,1,2,1}%
  \or
    \hdrfplfourconfig{2,6,4,3,6,1,3,4,3,4,5,3,4,3,4,5}\hskip0.63cm%
    \hdrfplfourconfig{2,1,5,3,6,4,2,4,3,6,1,3,4,3,4,5}\hskip0.63cm%
    \hdrfplfourconfig{2,1,5,3,6,4,5,5,3,6,4,2,4,3,6,1}%
  \or
    \hdrfplfourconfig{2,1,5,3,1,2,4,5,5,3,1,2,4,5,2,1}\hskip0.63cm%
    \hdrfplfourconfig{2,1,5,3,1,5,2,4,5,5,3,6,4,5,2,1}\hskip0.63cm%
    \hdrfplfourconfig{2,1,5,3,6,4,2,4,3,6,6,6,4,3,6,1}\hskip0.63cm%
    \hdrfplfourconfig{2,6,4,3,6,1,3,4,3,4,2,6,4,3,6,1}\hskip0.63cm%
    \hdrfplfourconfig{2,1,5,3,6,4,5,5,3,1,5,2,4,5,2,1}\hskip0.63cm%
    \hdrfplfourconfig{2,6,4,3,6,6,6,4,3,1,3,6,4,5,2,1}\hskip0.63cm%
    \hdrfplfourconfig{2,1,5,3,6,4,2,4,3,1,3,6,4,5,2,1}%
  \or
    \hdrfplfourconfig{2,1,5,3,1,2,4,5,2,6,1,2,6,1,2,1}\hskip0.63cm%
    \hdrfplfourconfig{2,1,5,3,1,5,2,4,2,4,3,6,6,1,2,1}\hskip0.63cm%
    \hdrfplfourconfig{2,1,5,3,1,5,2,4,5,2,6,6,4,3,6,1}%
  \or
    \hdrfplfourconfig{2,1,5,3,1,5,5,5,5,5,5,2,4,5,2,1}%
  \or
    \hdrfplfourconfig{2,1,5,3,1,5,5,5,2,4,5,2,6,1,2,1}\hskip0.63cm%
    \hdrfplfourconfig{2,1,5,3,1,5,5,5,5,2,4,2,4,3,6,1}\hskip0.63cm%
    \hdrfplfourconfig{2,1,5,3,1,5,2,4,5,2,1,3,4,3,4,5}%
  \or
    \hdrfplfourconfig{5,3,4,3,4,5,3,4,3,4,5,3,4,3,4,5}%
  \or
    \hdrfplfourconfig{2,6,4,3,1,3,6,4,5,5,3,6,4,5,2,1}\hskip0.63cm%
    \hdrfplfourconfig{5,3,4,3,4,2,6,4,3,1,3,6,4,5,2,1}\hskip0.63cm%
    \hdrfplfourconfig{5,3,4,3,4,5,3,4,3,4,2,6,4,3,6,1}%
  \or
    \hdrfplfourconfig{2,6,4,3,1,3,6,4,5,2,6,6,4,3,6,1}\hskip0.63cm%
    \hdrfplfourconfig{5,3,4,3,4,2,6,4,3,6,6,6,4,3,6,1}\hskip0.63cm%
    \hdrfplfourconfig{2,6,4,3,1,3,6,4,2,4,3,6,6,1,2,1}%
  \or
    \hdrfplfourconfig{5,3,4,3,4,2,1,5,3,1,5,2,4,5,2,1}\hskip0.63cm%
    \hdrfplfourconfig{2,1,2,6,1,5,3,1,5,5,5,2,4,5,2,1}\hskip0.63cm%
    \hdrfplfourconfig{2,6,4,3,1,3,1,5,5,5,5,2,4,5,2,1}%
  \or
    \hdrfplfourconfig{5,3,4,3,4,2,1,5,3,6,4,2,4,3,6,1}\hskip0.63cm%
    \hdrfplfourconfig{5,3,4,3,4,2,6,4,3,6,1,3,4,3,4,5}\hskip0.63cm%
    \hdrfplfourconfig{2,6,4,3,1,3,1,5,5,2,4,2,4,3,6,1}\hskip0.63cm%
    \hdrfplfourconfig{2,6,4,3,1,3,6,4,5,2,1,3,4,3,4,5}\hskip0.63cm%
    \hdrfplfourconfig{2,6,4,3,1,3,1,5,2,4,5,2,6,1,2,1}\hskip0.63cm%
    \hdrfplfourconfig{2,1,2,6,1,5,3,1,5,2,4,2,4,3,6,1}\hskip0.63cm%
    \hdrfplfourconfig{2,1,2,6,1,5,3,1,2,4,5,2,6,1,2,1}%
  \fi}

\newcommand{\hdrsixvertexfplglyph}[5]{%
  \begin{scope}[shift={(#1,0)}]
    \draw[line width=0.65pt,#2=0.5] (-0.62,0) -- (0,0);
    \draw[line width=0.65pt,#3=0.5] (0,0) -- (0.62,0);
    \draw[line width=0.65pt,#4=0.5] (0,-0.62) -- (0,0);
    \draw[line width=0.65pt,#5=0.5] (0,0) -- (0,0.62);
  \end{scope}}

\newcommand{\hdrsixvertexfpl}{%
  \begin{tikzpicture}[x=1cm,y=1cm,baseline=(current bounding box.center)]
    \useasboundingbox (-1.05,-2.98) rectangle (9.58,0.65);
    \foreach \x/\leftstyle/\rightstyle/\lowerstyle/\upperstyle in {
      0/hdr six arrow/hdr six arrow/hdr six arrow/hdr six arrow,
      1.8/hdr six invarrow/hdr six invarrow/hdr six invarrow/hdr six invarrow,
      3.6/hdr six invarrow/hdr six invarrow/hdr six arrow/hdr six arrow,
      5.4/hdr six arrow/hdr six arrow/hdr six invarrow/hdr six invarrow,
      7.2/hdr six invarrow/hdr six arrow/hdr six arrow/hdr six invarrow,
      9/hdr six arrow/hdr six invarrow/hdr six invarrow/hdr six arrow}{%
        \hdrsixvertexfplglyph{\x}{\leftstyle}{\rightstyle}{\lowerstyle}{\upperstyle}}
    \node[anchor=east] at (-0.72,-1.25) {odd};
    \node[anchor=east] at (-0.72,-2.50) {even};
    \foreach \x/\state in {0/1,1.8/2,3.6/3,5.4/4,7.2/5,9/6}{%
      \hdrfpllocalstate{\x}{-1.25}{\state}}
    \foreach \x/\state in {0/2,1.8/1,3.6/4,5.4/3,7.2/6,9/5}{%
      \hdrfpllocalstate{\x}{-2.50}{\state}}
  \end{tikzpicture}}

\newcommand{\hdryangbaxter}{%
  \begin{tikzpicture}[x=0.62cm,y=0.62cm,baseline=(current bounding box.center)]
    \useasboundingbox (0,0) rectangle (7.25,2.75);
    \begin{scope}[line width=0.45pt]
      \draw (0.45,2.05) -- (3.15,2.05);
      \draw (1.10,0) -- (1.10,2.75);
      \draw (0.45,0) -- (3.15,2.75);
      \draw (4.45,0) -- (7.15,2.75);
      \draw (6.50,0) -- (6.50,2.75);
      \draw (4.45,0.68) -- (7.15,0.68);
    \end{scope}
    \node[font=\tiny,anchor=east] at (0.18,2.02) {$1$};
    \node[font=\tiny,anchor=east] at (0.18,0.10) {$2$};
    \node[font=\tiny,anchor=north] at (1.40,0.12) {$3$};
    \node at (3.78,1.36) {$=$};
    \node[font=\tiny,anchor=east] at (4.20,0.69) {$1$};
    \node[font=\tiny,anchor=east] at (4.20,0.10) {$2$};
    \node[font=\tiny,anchor=north] at (6.82,0.12) {$3$};
  \end{tikzpicture}}

\newcommand{\hdrunitarity}{%
  \begin{tikzpicture}[x=0.75cm,y=0.38cm,baseline=(current bounding box.center)]
    \useasboundingbox (0,0) rectangle (8,3);
    \begin{scope}[line width=0.45pt]
      \draw (0.35,2.78) -- (1.08,1.50)
        .. controls (1.75,0.22) and (2.70,0.22) .. (3.38,1.50)
        -- (4.10,2.78);
      \draw (0.35,0.22) -- (1.08,1.50)
        .. controls (1.75,2.78) and (2.70,2.78) .. (3.38,1.50)
        -- (4.10,0.22);
      \draw (5.35,2.78) -- (7.95,2.78);
      \draw (5.35,0.22) -- (7.95,0.22);
    \end{scope}
    \node[font=\tiny,anchor=east] at (0.20,2.76) {$1$};
    \node[font=\tiny,anchor=east] at (0.20,0.24) {$2$};
    \node at (4.68,1.50) {$=$};
    \node[font=\tiny,anchor=east] at (5.18,2.76) {$1$};
    \node[font=\tiny,anchor=east] at (5.18,0.24) {$2$};
  \end{tikzpicture}}

\newcommand{\hdrsixvertexphase}{%
  \begin{tikzpicture}[x=1.25cm,y=1.25cm,baseline=(current bounding box.center)]
    \useasboundingbox (0,0) rectangle (5.55,4.65);
    \begin{scope}[line width=0.65pt]
      \draw[->] (0.75,0.60) -- (4.95,0.60);
      \draw[->] (0.75,0.60) -- (0.75,4.28);
      \draw (0.75,1.85) -- (2.00,0.60) -- (3.98,2.63);
      \draw (0.75,1.85) -- (2.95,4.08);
    \end{scope}
    \begin{scope}[line width=0.75pt,dash pattern=on 4pt off 3pt]
      \foreach \deltaval in {0,0.5,0.75}{%
        \draw plot[domain=0:90,samples=61,smooth,variable=\phaseangle]
          ({0.75+1.25*cos(\phaseangle)/sqrt(1-\deltaval*sin(2*\phaseangle))},
           {0.60+1.25*sin(\phaseangle)/sqrt(1-\deltaval*sin(2*\phaseangle))});
      }
    \end{scope}
    \node[anchor=east] at (0.58,4.20) {$a/c$};
    \node[anchor=west] at (5.02,0.48) {$b/c$};
    \node[anchor=west] at (0.02,0.23) {$\Delta=-\infty$};
    \node[anchor=north] at (2.00,0.52) {$1$};
    \node[anchor=east] at (0.66,1.85) {$1$};
    \node[font=\large] at (1.30,3.15) {F};
    \node[font=\large] at (1.12,1.03) {AF};
    \node[font=\large] at (2.22,2.07) {D};
    \node[font=\large] at (3.48,1.25) {F};
    \node[rotate=45] at (2.27,3.10) {$\Delta=1$};
    \node[rotate=45] at (3.38,1.72) {$\Delta=1$};
    \node[rotate=-45] at (1.45,1.38) {$\Delta=-1$};
  \end{tikzpicture}}

\newcommand{\hdrsixvertexglyph}[6]{%
  \begin{scope}[shift={(#1,#2)}]
    \draw[line width=0.6pt,#3=0.5] (-0.43,0) -- (0,0);
    \draw[line width=0.6pt,#4=0.5] (0,0) -- (0.43,0);
    \draw[line width=0.6pt,#5=0.5] (0,-0.43) -- (0,0);
    \draw[line width=0.6pt,#6=0.5] (0,0) -- (0,0.43);
  \end{scope}}

\newcommand{\hdrnilptile}[1]{%
  \draw[thin,densely dotted] ({#1-0.45},1.00) rectangle ({#1+0.45},1.90);}
\newcommand{\hdrdominotile}[1]{%
  \draw[thin,densely dotted] ({#1-0.45},-0.20) rectangle ({#1+0.45},0.70);}

\newcommand{\hdrsixvertexnilpdomino}{%
  \begin{tikzpicture}[x=1.9cm,y=1.8cm,baseline=(current bounding box.center)]
    \useasboundingbox (-0.65,-0.35) rectangle (6.96,3.05);
    \node[anchor=east] at (-0.51,2.55) {(a)};
    \node[anchor=east] at (-0.51,1.45) {(b)};
    \node[anchor=east] at (-0.51,0.25) {(c)};
    \hdrsixvertexglyph{0}{2.55}{hdr six arrow}{hdr six arrow}{hdr six arrow}{hdr six arrow}
    \hdrsixvertexglyph{1}{2.55}{hdr six invarrow}{hdr six invarrow}{hdr six invarrow}{hdr six invarrow}
    \hdrsixvertexglyph{2}{2.55}{hdr six invarrow}{hdr six invarrow}{hdr six arrow}{hdr six arrow}
    \hdrsixvertexglyph{3}{2.55}{hdr six arrow}{hdr six arrow}{hdr six invarrow}{hdr six invarrow}
    \hdrsixvertexglyph{4.59}{2.55}{hdr six invarrow}{hdr six arrow}{hdr six arrow}{hdr six invarrow}
    \hdrsixvertexglyph{6.18}{2.55}{hdr six arrow}{hdr six invarrow}{hdr six invarrow}{hdr six arrow}
    \foreach \x in {0,1,2,3,4,5.18,6.18}{\hdrnilptile{\x}\hdrdominotile{\x}}
    \begin{scope}[line width=1.45pt,line cap=butt,line join=miter]
      \draw (-0.45,1.45) -- (0,1.45) -- (0,1.90);
      \draw (0,1.00) -- (0.45,1.45);
      \draw (2,1.00) -- (2,1.90);
      \draw (2.55,1.45) -- (3.45,1.45);
      \draw (4,1.00) -- (4.45,1.45);
      \draw (5.18,1.00) -- (5.18,1.45) -- (5.63,1.45);
      \draw (5.73,1.45) -- (6.18,1.45) -- (6.18,1.90);
      \draw (-0.45,-0.20) -- (0.45,0.70);
      \draw (-0.45,0.70) -- (0,0.25);
      \draw (0.55,-0.20) -- (1.45,0.70);
      \draw (1,0.25) -- (1.45,-0.20);
      \draw (1.55,0.70) -- (2.45,-0.20);
      \draw (2,0.25) -- (2.45,0.70);
      \draw (2.55,0.70) -- (3.45,-0.20);
      \draw (2.55,-0.20) -- (3,0.25);
      \draw (3.55,-0.20) -- (4.45,0.70);
      \draw (4.73,0.70) -- (5.63,-0.20);
      \draw (5.73,-0.20) -- (6.63,0.70);
      \draw (5.73,0.70) -- (6.63,-0.20);
    \end{scope}
    \node[font=\tiny,fill=white,inner sep=0.5pt] at (4.59,1.45) {or};
    \node[font=\tiny,fill=white,inner sep=0.5pt] at (4.59,0.25) {or};
  \end{tikzpicture}}

\tikzset{
  hdr five frame/.style={draw=gray!78,densely dotted,line width=0.42pt,
    line cap=round,line join=round},
  hdr five guide/.style={draw=gray!78,dash pattern=on 3pt off 3pt,
    line width=0.43pt,line cap=butt,line join=miter},
  hdr five path/.style={draw=black,line width=1.45pt,
    line cap=butt,line join=miter},
  hdr five region/.style={draw=gray!78,densely dotted,line width=0.40pt,
    line cap=round,line join=round}
}

\newcommand{\hdrfivevertexfigure}{%
  \begin{tikzpicture}[x=0.00128cm,y=-0.00128cm,
      baseline=(current bounding box.center)]
    \useasboundingbox (1750,4650) rectangle (10850,8150);

    \foreach \x in {2100,3900,5700,7500,9300}{%
      \draw[hdr five frame] (\x,4800) rectangle ({\x+1200},6000);
      \draw[hdr five guide] ({\x+600},4800)--({\x+600},6000);
      \draw[hdr five guide] (\x,5400)--({\x+1200},5400);}
    \draw[hdr five path] (4500,4800)--(4500,6000);
    \draw[hdr five path] (5700,5400)--(6900,5400);
    \draw[hdr five path] (8100,6000)--(8100,5400)--(8700,5400);
    \draw[hdr five path] (9300,5400)--(9900,5400)--(9900,4800);

    \path[hdr five region,fill=hdrpppink]
      (2100,6600)--(1800,8100)--(3300,7800)--(3600,6300)--cycle;
    \path[hdr five region,fill=hdrppgreen]
      (5400,6300)--(5100,7800)--(3900,6600)--cycle;
    \path[hdr five region,fill=hdrppgreen]
      (3900,6600)--(3600,8100)--(5100,7800)--cycle;
    \path[hdr five region,fill=hdrppblue]
      (7200,6300)--(6900,7800)--(5700,6600)--cycle;
    \path[hdr five region,fill=hdrppblue]
      (5700,6600)--(5400,8100)--(6900,7800)--cycle;
    \path[hdr five region,fill=hdrppblue]
      (9000,6300)--(8700,7800)--(7500,6600)--cycle;
    \path[hdr five region,fill=hdrppgreen]
      (7500,6600)--(7200,8100)--(8700,7800)--cycle;
    \path[hdr five region,fill=hdrppgreen]
      (10800,6300)--(10500,7800)--(9300,6600)--cycle;
    \path[hdr five region,fill=hdrppblue]
      (9300,6600)--(9000,8100)--(10500,7800)--cycle;

    \draw[hdr five guide] (1950,7425)--(2400,7500)--(2475,7950);
    \draw[hdr five path] (2400,7500)--(3000,6900);
    \draw[hdr five guide] (2925,6450)--(3000,6900)--(3450,6975);

    \draw[hdr five guide] (5250,6975)--(4800,6900)
      --(4200,7500)--(3750,7425);
    \draw[hdr five path] (4800,6900)--(4725,6450);
    \draw[hdr five path] (4200,7500)--(4275,7950);

    \draw[hdr five guide] (6525,6450)--(6600,6900)
      --(6000,7500)--(6075,7950);
    \draw[hdr five path] (7050,6975)--(6600,6900);
    \draw[hdr five path] (6000,7500)--(5550,7425);

    \draw[hdr five guide] (8325,6450)--(8400,6900)
      --(7800,7500)--(7350,7425);
    \draw[hdr five path] (8400,6900)--(8850,6975);
    \draw[hdr five path] (7800,7500)--(7875,7950);

    \draw[hdr five guide] (10650,6975)--(10200,6900)
      --(9600,7500)--(9675,7950);
    \draw[hdr five path] (10200,6900)--(10125,6450);
    \draw[hdr five path] (9600,7500)--(9150,7425);
  \end{tikzpicture}}

\newcommand{\hdrfivevertexdualfigure}{%
  \begin{tikzpicture}[x=0.00128cm,y=-0.00128cm,
      baseline=(current bounding box.center)]
    \useasboundingbox (-50,4650) rectangle (9050,8150);

    \foreach \x in {300,2100,3900,5700,7500}
      \draw[hdr five frame] (\x,4800) rectangle ({\x+1200},6000);
    \draw[hdr five guide] (300,5400)--(1500,5400);
    \draw[hdr five path] (300,5400)--(900,4800);
    \draw[hdr five path] (900,6000)--(1500,5400);

    \draw[hdr five guide] (2700,4800)--(2100,5400)
      --(3300,5400)--(2700,6000);

    \draw[hdr five guide] (3900,5400)--(4500,4800);
    \draw[hdr five guide] (4500,6000)--(5100,5400);
    \draw[hdr five path] (3900,5400)--(5100,5400);

    \draw[hdr five guide] (6300,4800)--(5700,5400)--(6900,5400);
    \draw[hdr five path] (6300,6000)--(6900,5400);

    \draw[hdr five guide] (7500,5400)--(8700,5400)--(8100,6000);
    \draw[hdr five path] (7500,5400)--(8100,4800);

    \path[hdr five region,fill=hdrppgreen]
      (0,6300)--(300,7800)--(1500,6600)--cycle;
    \path[hdr five region,fill=hdrppgreen]
      (1500,6600)--(1800,8100)--(300,7800)--cycle;
    \path[hdr five region,fill=hdrpppink]
      (1800,6300)--(2100,7800)--(3300,6600)--cycle;
    \path[hdr five region,fill=hdrpppink]
      (3300,6600)--(3600,8100)--(2100,7800)--cycle;
    \path[hdr five region,fill=hdrppblue]
      (3600,6300)--(5100,6600)--(5400,8100)--(3900,7800)--cycle;
    \path[hdr five region,fill=hdrpppink]
      (5400,6300)--(5700,7800)--(6900,6600)--cycle;
    \path[hdr five region,fill=hdrppgreen]
      (6900,6600)--(7200,8100)--(5700,7800)--cycle;
    \path[hdr five region,fill=hdrppgreen]
      (7200,6300)--(7500,7800)--(8700,6600)--cycle;
    \path[hdr five region,fill=hdrpppink]
      (8700,6600)--(9000,8100)--(7500,7800)--cycle;

    \draw[hdr five guide] (135,6990)--(600,6900)
      --(1200,7500)--(1665,7418);
    \draw[hdr five path] (683,6430)--(601,6895);
    \draw[hdr five path] (1195,7501)--(1120,7981);

    \draw[hdr five guide] (2505,6443)--(2400,6900)
      --(3000,7500)--(2895,7950);
    \draw[hdr five path] (2412,6901)--(1939,6976);
    \draw[hdr five path] (2994,7491)--(3466,7416);

    \draw[hdr five guide] (4275,6450)--(4200,6900)--(3750,6975);
    \draw[hdr five path] (4200,6900)--(4800,7500);
    \draw[hdr five guide] (5250,7425)--(4800,7500)--(4725,7950);

    \draw[hdr five guide] (6090,6420)--(6000,6900)
      --(6600,7500)--(7072,7388);
    \draw[hdr five path] (5988,6874)--(5545,6957);
    \draw[hdr five path] (6528,7959)--(6603,7501);

    \draw[hdr five guide] (7350,7005)--(7800,6900)
      --(8400,7500)--(8333,7958);
    \draw[hdr five path] (7812,6908)--(7887,6443);
    \draw[hdr five path] (8400,7500)--(8857,7425);
  \end{tikzpicture}}

\newcommand{\hdrnilpbasicpattern}{%
  \begin{tikzpicture}[x=1.65cm,y=1.65cm,baseline=(current bounding box.center)]
    \useasboundingbox (0,0) rectangle (1,1);
    \draw[thin,densely dotted] (0,0) rectangle (1,1);
    \draw[line width=1.35pt] (0,0.5) -- (1,0.5);
    \draw[line width=1.35pt] (0.5,0) -- (0.5,1);
    \draw[line width=1.35pt] (0.5,0) -- (1,0.5);
    \node at (0.25,0.63) {$\alpha$};
    \node at (0.63,0.87) {$\beta$};
    \node at (0.76,0.63) {$\gamma$};
    \node at (0.37,0.24) {$\epsilon$};
    \node at (0.84,0.16) {$\delta$};
  \end{tikzpicture}}

\newcommand{\hdrexampledomino}{%
  \begin{tikzpicture}[x=1.58cm,y=1.58cm,baseline=(current bounding box.center)]
    \useasboundingbox (0,0) rectangle (2,2);
    \begin{scope}[thin,densely dotted]
      \draw (0,0) rectangle (1,1);
      \draw (0,1) rectangle (1,2);
      \draw (1,1) rectangle (2,2);
    \end{scope}
    \filldraw[fill=red!13,line width=1.25pt,line join=miter]
      (0,1) -- (1,2) -- (1.50,1.50) -- (0.50,0.50) -- cycle;
    \begin{scope}[thin,densely dotted]
      \draw (0,1) -- (1,1);
      \draw (1,1) -- (1,2);
    \end{scope}
    \begin{scope}[line width=1.25pt,line cap=butt]
      \draw (0,2) -- (0.50,1.50);
      \draw (0,0) -- (0.50,0.50) -- (1,0);
      \draw (1.50,1.50) -- (2,1);
    \end{scope}
  \end{tikzpicture}}

\tikzset{
  hdr kagome tiling/.style={black,line width=0.6pt,line cap=butt,line join=miter},
  hdr kagome spectral/.style={line width=0.45pt,dash pattern=on 2pt off 2pt,line cap=butt},
  hdr kagome alpha/.style={hdr kagome spectral,red!85!black},
  hdr kagome beta/.style={hdr kagome spectral,green!55!black},
  hdr kagome gamma/.style={hdr kagome spectral,blue},
  hdr kagome label/.style={font=\scriptsize,fill=white,inner sep=0.7pt}
}

\newcommand{\hdrkagomespectrallines}{%
  \begin{scope}[hdr kagome alpha]
    \draw (-0.5,0.75) -- (3,-1);
    \draw (-0.5,1.75) -- (4,-0.5);
    \draw (-0.5,2.75) -- (5,0);
    \draw (0,3.5) -- (6,0.5);
    \draw (1,4) -- (6.5,1.25);
    \draw (2,4.5) -- (6.5,2.25);
    \draw (3,5) -- (6.5,3.25);
  \end{scope}
  \begin{scope}[hdr kagome beta]
    \draw (-0.5,2.25) -- (5,5);
    \draw (-0.5,1.25) -- (6,4.5);
    \draw (-0.5,0.25) -- (6.5,3.75);
    \draw (0,-0.5) -- (6.5,2.75);
    \draw (1,-1) -- (6.5,1.75);
  \end{scope}
  \begin{scope}[hdr kagome gamma]
    \draw (0.5,-0.75) -- (0.5,3.75);
    \draw (1.5,-1.25) -- (1.5,4.25);
    \draw (2.5,-1.25) -- (2.5,4.75);
    \draw (3.5,-0.75) -- (3.5,5.25);
    \draw (4.5,-0.25) -- (4.5,5.25);
    \draw (5.5,0.25) -- (5.5,4.75);
  \end{scope}}

\newcommand{\hdrkagomecommontiling}{%
  \draw[hdr kagome tiling]
    (0,3) -- (0,0) -- (2,-1) -- (6,1) -- (6,4) -- (4,5) -- cycle;
  \begin{scope}[hdr kagome tiling]
    \draw (2,3) -- (3,3.5) -- (3,2.5) -- (2,2);
    \draw (2,1) -- (3,0.5);
    \draw (3,0.5) -- (6,2);
    \draw (5,1.5) -- (5,0.5);
    \draw (2,1) -- (3,1.5) -- (4,1);
    \draw (3,3.5) -- (2,4);
    \draw (4,2) -- (6,3);
    \draw (5,3.5) -- (5,2.5);
    \draw (4,2) -- (4,1);
    \draw (5,1.5) -- (5,2.5);
    \draw (3,2.5) -- (3,1.5);
    \draw (3,4.5) -- (5,3.5);
    \draw (5,4.5) -- (3,3.5);
    \draw (3,2.5) -- (4,3) -- (5,2.5);
    \draw (3,2.5) -- (4,2);
    \draw (4,3) -- (4,4);
    \draw (5,3.5) -- (6,4);
  \end{scope}}

\newcommand{\hdrkagomelefttiling}{%
  \begin{scope}[hdr kagome tiling]
    \draw (0,2) -- (1,2.5) -- (1,3.5);
    \draw (0,2) -- (1,1.5) -- (2,2) -- (1,2.5);
    \draw (0,1) -- (1,0.5) -- (1,1.5);
    \draw (1,0.5) -- (1,-0.5);
    \draw (1,3.5) -- (2,3) -- (2,2);
    \draw (3,0.5) -- (3,-0.5);
    \draw (1,0.5) -- (2,1) -- (2,2);
    \draw (1,-0.5) -- (2,0) -- (2,1);
    \draw (2,0) -- (3,-0.5);
  \end{scope}}

\newcommand{\hdrkagomerighttiling}{%
  \begin{scope}[hdr kagome tiling]
    \draw (1,0.5) -- (2,0);
    \draw (1,2.5) -- (2,2);
    \draw (1,1.5) -- (2,1);
    \draw (2,2) -- (2,1);
    \draw (0,2) -- (1,2.5) -- (1,3.5);
    \draw (0,1) -- (1,1.5);
    \draw (3,0.5) -- (3,-0.5);
    \draw (2,-1) -- (2,0) -- (3,0.5);
    \draw (0,1) -- (1,0.5) -- (1,-0.5);
    \draw (1,0.5) -- (2,1);
    \draw (1,1.5) -- (1,2.5);
  \end{scope}}

\newcommand{\hdrkagomedimensions}{%
  \begin{scope}[line width=0.45pt,>=latex]
    \draw[<->] (-0.25,-0.5) -- (1.75,-1.5)
      node[midway,below=2pt,fill=white,inner sep=0.7pt] {$a$};
    \draw[<->] (-0.75,0) -- (-0.75,3)
      node[midway,left=2pt,fill=white,inner sep=0.7pt] {$c$};
    \draw[<->] (-0.25,3.5) -- (3.75,5.5)
      node[midway,above=2pt,fill=white,inner sep=0.7pt] {$b$};
  \end{scope}}

\newcommand{\hdrkagomeparameterlabels}{%
  \foreach \x/\y/\i in {0.5/3.75/1,1.5/4.25/2,2.5/4.75/3,
                           3.5/5.25/4,4.5/5.25/5,5.5/4.75/6}{
    \node[hdr kagome label,text=blue,anchor=south] at (\x,{\y+0.08}) {$\gamma_{\i}$};
  }
  \node[hdr kagome label,text=red!85!black,anchor=north] at (3,-1.08) {$\alpha_1$};
  \node[hdr kagome label,text=red!85!black,anchor=north] at (4,-0.58) {$\alpha_2$};
  \node[hdr kagome label,text=red!85!black,anchor=north] at (5,-0.08) {$\alpha_3$};
  \node[hdr kagome label,text=red!85!black,anchor=west] at (6.10,0.5) {$\alpha_4$};
  \node[hdr kagome label,text=red!85!black,anchor=west] at (6.60,1.25) {$\alpha_5$};
  \node[hdr kagome label,text=red!85!black,anchor=west] at (6.60,2.25) {$\alpha_6$};
  \node[hdr kagome label,text=red!85!black,anchor=west] at (6.60,3.25) {$\alpha_7$};
  \node[hdr kagome label,text=green!55!black,anchor=south west] at (5.08,5.08) {$\beta_1$};
  \node[hdr kagome label,text=green!55!black,anchor=west] at (6.10,4.5) {$\beta_2$};
  \node[hdr kagome label,text=green!55!black,anchor=west] at (6.60,3.75) {$\beta_3$};
  \node[hdr kagome label,text=green!55!black,anchor=west] at (6.60,2.75) {$\beta_4$};
  \node[hdr kagome label,text=green!55!black,anchor=west] at (6.60,1.75) {$\beta_5$};}

\newcommand{\hdrkagomepanelone}{%
  \begin{tikzpicture}[x=0.65cm,y=0.65cm,baseline=(current bounding box.center)]
    \useasboundingbox (-1.35,-2.08) rectangle (7.65,5.90);
    \hdrkagomespectrallines
    \hdrkagomecommontiling
    \hdrkagomelefttiling
    \hdrkagomedimensions
    \hdrkagomeparameterlabels
    \node at (3,-1.90) {(a)};
  \end{tikzpicture}}

\newcommand{\hdrkagomepaneltwo}{%
  \begin{tikzpicture}[x=0.65cm,y=0.65cm,baseline=(current bounding box.center)]
    \useasboundingbox (-1.35,-2.08) rectangle (7.65,5.90);
    \filldraw[hdr kagome tiling,fill=red!12]
      (0,1) -- (0,0) -- (2,-1) -- (6,1) -- (6,2) -- (2,0) -- cycle;
    \hdrkagomespectrallines
    \hdrkagomecommontiling
    \hdrkagomerighttiling
    \hdrkagomedimensions
    \hdrkagomeparameterlabels
    \node at (3,-1.90) {(b)};
  \end{tikzpicture}}

\newcommand{\hdrcirclepattern}[2][1cm]{%
  \linkpattern[shape=hdrcircle,numbered,centered,unit=#1]{#2}}
\newcommand{\hdrlinepattern}[2][0.78cm]{%
  \linkpattern[numbered,centered,unit=#1]{#2}}
\newcommand{\hdrleadingmonomialpattern}{%
  \linkpattern[numbered,unit=0.62cm,
    tikzoptions={baseline=-\the\fontdimen22\textfont2}]{1/6,2/3,4/5,7/8}}
\newcommand{\hdrcirclepatternplain}[2][1cm]{%
  \linkpattern[shape=hdrcircle,centered,unit=#1]{#2}}

\newcommand{\hdrlpthree}[2][1cm]{%
  \ifcase#2\relax
  \or\hdrcirclepattern[#1]{1/6,2/5,3/4}%
  \or\hdrcirclepattern[#1]{1/6,2/3,4/5}%
  \or\hdrcirclepattern[#1]{1/2,3/6,4/5}%
  \or\hdrcirclepattern[#1]{1/4,2/3,5/6}%
  \or\hdrcirclepattern[#1]{1/2,3/4,5/6}%
  \fi}
\newcommand{\hdrlinepthree}[2][0.78cm]{%
  \ifcase#2\relax
  \or\hdrlinepattern[#1]{1/6,2/5,3/4}%
  \or\hdrlinepattern[#1]{1/6,2/3,4/5}%
  \or\hdrlinepattern[#1]{1/2,3/6,4/5}%
  \or\hdrlinepattern[#1]{1/4,2/3,5/6}%
  \or\hdrlinepattern[#1]{1/2,3/4,5/6}%
  \fi}

\newcommand{\hdrminilpthree}[1]{%
  \begingroup
  \linkpatternvertexradius=0.45\linkpatternvertexradius
  \hdrlpthree[0.40cm]{#1}%
  \endgroup}

\newcommand{\hdrlpfour}[2][1cm]{%
  \ifcase#2\relax
  \or\hdrcirclepattern[#1]{1/8,2/7,3/6,4/5}%
  \or\hdrcirclepattern[#1]{1/8,2/7,3/4,5/6}%
  \or\hdrcirclepattern[#1]{1/8,2/3,4/7,5/6}%
  \or\hdrcirclepattern[#1]{1/8,2/5,3/4,6/7}%
  \or\hdrcirclepattern[#1]{1/8,2/3,4/5,6/7}%
  \or\hdrcirclepattern[#1]{1/2,3/8,4/7,5/6}%
  \or\hdrcirclepattern[#1]{1/2,3/8,4/5,6/7}%
  \or\hdrcirclepattern[#1]{1/4,2/3,5/8,6/7}%
  \or\hdrcirclepattern[#1]{1/2,3/4,5/8,6/7}%
  \or\hdrcirclepattern[#1]{1/6,2/5,3/4,7/8}%
  \or\hdrcirclepattern[#1]{1/6,2/3,4/5,7/8}%
  \or\hdrcirclepattern[#1]{1/2,3/6,4/5,7/8}%
  \or\hdrcirclepattern[#1]{1/4,2/3,5/6,7/8}%
  \or\hdrcirclepattern[#1]{1/2,3/4,5/6,7/8}%
  \fi}
\newcommand{\hdrlinepfour}[2][0.62cm]{%
  \ifcase#2\relax
  \or\hdrlinepattern[#1]{1/8,2/7,3/6,4/5}%
  \or\hdrlinepattern[#1]{1/8,2/7,3/4,5/6}%
  \or\hdrlinepattern[#1]{1/8,2/3,4/7,5/6}%
  \or\hdrlinepattern[#1]{1/8,2/5,3/4,6/7}%
  \or\hdrlinepattern[#1]{1/8,2/3,4/5,6/7}%
  \or\hdrlinepattern[#1]{1/2,3/8,4/7,5/6}%
  \or\hdrlinepattern[#1]{1/2,3/8,4/5,6/7}%
  \or\hdrlinepattern[#1]{1/4,2/3,5/8,6/7}%
  \or\hdrlinepattern[#1]{1/2,3/4,5/8,6/7}%
  \or\hdrlinepattern[#1]{1/6,2/5,3/4,7/8}%
  \or\hdrlinepattern[#1]{1/6,2/3,4/5,7/8}%
  \or\hdrlinepattern[#1]{1/2,3/6,4/5,7/8}%
  \or\hdrlinepattern[#1]{1/4,2/3,5/6,7/8}%
  \or\hdrlinepattern[#1]{1/2,3/4,5/6,7/8}%
  \fi}

\newcommand{\hdrlptwo}[2][0.77cm]{%
  \ifcase#2\relax
  \or\hdrcirclepattern[#1]{1/4,2/3}%
  \or\hdrcirclepattern[#1]{1/2,3/4}%
  \or\hdrcirclepattern[#1]{1/3,2/4}
  \fi}

\newcommand{\hdrlpfourexampleone}[1][1.08cm]{%
  \hdrcirclepattern[#1]{1/7,2/5,3/6,4/8}}
\newcommand{\hdrlpfourexampletwo}[1][1.08cm]{%
  \hdrcirclepattern[#1]{1/5,2/7,3/8,4/6}}

\newcommand{\hdrlinepairnested}[1][0.63cm]{%
  \hdrlinepattern[#1]{1/4,2/3}}
\newcommand{\hdrlinepairadjacent}[1][0.63cm]{%
  \hdrlinepattern[#1]{1/2,3/4}}
\newcommand{\hdrlinepairnestedaa}[1][0.63cm]{%
  \linkpattern[numbered,centered,unit=#1]{1/4/[arrow],2/3/[arrow]}}
\newcommand{\hdrlinepairnestedai}[1][0.63cm]{%
  \linkpattern[numbered,centered,unit=#1]{1/4/[arrow],2/3/[invarrow]}}
\newcommand{\hdrlinepairnestedia}[1][0.63cm]{%
  \linkpattern[numbered,centered,unit=#1]{1/4/[invarrow],2/3/[arrow]}}
\newcommand{\hdrlinepairnestedii}[1][0.63cm]{%
  \linkpattern[numbered,centered,unit=#1]{1/4/[invarrow],2/3/[invarrow]}}
\newcommand{\hdrlinepairadjacentaa}[1][0.63cm]{%
  \linkpattern[numbered,centered,unit=#1]{1/2/[arrow],3/4/[arrow]}}
\newcommand{\hdrlinepairadjacentai}[1][0.63cm]{%
  \linkpattern[numbered,centered,unit=#1]{1/2/[arrow],3/4/[invarrow]}}
\newcommand{\hdrlinepairadjacentia}[1][0.63cm]{%
  \linkpattern[numbered,centered,unit=#1]{1/2/[invarrow],3/4/[arrow]}}
\newcommand{\hdrlinepairadjacentii}[1][0.63cm]{%
  \linkpattern[numbered,centered,unit=#1]{1/2/[invarrow],3/4/[invarrow]}}

\newcommand{\hdrmaximallycrossed}[1][1.1cm]{%
  \hdrcirclepatternplain[#1]{1/9,2/10,3/11,4/12,5/13,6/14,7/15,8/16}}

\newcommand{\hdrthreelittlearchescircle}{%
  \begin{tikzpicture}[x=2.3cm,y=2.3cm,baseline=(current bounding box.center)]
    \draw (0,0) circle[radius=0.41];
    \begin{scope}[blue,line width=0.9pt]
      \draw (0.140228,-0.385274) .. controls (0.052320,-0.143749)
        and (-0.150650,-0.026564) .. (-0.403771,-0.071196);
      \draw (-0.140228,-0.385274) .. controls (-0.106758,-0.293314)
        and (-0.200639,-0.239112) .. (-0.263543,-0.314078);
      \draw (0.263543,0.314078) .. controls (0.098330,0.117185)
        and (0.098330,-0.117185) .. (0.263543,-0.314078);
      \draw (0.403771,0.071196) .. controls (0.307397,0.054202)
        and (0.307397,-0.054202) .. (0.403771,-0.071196);
      \draw (-0.403771,0.071196) .. controls (-0.150650,0.026564)
        and (0.052320,0.143749) .. (0.140228,0.385274);
      \draw (-0.263543,0.314078) .. controls (-0.200639,0.239112)
        and (-0.106758,0.293314) .. (-0.140228,0.385274);
      \foreach \p in {(0.28,0),(0.21,0),(-0.14,0.242),(-0.105,0.182),
          (-0.14,-0.242),(-0.105,-0.182)}
        \fill \p circle[radius=1.15pt];
    \end{scope}
    \foreach \p in {(-0.403771,-0.071196),(-0.263543,-0.314078),
        (-0.140228,-0.385274),(0.140228,-0.385274),(0.263543,-0.314078),
        (0.403771,-0.071196),(0.403771,0.071196),(0.263543,0.314078),
        (0.140228,0.385274),(-0.140228,0.385274),(-0.263543,0.314078),
        (-0.403771,0.071196)}
      \node[/linkpattern/endpoint] at \p {};
    \node at (0.245,0.075) {$a$};
    \node at (-0.07,0.25) {$b$};
    \node at (-0.09,-0.27) {$c$};
  \end{tikzpicture}}

\newcommand{\hdrthreelittlearchesline}{%
  \begin{tikzpicture}[x=7.05cm,y=5.8cm,baseline=(current bounding box.center)]
    \draw (0.027778,0) -- (0.972222,0);
    \begin{scope}[blue,line width=0.9pt]
      \draw (0.027778,0) .. controls (0.132222,0.261111)
        and (0.867778,0.261111) .. (0.972222,0);
      \draw (0.138889,0) .. controls (0.212222,0.183333)
        and (0.787778,0.183333) .. (0.861111,0);
      \draw (0.194444,0) .. controls (0.238889,0.111111)
        and (0.427778,0.111111) .. (0.472222,0);
      \draw (0.305556,0) .. controls (0.318889,0.033333)
        and (0.347778,0.033333) .. (0.361111,0);
      \draw (0.527778,0) .. controls (0.572222,0.111111)
        and (0.761111,0.111111) .. (0.805556,0);
      \draw (0.638889,0) .. controls (0.652222,0.033333)
        and (0.681111,0.033333) .. (0.694444,0);
      \foreach \p in {(0.5,0.16),(0.5,0.18),(0.333,0.04),
          (0.333,0.06),(0.667,0.04),(0.667,0.06)}
        \fill \p circle[radius=1.15pt];
    \end{scope}
    \foreach \x in {0.027778,0.138889,0.194444,0.305556,0.361111,
        0.472222,0.527778,0.638889,0.694444,0.805556,0.861111,0.972222}
      \node[/linkpattern/endpoint] at (\x,0) {};
    \node at (0.53,0.16) {$a$};
    \node at (0.363,0.055) {$b$};
    \node at (0.697,0.055) {$c$};
    \node[anchor=north,inner sep=0pt,font=\scriptsize]
      at (0.027778,-0.016) {$\frac{\gamma_1}{q^2}$};
    \node[anchor=north,inner sep=0pt,font=\scriptsize]
      at (0.166667,-0.016) {$\cdots$};
    \node[anchor=north,inner sep=0pt,font=\scriptsize,xshift=-1.5pt]
      at (0.305556,-0.016) {$\frac{\gamma_{a+b}}{q^2}$};
    \node[anchor=north,inner sep=0pt,font=\scriptsize,xshift=1.5pt]
      at (0.361111,-0.016) {$\alpha_1$};
    \node[anchor=north,inner sep=0pt,font=\scriptsize]
      at (0.5,-0.016) {$\cdots$};
    \node[anchor=north,inner sep=0pt,font=\scriptsize,xshift=-4pt]
      at (0.638889,-0.016) {$\alpha_{b+c}$};
    \node[anchor=north,inner sep=0pt,font=\scriptsize,xshift=4pt]
      at (0.694444,-0.016) {$q^2\beta_1$};
    \node[anchor=north,inner sep=0pt,font=\scriptsize]
      at (0.833333,-0.016) {$\cdots$};
    \node[anchor=north,inner sep=0pt,font=\scriptsize]
      at (0.972222,-0.016) {$q^2\beta_{a+c}$};
  \end{tikzpicture}}

\newcommand{\hdrtransferrowexample}{%
  \begin{tikzpicture}[x=2.28cm,y=2.28cm,baseline=(current bounding box.center)]
    \draw (0,0) circle[radius=0.41] circle[radius=0.71];
    \foreach \a in {0,45,...,315}
      \draw (\a:0.41) -- (\a:0.71);
    \begin{scope}[/linkpattern/edge]
      \draw (0.156900,0.378791) .. controls (0.071753,0.173227)
        and (-0.173227,-0.071753) .. (-0.378791,-0.156900);
      \draw (0.156900,-0.378791) .. controls (0.104368,-0.251967)
        and (-0.104368,-0.251967) .. (-0.156900,-0.378791);
      \draw (0.378791,0.156900) .. controls (0.251967,0.104368)
        and (0.251967,-0.104368) .. (0.378791,-0.156900);
      \draw (-0.378791,0.156900) .. controls (-0.251967,0.104368)
        and (-0.104368,0.251967) .. (-0.156900,0.378791);
      \draw (-0.378791,-0.156900) .. controls (-0.517373,-0.214303)
        and (-0.517373,-0.214303) .. (-0.395980,-0.395980);
      \draw (-0.655954,-0.271705) .. controls (-0.517373,-0.214303)
        and (-0.517373,-0.214303) .. (-0.560000,0);
      \draw (-0.156900,-0.378791) .. controls (-0.214303,-0.517373)
        and (-0.214303,-0.517373) .. (0,-0.560000);
      \draw (-0.271705,-0.655954) .. controls (-0.214303,-0.517373)
        and (-0.214303,-0.517373) .. (-0.395980,-0.395980);
      \draw (0.156900,-0.378791) .. controls (0.214303,-0.517373)
        and (0.214303,-0.517373) .. (0,-0.560000);
      \draw (0.271705,-0.655954) .. controls (0.214303,-0.517373)
        and (0.214303,-0.517373) .. (0.395980,-0.395980);
      \draw (0.378791,-0.156900) .. controls (0.517373,-0.214303)
        and (0.517373,-0.214303) .. (0.560000,0);
      \draw (0.655954,-0.271705) .. controls (0.517373,-0.214303)
        and (0.517373,-0.214303) .. (0.395980,-0.395980);
      \draw (0.378791,0.156900) .. controls (0.517373,0.214303)
        and (0.517373,0.214303) .. (0.395980,0.395980);
      \draw (0.655954,0.271705) .. controls (0.517373,0.214303)
        and (0.517373,0.214303) .. (0.560000,0);
      \draw (0.156900,0.378791) .. controls (0.214303,0.517373)
        and (0.214303,0.517373) .. (0.395980,0.395980);
      \draw (0.271705,0.655954) .. controls (0.214303,0.517373)
        and (0.214303,0.517373) .. (0,0.560000);
      \draw (-0.156900,0.378791) .. controls (-0.214303,0.517373)
        and (-0.214303,0.517373) .. (-0.395980,0.395980);
      \draw (-0.271705,0.655954) .. controls (-0.214303,0.517373)
        and (-0.214303,0.517373) .. (0,0.560000);
      \draw (-0.378791,0.156900) .. controls (-0.517373,0.214303)
        and (-0.517373,0.214303) .. (-0.560000,0);
      \draw (-0.655954,0.271705) .. controls (-0.517373,0.214303)
        and (-0.517373,0.214303) .. (-0.395980,0.395980);
    \end{scope}
    \foreach \p in {(-0.378791,-0.156900),(-0.156900,-0.378791),
        (0.156900,-0.378791),(0.378791,-0.156900),(0.378791,0.156900),
        (0.156900,0.378791),(-0.156900,0.378791),(-0.378791,0.156900),
        (-0.655954,-0.271705),(-0.271705,-0.655954),(0.271705,-0.655954),
        (0.655954,-0.271705),(0.655954,0.271705),(0.271705,0.655954),
        (-0.271705,0.655954),(-0.655954,0.271705)}
      \node[/linkpattern/endpoint] at \p {};
    \begin{scope}[every node/.append style={font=\scriptsize,inner sep=2pt}]
      \node[anchor=east] at (-0.735,-0.315) {$1$};
      \node[anchor=north] at (-0.300,-0.735) {$2$};
      \node[anchor=north] at (0.300,-0.735) {$3$};
      \node[anchor=west] at (0.735,-0.315) {$4$};
      \node[anchor=west] at (0.735,0.315) {$5$};
      \node[anchor=south] at (0.300,0.735) {$6$};
      \node[anchor=south] at (-0.300,0.735) {$7$};
      \node[anchor=east] at (-0.735,0.315) {$8$};
    \end{scope}
  \end{tikzpicture}}

\newcommand{\hdrcylinderpaths}{%
  \draw (-1.19961,0.92114)
    .. controls (-1.22823,1.11696) and (-1.09182,1.28090) .. (-0.89510,1.30245)
    .. controls (-0.70390,1.33544) and (-0.55465,1.16547) .. (-0.59949,0.97670);
  \draw (0.00000,0.99450)
    .. controls (-0.06778,1.19376) and (0.09023,1.33926) .. (0.29840,1.33822)
    .. controls (0.59550,1.34218) and (0.57222,1.47051) .. (0.59949,1.67290)
    .. controls (0.61126,1.89134) and (0.52001,2.05299) .. (0.29840,2.03437)
    .. controls (0.10146,1.98277) and (0.00622,2.19347) .. (0.00000,2.38680)
    .. controls (0.03506,2.57979) and (-0.08838,2.74975) .. (-0.29840,2.73052)
    .. controls (-0.46036,2.74757) and (-0.53445,2.88589) .. (-0.59949,3.06520);
  \draw (0.59949,0.97670)
    .. controls (0.59138,1.17232) and (0.69027,1.30189) .. (0.89510,1.30245)
    .. controls (1.07482,1.25976) and (1.21964,1.43372) .. (1.19961,1.61734)
    .. controls (1.20212,1.79068) and (1.31132,1.95065) .. (1.49180,1.92271)
    .. controls (1.66643,1.91660) and (1.84831,2.04360) .. (1.80184,2.21204)
    .. controls (1.75750,2.40794) and (1.90752,2.57790) .. (2.08850,2.49078)
    .. controls (2.27562,2.42060) and (2.39364,2.25444) .. (2.37003,2.05600)
    .. controls (2.37435,1.85355) and (2.45564,1.73596) .. (2.61500,1.61103)
    .. controls (2.76818,1.58252) and (2.89998,1.69573) .. (2.84146,1.84014)
    .. controls (2.76204,1.95503) and (2.87630,2.08975) .. (3.01290,2.06063)
    .. controls (3.09548,2.05913) and (3.14288,2.14345) .. (3.14535,2.23002);
  \draw (-1.74805,0.83104)
    .. controls (-1.80689,1.02155) and (-1.68568,1.18002) .. (-1.49180,1.22656)
    .. controls (-1.30000,1.21633) and (-1.17264,1.42717) .. (-1.19961,1.61734)
    .. controls (-1.21114,1.80719) and (-1.07961,1.95243) .. (-0.89510,1.99860)
    .. controls (-0.64840,1.93312) and (-0.62685,1.83121) .. (-0.59949,1.67290)
    .. controls (-0.60735,1.48202) and (-0.48786,1.31369) .. (-0.29840,1.33822)
    .. controls (-0.12495,1.35698) and (-0.00308,1.51627) .. (0.00000,1.69070)
    .. controls (-0.02620,1.86972) and (-0.12493,1.98299) .. (-0.29840,2.03437)
    .. controls (-0.46990,2.01737) and (-0.63910,2.20128) .. (-0.59949,2.36900)
    .. controls (-0.63029,2.52841) and (-0.73794,2.65404) .. (-0.89510,2.69475)
    .. controls (-1.08728,2.67565) and (-1.20801,2.50638) .. (-1.19961,2.31344)
    .. controls (-1.18138,2.13651) and (-1.31663,1.95354) .. (-1.49180,1.92271) 
    .. controls (-1.65188,1.93212) and (-1.74972,2.06849) .. (-1.74805,2.22334)
    .. controls (-1.79011,2.35002) and (-1.85613,2.50395) .. (-1.98900,2.51673)
    .. controls (-2.14199,2.45227) and (-2.27566,2.25940) .. (-2.25145,2.09516)
    .. controls (-2.22461,1.90657) and (-2.30482,1.72293) .. (-2.48630,1.66502)
    .. controls (-2.64696,1.57084) and (-2.71893,1.40612) .. (-2.70427,1.22047)
    .. controls (-2.70432,1.10251) and (-2.59698,0.92810) .. (-2.48630,0.96887)
    .. controls (-2.29913,1.00558) and (-2.15853,0.86943) .. (-2.25145,0.70286);
  \draw (1.19961,0.92114)
    .. controls (1.13506,1.11400) and (1.29506,1.27813) .. (1.49180,1.22656)
    .. controls (1.66607,1.14431) and (1.84066,1.32719) .. (1.80184,1.51594)
    .. controls (1.76919,1.69246) and (1.90925,1.80428) .. (2.08850,1.79463)
    .. controls (2.28372,1.71165) and (2.40129,1.56970) .. (2.37003,1.35990)
    .. controls (2.36413,1.17625) and (2.45129,0.99829) .. (2.61500,0.91488)
    .. controls (2.77520,0.83942) and (2.89994,0.61498) .. (2.84146,0.44784);
  \draw (1.80184,0.81974)
    .. controls (1.73797,1.02007) and (1.89253,1.17470) .. (2.08850,1.09848)
    .. controls (2.27838,1.03930) and (2.41384,0.85769) .. (2.37003,0.66370);
  \draw (3.1824,0.65) .. controls (3.14,0.75) .. (3.10518,0.89507)
    .. controls (3.10545,1.05236) and (3.07319,1.21873) .. (3.01290,1.36448)
    .. controls (2.87072,1.45584) and (2.83082,1.27179) .. (2.84146,1.14404)
    .. controls (2.82504,0.96792) and (2.88961,0.79515) .. (3.01290,0.66833)
    .. controls (3.12763,0.57060) and (3.17129,0.40804) .. (3.11606,0.26783);
  \draw (0.89510,1.99860)
    .. controls (1.07333,2.01355) and (1.17352,2.13649) .. (1.19961,2.31344)
    .. controls (1.18375,2.49701) and (1.07456,2.65296) .. (0.89510,2.69475)
    .. controls (0.70222,2.77446) and (0.59441,2.57030) .. (0.59949,2.36900)
    .. controls (0.59644,2.14807) and (0.66826,1.98584) .. (0.89510,1.99860)
    -- cycle;
  \draw (-1.98900,1.12443)
    .. controls (-1.82399,1.19879) and (-1.75563,1.37329) .. (-1.74805,1.52724)
    .. controls (-1.73803,1.68227) and (-1.83117,1.79356) .. (-1.98900,1.82058)
    .. controls (-2.16982,1.75253) and (-2.25981,1.59207) .. (-2.25145,1.39906)
    .. controls (-2.25535,1.23322) and (-2.15484,1.09052) .. (-1.98900,1.12443)
    -- cycle;
  \draw (-2.70427,0.52427)
    .. controls (-2.63334,0.66339) and (-2.72950,0.79046) .. (-2.88410,0.76848)
    .. controls (-3.06761,0.65339) and (-3.15292,0.48256) .. (-3.08018,0.27852);
  \draw (-3.1824,0.58619)
    .. controls (-3.11561,0.71805) and (-3.07175,0.85420) .. (-3.04980,0.99920)
    .. controls (-3.06813,1.17986) and (-3.00391,1.32817) .. (-2.88410,1.46463)
    .. controls (-2.77097,1.57043) and (-2.70540,1.76168) .. (-2.70427,1.91657)
    .. controls (-2.70216,2.04393) and (-2.74931,2.14310) .. (-2.88410,2.16078)
    .. controls (-2.95980,2.12206) and (-3.09007,2.16926) .. (-3.10947,2.30073);
  \draw (-3.1824,1.24243)
    .. controls (-3.08160,1.37851) and (-3.02239,1.54643) .. (-3.02065,1.71979)
    .. controls (-2.99186,1.81322) and (-3.05970,1.91334) .. (-3.1824,1.89874);
  \draw (3.1824,1.29290)
    .. controls (3.09873,1.38356) and (3.07603,1.50377) .. (3.11608,1.61575)
    .. controls (3.12154,1.64808) and (3.14460,1.67004) .. (3.1824,1.68008);
  \draw (-2.70427,2.61277)
    .. controls (-2.71517,2.47531) and (-2.62408,2.35552) .. (-2.48630,2.36117)
    .. controls (-2.30421,2.43408) and (-2.21400,2.59882) .. (-2.25145,2.79136);
  \draw (-1.19961,3.00964)
    .. controls (-1.18234,2.82214) and (-1.31208,2.67503) .. (-1.49180,2.61886)
    .. controls (-1.65286,2.64643) and (-1.68648,2.76818) .. (-1.74805,2.91954);
  \draw (0.00000,3.08300)
    .. controls (0.02450,2.85665) and (0.15490,2.77222) .. (0.29840,2.73052)
    .. controls (0.48401,2.74429) and (0.54531,2.88715) .. (0.59949,3.06520);
  \draw (1.19961,3.00964)
    .. controls (1.25238,2.80952) and (1.33690,2.70339) .. (1.49180,2.61886)
    .. controls (1.66090,2.60892) and (1.73104,2.75436) .. (1.80184,2.90824);
  \draw (2.37003,2.75220)
    .. controls (2.40983,2.51079) and (2.50061,2.41574) .. (2.61500,2.30718)
    .. controls (2.69874,2.21846) and (2.91097,2.34193) .. (2.84146,2.53634);}
\newcommand{\hdrloopcylinder}{%
  \begin{tikzpicture}[x=1cm,y=1cm,baseline=(current bounding box.center)]
    \useasboundingbox (-3.38,-1.05) rectangle (3.38,4.82);
    \begin{scope}[line width=0.3pt]
      \draw (0,0) ellipse[x radius=3.1824,y radius=0.9945];
      \foreach \y in {0,0.6962,1.3923,2.0885}
        \draw (-3.1824,\y) arc[start angle=-180,end angle=-360,
          x radius=3.1824,y radius=0.9945];
      \foreach \x in {-2.8841,-2.4863,-1.9890,-1.4918,-0.8951,-0.2984,
          0.2984,0.8951,1.4918,2.0885,2.6150,3.0129}{%
        \pgfmathsetmacro{\hdrbase}{0.9945*sqrt(1-(\x/3.1824)^2)}%
        \draw (\x,\hdrbase) -- (\x,{\hdrbase+2.0885});}
      \draw (-3.1824,0) -- (-3.1824,4.376)
        (3.1824,0) -- (3.1824,4.376);
    \end{scope}
    \begin{scope}[blue,line width=1.1pt,line cap=butt,line join=round]
      \clip (-3.1824,-1.1) rectangle (3.1824,4.4);
      \hdrcylinderpaths
    \end{scope}
    \foreach \x/\y in {
      0.59949/0.97670,1.19961/0.92114,1.80184/0.81974,2.37003/0.66370,
      2.84146/0.44784,3.0962/0.2188,3.1559/-0.0994,2.9968/-0.3381,
      2.6586/-0.5569,2.1813/-0.7465,1.6535/-0.8659,1.0568/-0.9454,
      0.3912/-0.9945,-0.3050/-0.9945,-0.9123/-0.9746,-1.4785/-0.8858,
      -2.0659/-0.7558,-2.5632/-0.5874,-2.9106/-0.3978,-3.1201/-0.1989,
      -3.0604/0.2586,-2.70427/0.52427,-2.25145/0.70286,-1.74805/0.83104,
      -1.19961/0.92114,-0.59949/0.97670,0.00000/0.99450}
      \node[/linkpattern/endpoint] at (\x,\y) {};
    \foreach \x/\y in {-1.5912/3.5802,-1.5912/3.9780,-1.5912/4.3758,
        0/3.7134,0/4.1112,0/4.5090,
        1.5912/3.5802,1.5912/3.9780,1.5912/4.3758}
      \fill (\x,\y) circle[radius=0.8pt];
    \node[font=\small] at (-1.19961,0.675) {$2n$};
    \node[font=\small] at (-0.59949,0.725) {$1$};
    \node[font=\small] at (0,0.742) {$2$};
    \node[font=\small] at (0.59949,0.725) {$3$};
    \node[font=\small] at (1.19961,0.675) {$4$};
    \node[rotate=10,font=\small] at (-2.0885,0.3978) {$\cdots$};
    \node[rotate=-10,font=\small] at (2.0885,0.4973) {$\cdots$};
  \end{tikzpicture}}

\newcommand{\hdraffinewindingpaths}[5]{%
  \begin{scope}
    \tikzset{hdr affine strand/.style={blue,line width=0.9pt,line cap=butt,
      preaction={draw=white,line width=5.4pt,line cap=butt}}}
    \foreach \x in {5,...,0}{%
      \let\hdrwindingheight\relax
      \ifnum\x=#2\relax\def\hdrwindingheight{#3}\fi
      \ifnum\x=#4\relax\def\hdrwindingheight{#5}\fi
      \ifx\hdrwindingheight\relax
        \draw[hdr affine strand] (\x,#1) -- (\x,-#1);
      \else
        \draw[hdr affine strand] (\x,#1) -- (\x,{\hdrwindingheight+0.34})
          .. controls (\x,{\hdrwindingheight+0.10})
          and ({\x-0.10},\hdrwindingheight)
          .. ({\x-0.34},\hdrwindingheight) -- (-0.85,\hdrwindingheight);
        \draw[hdr affine strand] (5.55,\hdrwindingheight) -- ({\x+0.34},\hdrwindingheight)
          .. controls ({\x+0.10},\hdrwindingheight)
          and (\x,{\hdrwindingheight-0.10})
          .. (\x,{\hdrwindingheight-0.34}) -- (\x,-#1);
      \fi}
  \end{scope}}
\newcommand{\hdraffineyi}{%
  \begin{tikzpicture}[x=0.80cm,y=0.86cm,baseline=(current bounding box.center)]
    \useasboundingbox (-0.90,-1.03) rectangle (5.62,1.03);
    \hdraffinewindingpaths{0.78}{2}{0}{-1}{0}
    \node[anchor=north,font=\scriptsize] at (2,-0.82) {$i$};
  \end{tikzpicture}}
\newcommand{\hdraffinecommuting}{%
  \begin{tikzpicture}[x=0.72cm,y=0.72cm,baseline=(current bounding box.center)]
    \useasboundingbox (-0.90,-1.94) rectangle (13.52,1.94);
    \hdraffinewindingpaths{1.66}{2}{-0.88}{3}{0.88}
    \node[anchor=north,font=\scriptsize] at (2,-1.72) {$i$};
    \node[anchor=north,font=\scriptsize] at (3,-1.72) {$j$};
    \node at (6.35,0) {$=$};
    \begin{scope}[shift={(7.55,0)}]
      \hdraffinewindingpaths{1.66}{2}{0.88}{3}{-0.88}
      \node[anchor=north,font=\scriptsize] at (2,-1.72) {$i$};
      \node[anchor=north,font=\scriptsize] at (3,-1.72) {$j$};
    \end{scope}
  \end{tikzpicture}}

\newcommand{\hdraffineyeigenvector}{%
  \begin{tikzpicture}[x=2.15cm,y=2.15cm,baseline=(current bounding box.center)]
    \useasboundingbox (-0.78,-0.78) rectangle (0.78,0.78);
    \draw (0,0) circle[radius=0.41] circle[radius=0.71];
    \begin{scope}[blue,line width=0.9pt,line cap=butt]
      \foreach \a/\b in {202.5/247.5,292.5/337.5,22.5/67.5,112.5/157.5}
        \draw (\a:0.41) .. controls (\a:0.272728) and (\b:0.272728) .. (\b:0.41);
      \draw (202.5:0.71) -- (202.5:0.41)
        (247.5:0.71) -- (247.5:0.41);
      \foreach \a in {337.5,22.5,67.5,112.5,157.5}
        \draw (\a:0.41) -- (\a:0.51) (\a:0.61) -- (\a:0.71);
      \draw (292.5:0.41) .. controls (292.5:0.49) and (0.18955,-0.53158) .. (282.5:0.56)
        arc[start angle=282.5,end angle=252.5,radius=0.56];
      \draw (242.5:0.56) arc[start angle=242.5,end angle=207.5,radius=0.56];
      \draw (197.5:0.56) arc[start angle=197.5,end angle=-57.5,radius=0.56]
        .. controls (0.22498,-0.52065) and (292.5:0.64) .. (292.5:0.71);
    \end{scope}
    \foreach \a in {22.5,67.5,...,337.5}
      \node[/linkpattern/endpoint] at (\a:0.71) {};
    \node[anchor=east,font=\scriptsize] at (202.5:0.77) {$1$};
    \node[anchor=north,font=\scriptsize] at (292.5:0.77) {$i$};
  \end{tikzpicture}}

\newcommand{\hdraffinecrossing}{%
  \begin{tikzpicture}[x=0.72cm,y=0.72cm,baseline=(current bounding box.center)]
    \draw[blue,line width=0.9pt] (-0.42,-0.34) -- (0.42,0.50);
    \draw[blue,line width=0.9pt] (-0.42,0.50) -- (-0.10,0.18);
    \draw[blue,line width=0.9pt] (0.10,-0.02) -- (0.42,-0.34);
    \node[anchor=north east,font=\scriptsize] at (-0.42,-0.36) {$i$};
    \node[anchor=north west,font=\scriptsize] at (0.42,-0.36) {$i+1$};
    \path ($ (0,0.08)!-1!(current bounding box.south) $);
  \end{tikzpicture}}

\newcommand{\hdrlinkpatterntoyoung}{%
  \begin{tikzpicture}[x=0.55cm,y=0.55cm,baseline=(current bounding box.center)]
    \filldraw[fill=red!12,thin]
      (0.5,0) -- (3.5,3) -- (5.5,1) -- (4.5,0) --
      (3.5,1) -- (1.5,-1) -- cycle;
    \draw[thin] (1.5,1) -- (2.5,0)
      (2.5,2) -- (3.5,1)
      (3.5,1) -- (4.5,2);
    \draw[line width=1pt]
      (-0.5,-1) -- (0.5,0) -- (1.5,-1) -- (3.5,1) --
      (4.5,0) -- (5.5,1) -- (7.5,-1);
    \begin{scope}[blue,line width=0.9pt]
      \draw (0,3.5) -- (0,-0.5) -- (1,-0.5) -- (1,3.5);
      \draw (2,3.5) -- (2,-0.5) -- (7,-0.5) -- (7,3.5);
      \draw (3,3.5) -- (3,0.5) -- (4,0.5) -- (4,3.5);
      \draw (5,3.5) -- (5,0.5) -- (6,0.5) -- (6,3.5);
    \end{scope}
    \foreach \x in {0,...,7}
      \node[/linkpattern/endpoint] at (\x,3.5) {};
  \end{tikzpicture}}

\newcommand{\hdryoungweight}{%
  \begin{tikzpicture}[x=0.55cm,y=0.55cm,baseline=(current bounding box.center)]
    \filldraw[fill=red!12,thin]
      (2,2) -- (5,5) -- (7,3) -- (6,2) --
      (5,3) -- (3,1) -- cycle;
    \filldraw[fill=cyan!35,thin]
      (1,1) -- (2,2) -- (3,1) -- (5,3) -- (6,2) --
      (7,3) -- (9,1) -- (8,0) -- (7,1) -- (6,0) --
      (5,1) -- (4,0) -- (3,1) -- (2,0) -- cycle;
    \draw[thin] (3,3) -- (5,1)
      (4,4) -- (5,3)
      (5,3) -- (6,4)
      (5,1) -- (6,2) -- (7,1) -- (8,2);
    \draw[line width=1pt]
      (0,0) -- (2,2) -- (3,1) -- (5,3) --
      (6,2) -- (7,3) -- (10,0);
    \node at (2,1) {$t$};
    \node at (4,1) {$t$};
    \node at (6,1) {$t$};
    \node at (8,1) {$t$};
    \node at (5,2) {$t^{-1}$};
    \node at (7,2) {$t^{-1}$};
  \end{tikzpicture}}

\newcommand{\hdrwiringframe}[1]{%
  \draw[thin] (0,0) rectangle (4,#1);
  \foreach \x/\lab in {0.5/1,1.5/2,2.5/3,3.5/4}{
    \node[anchor=north,font=\scriptsize] at (\x,-0.13) {$\lab$};
    \node[anchor=south,font=\scriptsize] at (\x,{#1+0.13}) {$\lab$};
  }}
\newcommand{\hdrwiringendpoints}[1]{%
  \foreach \x in {0.5,1.5,2.5,3.5}{
    \node[/linkpattern/endpoint] at (\x,0) {};
    \node[/linkpattern/endpoint] at (\x,#1) {};
  }}
\newcommand{\hdrnilheckeone}{%
  \begin{tikzpicture}[x=0.88cm,y=0.88cm,baseline=(current bounding box.center)]
    \hdrwiringframe{2}
    \draw[thin] (0,1) -- (4,1);
    \begin{scope}[blue,line width=0.9pt]
      \draw (1.5,2) -- (0.5,1) -- (0.5,0);
      \draw (0.5,2) -- (1.5,1) -- (3.5,0);
      \draw (1.5,0) -- (3.5,2);
      \draw (2.5,2) -- (3.5,1) -- (2.5,0);
    \end{scope}
    \hdrwiringendpoints{2}
  \end{tikzpicture}}
\newcommand{\hdrnilhecketwo}{%
  \begin{tikzpicture}[x=0.88cm,y=0.88cm,baseline=(current bounding box.center)]
    \hdrwiringframe{1}
    \begin{scope}[blue,line width=0.9pt]
      \draw (0.5,1) -- (3.5,0);
      \draw (1.5,1) -- (0.5,0);
      \draw (1.5,0) -- (3.5,1);
      \draw (2.5,1) -- (2.5,0);
    \end{scope}
    \hdrwiringendpoints{1}
  \end{tikzpicture}}
\newcommand{\hdrnilheckethree}{%
  \begin{tikzpicture}[x=0.88cm,y=0.88cm,baseline=(current bounding box.center)]
    \hdrwiringframe{2}
    \draw[thin] (0,1) -- (4,1);
    \begin{scope}[blue,line width=0.9pt]
      \draw (1.5,0) -- (0.5,1) -- (0.5,2);
      \draw (2.5,0) -- (1.5,1) -- (3.5,2);
      \draw (0.5,0) -- (2.5,1) -- (1.5,2);
      \draw (3.5,0) -- (3.5,1) -- (2.5,2);
    \end{scope}
    \hdrwiringendpoints{2}
  \end{tikzpicture}}

\newcommand{\hdrperiodicblockframe}{%
  \draw (0,0) -- (2,-2) -- (4,-2) -- (2,0) -- cycle;
  \draw[dashed] (0,0) -- (-0.5,0.5)
    (2,0) -- (1.5,0.5)
    (2,-2) -- (2.5,-2.5)
    (4,-2) -- (4.5,-2.5);}
\newcommand{\hdrupperlowerblock}{%
  \begin{tikzpicture}[x=0.64cm,y=0.64cm,baseline=(current bounding box.center)]
    \hdrperiodicblockframe
    \draw (2,0) -- (2,-2);
    \node at (1.333,-0.667) {$U$};
    \node at (2.667,-1.333) {$L$};
  \end{tikzpicture}}
\newcommand{\hdrbrauerblock}{%
  \begin{tikzpicture}[x=0.64cm,y=0.64cm,baseline=(current bounding box.center)]
    \hdrperiodicblockframe
    \draw (1,0) -- (1,-1) -- (2,-1)
      (2,-1) -- (3,-1) -- (3,-2)
      (2,0) -- (2,-2);
    \node at (0.667,-0.333) {$0$};
    \node at (1.5,-0.5) {$X$};
    \node at (1.667,-1.333) {$0$};
    \node at (2.333,-0.667) {$\star$};
    \node at (2.5,-1.5) {$Y$};
    \node at (3.333,-1.667) {$\star$};
  \end{tikzpicture}}

\tikzset{hdr pipe/.style={line width=1.4pt,line cap=butt,line join=round}}

\newcommand{\hdrcolouredelbow}[2]{%
  \draw[#1,hdr pipe] (0,0.5) .. controls (0,0.2) and (-0.2,0) .. (-0.5,0);
  \draw[#2,hdr pipe] (0,-0.5) .. controls (0,-0.2) and (0.2,0) .. (0.5,0);}
\newcommand{\hdrcolouredcrossing}[2]{%
  \draw[#1,hdr pipe] (0,0.5) -- (0,-0.5);
  \draw[#2,hdr pipe] (-0.5,0) -- (0.5,0);}
\newcommand{\hdrcolouredboundaryelbow}[1]{%
  \draw[#1,hdr pipe] (0,0.5) .. controls (0,0.2) and (-0.2,0) .. (-0.5,0);}
\newcommand{\hdrmulticolouredpipedream}{%
  \begin{tikzpicture}[x=0.70cm,y=0.70cm,baseline=(current bounding box.center)]
    \draw[thin] (0,0) -- (0,-4) -- (4,0) -- cycle;
    \foreach \i in {1,2,3}{
      \draw[thin] (\i,0) -- (\i,{\i-4});
      \draw[thin] (0,-\i) -- ({4-\i},-\i);
    }
    \foreach \i in {1,...,4}{
      \node[font=\scriptsize] at ({\i-0.5},0.30) {$\i$};
      \node[font=\scriptsize] at (-0.30,{0.5-\i}) {$\i$};
    }
    \def\hdrgreen{green!70!black}
    \def\hdrorange{orange!90!black}
    \begin{scope}[shift={(0.5,-0.5)}]\hdrcolouredcrossing{red}{\hdrgreen}\end{scope}
    \begin{scope}[shift={(1.5,-0.5)}]\hdrcolouredcrossing{\hdrorange}{\hdrgreen}\end{scope}
    \begin{scope}[shift={(2.5,-0.5)}]\hdrcolouredcrossing{blue}{\hdrgreen}\end{scope}
    \begin{scope}[shift={(3.5,-0.5)}]\hdrcolouredboundaryelbow{\hdrgreen}\end{scope}
    \begin{scope}[shift={(0.5,-1.5)}]\hdrcolouredelbow{red}{\hdrorange}\end{scope}
    \begin{scope}[shift={(1.5,-1.5)}]\hdrcolouredelbow{\hdrorange}{blue}\end{scope}
    \begin{scope}[shift={(2.5,-1.5)}]\hdrcolouredboundaryelbow{blue}\end{scope}
    \begin{scope}[shift={(0.5,-2.5)}]\hdrcolouredcrossing{\hdrorange}{blue}\end{scope}
    \begin{scope}[shift={(1.5,-2.5)}]\hdrcolouredboundaryelbow{blue}\end{scope}
    \begin{scope}[shift={(0.5,-3.5)}]\hdrcolouredboundaryelbow{\hdrorange}\end{scope}
  \end{tikzpicture}}
\newcommand{\hdrcolouredpipedream}{%
  \begin{tikzpicture}[x=0.70cm,y=0.70cm,baseline=(current bounding box.center)]
    \draw[thin] (0,0) -- (0,-5) -- (5,0) -- cycle;
    \foreach \i in {1,...,4}{
      \draw[thin] (\i,0) -- (\i,{\i-5});
      \draw[thin] (0,-\i) -- ({5-\i},-\i);
    }
    \foreach \i in {1,...,5}{
      \node[font=\scriptsize] at ({\i-0.5},0.30) {$\i$};
      \node[font=\scriptsize] at (-0.30,{0.5-\i}) {$\i$};
    }
    \def\hdrgreen{green!70!black}
    \begin{scope}[shift={(0.5,-0.5)}]\hdrcolouredcrossing{red}{\hdrgreen}\end{scope}
    \begin{scope}[shift={(1.5,-0.5)}]\hdrcolouredelbow{\hdrgreen}{\hdrgreen}\end{scope}
    \begin{scope}[shift={(2.5,-0.5)}]\hdrcolouredelbow{\hdrgreen}{red}\end{scope}
    \begin{scope}[shift={(3.5,-0.5)}]\hdrcolouredelbow{red}{\hdrgreen}\end{scope}
    \begin{scope}[shift={(4.5,-0.5)}]\hdrcolouredboundaryelbow{\hdrgreen}\end{scope}
    \begin{scope}[shift={(0.5,-1.5)}]\hdrcolouredcrossing{red}{\hdrgreen}\end{scope}
    \begin{scope}[shift={(1.5,-1.5)}]\hdrcolouredelbow{\hdrgreen}{red}\end{scope}
    \begin{scope}[shift={(2.5,-1.5)}]\hdrcolouredelbow{red}{\hdrgreen}\end{scope}
    \begin{scope}[shift={(3.5,-1.5)}]\hdrcolouredboundaryelbow{\hdrgreen}\end{scope}
    \begin{scope}[shift={(0.5,-2.5)}]\hdrcolouredcrossing{red}{\hdrgreen}\end{scope}
    \begin{scope}[shift={(1.5,-2.5)}]\hdrcolouredcrossing{red}{\hdrgreen}\end{scope}
    \begin{scope}[shift={(2.5,-2.5)}]\hdrcolouredboundaryelbow{\hdrgreen}\end{scope}
    \begin{scope}[shift={(0.5,-3.5)}]\hdrcolouredelbow{red}{red}\end{scope}
    \begin{scope}[shift={(1.5,-3.5)}]\hdrcolouredboundaryelbow{red}\end{scope}
    \begin{scope}[shift={(0.5,-4.5)}]\hdrcolouredboundaryelbow{red}\end{scope}
  \end{tikzpicture}}

\newcommand{\hdrplaquette}[3][0.64cm]{%
  \begingroup
  \setlength{\loopcellsize}{#1}%
  \def\linkpatternedgecolor{#2}%
  \tikz[baseline=(current bounding box.center)]{\plaq{#3}}%
  \endgroup}
\newcommand{\hdrblankplaquette}{\hdrplaquette{blue}{}}
\newcommand{\hdrturnone}{\hdrplaquette{blue}{b}}
\newcommand{\hdrturntwo}{\hdrplaquette{blue}{a}}
\newcommand{\hdrcrossingplaquette}{\hdrplaquette{blue}{c}}
\newcommand{\hdrpipedreamturn}{%
  \begingroup
  \tikzset{/linkpattern/edge/.append style={hdr pipe}}%
  \hdrturntwo
  \endgroup}
\newcommand{\hdrpipedreamcrossing}{%
  \begingroup
  \tikzset{/linkpattern/edge/.append style={hdr pipe}}%
  \hdrcrossingplaquette
  \endgroup}

\newcommand{\hdrorientedturnone}{%
  \tikz[x=1cm,y=1cm,baseline=(current bounding box.center)]{
    \useasboundingbox (-0.5,-0.5) rectangle (0.5,0.5);
    \draw[line width=0.65pt,hdr six invarrow=0.18]
      (0,0.5) .. controls (0,0.2) and (-0.2,0) .. (-0.5,0);
    \draw[line width=0.65pt,hdr six arrow=0.82]
      (0,-0.5) .. controls (0,-0.2) and (0.2,0) .. (0.5,0);
  }}
\newcommand{\hdrorientedturntwo}{%
  \tikz[x=1cm,y=1cm,baseline=(current bounding box.center)]{
    \useasboundingbox (-0.5,-0.5) rectangle (0.5,0.5);
    \draw[line width=0.65pt,hdr six arrow=0.82]
      (0,0.5) .. controls (0,0.2) and (-0.2,0) .. (-0.5,0);
    \draw[line width=0.65pt,hdr six arrow=0.82]
      (0,-0.5) .. controls (0,-0.2) and (0.2,0) .. (0.5,0);
  }}

\newif\ifhdrcrossingcell
\newcommand{\hdrnumericlabel}[1]{#1}
\newcommand{\hdrxlabel}[1]{x_{#1}}
\newcommand{\hdrylabel}[1]{y_{#1}}
\newcommand{\hdrpipedreambase}[5]{%
  \begingroup
  \def\linkpatternedgecolor{blue}%
  \tikzset{/linkpattern/edge/.append style={hdr pipe}}%
  \begin{tikzpicture}[x=#1,y=#1,baseline=(current bounding box.center)]
    \draw[thin] (0,0) -- (0,-4) -- (4,0) -- cycle;
    \foreach \i in {1,2,3}{
      \draw[thin] (\i,0) -- (\i,{\i-4});
      \draw[thin] (0,-\i) -- ({4-\i},-\i);
    }
    \foreach \i in {1,...,4}{
      \node[font=\scriptsize] at ({\i-0.5},0.30) {$#4{\i}$};
      \node[font=\scriptsize] at (-0.30,{0.5-\i}) {$#5{\i}$};
    }
    \foreach \r in {1,...,4}{
      \pgfmathtruncatemacro{\hdrlastcolumn}{4-\r}
      \ifnum\hdrlastcolumn>0
        \foreach \c in {1,...,\hdrlastcolumn}{
          \begin{scope}[shift={({\c-0.5},{0.5-\r})}]
            \hdrcrossingcellfalse
            \ifnum#2=2 \hdrcrossingcelltrue\fi
            \pgfmathtruncatemacro{\hdrcellnumber}{10*\r+\c}
            \ifnum#2=3
              \ifnum\hdrcellnumber=11 \hdrcrossingcelltrue\fi
              \ifnum\hdrcellnumber=12 \hdrcrossingcelltrue\fi
              \ifnum\hdrcellnumber=13 \hdrcrossingcelltrue\fi
              \ifnum\hdrcellnumber=31 \hdrcrossingcelltrue\fi
            \fi
            \ifnum#2=4
              \ifnum\hdrcellnumber=11 \hdrcrossingcelltrue\fi
              \ifnum\hdrcellnumber=12 \hdrcrossingcelltrue\fi
              \ifnum\hdrcellnumber=13 \hdrcrossingcelltrue\fi
              \ifnum\hdrcellnumber=22 \hdrcrossingcelltrue\fi
            \fi
            \ifnum#2>0
              \ifhdrcrossingcell\plaqc\else\plaqa\fi
            \fi
          \end{scope}
        }
      \fi
      \pgfmathsetmacro{\hdrdiagonalcolumn}{4.5-\r}
      \begin{scope}[shift={(\hdrdiagonalcolumn,{0.5-\r})}]
        \plaqj
      \end{scope}
    }
  \end{tikzpicture}%
  \endgroup}
\newcommand{\hdrpipedream}[2][0.75cm]{%
  \hdrpipedreambase{#1}{#2}{}{\hdrnumericlabel}{\hdrnumericlabel}}
\newcommand{\hdrpipedreamvariables}[1][0.75cm]{%
  \hdrpipedreambase{#1}{0}{}{\hdrylabel}{\hdrxlabel}}

\newcommand{\hdrpipeendpoints}[1][0.75cm]{%
  \begin{tikzpicture}[x=#1,y=#1,baseline=(current bounding box.center)]
    \draw[thin] (0,0) -- (0,-4) -- (4,0) -- cycle;
    \foreach \i in {1,...,4}{
      \node[/linkpattern/endpoint] at ({\i-0.5},0) {};
      \node[/linkpattern/endpoint] at (0,{0.5-\i}) {};
      \node[font=\scriptsize] at ({\i-0.5},0.30) {$\i'$};
      \node[font=\scriptsize] at (-0.30,{0.5-\i}) {$\i$};
    }
  \end{tikzpicture}}

\newcommand{\hdrnilheckeidentity}[1][0.75cm]{%
  \begin{tikzpicture}[x=#1,y=#1,baseline=(current bounding box.center)]
    \draw[thin] (0,0) rectangle (4,-1);
    \foreach \i in {1,...,4}{
      \node[/linkpattern/endpoint] at ({\i-0.5},0) {};
      \node[/linkpattern/endpoint] at ({\i-0.5},-1) {};
      \node[font=\scriptsize] at ({\i-0.5},0.30) {$\i'$};
      \node[font=\scriptsize] at ({\i-0.5},-1.30) {$\i$};
    }
  \end{tikzpicture}}

\newcommand{\hdrTLgenerator}{%
  \begingroup
  \tikzset{/linkpattern/vertex/.style={draw=none,fill=none,inner sep=0pt}}%
  \begin{tikzpicture}[baseline=(current bounding box.center)]
    \linkpattern[tikzstarted,tangle,axis=false,unit=0.65cm,height=0.42]{1/2,1'/2'}
    \node[font=\scriptsize] at (0.65cm,-0.49cm) {$i$};
    \node[font=\scriptsize] at (1.30cm,-0.49cm) {$i+1$};
  \end{tikzpicture}%
  \endgroup}

\newcommand{\hdrTLactionone}{%
  \begingroup
  \tikzset{/linkpattern/vertex/.style={draw=none,fill=none,inner sep=0pt}}%
  \begin{tikzpicture}[baseline=(current bounding box.center)]
    \linkpattern[tikzstarted,unit=0.78cm]{1/6,2/3,4/5}
    \linkpattern[tikzstarted,tangle,axis=false,unit=0.78cm,height=0.5,
      pos={(0,-0.39cm)}]{1/2,1'/2',3/3',4/4',5/5',6/6'}
    \draw (0.78cm,0) -- (4.68cm,0);
    \draw (0.78cm,-0.78cm) -- (4.68cm,-0.78cm);
    \foreach \i in {1,...,6}{
      \node[/linkpattern/endpoint] at ({0.78cm*\i},-0.78cm) {};
      \node[font=\scriptsize] at ({0.78cm*\i},-1.03cm) {$\i$};
    }
  \end{tikzpicture}%
  \endgroup}

\newcommand{\hdrTLactiontwo}{%
  \begingroup
  \tikzset{/linkpattern/vertex/.style={draw=none,fill=none,inner sep=0pt}}%
  \begin{tikzpicture}[baseline=(current bounding box.center)]
    \linkpattern[tikzstarted,unit=0.78cm]{1/6,2/3,4/5}
    \linkpattern[tikzstarted,tangle,axis=false,unit=0.78cm,height=0.5,
      pos={(0,-0.39cm)}]{2/3,2'/3',1/1',4/4',5/5',6/6'}
    \draw (0.78cm,0) -- (4.68cm,0);
    \draw (0.78cm,-0.78cm) -- (4.68cm,-0.78cm);
    \foreach \i in {1,...,6}{
      \node[/linkpattern/endpoint] at ({0.78cm*\i},-0.78cm) {};
      \node[font=\scriptsize] at ({0.78cm*\i},-1.03cm) {$\i$};
    }
  \end{tikzpicture}%
  \endgroup}

\renewcommand\smallboxes{\setlength{\cellsize}{7pt}\def\boxformat{\scriptstyle}}
\let\emptycell\missingcell
\def\boxformat{}
\long\def\rem#1{}
\newcommand\F{\mathcal{F}}
\title[6v, loops and tilings: integrability, combinatorics]{Six-Vertex, Loop and Tiling Models:\\ Integrability and Combinatorics}
\author[P.~Zinn-Justin]{Paul~Zinn-Justin}
\address{LPTMS (CNRS, UMR 8626), Universit\'e Paris-Sud\\91405 Orsay Cedex, France}
\address{LPTHE (CNRS, UMR 7589), Universit\'e Pierre et Marie Curie\\75252 Paris Cedex, France}
\email{pzinn@lpthe.jussieu.fr}
\hypersetup{
  pdftitle={Six-Vertex, Loop and Tiling Models: Integrability and Combinatorics},
  pdfauthor={Paul Zinn-Justin}
}
\thanks{\textit{Acknowledgements.}
I would like to thank A.~Knutson, J.-M.~Maillet, F.~Smirnov who
accepted to be members of my jury, and especially V.~Pasquier, N.~Reshetikhin
and X.~Viennot who agreed to referee this manuscript. I would also like to
specially thank
J.-B.~Zuber, who besides collaborating with me on numerous occasions,
has regularly advised me in my work and provided guidance during the preparation
of this habilitation.
\par\smallskip
I would like to thank my colleagues at my former laboratory (LPTMS Orsay)
and new one (LPTHE Jussieu), especially their directors S.~Ouvry
and O.~Babelon who welcomed me in their labs and were always open for
discussions.
\par\smallskip
My thanks to all my collaborators, here listed in the order of the number
of co-publications (and alphabetically in case of ties):
P.~Di Francesco, with whom I have had the pleasure to have a most fruitful
collaboration, which is the basis for a good part of the results
in the manuscript; J.~Jacobsen, J.-B.~Zuber; V.~Korepin,
N.~Andrei, J.~de Gier, T.~Fonseca,
V.~Kazakov, A.~Knutson (the only person
with whom I wrote a paper without ever meeting
in person), P.~Pyatov, A.~Razumov, G.~Schaeffer, Yu.~Stroganov.
My interaction with all of them has been influential in my scientific
career.
\par\smallskip
Finally, I would like to thank C.~Le Vaou whose invaluable help with the
administrative issues during all these years at LPTMS made it possible
for me to concentrate on my research; and my family and friends for
their support during the writing process.
\par\smallskip
My work was supported by
European Union Marie Curie Research Training Networks
``ENRAGE'' MRTN-CT-2004-005616, ``ENIGMA'' MRT-CT-2004-5652,
European Science Foundation program ``MISGAM''
and Agence Nationale de la Recherche program ``GIMP'' ANR-05-BLAN-0029-01.}
\numberwithin{equation}{section}
\newtheorem*{conj*}{Theorem}

\begin{document}
\begin{abstract}
This is a review (including some background material) of the author's work and related activity
on certain exactly solvable statistical 
models in two dimensions, including the six-vertex model, loop models and lozenge tilings. Applications to enumerative
combinatorics and to algebraic geometry are described.
\end{abstract}
\maketitle

\tableofcontents
\vfill\eject

\section*{Introduction}
Exactly solvable (integrable) two-dimensional lattice 
statistical models have played an important role in theoretical physics:
starting with Onsager's solution of the Ising model, they have provided non-trivial examples of critical phenomena
in two dimensions and given examples of lattice realizations 
of various known conformal field theories and of their perturbations.
In all these physical applications, one is interested in the thermodynamic limit where the size of the system tends
to infinity and where details of the lattice become irrelevant. In the most basic scenario, one considers the infra-red
limit and recovers this way conformal invariance.

On the other hand, combinatorics is the study of discrete structures in mathematics. Combinatorial
properties of integrable models will thus be uncovered by taking a different point of view on them which considers them
on finite lattices and emphasizes their
discrete properties. The purpose of this text is to show that the same methods and concepts of quantum integrability
lead to non-trivial combinatorial results. The latter may be of intrinsic mathematical interest, in some cases
proving, reproving, extending statements found in the literature. They may also lead back to physics
by taking appropriate scaling limits. 

Let us be more specific on the kind of applications we have in mind. First and foremost comes the connection to 
enumerative combinatorics.
For us the story begins in 1996, when Kuperberg showed how to enumerate alternating sign matrices using Izergin's formula
for the six-vertex model. The observation that alternating sign matrices are nothing but configurations of the six-vertex
model in disguise, paved the way to a fruitful interaction between two subjects which were disjoint until then:
(i) the study of alternating sign matrices, which began in the early eighties after their
definition by Mills, Robbins, and Rumsey in relation to Dodgson condensation, and whose enumerative properties
were studied in the following years, displaying remarkable connections with a much older class of combinatorial
objects, namely plane partitions; and (ii) the study of the six-vertex model, one of the most fundamental
solvable statistical models in two dimensions, which was undertaken in the sixties and has remained at the center
of the activity around quantum integrable models ever since.

One of the most noteworthy recent chapters in this continuing story is the Razumov--Stroganov conjecture, in 2001,
which emerged out of a collective effort by combinatorialists and physicists to understand the connection between
the aforementioned objects and another class of statistical models, namely loop models. The work of the author
was mostly a byproduct of various attempts to understand (and possibly prove) this conjecture. A large part of this
manuscript is dedicated to reviewing these questions.

Another interesting, related application is to algebraic combinatorics, due to the appearance of certain families of polynomials
in quantum integrable models. In the context of the Razumov--Stroganov conjecture, they were introduced by Di Francesco and Zinn-Justin
in 2004, but their true meaning was only clarified subsequently by Pasquier, creating a connection to representation theory
of affine Hecke algebras and to previously studied classes of polynomials such as Macdonald polynomials. These polynomials
satisfy relations which are typically studied in algebraic combinatorics, e.g.,\ involving divided difference operators.
The use of specific bases of spaces of polynomials, which is necessary for a combinatorial interpretation, connects
to the theory of canonical bases and the work of Kazhdan and Lusztig.

Finally, an exciting and fairly new aspect in this study of integrable models is to try to find an algebro-geometric interpretation
of some of the objects and of the relations that they satisfy. The most naive version of it would be to relate the integer numbers
that appear in our models to problems in enumerative geometry, so that they become intersection numbers for certain algebraic
varieties. A more sophisticated version involves equivariant cohomology or K-theory, which typically leads to polynomials
instead of integers. The connection between integrable models and certain classes of polynomials with geometric meaning
is not entirely new, and the work that will be described here bears some resemblance, as will be reminded here, to
that of Fomin and Kirillov on Schubert and Grothendieck polynomials. However there are also novelties, including 
the use of the multidegree technology of Knutson et al, and we apply these ideas to a broad class of models,
resulting in new formulas for known algebraic varieties such as orbital varieties and the commuting variety,
as well as in the discovery of new geometric objects, such as the Brauer loop scheme.

The presentation that follows, though based on the articles of the author, is meant to be essentially self-contained. It is aimed at researchers and graduate students in mathematical physics or in combinatorics
with an interest in exactly solvable statistical models.
For simplicity, the integrable models that are defined are based on the underlying affine quantum group
$U_q(\widehat{\mathfrak{sl}(2)})$, with the notable exception of the discussion of the Brauer loop model
in the last section.
Furthermore, only the spin $1/2$ representation and periodic boundary conditions are considered.
There are interesting generalizations to higher rank, higher spin and to other boundary conditions
of some of these results, on which the author has worked, 
but for these the reader is referred to the literature.

The plan of this manuscript is the following. In section \ref{secff}, we discuss free fermionic methods.
Though free fermions in two dimensions may seem like an excessively simple physical model, they already provide a wealth
of combinatorial formulae. In fact they have become extremely popular in the recent mathematical literature. 
We shall apply the basic formalism of free fermions to introduce Schur functions, and then spend some time
reviewing the properties of the latter, because they will reappear many times in our discussion. 
We shall then briefly discuss the application to the enumeration of
plane partitions. Section \ref{sec6v} covers the six-vertex model, and in particular the six-vertex model with
domain-wall boundary conditions. We shall discuss its quantum integrability, which is the root of its
exact solvability. Then we shall apply it to the enumeration of alternating sign matrices.
In section \ref{seclooprs}, we shall discuss statistical models of loops, their interrelations with the six-vertex model,
their combinatorial properties and formulate the Razumov--Stroganov conjecture. Section \ref{secqkz} introduces the
quantum Knizhnik--Zamolodchikov equation, which will be used to reconnect some of the objects discussed previously.
The last section, \ref{secgeom}, will be devoted to a brief review of the current status on the geometric reinterpretation
of some of the concepts above, focusing on the central role of the quantum Knizhnik--Zamolodchikov equation.

\section{Free fermionic methods}\label{secff}
As mentioned above, we want to spend some time defining a typical free fermionic model and
to apply it to rederive some useful formulae for Schur functions, which will be needed later.
We shall also need some formulae concerning the enumeration of plane partitions, which will appear
at the end of this section.

\subsection{Definitions}
\subsubsection{Operators and Fock space}
Consider a fermionic operator $\psi(z)$:
\begin{equation}\label{defpsi}
\psi(z)=\sum_{k\in \mathbb{Z}+\oh} \psi_{k} z^{-k-\oh},\qquad
\psis(z)=\sum_{k\in \mathbb{Z}+\oh} \psis_{-k} z^{-k-\oh}
\end{equation}
with anti-commutation relations
\begin{equation}
[\psis_r,\psi_s]_+=\delta_{rs}
\qquad
[\psi_r,\psi_s]_+=[\psis_r,\psis_s]_+=0
\end{equation}
$\psi(z)$ and $\psis(z)$
should be thought of as generating series for the $\psi_k$
and $\psis_k$, so that $z$ is just a formal variable.
What we have here is 
a complex (charged) fermion, with particles, and anti-particles
which can be identified with holes in the Dirac sea.
These fermions are one-dimensional, in the sense that their states
are indexed by (half-odd-)integers; $\psis_k$ creates a particle
(or destroys a hole) at location $k$, whereas $\psi_k$ destroys a particle
(creates a hole) at location $k$.

We shall explicitly build the Fock space $\F$ and the representation
of the fermionic operators now. Start from a vacuum $\ket{0}$
which satisfies
\begin{equation}
\psi_k\ket{0}=0 \quad k>0,
\qquad
\psis_k\ket{0}=0\quad k<0
\end{equation}
that is, it is a Dirac sea filled up to location 0:
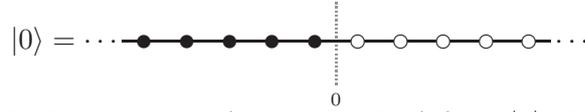
\[
\ket{0}=\cdots
\vcenterbox{\hdrdiracsea{0}}
\cdots\]

Then any state can be built by action of the $\psi_k$ and $\psis_k$
from $\ket{0}$. In particular one can define more general vacua at
level $\ell\in\mathbb{Z}$:
\begin{equation}
\ket{\ell}=\begin{cases}
\psis_{\ell-\oh}\psis_{\ell-\frac{3}{2}}\cdots \psis_\oh \ket{0}&\ell>0\\
\psi_{\ell+\oh}\psi_{\ell+\frac{3}{2}}\cdots \psi_{-\oh} \ket{0}&\ell<0
\end{cases}
=\cdots
\vcenterbox{\hdrdiracsea{\ell}}
\cdots
\end{equation}
which will be useful in what follows. They satisfy
\begin{equation}
\psi_k\ket{\ell}=0 \quad k>\ell,
\qquad
\psis_k\ket{\ell}=0 \quad k<\ell
\end{equation}
More generally, define a {\em partition}\/ to be a weakly decreasing finite sequence of non-negative integers:
$\lambda_1\ge\lambda_2\ge\cdots\ge\lambda_n\ge 0$. We usually represent partitions as
{\em Young diagrams} (also called Ferrers diagrams): 
for example $\lambda=(5,2,1,1)$ is depicted as
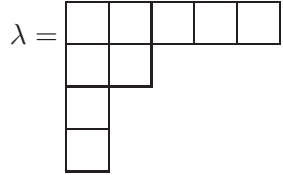
\[
\lambda=\tableau{&&&&\\&\\\\\\}
\]
To each partition $\lambda=(\lambda_1,\ldots,\lambda_n)$ we associate
the following state in $\F_\ell$:
\begin{equation}\label{defstate}
\ket{\lambda;\ell}=\psis_{\ell+\lambda_1-\oh}\psis_{\ell+\lambda_2-\frac{3}{2}}\cdots\psis_{\ell+\lambda_n-n+\oh}\ket{\ell-n}
\end{equation}
Note the important property that if one ``pads'' a partition with extra zeroes, then the corresponding state
remains unchanged.
In particular for the empty diagram $\varnothing$, $\ket{\varnothing;\ell}=\ket{\ell}$.
For $\ell=0$ we just write $\ket{\lambda;0}=\ket{\lambda}$.

This definition has the following nice graphical interpretation: the state $\ket{\lambda;\ell}$ can be described
by numbering the edges of the boundary of the Young diagram, in such a way that the main diagonal passes between
$\ell-\oh$ and $\ell+\oh$; then the occupied (resp.\ empty) sites correspond to vertical (resp.\ horizontal) edges.
With the example above and $\ell=0$, we find (only the occupied sites are numbered for clarity)
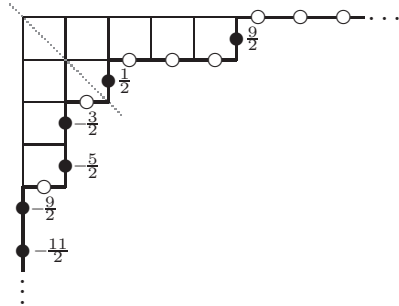
\[
\begin{tikzpicture}[x=\cellsize,y=\cellsize,baseline=(current bounding box.south)]
\node[tableau] at (0,0)
 {&&&&&\missingcell \scriptscriptstyle\!\!\!\frac{9}{2}
   &\missingcell&\missingcell&\hdotscell\\
  &&\missingcell\scriptscriptstyle\!\!\!\frac{1}{2}\\
  &\missingcell\scriptscriptstyle-\!\frac{3}{2}\\
  &\missingcell\scriptscriptstyle-\!\frac{5}{2}\\
  \missingcell {\scriptscriptstyle-\!\frac{9}{2}}\\
  \missingcell {\scriptscriptstyle\,\,\,-\!\frac{11}{2}}\\
  \vdotscell\\};
\draw[gray,dotted] (-0.3,1.3) -- (2.3,-1.3);
\draw[thick] (0,-5) -- (0,-3) -- (1,-3) -- (1,-1) --
 (2,-1) -- (2,0) -- (5,0) -- (5,1) -- (8,1);
\foreach \p in {(0,-4.5),(0,-3.5),(1,-2.5),(1,-1.5),(2,-0.5),(5,0.5)}
 \fill \p circle[radius=0.15];
\foreach \p in {(7.5,1),(6.5,1),(5.5,1),(4.5,0),(3.5,0),(2.5,0),(1.5,-1),(0.5,-3)}
 \draw[fill=white] \p circle[radius=0.15];
\end{tikzpicture}
\]

The $\ket{\lambda;\ell}$, where $\lambda$ runs over all possible partitions (two partitions
being identified if they are obtained from each other by adding or removing zero parts), form an orthonormal 
basis of a subspace of $\F$ which we denote by $\F_\ell$. $\psi_k$ and $\psis_k$ are Hermitean conjugate of each other.

Note that \eqref{defstate} fixes our sign convention of the states. In particular, this implies
that when one acts with $\psi_k$ (resp.\ $\psis_k$) on a state $\ket{\lambda}$ with a particle
(resp.\ a hole) at $k$, one produces a new state $\ket{\lambda'}$ with the particle
removed (resp.\ added) at $k$ times $-1$ to the power the number of particles to the right of $k$.

The states $\lambda$ can also be produced from the vacuum by acting with $\psi$ to create holes;
paying attention to the sign issue, we find
\begin{equation}\label{defstatedual}
\ket{\lambda;\ell}=(-1)^{|\lambda|}\psi_{\ell-\lambda'_1+\oh}\cdots\psi_{\ell-\lambda'_m+m-\oh}\ket{\ell+m}
\end{equation}
where the $\lambda'_i$ are the lengths of the columns of $\lambda$,
$|\lambda|$ is the number of boxes of $\lambda$
and $m=\lambda_1$. This formula is formally identical to \eqref{defstate} if we renumber the
states from right to left, exchange $\psi$ and $\psis$, and replace $\lambda$ with its {\em transpose} diagram
$\lambda'$ (this property is graphically clear).
So the particle--hole duality translates into transposition of Young diagrams.

Finally, introduce the normal ordering with respect to the vacuum $\ket{0}$:
\begin{equation}
\normord{\psis_j \psi_k}=-\normord{\psi_k \psis_j}=
\begin{cases}
\psis_j \psi_k&j>0\\
-\psi_k \psis_j&j<0
\end{cases}
\end{equation}
which allows us to get rid of trivial infinite quantities.

\subsubsection{\texorpdfstring{$\mathfrak{gl}(\infty)$ and $\widehat{\mathfrak{gl}(1)}$}{gl(infinity) and affine gl(1)} action}\label{secglact}
The operators $\psis(z)\psi(w)$ give rise
to the Schwinger representation
of $\mathfrak{gl}(\infty)$ on $\F$, whose usual basis is 
the $\normord{\psis_r \psi_s}$, $r,s\in\mathbb{Z}+\oh$, and the identity. 
In the first quantized picture this representation is simply the natural
action of $\mathfrak{gl}(\infty)$ on the one-particle Hilbert space $\mathbb{C}^{\mathbb{Z}+\oh}$ and
exterior products thereof.
The electric charge $J_0=\sum_r \normord{\psis_r \psi_r}$
is a conserved number and classifies the irreducible representations of 
$\mathfrak{gl}(\infty)$ inside $\F$. The
highest weight vectors are precisely our vacua $\ket{\ell}$, 
$\ell\in\mathbb{Z}$, so that $\F=\oplus_{\ell\in\mathbb{Z}} \F_\ell$ with $\F_\ell$ the subspace
in which $J_0=\ell$.

The $\mathfrak{gl}(1)$ current
\begin{equation}
j(z)=\normord{\psis(z)\psi(z)}=\sum_{n\in\mathbb{Z}} J_n z^{-n-1}
\end{equation}
with $J_n=\sum_r \normord{\psis_{r-n} \psi_r}$ forms a representation of the $\widehat{\mathfrak{gl}(1)}$ (Heisenberg)
sub-algebra of $\mathfrak{gl}(\infty)$:
\begin{equation}\label{heis}
[J_m,J_n]=m \delta_{m,-n}
\end{equation}
The central-extension generator of $\widehat{\mathfrak{gl}(1)}$ is represented by the identity in \eqref{heis}.
The $\F_\ell$ are isomorphic as representations of the oscillator Heisenberg subalgebra generated by
the $J_n$ with $n\ne0$ and the identity, but not when the zero mode $J_0$ is included.

Note that positive modes commute among themselves. This allows us to define the general ``Hamiltonian''
\begin{equation}
H[t]=\sum_{q=1}^\infty t_q J_q
\end{equation}
where $t=(t_1,\ldots,t_q,\ldots)$ is a set of parameters (``times'').

The $J_q$, $q>0$, displace one of the fermions
$q$ steps to the left, or one of the holes $q$ steps to the right.
This is expressed by
the formulae describing the time evolution of the fermionic fields:
\begin{equation}\label{commrel}
\begin{split}
\e{H[t]}\psi(z)\e{-H[t]}
&=\e{-\sum_{q=1}^\infty t_q z^q}\psi(z)\\
\e{H[t]}\psis(z)\e{-H[t]}
&=\e{+\sum_{q=1}^\infty t_q z^q}\psis(z)
\end{split}
\end{equation}
(proof: compute $[J_q,\psi^{[\star]}(z)]=\pm z^q \psi^{[\star]}(z)$ and exponentiate). Of course, similarly,
$J_{-q}$, $q>0$, moves one fermion $q$ steps to the right, or one hole
$q$ steps to the left.
(in what follows, we shall reverse the natural convention and read
expressions from left to right, in which case one should say that
acting on bras, the $J_q$ displace fermions $q$ steps to the right
and holes $q$ steps to the left, and vice versa for $J_{-q}$).

\subsubsection{Bosonization}
The current $j(z)$ can also be thought of as the derivative of a bosonic
field: $j(z)=\frac{\der}{\der z}\varphi(z)$ where 
\[
\varphi(z)=-\frac{\der}{\der J_0}+J_0\log z-\sum_{n\ne 0}\frac{1}{n} J_n z^{-n}
\]
$\der/\der J_0$ is the formal conjugate variable of $J_0$;
only its exponential makes sense as an operator on $\F$, acting as
$\exp \pm\der/\der J_0 \ket{\lambda;\ell}=\ket{\lambda;\ell\mp 1}$.
This means that we can interpret each $\F_\ell$ as a bosonic Fock space
(with no action of the zero mode); in the correspondence with symmetric
functions to be introduced right below, the fermionic basis
of section~\ref{secglact} is the basis of Schur functions, whereas the
bosonic basis obtained by acting with products of 
$J_{-q}$ on $\ket{0}$
(or the $J_q$ on $\bra{0}$) is the basis of (products of) power sums.

Furthermore, one can show that the fermionic fields themselves can be
recovered from the bosonic field $\varphi(z)$:
\[
\psi(z)=\normord{e^{-\varphi(z)}}\qquad
\psi^\star(z)=\normord{e^{\varphi(z)}}
\]
\goodbreak
\subsection{Schur functions}\label{ffschur}
\subsubsection{Free fermionic definition}
It is known that the map $\ket{\Phi}\mapsto \bra{\ell}\e{H[t]}\ket{\Phi}$ is an isomorphism from $\F_\ell$ to
the space of polynomials in an infinite number of variables $t_1,t_2,\ldots$. Thus, we obtain a basis of the latter
as follows:
for a given Young diagram $\lambda$, define the {\em Schur function} $s_\lambda[t]$ by
\begin{equation}\label{defschur}
s_\lambda[t]=\bra{\ell}\e{H[t]}\ket{\lambda;\ell}
\end{equation}
(by translational invariance it is in fact independent of $\ell$).
In the language of Schur functions, the $t_q$ are (up to a conventional factor $1/q$) the {\em power sums},
see section \ref{fqformula} below.

We provide here various expressions of $s_\lambda[t]$ using the free fermionic formalism.
In fact, many of the methods used
are equally applicable to the following more general quantity:
\begin{equation}\label{defskewschur}
s_{\lambda/\mu}[t]=\bra{\mu;\ell}\e{H[t]}\ket{\lambda;\ell}
\end{equation}
where $\lambda$ and $\mu$ are two partitions. It is easy to see that in order for $s_{\lambda/\mu}[t]$
to be non-zero, $\mu\subset\lambda$ as Young diagrams; in this case $s_{\lambda/\mu}$ is known
as the {\em skew Schur function} associated to the skew Young diagram $\lambda/\mu$. The latter
is depicted as the complement of $\mu$ inside $\lambda$. This is appropriate because
skew Schur functions factorize in terms of the connected components of the skew Young diagram $\lambda/\mu$.

\goodbreak{\em Examples:} $s_{\smallboxes\tableau{\\}}=t_1$,
$s_{\smallboxes\tableau{\\\\}}=\frac{1}{2}t_1^2-t_2$,
$s_{\smallboxes\tableau{&\\}}=\frac{1}{2}t_1^2+t_2$,
$s_{\smallboxes\tableau{&\\\\}}=\frac{1}{3}t_1^3-t_3$.

$s_{\smallboxes\tableau{\emptycell&\\\\}}=s_{\smallboxes\tableau{\\}}^2=t_1^2$,
$s_{\smallboxes\tableau{\emptycell&&\\&\\}}=\frac{5}{24}t_1^4+\frac{1}{2}t_1^2t_2+\frac{1}{2}t_2^2-t_1t_3-t_4$.

\subsubsection{Wick theorem and Jacobi--Trudi identity}
First, we apply the Wick theorem. Consider 
as the definition of the time evolution of fermionic fields:
\begin{equation}
\begin{split}
\psi_k[t]&=\e{H[t]}\psi_k\e{-H[t]}\\
\psis_k[t]&=\e{H[t]}\psis_k\e{-H[t]}\\
\end{split}
\end{equation}
In fact, \eqref{commrel} gives us the ``solution'' of the equations of motion
in terms of the generating series $\psi(z)$, $\psis(z)$.

Noting that the Hamiltonian is quadratic in the fields, we now state
the Wick theorem:
\begin{equation}
\bra{\ell}\psi_{i_1}[0]\cdots\psi_{i_n}[0]
\psis_{j_1}[t]\ldots\psis_{j_n}[t]\ket{\ell}
=
\det_{1\le p,q\le n} \bra{\ell}\psi_{i_p}[0]\psis_{j_q}[t]\ket{\ell}
\end{equation}

Next, start from the expression \eqref{defskewschur} of $s_{\lambda/\mu}[t]$: padding with zeroes $\lambda$ or $\mu$ so that
they have the same number of parts $n$, we can write
\[
s_{\lambda/\mu}[t]=\bra{-n}
\psi_{\mu_n-n+\oh}\cdots\psi_{\mu_1-\oh}
\e{H[t]}
\psis_{\lambda_1-\oh}\cdots\psis_{\lambda_n-n+\oh}
\ket{-n}
\]
and apply the Wick theorem to find:
\[
s_{\lambda/\mu}[t]=
\det_{1\le p,q\le n} \bra{-n}\psi_{\mu_p-p+\oh}\e{H[t]}\psis_{\lambda_q-q+\oh}\ket{-n}
\]
It is easy to see that $\bra{-n}\psi_i \e{H[t]} \psis_j\ket{-n}$
does not depend on $n$ and thus only depends on $j-i$. Let us denote it
\begin{equation}\label{gencomp}
h_k[t]=\bra{1}\e{H[t]} \psis_{k+\oh}\ket{0}
\qquad
\sum_{k\ge 0} h_k[t] z^k = \bra{1}\e{H[t]} \psis(z)\ket{0} = \e{\sum_{q\ge 1} t_q z^q}
\end{equation}
($k=j-i$; note that $h_k[t]=0$ for $k<0$).

The final formula we obtain is
\begin{equation}\label{skewgiamb}
s_{\lambda/\mu}[t]=
\det_{1\le p,q\le n} \left(h_{\lambda_q-\mu_p-q+p}[t]\right)
\end{equation}
or, for regular Schur functions,
\begin{equation}\label{giamb}
s_{\lambda}[t]=
\det_{1\le p,q\le n} \left(h_{\lambda_q-q+p}[t]\right)
\end{equation}
This is known as the Jacobi--Trudi identity.

By using ``particle--hole duality'', we can find a dual form of this identity.
We describe our states in terms of hole positions, parametrized by the lengths of the columns $\lambda'_p$ and
$\mu'_q$, according to \eqref{defstatedual}:
\[
s_{\lambda/\mu}[t]=(-1)^{|\lambda|+|\mu|}\bra{m}\psis_{-\mu'_m+m-\oh}\cdots\psis_{-\mu'_1+\oh}
\e{H[t]}
\psi_{-\lambda'_1+\oh}\cdots\psi_{-\lambda'_m+m-\oh}\ket{m}
\]
Again the Wick theorem applies. It expresses $s_{\lambda/\mu}$ in terms of
$\bra{m}\psis_i\e{H[t]}\psi_j\ket{m}$, a two-point function that only depends
on $i-j=k$ and is given by
\begin{equation}\label{genele}
e_k[t]=(-1)^k\bra{-1}\e{H[t]}\psi_{-k+\oh}\ket{0}
\qquad
\sum_{k\ge 0} e_k[t] z^k 
=\bra{-1}\e{H[t]}\psi(-z)\ket{0}
=\e{\sum_{q\ge 1}(-1)^{q-1} t_q z^q}
\end{equation}

The final formula takes the form
\begin{equation}\label{skewdualgiamb}
s_{\lambda/\mu}[t]=
\det_{1\le p,q\le n} \left(e_{\lambda'_q-\mu'_p-q+p}[t]\right)
\end{equation}
or, for regular Schur functions,
\begin{equation}
s_{\lambda}[t]=\label{dualgiamb}
\det_{1\le p,q\le n} \left( e_{\lambda'_q-q+p}[t]\right)
\end{equation}
This is the dual Jacobi--Trudi identity, also known as Von N\"agelsbach--Kostka identity.

\subsubsection{Weyl formula}\label{fqformula}
In the following sections \ref{fqformula}--\ref{secnilp},
we shall fix an integer $n$ and consider the following change of variable (this is essentially the
Miwa transformation \cite{Miwa-transf}) $t_q=\frac{1}{q}\sum_{i=1}^n x_i^q$. The Schur function becomes
a symmetric polynomial of these variables $x_i$, which we denote by $s_\lambda(x_1,\ldots,x_n)$, and we now
derive a different (first quantized) formula for it.

Due to obvious translational invariance of all the operators involved, we may as well set $\ell=n$.
Use the definition \eqref{defstate} of $\ket{\lambda}$ and the commutation relations \eqref{commrel}
to rewrite the left hand side as
\[
\bra{n}\e{H[t]}\ket{\lambda;n}=\e{\sum_{q\ge1} t_q\sum_{i=1}^n z_i^q} \bra{n}\psis(z_1)\psis(z_2)\cdots\psis(z_n)\ket{0}
\big|_{z_1^{n+\lambda_1-1}z_2^{n+\lambda_2-2}\ldots z_n^{\lambda_n}}
\]
where $\big|_{\ldots}$ means picking one term in a generating series.

We can easily evaluate the remaining bra-ket to be: (we now use the $\ell=0$ notation for the l.h.s.)
\[
\bra{0}\e{H[t]}\ket{\lambda}=\e{\sum_{q\ge1} t_q\sum_{i=1}^n z_i^q} 
\prod_{1\le i<j\le n} (z_i-z_j)
\big|_{z_1^{n+\lambda_1-1}z_2^{n+\lambda_2-2}\ldots z_n^{\lambda_n}}
\]
Now write $t_q=\frac{1}{q}\sum_{j=1}^n x_j^q$ and note that $\e{\sum_{q\ge1} t_q \sum_{i=1}^n z_i^q}=
\prod_{i,j=1}^n (1-z_i x_j)^{-1}$. We recognize (part of) the Cauchy determinant:
\[
\bra{0}\e{H[t]}\ket{\lambda}=
\frac{\det_{1\le i,j\le n}(1-x_i z_j)^{-1}}{\prod_{i<j}(x_i-x_j)}
\big|_{z_1^{n+\lambda_1-1}z_2^{n+\lambda_2-2}\cdots z_n^{\lambda_n}}
\]
At this stage we can just expand separately each column of the matrix $(1-x_iz_j)^{-1}$ to pick the right power of $z_j$;
we find:
\begin{equation}\label{defweyl}
\bra{0}\e{H[t]}\ket{\lambda}=
\frac{\det_{1\le i,j\le n}(x_i^{\lambda_j+n-j})}{\prod_{i<j}(x_i-x_j)}
\end{equation}

{\em Remark 1:} defined in terms of a fixed number $n$ of variables, as in \eqref{defweyl},
$s_\lambda(x_1,\ldots,x_n)$ has the following group-theoretic interpretation.
The polynomial irreducible representations of $GL(n)$ are known to be indexed by partitions with at most $n$ rows.
Then $s_\lambda(x_1,\ldots,x_n)$ is the character of representation $\lambda$ evaluated at the diagonal
matrix $\mathrm{diag}(x_1,\ldots,x_n)$. Hence,
the dimension of $\lambda$ as a $GL(n)$ representation is given by
$s_\lambda(\underbrace{1,\ldots,1}_n)=\prod_{1\le i<j\le n}(\lambda_i-i-\lambda_j+j)/(j-i)$. 

{\em Remark 2:} the more general Miwa transformation allows for coefficients:
$t_q=\frac{1}{q}\sum_{i=1}^n \alpha_i x_i^q$. In particular
if we use minus signs, we get the notion of {\em plethystic negation}.  
Combining it with the usual negation
of variables is equivalent to transposing Young diagrams: indeed it amounts to exchanging the $h_k[t]$ and the
$e_k[t]$. In other words,
\[
s_\lambda[t]=s_{\lambda'}[-\epsilon t]\qquad -\epsilon t_q := (-1)^{q-1} t_q
\]
More generally, one defines the {\em supersymmetric Schur function} $s_\lambda(x_1,\ldots,x_n/y_1,\ldots,y_m)$
to be equal to $s_\lambda[t]$ where $t_q=\frac{1}{q}(\sum_{i=1}^n x_i^q - \sum_{i=1}^m (-y_i)^q)$.

\subsubsection{Schur functions and lattice fermions}\label{secschurlat}
Note that the change of variables $t_q=\frac{1}{q}\sum_{j=1}^n x_j^q$ allows us to write
\[
\e{H[t]}=
\prod_{i=1}^n \e{\varphi_+(x_i^{-1})}\qquad \varphi_+(z)=\sum_{q\ge1}\frac{z^{-q}}{q} J_q
\]
So we can think of the ``time evolution'' as a series of discrete steps represented
by commuting operators $\exp\varphi_+(x_i^{-1})$. In the language of statistical mechanics,
these are transfer matrices (and the existence of a
one-parameter family of commuting transfer matrices $\exp\varphi_+(z)$
is of course related to the integrability of the model).
We now show that they have a very simple meaning in terms of lattice fermions.

\begin{figure}
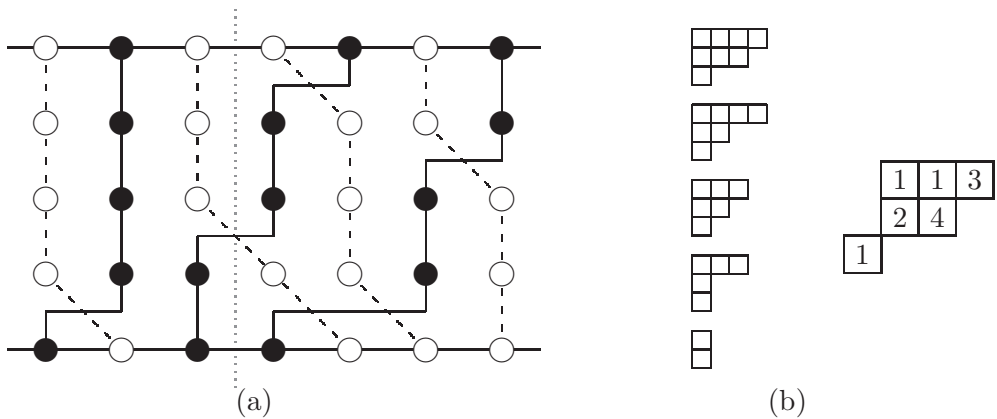

\hdrlatticefermionfigure
\caption{A lattice fermion configuration and the corresponding (skew) SSYT.}
\label{figlatferm}
\end{figure}

Consider a two-dimensional square lattice, one direction being our space $\mathbb{Z}+\oh$
and one direction being time. In what follows we shall read
expressions from left to right (i.e.,\ think of operators as acting successively
on bras) and draw pictures from bottom to top (i.e.,\ time flowing upwards).
The rule to go from one step to the next according
to the evolution operator $\exp\varphi_+(z)$
can be formulated either in terms of particles or in terms of holes: 
\begin{itemize}
\item Each particle can go straight or hop to the right as long as it does not reach the (original) location of the next particle.
Each step to the right is given a weight of $x$.
\item Each hole can only go straight or one step to the left as long as it does not bump into its neighbor. 
Each step to the left is given a weight
of $x$.
\end{itemize}
Obviously the second description is simpler.
An example of a possible evolution of the system with given initial and final states is shown
on Fig.~\ref{figlatferm}.

The proof of these rules consists of computing explicitly $\bra{\mu}\e{\varphi_+(x^{-1})}\ket{\lambda}$
by applying say \eqref{skewdualgiamb} for $t_q=\frac{1}{q}x^q$, and noting
that in this case, according to \eqref{genele}, $e_n[t]=0$ for $n>1$.
This strongly constrains the possible transitions and produces the description above.

\subsubsection{Relation to semistandard Young tableaux}
A semistandard
Young tableau (SSYT) of shape $\lambda$ is a filling of the Young diagram of $\lambda$
with elements of some ordered alphabet, in such a way that rows are weakly increasing and columns are strictly
increasing.

We shall use here the alphabet $\{1,2,\ldots,n\}$. For example with $\lambda=(5,2,1,1)$
one possible SSYT with $n\ge5$ is:
\[
\tableau{1&2&4&5&5\\3&3\\4\\5\\}
\]

It is useful to think of Young tableaux as time-dependent Young diagrams where the number
indicates the step at which a given box was created. Thus, with the same example, we get
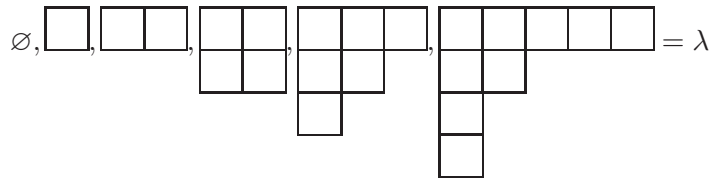
\[
\varnothing, \tableau{\\}, \tableau{&\\}, \tableau{&\\&\\}, \tableau{&&\\&\\\\},
\tableau{&&&&\\&\\\\\\}=\lambda
\]

So a Young tableau is nothing but a statistical configuration of our lattice fermions,
where the initial state is the vacuum. Similarly, a skew SSYT is a filling of a skew Young diagram
with the same rules; it corresponds to a statistical configuration of lattice fermions with arbitrary
initial and final states.
The correspondence is exemplified on Fig.~\ref{figlatferm}.
Equivalently, trajectories of particles can be read row by row in the SSYT,
while trajectories of holes are read column by column, the numbers
corresponding to time steps of horizontal moves.

Each extra box corresponds to a step to the right for particles or to the left for holes.
The initial and final states are $\varnothing$ and $\lambda$, which is the case for Schur functions,
cf.\ \eqref{defschur}.
We conclude that the following formula holds:
\begin{equation}\label{defschurb}
s_\lambda(x_1,\ldots,x_n)=\sum_{T\in \mathrm{SSYT}(\lambda,n)} \prod_{b\ \text{box of }T} x_{T_b}
\end{equation}
This is often taken as a definition of Schur functions. 
It is explicitly stable with respect to $n$ in the sense that
$s_\lambda(x_1,\ldots,x_n,0,\ldots,0)=s_\lambda(x_1,\ldots,x_n)$. It is however not obvious 
from it that $s_\lambda$ is symmetric by permutation of its variables. This fact
is a manifestation of the underlying free fermionic (``integrable'') behavior.
Of course an identical formula holds for the more general case of skew Schur functions.

\subsubsection{Nonintersecting lattice paths and the Lindstr\"om--Gessel--Viennot formula}\label{secnilp}
\begin{figure}
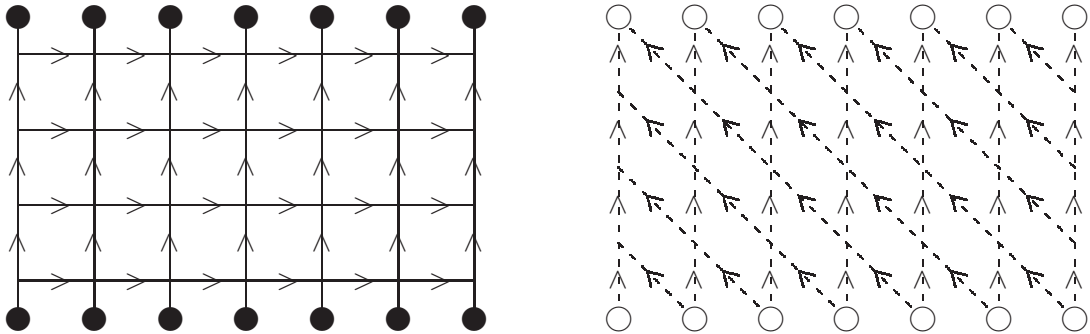

\hbox{\hdrnilpparticlegraph\qquad\hdrnilpholegraph}
\caption{Underlying directed graphs for particles and holes.}
\label{NILPgraph}
\end{figure}

The rules of evolution given in section \ref{secschurlat} strongly suggest
the following explicit description of the lattice fermion configurations.
Consider the directed graphs of Fig.~\ref{NILPgraph} (the graphs are in principle infinite to the
left and right, but any given bra-ket evaluation only involves a finite number of particles and holes and
therefore the graphs can be truncated to a finite part). Consider {\em nonintersecting lattice paths}
(NILPs) on these graphs: they are paths with given starting points (at the bottom) and given ending points 
(at the top), which follow the edges of the graph respecting the orientation of the arrows, 
and which are not allowed to touch at any vertices. One can check that the trajectories of holes and
particles following the rules described in section \ref{secschurlat} are exactly the most general NILPs
on these graphs.

In this context, the Jacobi--Trudi identity \eqref{giamb} becomes a consequence of the so-called 
Lindstr\"om--Gessel--Viennot formula \cite{Lind,GV}. This formula expresses, as a determinant,
the weighted sum of NILPs on a general directed acyclic graph from starting locations
$i_1,\ldots,i_n$ to ending locations $j_1,\ldots,j_n$. The weight of a path is the product
of the weights of its edges:
\begin{equation}\label{LGV}
N(i_1,\ldots,i_n;j_1,\ldots,j_n)=
\det_{p,q} N(i_p;j_q)
\end{equation}
More precisely, in Lindstr\"om's formula, sets of NILPs such that the path starting from $i_k$ ends at $j_{w(k)}$ get
an extra sign which is that of the permutation $w$.
This is nothing but the Wick theorem once again (but with fermions
living on a general graph), and from this point of view is a simple exercise in Grassmannian Gaussian integrals.
In the special case of a planar graph with appropriate starting points
(no paths are possible between them) and ending points, only one permutation, say the identity up to relabelling, contributes.

In order to use this formula, one only needs to compute $N(i;j)$, the weighted sum of paths from $i$ to $j$. 
Let us do so in our problem.

In the case of particles (left graph), numbering the initial and final points from left to right,
we find that the weighted sum of paths from $i$ to $j$, where a weight $x_k$ is given to each right move
at time-step $k$, only depends on $j-i$; if we denote it by $h_{j-i}(x_1,\ldots,x_n)$, we have the obvious
generating series formula
\[
\sum_{k\ge 0} h_k(x_1,\ldots,x_n) z^k=\prod_{i=1}^n \frac{1}{1-z x_i}
\]
Note that this formula coincides with the alternate definition \eqref{gencomp} of $h_k[t]$ if we set as usual
$t_q=\frac{1}{q}\sum_{i=1}^n x_i^q$. Thus, applying the LGV formula \eqref{LGV} and choosing the correct
initial and final points for Schur functions or skew Schur functions, we recover immediately
(\ref{skewgiamb},\ref{giamb}).

In the case of holes (right graph), numbering the initial and final points from right to left,
we find once again that the weighted sum of paths from $i$ to $j$, 
where a weight $x_k$ is given to each left move
at time-step $k$, only depends on $j-i$; if we denote it by $e_{j-i}(x_1,\ldots,x_n)$, we have the
equally obvious generating series formula
\[
\sum_{k\ge 0} e_k(x_1,\ldots,x_n) z^k=\prod_{i=1}^n (1+z x_i)
\]
which coincides with \eqref{genele}, thus allowing us to recover (\ref{skewdualgiamb},\ref{dualgiamb}).

\subsubsection{Relation to standard Young tableaux}\label{secSYT}
A standard Young tableau (SYT) of shape $\lambda$ is a filling of the Young diagram of $\lambda$
with elements of some ordered alphabet, in such a way that both rows and columns are strictly
increasing. There is no loss of generality in assuming that the alphabet is $\{1,\ldots,n\}$,
where $n=|\lambda|$ is the number of boxes of $\lambda$. For example, 
\[
\tableau{1&2&6&8&9\\3&4\\5\\7\\}
\]
is a SYT of shape $(5,2,1,1)$.

\begin{figure}
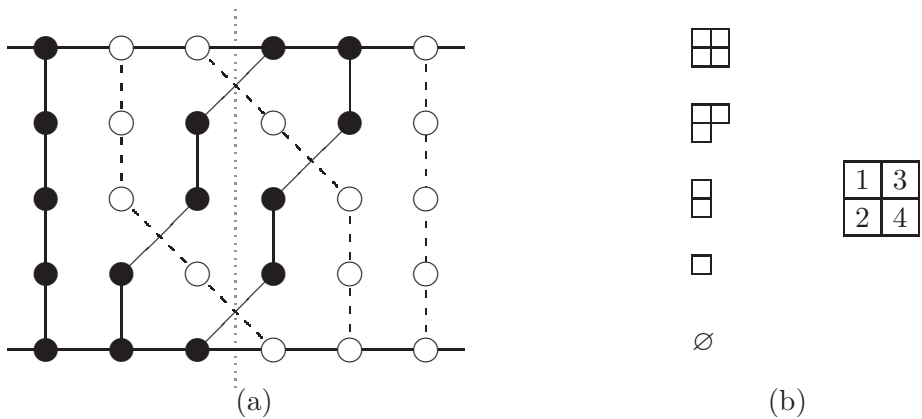

\hdrlatticefermionstandardfigure
\caption{A lattice fermion configuration and the corresponding SYT.}
\label{figlatfermb}
\end{figure}

Standard Young tableaux are connected to the representation theory of the symmetric group; 
the number of such tableaux 
with given shape $\lambda$ is the dimension of $\lambda$ as an irreducible representation
of the symmetric group, which is up to a factor $n!$ the evaluation of the Schur function $s_\lambda$ at $t_q=\delta_{1q}$. 
Indeed, in this case one has $H[t]=J_1$, and there is only one term contributing to 
the bra-ket $\bra{0}\e{H[t]}\ket{\lambda}$
in the expansion of the exponential:
\[
s_\lambda[\delta_{1\cdot}]=\frac{1}{n!}\bra{0} J_1^n \ket{\lambda}
\]
In terms of lattice fermions, $J_1$ has a direct interpretation as the transfer matrix for one particle
hopping one step to the left.
As the notion of SYT is invariant by transposition, particles and holes play a symmetric role so that the evolution can be summarized by
either of the two rules:
\begin{itemize}
\item Exactly one particle moves one step to the right in such a way that it does not bump into its neighbor;
all the other particles go straight.
\item Exactly one hole moves one step to the left in such a way that it does not bump into its neighbor;
all the other holes go straight.
\end{itemize}
An example of such a configuration is given on Fig.~\ref{figlatfermb}.

{\em Remark:} more generally, the Frobenius formula states that
the coefficient of $\prod_q t_q^{\alpha_q}$ in $s_\lambda[t]$ is 
$\chi_\lambda(\alpha)/\prod_q \alpha_q!$, where
$\chi_\lambda(\alpha)$ is the character
of the symmetric group for representation $\lambda$ evaluated at the
conjugacy class with $\alpha_q$ cycles of length $q$.
Thus, $\chi_\lambda(\alpha)=\bra{0}\prod_q J_q^{\alpha_q}\ket{\lambda}$.
This formula is usually written in terms of Young diagrams and known as
the Murnaghan--Nakayama formula for characters of the 
symmetric group.

\subsubsection{Cauchy formula}
As an additional remark, consider the commutation of $\e{H[t]}$ and $\e{H^\star[u]}$,
where $H^\star[u]$, the transpose of $H[u]$, is obtained from it by replacing $J_q$ with $J_{-q}$.
Using the Baker--Campbell--Hausdorff
formula and the commutation relations \eqref{heis}
we find
\[
\e{H[t]}\e{H^\star[u]}
=\e{\sum_{q\ge1}q t_q u_q}
\e{H^\star[u]}\e{H[t]}
\]
or equivalently
\[
\e{\varphi_+(x^{-1})}\e{\varphi_-(y)}
=\frac{1}{1-x y}\e{\varphi_-(y)}\e{\varphi_+(x^{-1})}.
\]
Here $\varphi_\pm(z)=\sum_{q\ge1}\frac{z^{\mp q}}{q}J_{\pm q}$ are the positive
and negative modes of the bosonic field
$\varphi(z)=-\der/\der J_0+J_0\log z-\varphi_+(z)+\varphi_-(z)$.

If we now use the fact that the $\ket{\lambda}$ form a basis of $\F_0$, we obtain the Cauchy formula:
\begin{equation}\label{cauchy}
\bra{0}\e{H[t]}\e{H^\star[u]}\ket{0}=
\sum_\lambda s_\lambda[t]s_\lambda[u]=\prod_{i,j}(1-x_i y_j)^{-1}=\e{\sum_{q\ge1}q t_q u_q}
\end{equation}
with $t_q=\frac{1}{q}\sum_{i=1}^n x_i^q$,
$u_q=\frac{1}{q}\sum_{i=1}^n y_i^q$.

\subsection{Application: enumeration of plane partitions}\label{secpp}
{\em Plane partitions}\/ are a well-known class of combinatorial objects.
The name originates from the way they were first introduced \cite{MacM} as two-dimensional
generalizations of partitions; here we shall directly define plane partitions graphically.
Their study has a long history in mathematics, with a renewal of interest
in the eighties \cite{Stan} in combinatorics, and more recently in mathematical physics \cite{OR}.\looseness=-1

\begin{figure}
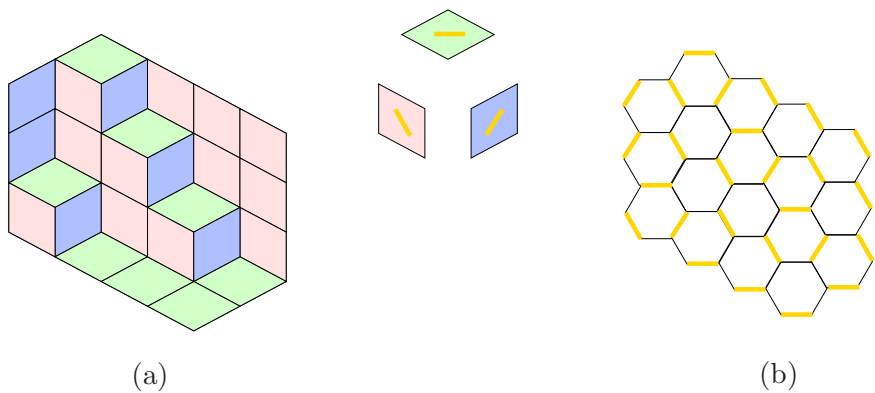

\hdrplanepartitionfigure
\caption{(a) A plane partition of size $2\times3\times4$. (b) The corresponding dimer configuration.}\label{figpp}
\end{figure}

\subsubsection{Definition}
Intuitively, plane partitions are pilings of boxes (cubes) in the corner of a room, subject to the constraints of gravity.
An example is given on Fig.~\ref{figpp}(a). Typically, we ask for the cubes to be contained inside a 
bigger box (parallelepiped) of given sizes.

Alternatively, one can project the picture onto a two-dimensional plane (which is inevitably what we do when we draw
the picture on paper) and the result is a tiling of a region of the plane by lozenges (rhombi with 60/120 degrees angles) 
of three possible orientations,
as shown on the right of the figure. If the cubes are inside a parallelepiped of size $a\times b\times c$, then,
possibly drawing the walls of the room as extra tiles, we obtain a lozenge tiling of a hexagon with sides $a,b,c$,
which is the situation we consider now.

Note that each lozenge is the union of two adjacent triangles which live on an underlying fixed triangular lattice.
So this is a statistical model on a regular lattice. In fact, we can identify it with a model of 
{\em dimers}\/ living on the dual lattice, that is the honeycomb lattice. Each lozenge corresponds to an occupied
edge, see Fig.~\ref{figpp}(b). Dimer models have a long history of their own (most notably,
Kasteleyn's formula \cite{Kast} is the standard route to their exact solution, which we do not use here), 
which we cannot possibly review here.

\subsubsection{MacMahon formula}
\begin{figure}
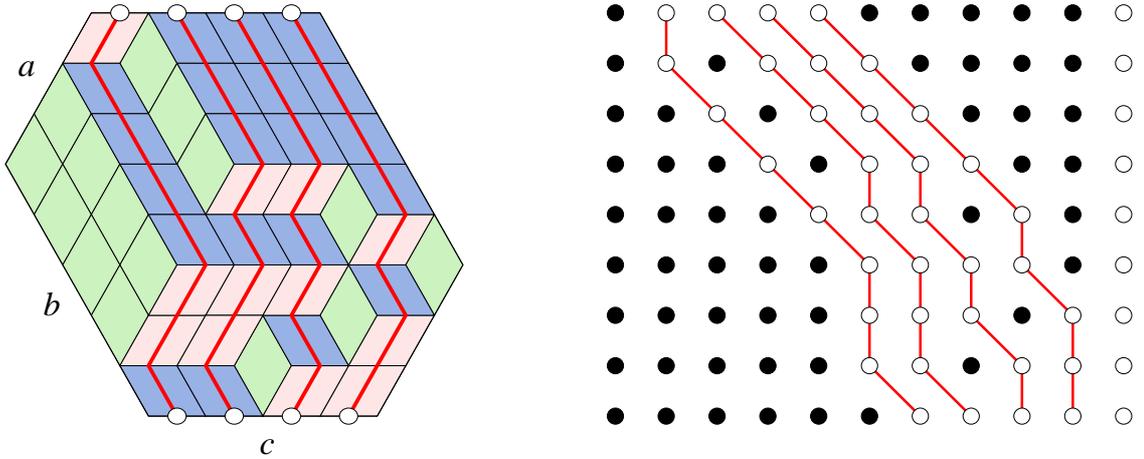

\hdrplanepartitionnilpfigure
\caption{NILPs corresponding to a plane partition.}\label{figppnilp}
\end{figure}

In order to display the free fermionic nature of plane partitions, we shall consider the following operation.
In the 3D view, consider slices of the piling of boxes by hyperplanes parallel to two of the three axes and such
that they are located half-way between successive rows of cubes. In the 2D view, this corresponds to selecting two
orientations among the three orientations of the lozenges and building paths out of these. Fig.~\ref{figppnilp}
shows on the left the result of such an operation:
a set of lines going from one side to the opposite side of the hexagon. They are by definition nonintersecting
and can only move in two directions. Conversely, any set of such NILPs produces a plane partition.

At this stage one can apply the LGV formula. But there is no need since this is actually the case already considered
in section 1.3.4. Compare Figs.~\ref{figppnilp} and \ref{figlatferm}: the trajectories of holes are exactly our paths 
(the trajectories of particles form another
set of NILPs corresponding to another choice of two orientations of lozenges).
If we attach a weight of $x_i$ to each blue lozenge at step $i$, we find that the weighted enumeration of plane
partitions in an $a\times b\times c$ box is given by:
\[
N_{a,b,c}(x_1,\ldots,x_{a+b})
=\bra{0}\e{H[t]}\ket{b\times c}=s_{b\times c}(x_1,\ldots,x_{a+b})
\]
where $b\times c$ is the rectangular Young diagram with height $b$ and width $c$.
In particular the unweighted enumeration is the dimension of the
Young diagram $b\times c$ as a $GL(a+b)$ representation:
\begin{equation}\label{maceq}
N_{a,b,c}
=\prod_{i=1}^a\prod_{j=1}^b\prod_{k=1}^c \frac{i+j+k-1}{i+j+k-2}
\end{equation}
which is the celebrated MacMahon formula. But the more general formula provides various refinements. For example,
one can assign a weight of $q$ to each cube in the 3D picture. It can be shown that this is
achieved by setting $x_i=q^{a+b-i}$ (up to a global power of $q$). This way we find the $q$-deformed formula
\[
N_{a,b,c}(q)
=\prod_{i=1}^a\prod_{j=1}^b\prod_{k=1}^c \frac{1-q^{i+j+k-1}}{1-q^{i+j+k-2}}
\]
Many more formulae can be obtained in this formalism. The reader may for example prove that
\[
N_{a,b,c}=\sum_{\lambda: \lambda_1\le c}
s_\lambda(\underbrace{1,\ldots,1}_a)
s_\lambda(\underbrace{1,\ldots,1}_b)
\]
by writing it as a specialization of $\bra{0}\prod_{i=1}^a \varphi_+(1/x_i) P_c \prod_{i=1}^b \varphi_-(y_i) \ket{0}$ where $P_c$ projects onto Young diagrams with $\lambda_1\le c$ (this point of view is close to the one of \cite{OR},
and differs from the previous one in its view of the implicit
boundary conditions
outside the hexagon -- corner of a room rather than wall with a ledge)
(note that a bijective proof also exists, relating a plane partition
to a pair of semistandard Young tableaux of same shape);
or that
\[
N_{a,b,c}=\det(1+T_{c\times b}T_{b\times a}T_{a\times c})
\]
(where $T_{y\times x}$ is the matrix with $y$ rows and $x$ columns and entries $i \choose j$, $i=0,\ldots,y-1$,
$j=0,\ldots,x-1$), as well as investigate their possible refinements.
(for more formulae similar to the last one, see \cite{artic30}). Finally, one can take the limit $a,b,c\to\infty$,
and by comparing the power of the factors $1-q^a$ in the numerator and the denominator, one finds another classical
formula
\[
N_{\infty,\infty,\infty}(q)=\prod_{n=1}^\infty (1-q^n)^{-n}
\]

Note that our description in terms of paths clearly breaks the threefold symmetry of the original hexagon.
It strongly suggests that one should be able to introduce {\em three}\/ series of parameters
to provide an even more refined counting of plane partitions.
With two sets of parameters, this is in fact known in the combinatorial literature and is related to so-called
double Schur functions (these will reappear in section \ref{secdoubleschur}).
The full three-parameter generalization is less well-known and appears in \cite{artic38}, as will be recalled 
in section \ref{fewlittle}.


{\em Remark}: as the name suggests, plane partitions are higher dimensional versions of partitions, that is of Young diagrams.
After all, each slice we have used to define our NILPs is also a Young diagram itself. However 
these Young diagrams should not be confused with the ones obtained from the NILPs by the correspondence of section \ref{ffschur}.

\subsubsection{Totally symmetric self-complementary plane partitions}
In the mathematical literature, many more complicated enumeration problems are addressed, see \cite{Stan}.
In particular, consider lozenge tilings of a hexagon of shape $2a\times 2a\times 2a$.
One notes that there is a group of transformations acting naturally on the set of configurations. We consider here
the dihedral group of order 12, which consists of rotations by $\pi/3$ and reflections about axes passing through
opposite corners of the hexagon or the midpoints of opposite edges. 
To each of its subgroups one can associate an enumeration problem.

\begin{figure}
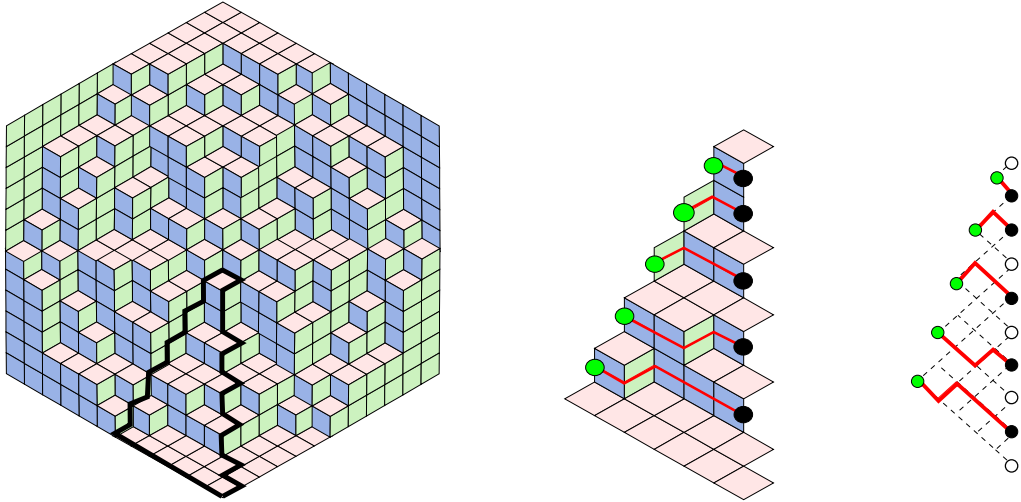

\hdrtsscppfigure
\caption{A TSSCPP and the associated NILP.}\label{figtsscppnilp}
\end{figure}

Here we discuss only the case of maximal symmetry, i.e.,\ the enumeration of plane partitions with the dihedral
symmetry. They are called in this case totally symmetric self-complementary plane partitions (TSSCPPs).
The fundamental domain is a twelfth of the hexagon, see Fig.~\ref{figtsscppnilp}. Inside this fundamental domain,
one can use the equivalence to NILPs by considering green and blue lozenges. However it is clear that the
resulting NILPs are not of the same type as those considered before for general plane partitions, for two reasons:
(i) the starting and ending points are not on parallel lines, and (ii) the endpoints are in fact free to lie anywhere on a vertical
line. However the LGV formula still holds. For future purposes we provide an integral formula for the counting
of TSSCPPs where a weight $\tau$ is attached to every blue lozenge in the fundamental domain \cite{artic41}.

Let us call $r_j$ the location of the endpoint of the $j^{\rm th}$ path, numbered from top to bottom starting at zero.
We first apply the LGV formula to write the number of NILPs with given endpoints to be $\det(N_{i,r_j})_{1\le i,j\le n-1}$
where $N_{i,r}=\tau^{2i-r-1}{i\choose 2i-r-1}=(1+\tau u)^i|_{u^{2i-r-1}}$. Next we sum over them and obtain
\begin{equation}
\begin{split}
N_n(\tau)&=\sum_{0\le r_1<r_2<\cdots<r_{n-1}} \det[(1+\tau u_i)^i u_i^{r_j}]\big|_{\prod_{i=1}^{n-1} u_i^{2i-1}}\\
&=\prod_{i=1}^{n-1} (1+\tau u_i)^i \sum_{0\le r_1<r_2<\cdots<r_{n-1}} \det(u_i^{r_j})\big|_{\prod_{i=1}^{n-1} u_i^{2i-1}}
\end{split}
\end{equation}
We recognize the numerator of a Schur function; the summation is simply over all Young diagrams with at most $n-1$ parts.
At this stage we use a classical summation formula, 
$\sum_\lambda s_\lambda(u_1,\ldots,u_{n-1})=\prod_{i=1}^{n-1}(1-u_i)^{-1}\prod_{1\le i<j\le n-1}(1-u_iu_j)^{-1}$,
to conclude that
\begin{equation}\label{tsscppcounting}
N_n(\tau)=\prod_{1\le i<j\le n-1} \frac{u_j-u_i}{1-u_i u_j} \prod_{i=1}^{n-1}\frac{(1+\tau u_i)^i}{1-u_i}
\big|_{\prod_{i=1}^{n-1} u_i^{2i-1}}
\end{equation}
where $\big|_{\prod_{i=1}^{n-1} u_i^{2i-1}}$ is now interpreted as picking the coefficient of a monomial in a power
series around zero.

This formula can be used to generate efficiently these polynomials by computer; in particular, we find the numbers
\[
N_n(1)=1,2,7,42,429\ldots
\]
which have only small prime factors. This allows us to conjecture a simple product form:
\[
N_n(1)=\prod_{i=0}^{n-1}\frac{(3i+1)!}{(n+i)!}=\frac{1!4!\ldots (3n-2)!}{n!(n+1)!\ldots(2n-1)!}
\]
which was in fact proven in \cite{And-TSSCPP}.
As a byproduct of what follows (sections \ref{secdwbc} and \ref{secintform}),
we shall obtain a (rather indirect) derivation of this evaluation.

\begin{figure}
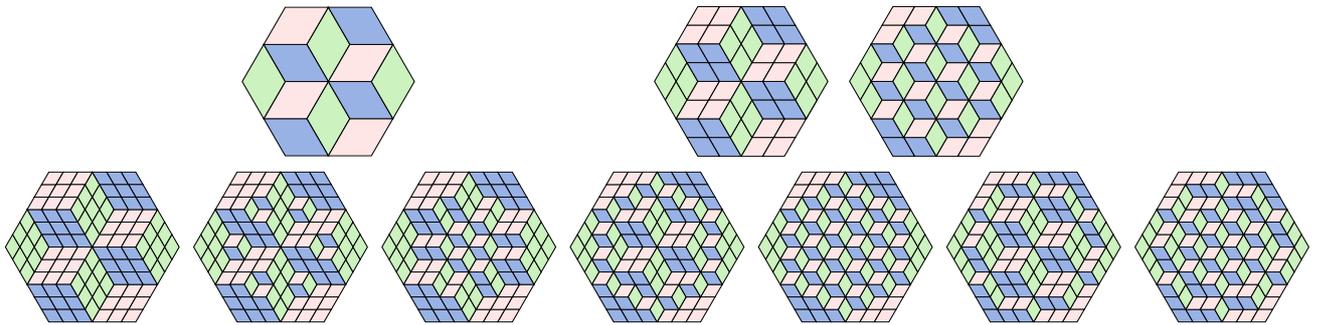

\hdrtsscppcatalogue
\caption{All TSSCPPs of size 1, 2, 3.}
\end{figure}

\subsection{Classical integrability}
The free fermionic Fock space is also important for the construction of solutions of {\em classically integrable}
hierarchies. We cannot possibly describe
these important ideas here, and refer the reader to \cite{JM-ff} and references therein for details. 
Since an explicit example will appear in section 2, let us simply say a few general words.
Recall the isomorphism $\Phi\mapsto\bra{\ell}\e{H[t]}\ket\Phi$ from $\F_\ell$ to the space of polynomials in the
variables $t_q$ (or equivalently to the space of symmetric functions if the $t_q$ are interpreted as power sums).
The resulting function will be a {\em tau-function} of the Kadomtsev--Petiashvili (KP) hierarchy (as a function
of the $t_q$) for appropriately chosen $\ket{\Phi}$. By appropriately chosen we mean the following.

In the first quantized picture, the essential property of free fermions is the possibility to write their wave function as a Slater
determinant; this amounts to considering states which are
exterior products of one-particle states. Geometrically this is interpreted as saying that
the state (defined up to multiplication by a scalar) really lives in a subspace of the full Hilbert space called a Grassmannian.
The equations defining this space (Pl\"ucker relations) are quadratic; these equations are differential
equations satisfied by $\bra{\ell}\e{H[t]}\ket{\Phi}$. They are Hirota's form of the equations defining the 
KP hierarchy. 

In section 2 we shall find ourselves in a slightly more elaborate setting, which results in the Toda lattice hierarchy.

\section{The six-vertex model}\label{sec6v}
The six-vertex model is an important model of classical statistical
mechanics in two dimensions, being the prototypical (vertex) integrable model. 
The ice model (infinite temperature limit of the six-vertex model) 
was solved by Lieb \cite{Lieb1} in 1967 by means of Bethe Ansatz, followed 
by several generalizations \cite{Lieb2,Lieb3,Lieb4}.
The solution of the most general six-vertex model was given by Sutherland \cite{Suth} in 1967.
The bulk free energy was calculated in these papers for periodic boundary conditions (PBC). 
Here our main interest will be in a different kind of boundary conditions, the so-called domain-wall boundary conditions.
But first we provide a brief review of the six-vertex model.

\begin{figure}
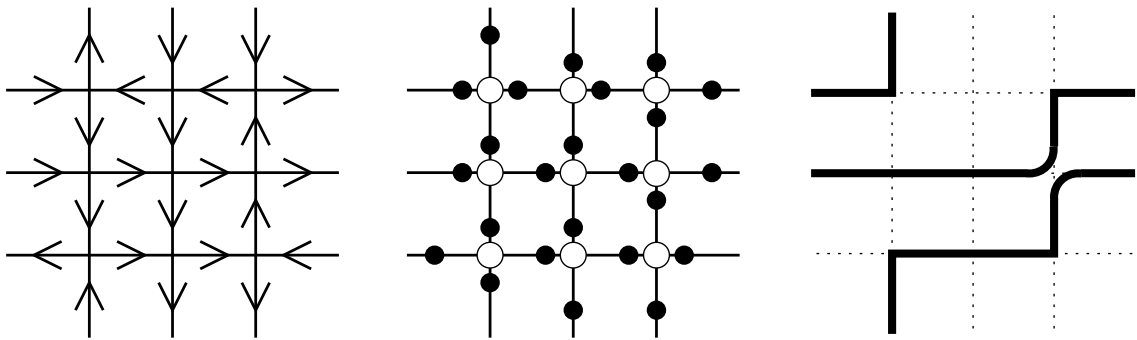

\hdrsixvertexconfiguration\qquad
\hdrsixvertexice\qquad
\hdrsixvertexpaths
\caption{A configuration of the six-vertex model.}\label{fig6vconfig}
\end{figure}
\subsection{Definition}
\subsubsection{Configurations}
The six-vertex model is defined on a (subset of the) 
square lattice by putting arrows (two possible directions) on each edge of the lattice, with the additional rule
that at each vertex, there are as many incoming arrows as outgoing ones. See Fig.~\ref{fig6vconfig} for an example,
and for two alternative descriptions: the ``square ice'' version in which arrows represent which oxygen atom
(sitting at each lattice vertex) the hydrogen ions (living on the edges) are closer to, with the ``ice rule''
that exactly two hydrogen ions are close to each oxygen atom; and the ``path'' version in which one considers
edges with right or up arrows as occupied, so that they form north-east going paths.
Around a given vertex, there are only 6 configurations of edges which respect the arrow conservation rule, see Fig.~\ref{6vweights}, hence the name of the model.

\subsubsection{Weights}
\begin{figure}
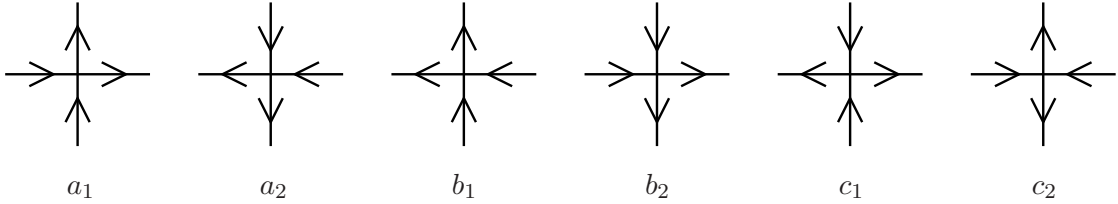

\hdrsixvertexweights
\caption{Weights of the six-vertex model.}\label{6vweights}
\end{figure}
The weights are assigned to the six
vertices, see Fig.~\ref{6vweights}.
Thus the partition function is given by
\[
Z=\sum_{\text{configurations}} \prod_{\text{vertex}} \text{(weight of the vertex)}
\]

An additional remark is useful. With any fixed boundary conditions, one can show that the difference between the
numbers of vertices of the two types $c$ is constant (independent of the configuration). This means that 
only the product $c^2=c_1c_2$ of their two weights matters. 

Let us denote similarly $a^2=a_1 a_2$ and $b^2=b_1 b_2$. One can write
\[
a_1=a e^{+E_x+E_y}
\qquad
a_2=a e^{-E_x-E_y}
\qquad
b_1=b e^{-E_x+E_y}
\qquad
b_2=b e^{+E_x-E_y}
\]
and consider that $a$, $b$, $c$ are the weights of the vertices, while $E_x$, $E_y$ are electric fields.
In what follows, we shall consider by default the model without any electric field,
where the Boltzmann weights are invariant by reversal of every arrow 
and $a_1=a_2=a$, $b_1=b_2=b$, $c_1=c_2=c$; and sometimes comment on the generalization to non-zero fields.

There is another way to formulate the partition function, using a transfer matrix. 
In order to set up a transfer matrix formalism, we first need
to specify the boundary conditions. Let us consider doubly periodic boundary conditions in the two directions of the lattice,
so that the model is defined on lattice of size $M\times L$ with the topology of a torus. Then one can write
\newcommand\tr{\mathop{\rm tr}\nolimits}
\[
Z=\tr T_L^{\,M}
\]
where $T_L$ is the $2^L\times 2^L$ transfer matrix which corresponds to a periodic strip of size $L$. Explicitly,
the indices of the matrix $T_L$ are sequences of $L$ up/down arrows.
$T_L$ can itself be expressed as a product of matrices which encode the vertex weights; in the case of
integrable models, we usually denote this matrix by the letter $R$:
\begin{equation}\label{6vR}
R=
\bordermatrix{
&\rightarrow\uparrow&\rightarrow\downarrow&\leftarrow\uparrow&\leftarrow\downarrow\cr
\rightarrow\uparrow&a&0&0&0\cr
\rightarrow\downarrow&0&b&c&0\cr
\leftarrow\uparrow&0&c&b&0\cr
\leftarrow\downarrow&0&0&0&a}
\end{equation}
Then we have
\begin{equation}
T_L=\tr_0 (R_{0L}\cdots R_{02}R_{01})=\cdots
\vcenterbox{\hdrtransfermatrixlattice}
\cdots
\end{equation}
where $R_{ij}$ means the matrix $R$ acting on the tensor product of $i^{\rm th}$ and $j^{\rm th}$ spaces, and $0$ is
an additional auxiliary space encoding the horizontal edges, as on the picture (note that the trace is on
the auxiliary space and graphically means that the horizontal line reconnects with itself). On the picture ``time''
flows upwards and to the right. 

The introduction of a (horizontally constant) 
vertical electric field amounts to multiplying the transfer matrix by an operator which commutes with it,
of the form $\e{E_y \Sigma^z}$ ($\Sigma^z$ being the number of up arrows minus the number of down arrows). 
More interestingly, adding a (vertically constant)
horizontal field amounts to twisting the periodic transfer matrix:
indeed, all the horizontal fields, using conservation of arrows at each vertex, can be moved to a single site, so that
the transfer matrix becomes, up to conjugation,
\begin{equation}\label{twistedtm}
T_L=\tr_0 (R_{0L}\cdots R_{02}R_{01}\Omega)
\end{equation}
where the twist $\Omega$ acts on the auxiliary space. 
For a constant field, it is of the form $\Omega=\e{LE_x\sigma^z}$. 
\goodbreak

\subsection{Integrability}
\subsubsection{Properties of the \texorpdfstring{$R$}{R}-matrix}
Let us now introduce the following parametrization of the weights:
\begin{equation}\label{6vwei}
\begin{split}
a&=q\,x-q^{-1}x^{-1}\\
b&=x-x^{-1}\\
c&=q-q^{-1}
\end{split}
\end{equation}
$x$, $q$ are enough to parametrize them up to global scaling. Instead of $q$ one often uses
\[
\Delta=\frac{a^2+b^2-c^2}{2ab}=\frac{q+q^{-1}}{2}
\]
In general, $q$ or $\Delta$ are fixed whereas $x$ is a variable parameter, called spectral parameter.
It can be thought itself as a ratio of two spectral parameters attached to the lines crossing at the vertex.

The matrix $R(x)$ then satisfies the following remarkable identity: (Yang--Baxter equation)
\begin{multline*}
R_{12}(x_2/x_1)R_{13}(x_3/x_1)R_{23}(x_3/x_2)=R_{23}(x_3/x_2)R_{13}(x_3/x_1)R_{12}(x_2/x_1)
\ 
\vcenterbox{\hdryangbaxter}
\end{multline*}
This is formally the same equation that is satisfied by $S$ matrices in an integrable field theory (field theory
with factorized scattering, i.e.,\ such that every $S$ matrix is a product of two-body $S$ matrices).

The $R$-matrix also satisfies the unitarity equation:
\[
R_{12}(x)R_{21}(x^{-1})=(q\,x-q^{-1}x^{-1})(q\,x^{-1}-q^{-1}x)
\qquad\vcenterbox{\hdrunitarity}
\]
with $x=x_2/x_1$.
The scalar function could of course be absorbed by appropriate normalization of $R$.

\subsubsection{Commuting transfer matrices}\label{commtm}
Consider now the transfer matrix as a function of the spectral parameter $x$, possibly with a twist:
\begin{equation}\label{tmparam}
T_L(x)=\tr_0 (R_{0L}(x)\cdots R_{02}(x)R_{01}(x)\Omega)
\end{equation}

Then using the Yang--Baxter equation repeatedly one obtains the relation
\[
[T_L(x),T_L(x')]=0
\]
We thus have an infinite family of commuting operators. In practice, for a finite chain $T_L(x)$ is a Laurent polynomial
of $x$ so there is a finite number of independent operators.

Note that we could have used the more general {\em inhomogeneous}\/ transfer matrix
\[
T_L(x;y_1,\ldots,y_L)=\tr_0 (R_{0L}(y_L/x)\cdots R_{02}(y_2/x) R_{01}(y_1/x)\Omega)
\]
where now we have spectral parameters $y_i$ attached to each vertical line $i$ and one more parameter $x$
attached to the auxiliary line. Then the same commutation relations hold
for fixed $y_i$ and variable $x$.

As is well-known, the commutation of the transfer matrices is only one relation in the Yang--Baxter algebra generated
by the so-called RTT relations. The latter lead to an exact solution of the model using Algebraic Bethe Ansatz 
\cite{faddeev}.

\subsection{Phase diagram}
The phase diagram of the six-vertex model in the absence of electric field 
is discussed in great detail in chapter 8 of \cite{Baxter}. It can be deduced from the exact solution of the model
using Bethe Ansatz after taking the thermodynamic limit.
The physical properties of the system
depend only on $\Delta=(q+q^{-1})/2$, $x$ playing the role of a lattice anisotropy parameter.
\begin{figure}
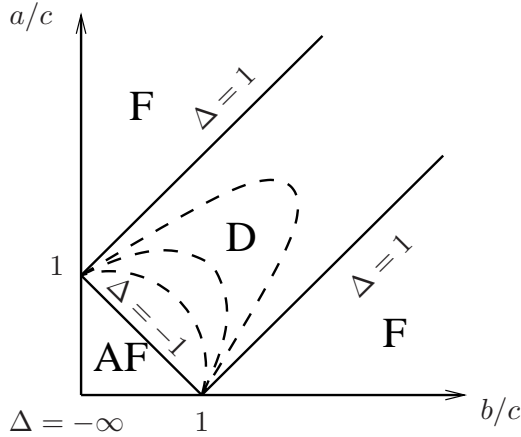

\hdrsixvertexphase
\caption{Phase diagram of the six-vertex model.}\label{fig6vphase}
\end{figure}
We distinguish three phases, see Fig.~\ref{fig6vphase}:
\begin{enumerate}
\item{$\Delta\ge 1$}: the ferroelectric phase.
This phase is non-critical. Furthermore, there are no local degrees of freedom: the system is frozen in regions
filled with one of the vertices of type $a$ or $b$ (i.e.,\ all arrows aligned), 
and no local changes (that respect arrow conservation) are possible.

\item{$\Delta< -1$}: the anti-ferroelectric phase.
This phase is non-critical. This time there is a finite correlation length. The ground state of the transfer matrix
corresponds to a state with zero polarization (in the limit $\Delta\to-\infty$, it is simply an alternation
of up and down arrows).

\item{$-1\le\Delta<1$}: the disordered phase.
This phase is critical. It possesses a continuum limit with conformal symmetry, and this limiting infra-red
Conformal Field Theory is well-known: it is simply the $c=1$ theory of a free boson on a circle with radius $R$
given by
$R^2=\frac{1}{2(1-\gamma/\pi)}$, $\Delta=-\cos\gamma$, $0<\gamma<\pi$.
\end{enumerate}

The phase diagram in the presence of an electric field is more complicated, though the basic division into the
three phases above remains.
See \cite{SB-6v,Nolden-6v} for a description.\footnote{Note that the discussion of the phase
diagram in \cite{artic20} is incomplete.}

\subsection{Free-fermion point}
Inside the disordered phase, there is a special point $\Delta=0$. We provide various representations
of the six-vertex model which display the free fermionic behavior of this region of parameter space.

\subsubsection{NILP representation}\label{NILP6v}
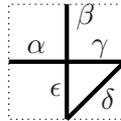
\begin{figure}
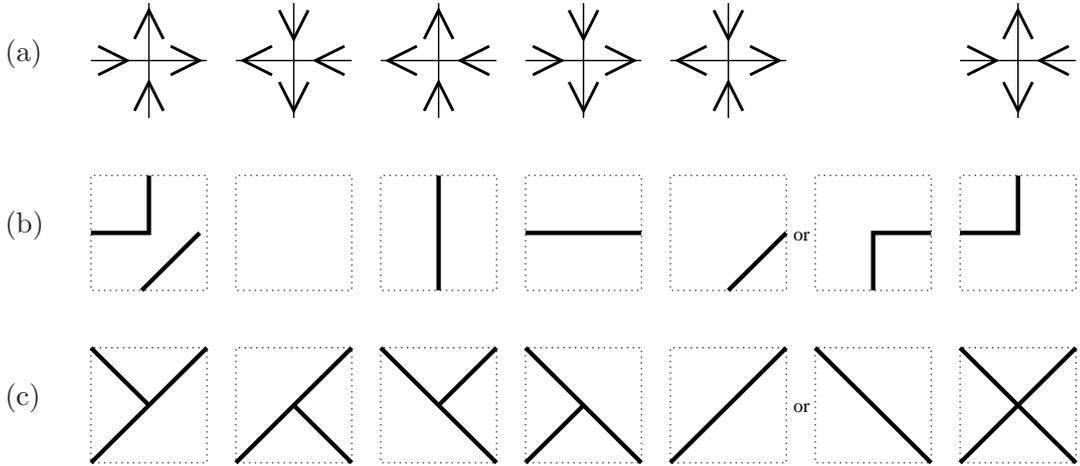

\hdrsixvertexnilpdomino
\caption{Correspondence between (a) ($\Delta=0)$ six-vertex (b) NILPs and (c) domino tilings.}\label{fig6vnilp}
\end{figure}

It is tempting to try to interpret the ``north-east going paths'' of Fig.~\ref{fig6vconfig} as nonintersecting
lattice paths. The problem is that they can touch at vertices. One way to fix it is to consider the slightly modified paths of Fig.~\ref{fig6vnilp}(b).
The rule is to replace each vertex of (a) with the corresponding 
dotted square of (b) and then patch together the latter to form the 
paths.\footnote{Going from (a) to (b) amounts to
combining the equivalences of \cite{Joh-arctic} and \cite{artic13}.}
Note that the correspondence is no longer one-to-one: each vertex of type $c_1$ corresponds to two possible local paths.

\goodbreak
The directed graph of the NILPs 
is the basic pattern {\hdrnilpbasicpattern} repeated, with paths moving upwards and
to the right, and with weights indicated on the edges.
Comparing the weights we get the relations
\begin{align*}
a_1&=\alpha\beta\delta & a_2&=1 \\
b_1&=\beta\epsilon & b_2&=\alpha\gamma\\
c_1&=\delta+\epsilon\gamma & c_2&=\alpha\beta
\end{align*}
Combining these we find that $a_1a_2+b_1b_2-c_1c_2=0$, so the correspondence only makes sense at $\Delta=0$
(and there are really only 4 parameters and not 5 as one might naively assume).

\subsubsection{Domino tilings}\label{domino}
There is also a prescription to turn six-vertex configurations into
{\em domino tilings} that is illustrated on Fig.~\ref{fig6vnilp}(c) \cite{artic13}.
As already mentioned, going from (b) to (c) is nothing but a slightly modified version of the bijection
of \cite{Joh-arctic} between NILPs and domino tilings.

In order to understand the correspondence of Boltzmann weights, note that patching together the pictures of
Fig.~\ref{fig6vnilp}(c) produces dominoes that span three dotted squares, for example
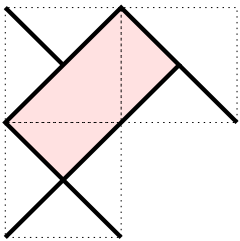
\[
\hdrexampledomino
\]
In particular, one half of the domino is contained inside one square.
This allows us to classify dominoes into four kinds, depending on which half of the square it occupies 
(these are called north-, west-, south-, and east-going in \cite{Joh-arctic}). 
Going back to Fig.~\ref{fig6vnilp}(c), we conclude that $a_1$, $a_2$, $b_1$, $b_2$ can be considered as the Boltzmann weights of
the four kinds of dominoes. Furthermore, we have the relations
\[
c_1=a_1a_2+b_1b_2\qquad c_2=1
\]
from which we derive as expected $a_1 a_2+b_1 b_2-c_1 c_2=0$.

Just as plane partitions are dimers on the honeycomb lattice, domino tilings can be considered equivalently
as dimers on the square lattice.

\subsubsection{Free fermionic five-vertex model}\label{sec5v}
The general {\em five-vertex model} is obtained by sending one of the $a$ or $b$ weights to zero
while all other weights remain finite;
in other words, one simply forbids one of the 6 types of vertices.
For a discussion of the general five-vertex model, see \cite{HWKK-5v}
(see also appendix A of \cite{NK-5v}).
In the first part of this section, we choose to
send both horizontal and vertical electric fields to minus infinity and $a$ to zero, 
in such a way that $a_1$ becomes zero.
In the representation in terms of north-east going paths, this amounts to forbidding crossings; 
however, these paths in general interact when they are close to each other.
The paths become NILPs (i.e.,\ they only interact through the Pauli principle) 
only if their weights are products over the edges, which implies that
$b_1 b_2=c_1 c_2$. This leads us back to the model of section \ref{NILP6v}, but with
$\delta$ sent to zero: what we get this way is the free fermionic five-vertex model,
first discussed in \cite{Wu-5v}.

\begin{figure}
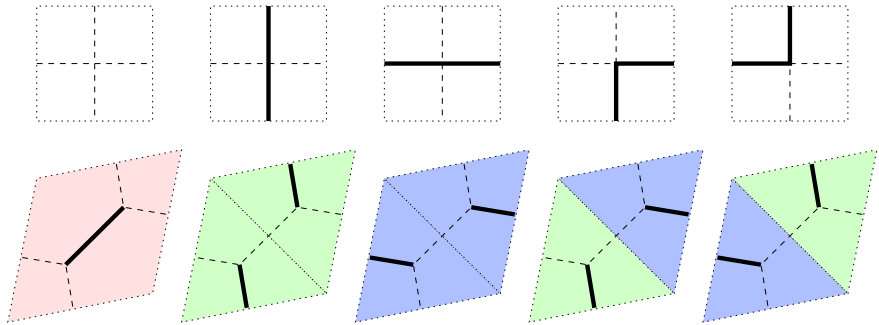

\hdrfivevertexfigure
\caption{From the five-vertex model to dimers or plane partitions.}\label{fig5v}
\end{figure}

If $\delta=0$ the NILPs of Fig.~\ref{fig6vnilp}(b) simply live on a regular square lattice, and
of course at this stage we recognize the transfer matrix discussed in section \ref{secschurlat}, and illustrated
on Fig.~\ref{figlatferm} (green lines). In section \ref{secpp} on plane partitions, it was also identified with the transfer matrix of 
lozenge tilings. To complete the circle of equivalences, we show on Fig.~\ref{fig5v} how to go from NILPs to either dimers
on the honeycomb lattice or lozenge tilings,
following Reshetikhin \cite{Resh-lectures}.

\begin{figure}
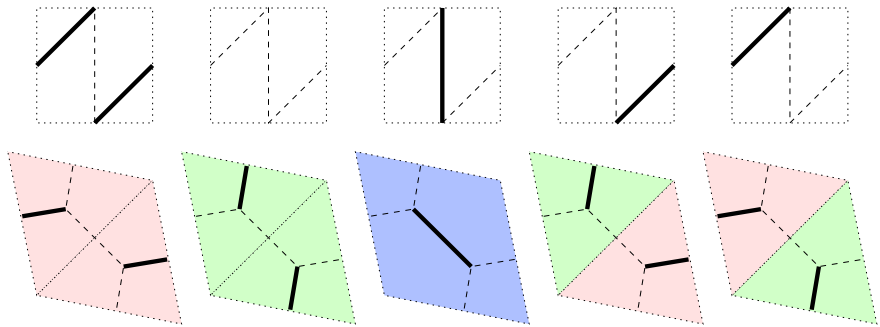

\hdrfivevertexdualfigure
\caption{From the five-vertex model to dimers or plane partitions, dual version.}\label{fig5v2}
\end{figure}

There is a second case which is worth mentioning (if only because it will reappear in section \ref{secdoubleschur}): 
suppose instead that we send $b_1$ to zero.
This time the north-east going paths cannot go straight north any more. In this case it is natural to redraw
all north-east moves with a right turn as straight lines (not just south-side-goes to east but also west-side-goes-to-north),
and this way we recognize again the green lines of Fig.~\ref{figlatferm}, with a slight modification: the whole picture is distorted
in such a way that each path moves one step further to the right (so that north steps become north-east steps).
If we want these paths to be NILPs, we reproduce the weights of section \ref{NILP6v} with $\epsilon=0$. Finally the correspondence
to lozenge tilings/dimers is illustrated on Fig.~\ref{fig5v2}.

Note a difference between the models of lozenge tilings corresponding to these two versions of the free fermionic five-vertex model:
the vertical spectral parameters flow north-east in the first picture, whereas they flow north-west in the second picture.
Ultimately, this is related to two possible inhomogeneous versions of Schur functions (double vs dual [double] Schur functions
in the language of \cite{Molev-coproduct}). See also the recent work \cite{artic46} where these lozenge tilings are embedded
in a more general square-triangle-rhombus tiling model.

\subsection{Domain-wall boundary conditions}\label{secdwbc}
Domain-wall boundary conditions (DWBC) are special boundary conditions which were originally introduced in order
to study correlation functions of the six-vertex model \cite{Kor}. However they are also interesting in their own right.
\subsubsection{Definition}
\begin{figure}
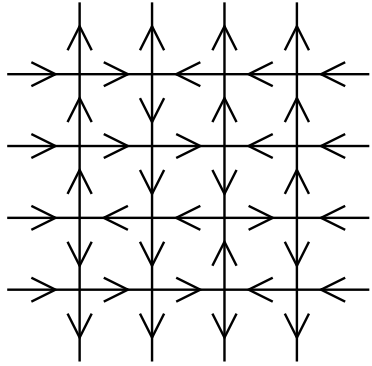

\hdrsixvertexdwbc
\caption{An example of a configuration with domain-wall boundary conditions.}\label{fig6vdwbc}
\end{figure}

DWBC are defined on a $n\times n$ square grid: all the external edges of the grid are fixed according to the rule that
vertical ones are outgoing and horizontal ones are incoming. An example is given on Fig.~\ref{fig6vdwbc}.

To each horizontal (resp.\ vertical) line one associates a spectral parameter $x_i$ (resp.\ $y_j$).
The partition function is thus:
\[
Z_n(x_1,\ldots,x_n;y_1,\ldots,y_n)=\sum_{\text{configurations}} \prod_{i,j=1}^n w(y_j/x_i)
\]
where $w=a,b,c$ depending on the type of vertex (cf.\ \eqref{6vwei}). 
Here we do not allow any electric field for the simple reason
that with DWBC (as with any fixed boundary conditions), using the same type of arguments as in the previous section,
one can push the effect of the field to the boundary, where it only contributes a constant to the partition function.

{\em Remark:} the (one-to-many) correspondence of section \ref{domino} sends DWBC six-vertex configurations to domino tilings
of the {\em Aztec diamond}\/ \cite{JPS}.

\subsubsection{Korepin's recurrence relations}\label{koreprec}
In \cite{Kor}, a way to compute $Z_n$ inductively was proposed. It is based on the following properties:
\begin{itemize}
\item $Z_1=q-q^{-1}$.
\item $Z_n(x_1,\ldots,x_n;y_1,\ldots,y_n)$ is a symmetric function of the $\{ x_i\}$ and of the $\{ y_i\}$ (separately).
This is a consequence of repeated application of the Yang--Baxter equation 
(or equivalently of one of the components of the so-called RTT relations):
\begin{multline*}
(q\,y_{i+1}/y_i -q^{-1}y_i/y_{i+1}) 
Z_n(\ldots, y_i,y_{i+1},\ldots)
=(q\,y_{i+1}/y_i -q^{-1}y_i/y_{i+1}) 
\vcenterbox{\hdrpermutationdiagram{0}{$\scriptstyle y_i$}{$\scriptstyle y_{i+1}$}}
\\
=\vcenterbox{\hdrpermutationdiagram{1}{$\scriptstyle y_i$}{$\scriptstyle y_{i+1}$}}
=\vcenterbox{\hdrpermutationdiagram{2}{$\scriptstyle y_i$}{$\scriptstyle y_{i+1}$}}
=\cdots
=\vcenterbox{\hdrpermutationdiagram{3}{$\scriptstyle y_i$}{$\scriptstyle y_{i+1}$}}
\\
=(q\,y_{i+1}/y_i -q^{-1}y_i/y_{i+1}) 
\vcenterbox{\hdrpermutationdiagram{0}{$\scriptstyle y_{i+1}$}{$\scriptstyle y_i$}}
=(q\,y_{i+1}/y_i-q^{-1}y_i/y_{i+1})
Z_n(\ldots,y_{i+1},y_i,\ldots)
\end{multline*}
and similarly for the $x_i$.
\item $Z_n$ multiplied by $x_i^{n-1}$ (resp.\ $y_i^{n-1}$) 
is a polynomial of degree at most $n-1$ in each variable $x_i^2$ (resp.\ $y_i^2$).
This is because (i) each variable say $x_i$ appears only on row $i$ (ii) $a$, $b$ are linear combinations of
$x_i^{-1}$, $x_i$ and $c$ is a constant and (iii) there is at least one vertex of type $c$ on each row/column.

\item The $Z_n$ obey the following recursion relation:
\begin{multline}\label{koreprecur}
Z_{n}(x_{1},\ldots,x_n;y_1=x_{1},\ldots,y_{n})\\=(q-q^{-1})
\prod_{i=2}^n(q\,x_1/x_i-q^{-1}x_{i}/x_1)
\prod_{j=2}^{n}(q\,y_j/x_1-q^{-1}x_1/y_{j})
Z_{n-1}(x_{2},\ldots,x_n;y_{2},\ldots,y_{n})
\end{multline}
Since $y_1=x_1$ implies $b(y_1/x_1)=0$,
by inspection all configurations with non-zero weights are of the
form shown on Fig.~\ref{recurs}. This results in the identity.
\begin{figure}
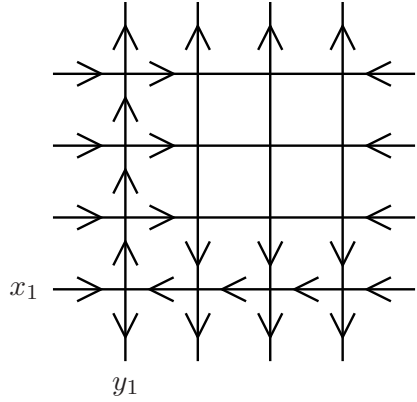

\centering
\hdrsixvertexrecurrence
\caption{Graphical proof of the recursion relation.}
\label{recurs}
\end{figure}
\end{itemize}

Note that by the symmetry property, Eq.~\eqref{koreprecur} fixes
$Z_n$ at $n$ distinct values of $y_{1}$: $x_i$, $i=1,\ldots,n$.
Since $Z_n$ is of degree $n-1$ in $y_{1}^2$, it is entirely determined by it.

\subsubsection{Izergin's formula}
Remarkably, there is a closed expression for $Z_n$ due to Izergin \cite{Iz-6v,ICK}.
It is a determinant formula:
\begin{equation}\label{izdet}
Z_n=\frac{\prod_{i,j=1}^n (x_j/y_i-y_i/x_j)(q\,x_j/y_i-q^{-1}y_i/x_j)}{\prod_{1\le i<j\le n}(x_i/x_j-x_j/x_i)(y_i/y_j-y_j/y_i)}
\det_{i,j=1\ldots n}\left(\frac{q-q^{-1}}{(x_j/y_i-y_i/x_j)(q\,x_j/y_i-q^{-1}y_i/x_j)}\right)
\end{equation}
The hard part lies in finding the formula, but once it is found,
it is a simple check to prove that it satisfies all the properties of the previous section.
The symmetry under interchange of variables is evident from the
structure of the formula, and the recurrence formula follows from looking at the zeroes outside
the determinant and the poles inside the determinant: indeed, they must compensate each other for the result
to be non-zero at say $y_1=x_1$, which immediately leads to expanding the determinant along the $x_1$ column and to
the recurrence.

\subsubsection{Relation to classical integrability and random matrices}
The Izergin determinant formula is curious because it involves a simple determinant, which reminds us of free fermionic
models. And indeed it turns out that it can be written in terms of free fermions, or equivalently that it provides
a solution to a hierarchy of classically integrable PDE, in the present case the two-dimensional Toda lattice
hierarchy. We cannot go into any detail here but provide a few remarks.

Consider a function of two sets of $n$ variables of the form
\begin{equation}\label{taun}
\tau_n(X,Y)=\frac{\det \phi(X,Y)}{\Delta(X)\Delta(Y)}
\end{equation}
where if $X=( x_1,\ldots,x_n)$, $Y=( y_1,\ldots, y_n )$, then
\[
\det \phi(X,Y)=\det_{i,j=1,\ldots,n} \phi(x_i, y_j)\qquad \Delta(X)=\prod_{1\le i<j\le n}
(x_i-x_j)
\]
Certainly Izergin's formula \eqref{izdet} is of this form (up to some
prefactors and to $x_i\to x_i^2$, $y_i\to y_i^2$);
but it is also the case of the Cauchy formula
\eqref{cauchy} (under the form of the Cauchy determinant)
and even of Weyl's formula \eqref{defweyl} (once divided by $\Delta(\lambda)$; though usually one considers it as a function
of the $x_i$ only, which results in a tau-function of the KP hierarchy only).
It also appears in problems of random matrices,
and in particular in the Harish Chandra--Itzykson--Zuber integral \cite{HC,IZ} (see \cite{artic21}).
$\tau_n$ is of course symmetric by permutation of variables in $X$, and in $Y$.

We now make use of the following set of bilinear determinant identities:
\begin{equation}\label{bildet}
\sum_{i=1}^{n+1} \det \phi(X-x_i,Y) \det \phi(X'+x_i,Y')=\sum_{j=1}^{m+1} \det \phi(X',Y'-y'_j)\det \phi(X,Y+y'_j)
\end{equation}
where $X=(x_1,\ldots,x_{n+1})$, $Y=(y_1,\ldots,y_n)$,
$X'=(x'_1,\ldots,x'_m)$, $Y'=(y'_1,\ldots,y'_{m+1})$.

Using as in section 1 the Miwa transformation:\footnote{Since we have only a finite number of variables, the
correct prescription is to consider a symmetric polynomial in $n$ variables as a linear combination of Schur functions
with at most $n$ rows.}
$t_q=\frac{1}{q}\sum_{i=1}^n x_i^q$,
$s_q=\frac{1}{q}\sum_{i=1}^n y_i^q$, and similarly for the primed variables,
we find after standard contour integration tricks that \eqref{bildet} can be rewritten in terms of the $\tau_n$ as
\begin{multline*}
\oint {d u\over 2\pi i u} u^{m-n}
\tau_m[t-[u],s]
\tau_{n+1}[t'+[u],s']
\e{\sum_{q\ge 1}(t_q-t'_q) u^q}\\
=\oint {d v\over 2\pi i v} v^{n-m}
\tau_n[t',s'-[v]]
\tau_{m+1}[t,s+[v]]
\e{\sum_{q\ge 1}(s'_q-s_q) v^q}
\end{multline*}
where $[u]=(u^{-1},u^{-2}/2,\ldots,u^{-q}/q,\ldots)$, and
where the integrals $\oint du/(2\pi i u)$ have the meaning of picking the constant term of a Laurent series.
Expanding this equation in powers of $t_q-t'_q$ and $s_q-s'_q$ results in an infinite set of partial differential equations
satisfied by the $\tau_n$.
They are the Hirota form of the {\em two-dimensional Toda lattice hierarchy}. Inside it, there are two copies of the 
KP hierarchy corresponding to varying only one set of variables (the $t_q$ or the $s_q$) and keeping $n$ fixed.
$\tau_n$ is the tau-function of the hierarchy.

In particular, if one expands to first order in $t_1-t'_1$ and sets $m=n-1$, we find
\begin{equation}\label{toda}
\tau_{n+1}\tau_{n-1}=\tau_n \frac{\der}{\der t_1}
\frac{\der}{\der s_1} \tau_n-
\frac{\der}{\der t_1} \tau_n
\frac{\der}{\der s_1} \tau_n
\end{equation}
which is a form of the {\em Toda lattice equation}.

There is another representation which is particularly useful for the homogeneous limit. Consider the Laplace (or Fourier, we
are working at a formal level) transform of $\phi$:
\[
\phi(x,y)=\iint d\mu(a,b)\, \e{xa+yb}
\]
Then one can write
\[
\tau_n(X,Y)=\frac{1}{n!\Delta(X)\Delta(Y)}\idotsint \prod_{i=1}^n d\mu(a_i,b_i) \det_{i,j=1,\ldots,n}(\e{x_i a_j}) \det_{i,j=1,\ldots,n}(\e{y_i b_j})
\]
This is formally identical to the partition function of a generalized two-matrix model with external fields for both matrices. 

Next, let us consider the homogeneous limit $\tau_n(x,y)$ of such a function $\tau_n(X,Y)$ where all $x_i$ tend to $x$ and all $y_i$ tend to $y$.
Noting that $\det(\e{x_i a_j})/\Delta(X)\sim c_n\, \e{x\sum_i a_i} \Delta(a)$
where $c_n=(\prod_{k=1}^{n-1}k!)^{-1}$
(by the usual trick of taking $x_i=x+i\epsilon$, $\epsilon\to 0$)
and similarly for the $y_i$, 
\[
\tau_n(x,y)=c_n c_{n+1}\idotsint \prod_{i=1}^n d\mu(a_i,b_i) \Delta(a)\e{x\sum_{i=1}^n a_i} \Delta(b)\e{y\sum_{j=1}^n b_j}
\]
This is a generalized two-matrix model with linear potentials. With an arbitrary potential, the partition function of such a model is known
to be a tau-function of the two-dimensional Toda lattice hierarchy. In fact, we have the following fermionic representation, with notations
similar to section \ref{secff}:
\[
\tau_n(x,y)\propto\bra{n,n}\e{x J_{+,1}+y J_{-,1}}
\Big(\int d\mu(a,b) \psi_+^{\star}(a)\psi_-^{\star}(b)\Big)^n
\ket{0,0}
\]

Since we only have here linear potentials, i.e.,\ the primary times (the ``$t_1$''),
we shall only recover the first equation of the hierarchy. Let us do so. First note the determinant formula
\[
\tau_n(x,y)=c_n^2 \det_{i,j=0,\ldots,n-1} \iint d\mu(a,b) a^i b^j \e{xa+yb}=c_n^2 \det_{i,j=0,\ldots,n-1}
\left(\frac{\der^i}{\der x^i}\frac{\der^j}{\der y^j} \phi(x,y)\right)
\]
Of course the latter form could have been derived directly from \eqref{taun}. Next apply to either of these expressions
the Desnanot--Jacobi determinant identity. More precisely, consider the matrix of size $n+1$ and write
that its determinant times the determinant of the sub-matrix of size $n-1$ with last two rows and columns removed equals
the difference of the two possible 
products of determinants of sub-matrices of size $n$ with one row, one column, removed and the other row, other column removed
(among the last two rows and columns). The result is:
\begin{equation}\label{todab}
n^2\tau_{n+1}\tau_{n-1}=\tau_n \frac{\der}{\der x}
\frac{\der}{\der y} \tau_n-
\frac{\der}{\der x} \tau_n
\frac{\der}{\der y} \tau_n
\end{equation}
(the factor of $n^2$ takes care of the $c_n^2$).

Finally, in the special case that $\phi(x,y)$ only depends on $x-y$, then the previous formulae simplify.
The measure $d\mu(a,b)$ is concentrated on $a=-b$ and we have
\[
\tau_n(X,Y)=
\frac{1}{n!\Delta(X)\Delta(Y)}\idotsint \prod_{i=1}^n d\mu(a_i) \det_{i,j=1,\ldots,n}(\e{x_i a_j}) \det_{i,j=1,\ldots,n}(\e{-y_i a_j})
\]
which is a generalized one-matrix model with external field. In the homogeneous limit,
$\tau_n(x,y)$ becomes a function of a single variable $t=x-y$ and we can write
\begin{equation}\label{onemm}
\tau_n(t)=c_n c_{n+1}\idotsint \prod_{i=1}^n d\mu(a_i) \Delta(a)^2\e{t\sum_{i=1}^n a_i}
\end{equation}
which is a one-matrix model with linear potential.
With an arbitrary potential, the partition function of such a model is known
to be a tau-function of the Toda chain hierarchy. Writing as before
\[
\tau_n(t)=c_n^2 \det_{i,j=0,\ldots,n-1} \int d\mu(a) a^{i+j} \e{t a}=c_n^2 \det_{i,j=0,\ldots,n-1}
\left(\frac{\der^{i+j}}{\der t^{i+j}} \phi(t)\right)
\]
and applying the Jacobi--Desnanot identity we obtain 
\begin{equation}\label{todachain}
n^2\tau_{n+1}\tau_{n-1}=\tau_n \tau''_n-\tau'_n{}^2
\end{equation}
which is a form of the {\em Toda chain equation}.

\subsubsection{Thermodynamic limit}
In \cite{artic12}, the Toda chain equation \eqref{todachain} 
above was used to derive the asymptotic behavior of the partition function of the six-vertex model
with DWBC in two of its three phases: ferroelectric and disordered. 
These are the two phases where we expect the limit $n\to\infty$ to be smooth, as we shall explain.
In this case, making the Ansatz that the free energy is extensive, we plug the asymptotic expansion 
$\tau_n\sim e^{-n^2 f}$ into the Toda chain equation and are left with a simple differential equation to solve:
\begin{equation}\label{todafree}
f''=\e{2f}
\end{equation}
In fact, this is the simplest reasonable Ansatz that is compatible with \eqref{todachain}, so
that any solution of \eqref{todachain} with a smooth large $n$ limit 
will be governed by the differential equation \eqref{todafree}.

This is how the computation of the bulk free energy of the six-vertex model with DWBC for $\Delta>-1$ is performed
in \cite{artic12}. It is much simpler than the computation for PBC and the result is given in terms of elementary functions,
and is thus different from that of PBC. Explicitly, we find 
\begin{equation}\label{thermo6v}
Z_n^{1/n^2}\to 
\begin{cases} 
\max(a,b)& \Delta>1\\
i\,a\,b\,\frac{\pi/\gamma}{\cos \pi t/\gamma}& |\Delta|<1,\ q=-\e{-i\gamma}\ (\gamma>0),\ x=\e{i(t+\gamma)}
\end{cases}
\end{equation}
Note the obvious interpretation of the result in the ferroelectric regime -- a
frozen phase where almost all arrows are aligned.

In the anti-ferroelectric phase $\Delta<-1$ one expects a less smooth large $n$ limit because the alternation of arrows
that is favored in this phase will interact with the boundaries. This statement is made more precise in \cite{artic13}, where
matrix model techniques are used to compute the large $n$ limit of \eqref{onemm} (which essentially boil down to an appropriate saddle
point analysis of it). And indeed one finds that $\log \tau_n$ has an oscillating term of order $1$, which explains that the simple
differential equation \eqref{todafree} cannot account for the asymptotic behavior.

In any case, in all phases one finds a result that is different from that of PBC. The explanation of this phenomenon is that
the six-vertex model suffers from a strong dependence on boundary conditions due to the constraints imposed by arrow
conservation. In particular there is no thermodynamic limit in the usual sense (i.e.,\ independently of
boundary conditions). 
In \cite{artic20} it was suggested more precisely that the six-vertex model undergoes 
spatial phase separation, similarly to plane partitions \cite{CLP} and other dimer models \cite{Kenyon-notes}.
In other words, even far from the boundary of the system, the system loses any translational-invariance and the physical behavior around a 
given point is a function of the local polarization: as such, the model can have several (possibly, all) of the three phases
coexisting in different regions.
This was motivated by some numerical evidence, as well as by the exact result at the free-fermion point 
$\Delta=0$, at which the arctic circle theorem \cite{JPS}
applies: the boundary between ferroelectric and disordered phases
is known exactly to be an ellipse (a circle for $a=b$) tangent to the four sides
of the square.
So the apparent simplicity of the computation of the bulk free energy for DWBC conceals a complicated physical picture.

Since then, there has been a considerable amount of work in this area. There has been more numerical work \cite{AR}.
Some of the results of \cite{artic12} have been proven rigorously and extended using sophisticated machinery in the series
of papers \cite{BF,BL,BL2} by Bleher et al.  
Finally, the curve separating phases has been studied in the work of Colomo and Pronko
\cite{CP,CP2},
and recently they proposed equations for this curve in the cases $a=b$, $\Delta=\pm 1/2$ \cite{CP3}.
The point $\Delta=1/2$ is of special interest: it corresponds to all weights equal, and is the original ice model. It is
also the subject of the next section.

\subsubsection{Application: alternating sign matrices}\label{ASM}
\begin{figure}
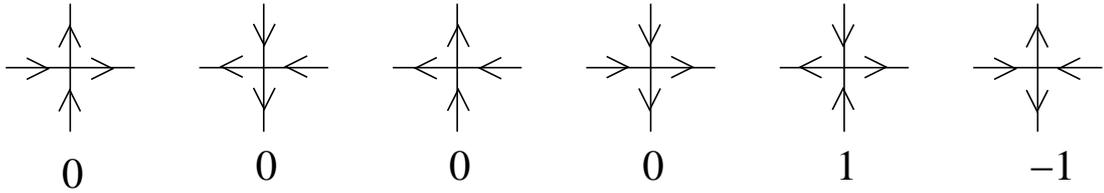

\hdrsixvertexasm
\caption{From six-vertex to ASMs.}\label{fig6vasm}
\end{figure}

{\em Alternating sign matrices}\/ are an important class of objects in modern combinatorics \cite{MRR-ASM}.
They are defined as follows. 
An alternating sign matrix (ASM) is a square matrix made of 0s, 1s and -1s 
such that if one ignores 0s,
1s and -1s alternate on each row and column starting and ending with 1s.
For example,
\[
\begin{matrix}
0&1&0&0\\
0&0&1&0\\
1&0&-1&1\\
0&0&1&0
\end{matrix}
\]
is an ASM of size 4. The enumeration of ASMs is a famous problem with a long history, see \cite{Bressoud}.
Here we simply note that ASMs are in fact in bijection with six-vertex model configurations with DWBC \cite{Kup-ASM}.
The correspondence is quite simple and is summarized on Fig.~\ref{fig6vasm}. For example, Fig.~\ref{fig6vdwbc}
becomes the $4\times 4$ ASM above.

We can therefore reinterpret the partition function of the six-vertex model with DWBC as a weighted enumeration of ASMs.
It is natural to set the weight of all zeroes to be equal ($a=b$), which leaves us with only one parameter $c/a$,
the weight of a $\pm 1$. In fact here we shall consider only the pure enumeration problem that is all weights
equal. We thus compute $\Delta=1/2$ and $q=\e{i\pi/3}$, and then $x_i=q^{-1}$, $y_j=1$ so that
the three weights are $w(y_j/x_i)=q-q^{-1}$.

At this stage there are several options. Either one tries to evaluate directly the formula \eqref{izdet};
since the determinant vanishes in the homogeneous limit where all the $x_i$ or $y_j$ coincide, this is a somewhat involved
computation and is the content of Kuperberg's paper \cite{Kup-ASM}.

There is however a much easier way, discovered independently by Stroganov \cite{Strog-IK} and Okada \cite{Oka}.
It consists of identifying $Z_n$ {\em at $q=\e{i\pi/3}$} with a Schur function. 
Consider the partition $\lambda^{(n)}=(n-1,n-1,n-2,n-2,\ldots,1,1)$, that is the Young diagram
\begin{equation}\label{defstairs}
\lambda^{(n)}=
\overbrace{
\tableau{
&&\hdotscell&\\
&&\hdotscell&\\
\vdotscell&\vdotscell&\vhdotscell\\
&\\
&\\
\\
\\}}^{n-1}
\end{equation}
$s_{\lambda^{(n)}}(z_1,\ldots,z_{2n})$ is a polynomial
of degree at most $n-1$ in each $z_i$ (use \eqref{defschurb}) and,
satisfies the following
\begin{equation}\label{recschur}
s_{\lambda^{(n)}}(z_1,\ldots,z_j=q^{-2} z_i,\ldots,z_{2n})=\prod_{\substack{k=1\\ k\ne i,j}}^{2n}
(z_k-q^{2}z_i) s_{\lambda^{(n-1)}}(z_1,\ldots,\hat z_i,\ldots,\hat z_j,\ldots,z_{2n})
\end{equation}
where the hat means that these variables are skipped 
(start from \eqref{defweyl}, find all the zeroes as $z_j=q^2z_i$ and then set $z_i=z_j=0$ to find what is left).

This looks similar to recursion relations \eqref{koreprecur}. After appropriate identification one finds:
\begin{multline*}
Z_n(x_1,\ldots,x_n;y_1,\ldots,y_n)\big|_{q=\e{i\pi/3}}=(-1)^{n(n-1)/2}
(q-q^{-1})^n \prod_{i=1}^n (q\,x_iy_i)^{-(n-1)}
\\
s_{\lambda^{(n)}}(q^{2}x_1^2,\ldots,q^{2}x_n^2,y_1^2,\ldots,y_n^2)
\end{multline*}
Note that $Z_n$ possesses at the point $q=\e{i\pi/3}$ an enhanced symmetry in the whole set of variables
$\{ q\,x_1,\ldots,q\,x_n,y_1,\ldots,y_n\}$. Finally, setting $x_i=q^{-1}$ and $y_j=1$ and remembering that
this will give a weight of $(q-q^{-1})^{n^2}$ to each ASM,
one concludes that the number of ASMs is given by
\[
A_n=3^{-n(n-1)/2}
s_{\lambda^{(n)}}(\underbrace{1,\ldots,1}_{2n})
=3^{-n(n-1)/2}\prod_{1\le i<j\le 2n}\frac{\lambda^{(n)}_i-i-\lambda^{(n)}_j+j}{j-i}
\]
Simplifying the product results in
\begin{equation}\label{An}
A_n=\prod_{i=0}^{n-1}\frac{(3i+1)!}{(n+i)!}=1,2,7,42,429\ldots
\end{equation}
which is a sequence of numbers we have encountered before! In fact, the first proof of 
formula \eqref{An}, due to Zeilberger \cite{Zeil-ASM}, amounts to showing (non-bijectively) that the number of ASMs
is the same as the number of TSSCPPs.

These are the ASMs of size 1, 2, 3 ($+=1$, $-=-1$):
\begin{gather*}
\begin{smallmatrix}+\end{smallmatrix}\\
\begin{smallmatrix}+&0\\0&+\end{smallmatrix}\qquad
\begin{smallmatrix}0&+\\+&0\end{smallmatrix}\\
\begin{smallmatrix}+&0&0\\0&+&0\\0&0&+\end{smallmatrix}
\qquad
\begin{smallmatrix}+&0&0\\0&0&+\\0&+&0\end{smallmatrix}
\qquad
\begin{smallmatrix}0&+&0\\+&0&0\\0&0&+\end{smallmatrix}
\qquad
\begin{smallmatrix}0&+&0\\0&0&+\\+&0&0\end{smallmatrix}
\qquad
\begin{smallmatrix}0&+&0\\+&-&+\\0&+&0\end{smallmatrix}
\qquad
\begin{smallmatrix}0&0&+\\+&0&0\\0&+&0\end{smallmatrix}
\qquad
\begin{smallmatrix}0&0&+\\0&+&0\\+&0&0\end{smallmatrix}
\end{gather*}

As a check, one can take $n\to\infty$ in \eqref{An}, and using Stirling's formula one finds
\[
A_n^{1/n^2}\to \frac{3\sqrt{3}}{4}
\]
This is to be compared with \eqref{thermo6v} for $\gamma=2\pi/3$, where we find $Z_n^{1/n^2}\to 9i/8$. The
two formulae agree considering $Z_n=(\sqrt{3}i/2)^{n^2} A_n$.

\section{Loop models and Razumov--Stroganov conjecture}\label{seclooprs}
\subsection{Definition of loop models}
Loop models are an important class of two-dimensional statistical lattice models.
They display a broad range of critical phenomena, and in fact many classical models
are equivalent to a loop model. The critical exponents, formulated in the language of loop models,
often acquire a simple geometric meaning; and many methods have been used to study their continuum limit,
including the Coulomb gas approach \cite{Nien-CG}, 
Conformal Field Theory \cite{BPZ} and more recently the Stochastic L\"owner Evolution \cite{LSW1,LSW2,LSW3}.
Here we are of course more interested in their properties on a finite lattice (in relation to
combinatorics) and in the use of integrable methods.

We shall introduce two classes of loop models on the square lattice, 
which turn out to be both closely related to the six-vertex model.
Then we shall discuss a very non-trivial connection between these two loop models
(the Razumov--Stroganov conjecture).

\begin{figure}
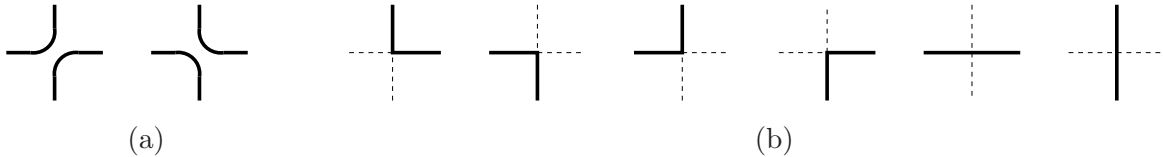

\hdrloopvertices
\caption{Vertices of (a) the CPL model and (b) the FPL model.}\label{figloops}
\end{figure}

Let us first discuss common features of the two models. Their configurations consist of loops living on the
edges of the square lattice. The most important feature is the {\em non-local}\/ Boltzmann weight produced
by assigning a fugacity of $\tau$ to closed loops. Here $\tau$ is a real parameter (usually called $n$,
due to the connection with the $O(n)$ model -- however for various reasons we avoid this notation here).
This can be supplemented by possible local weights for 
the various configurations around a given vertex, see Fig.~\ref{figloops}.

\subsubsection{Completely packed loops}\label{CPLdef}
Configurations of completely packed loops (CPL) 
consist of nonintersecting loops occupying {\em every}\/ edge of the square lattice,
which produces two possibilities at each vertex, represented on Fig.~\ref{figloops}(a).
Besides the weight $\tau$ of each closed loop, one can introduce a local weight of $u$ for one of the two types of 
CPL vertices, say NE/SW loops.

The model is known to be critical for $|\tau|<2$, and its continuum limit is described by
a theory with central charge $c=1-6\frac{\gamma^2}{\pi(\pi-\gamma)}$,
where $\tau=2\cos\gamma$, $0<\gamma<\pi$.

\subsubsection{Fully packed loops: FPL and \texorpdfstring{FPL$^2$}{FPL2} models}\label{FPLdef}
Configurations of fully packed loops (FPL) consist of nonintersecting loops such that there is exactly one loop
at each vertex, which results in the six possibilities described on Fig.~\ref{figloops}(b). One can then 
give a weight of $\tau$ to each closed loop (FPL model). 
However, one can do better: noting that the empty edges also form loops
(dashed lines on the figure), one can put them on the same footing as occupied edges and assign
them a fugacity too, say $\tilde\tau$. This more general model is usually called FPL$^2$ model.

One reason that the FPL$^2$ is interesting is the following: the FPL model is not integrable for $\tau\ne 1$
(the very special case $\tau=1$ is of interest to us and will be considered below). 
However, the FPL$^2$ is integrable for $\tau=\tilde\tau$,
that is if the two types of loops are given equal weights. This was shown using Coordinate Bethe Ansatz
in \cite{DCN} and then rederived using Algebraic Bethe Ansatz in \cite{artic28}.

For generic values of $\tau$, $\tilde\tau$, the Coulomb gas approach provides non-rigorous arguments to
identify the continuum field theory, see \cite{JK-FPL2}, and allows us to compute the central charge
to be $c=3-6\frac{\gamma^2}{\pi(\pi-\gamma)}-6\frac{\tilde\gamma^2}{\pi(\pi-\tilde\gamma)}$,
where $\tau=2\cos\gamma$ and $\tilde\tau=2\cos\tilde\gamma$, $0<\gamma,\tilde\gamma<\pi$. In particular the FPL model has central charge
$c=2-6\frac{\gamma^2}{\pi(\pi-\gamma)}$, which is one more than the corresponding CPL model.

We shall not discuss the possibility of adding local weights in detail.
Let us simply note that even if we impose rotational invariance of local Boltzmann weights, we can introduce an energy
cost for 90-degree turns of the loops, which amounts to giving them a certain amount of bending rigidity.
Such a model was studied numerically in \cite{JK-flory}.

\subsection{Equivalence to the six-vertex model and Temperley--Lieb algebra}
\subsubsection{From FPL to six-vertex}\label{FPLequiv}
\begin{figure}
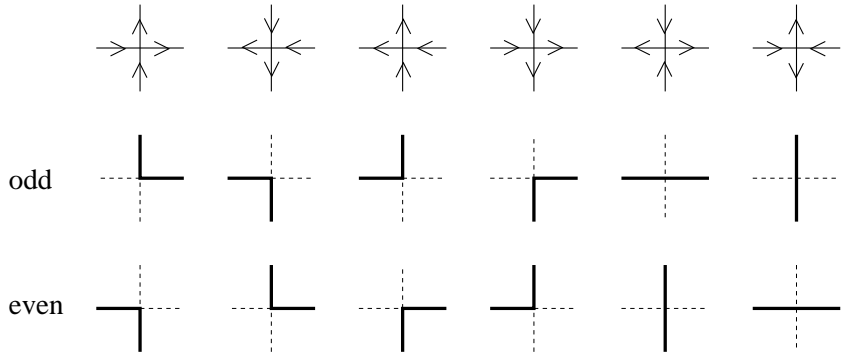

\hdrsixvertexfpl
\caption{From six-vertex to FPLs.}\label{fig6vfpl}
\end{figure}

The relation between the six-vertex model and the FPL model is rather limited, so we treat it first.
The limitation comes from the fact that one cannot assign an actual weight to the loops, so that we obtain
a $\tau=1$ model (with only local weights). 
The correspondence between configurations is one-to-one: starting from the six-vertex model side, 
one imposes that at every vertex, arrows pointing in the same direction should be in the same state (occupied
or empty) on the FPL side. This forces us to distinguish odd and even sub-lattices, and leads to the rules of Fig.~\ref{fig6vfpl}.

For rotational invariance of the FPL weights one should have $a=b$. 
$c/a$ then plays the role of rigidity parameter of the loops mentioned in section \ref{FPLdef}.

The rest of this section is devoted to the equivalence of CPL and six-vertex models.

\subsubsection{From CPL to six-vertex}\label{CPLequiv}
\begin{figure}
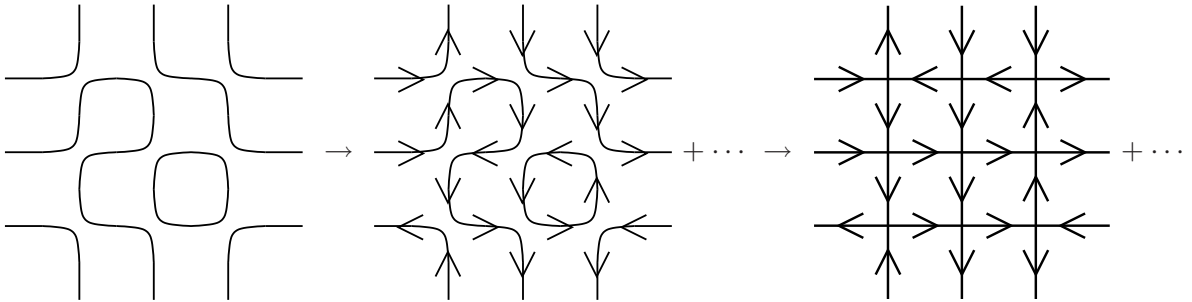

$\vcenterbox{\hdrunorientedloopconfiguration[1cm]}
\ \rightarrow\ 
\vcenterbox{\hdrorientedloopconfiguration[1cm]}
+\cdots
\ \rightarrow\ 
\vcenterbox{\hdrsixvertexconfiguration[1cm]}
+\cdots$
\caption{From CPLs to six-vertex.}\label{fig6vcpl}
\end{figure}
An example is shown on Fig.~\ref{fig6vcpl}.

Start from a CPL configuration. 
The (unoriented) loops carry a weight of $\tau$. A convenient way
to make the latter weight local is to turn unoriented loops into {\em oriented loops}: each configuration is now expanded
into $2^{\text{\# loops}}$ configurations with every possible orientation of the loops.
The weight of a 90-degree turn is chosen to be $\omega^{\pm 1/4}$, where $\tau=\omega+\omega^{-1}$.

Finally we forget about the original loops, retaining only the arrows. We note that the arrow conservation is
automatically satisfied around each vertex: we thus obtain one of the six-vertex configurations.
\begin{align*}
a
&=\vcenterbox{\hdrorientedturnone}
=\vcenterbox{\rotatebox{180}{\hdrorientedturnone}}
=u\\
b
&=\vcenterbox{\rotatebox{90}{\hdrorientedturnone}}
=\vcenterbox{\rotatebox{270}{\hdrorientedturnone}}
=1\\
c_1
&=\vcenterbox{\hdrorientedturntwo}
+\vcenterbox{\reflectbox{\hdrorientedturntwo}}
=u\,\omega^{1/2}+\omega^{-1/2}
\qquad
c_2
=\vcenterbox{\rotatebox{90}{\hdrorientedturntwo}}
+\vcenterbox{\rotatebox{90}{\reflectbox{\hdrorientedturntwo}}}
=\omega^{1/2}+u\,\omega^{-1/2}
\end{align*}
Note that if $u=1$ all weights become rotationally invariant and $a=b$, $c_1=c_2$.

Finally, one checks that the formula $\Delta=-\tau/2$ holds (equivalently $q=-\omega^{-1}$),
$u$ playing the role of spectral parameter. In particular the critical phase $|\Delta|<1$ corresponds to $|\tau|<2$.

{\em Remark}: this construction only works in the plane. On the cylinder or on the 
torus we have a problem: there are
non-contractible loops which according to the prescription above get a weight of $2$. 
This issue will reappear in the section \ref{secequiv} under the form of the twist.
It explains the discrepancy of central charges between the six-vertex model ($c=1$) and
CPL ($c<1$).

\subsubsection{Link patterns}\label{lpdef}
In order to understand this equivalence at the level of transfer matrices, one needs to introduce an appropriate space of states for the CPL model.
We now assume for simplicity that $L$ is even, $L=2n$.

Define a {\em link pattern}\/ of size $2n$ to be a
non-crossing pairing in a disk of $2n$ points lying on the boundary of the disk. Equivalently, we can map the disk to the upper half-plane
and ``flatten''
link patterns to pairings inside the upper half-plane of points on its boundary (a line). We shall switch from one description to the other
depending on what is more convenient. The points are labelled from $1$ to $2n$; in the half-plane, they are always ordered
from left to right, whereas in the disk the location of $1$ must be chosen, after which the labels increase counterclockwise.

Denote the set of link patterns of size $2n$ by $P_{2n}$. The number of such link
patterns is $c_n=\frac{(2n)!}{n!(n+1)!}$, the so-called Catalan number.

\goodbreak{\em Example:} in size $L=6$, there are 5 link patterns:
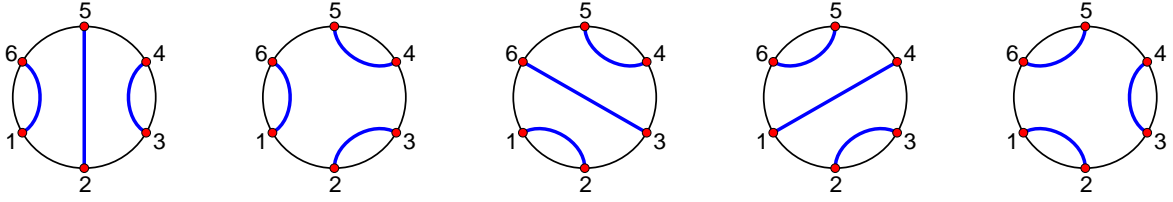
\[
\hdrlpthree{1}\qquad
\hdrlpthree{2}\qquad
\hdrlpthree{3}\qquad
\hdrlpthree{4}\qquad
\hdrlpthree{5}
\]

\subsubsection{Periodic boundary conditions and twist}\label{secequiv}
Suppose that we consider the CPL model with periodic boundary conditions in the horizontal direction, with a width of $L=2n$.
We can define a transfer matrix with indices living in the set of link patterns $P_{2n}$ as follows.
Consider appending a row of the CPL model to a link pattern; this way one produces a new link pattern:
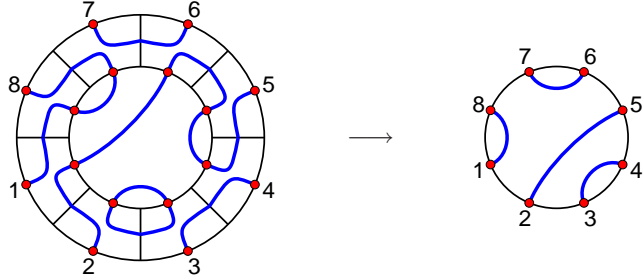
\[
\vcenterbox{\hdrtransferrowexample}
\qquad\longrightarrow\qquad
\vcenterbox{\hdrlpfour{4}}
\]
The transfer matrix $T_{\pi\pi'}$ is then the sum of weights of CPL rows such that the pattern $\pi'$ is turned into the pattern $\pi$.
The weights are calculated as follows: first one takes the product over each plaquette of the local weights; and then one multiplies
by $\tau$ to the power the number of loops that we have created.

What is the precise correspondence between the space of link patterns $\mathbb{C}P_{2n}$ and the space of spins $\mathbb{C}^{2^L}$ which
relates the transfer matrix of the six-vertex model and the newly defined one for the CPL model?

We start from the equivalence described in the section \ref{CPLequiv}. 
The basic idea is to orient the loops. So we start from a link pattern and add arrows to each
``loop'' (pairing of points). 
Forgetting about the original link pattern we obtain a collection of $2n$ up or down arrows, which form a state of the 6-vertex model in the transfer matrix
formalism. To assign weights it is convenient to think of the points as being on a straight line with the loops emerging perpendicularly: this way
each loop can only acquire a weight of $\omega^{\pm 1/2}$, depending on whether it is moving to the right or to the left. For example, in size $L=2n=4$,
\newcommand\ur[1]{\ \uparrow_{#1}\,}
\newcommand\dr[1]{\ \downarrow_{#1}\,}
\begin{align*}
\hdrlinepairnested&=
\omega\hdrlinepairnestedaa
&+&\hdrlinepairnestedai
&+&\hdrlinepairnestedia
&+&\omega^{-1}\hdrlinepairnestedii\\
&=\omega\ur1\ur2\dr3\dr4&+&\ur1\dr2\ur3\dr4&+&\dr1\ur2\dr3\ur4&+&\omega^{-1}\dr1\dr2\ur3\ur4\\
\hdrlinepairadjacent&=
\omega\hdrlinepairadjacentaa
&+&\hdrlinepairadjacentai
&+&\hdrlinepairadjacentia
&+&\omega^{-1}\hdrlinepairadjacentii\\
&=\omega\ur1\dr2\ur3\dr4&+&\ur1\dr2\dr3\ur4&+&\dr1\ur2\ur3\dr4&+&\omega^{-1}\dr1\ur2\dr3\ur4\\
\end{align*}
There is only one problem with this correspondence: it is not obviously compatible with periodic boundary conditions. We would like to identify a loop
from $i$ to $j$, $i<j$ and a loop from $j$ to $i+L$, $j<i+L$. This is only possible if we 
assume that $\uparrow_{i+L}=\omega \uparrow_i$, $\downarrow_{i+L}=\omega^{-1}\downarrow_i$,
i.e.,\ we impose {\em twisted boundary conditions}\/ on the six-vertex model. In the notations of \eqref{twistedtm} the twist
is $\Omega=\omega^{\sigma^z}$: it corresponds to an imaginary electric field.

This mapping from the space of link patterns (of dimension $c_n$) to that of sequences of arrows (of dimension $2^{2n}$) is
injective, so the space of link patterns is isomorphic to a certain subspace of $\mathbb{C}^{2^{2n}}$.
The claim, which we shall not prove in detail here but which is a natural consequence of the general formalism
is that the transfer matrix \eqref{twistedtm} of the six-vertex model with the twist defined above
leaves invariant this subspace and, once restricted to it, 
is identical to the transfer matrix of our loop model up to this isomorphism,
the correspondence of weights being the same as in section \ref{CPLequiv} (in particular, $\Delta=-\tau/2$).

The connection between CPL and six-vertex models is deep, in the sense that they are based on the same
algebraic structure, the Temperley--Lieb algebra and the associated solution of the Yang--Baxter equation,
but in different representations.
This is what we briefly discuss now.
These definitions will serve again when we study the quantum Knizhnik--Zamolodchikov equation (section \ref{secqkz}).

\newcommand{\TL}{${\rm TL}_L(\tau)$}
\subsubsection{Temperley--Lieb and Hecke algebras}\label{sectl}
The {\em Temperley--Lieb algebra}\/ of size $L$ and with parameter $\tau$ is given by generators $e_i$,
$i=1,\ldots,L-1$, and relations:
\begin{equation}\label{deftl}
e_i^2=\tau e_i\qquad e_i e_{i\pm 1}e_i=e_i\qquad e_i e_j=e_je_i\qquad |i-j|>1
\end{equation}
It is a quotient of the {\em Hecke algebra}, i.e.,\ the $e_i$ satisfy the less restrictive relations
\begin{equation}\label{defhecke}
e_i^2=\tau e_i\qquad e_i e_{i+1} e_i-e_i=e_{i+1}e_i e_{i+1}-e_{i+1}\qquad 
e_i e_j=e_j e_i\quad |i-j|>1
\end{equation}
Note that in Hecke (not Temperley--Lieb!), there is a symmetry $e_i \leftrightarrow \tau-e_i$.
The Hecke algebra is itself a quotient of the {\em braid group algebra}:
if $\tau=-(q+q^{-1})$ ($q$ is thus a free parameter),
then the $t_i=q^{-1/2}e_i+q^{1/2}$ satisfy the relations
\[
t_i t_{i+1}t_i = t_{i+1}t_i t_{i+1}\qquad t_i t_j = t_j t_i \quad |i-j|>1
\]

We are interested in two representations of the Temperley--Lieb algebra.

The first one is the representation on the ``space of spins'', that is the same space $(\mathbb{C}^2)^{\otimes L}$
of sequences of up/down arrows on which the six-vertex transfer matrix acts. It is given by making $e_i$
act on the $i^{\rm th}$ and $(i+1)^{\rm st}$ copies of $\mathbb{C}^2$ in the tensor product, with matrix
\begin{equation}\label{tlspin}
e_i=
\begin{pmatrix}0&0&0&0\\ 0&-q^{-1}&1&0\\ 0&1&-q&0\\ 0&0&0&0\end{pmatrix}
\end{equation}
It may be expressed in terms of Pauli matrices as
\begin{equation}\label{eipauli}
e_i={1\over2}\left(\sigma^x_i \sigma^x_{i+1}+\sigma^y_i\sigma^y_{i+1}
+\Delta (\sigma^z_i \sigma^z_{i+1}-1)+h(\sigma^z_{i+1}-\sigma^z_i)\right)
\end{equation}
where $h=\frac{1}{2}(q-q^{-1})$
and the $\sigma_i$ are the Pauli matrices at site $i$.

The second representation of Temperley--Lieb can be defined purely graphically.
We now assume as before that $L$ is even, $L=2n$.
In order to define the action of Temperley--Lieb generators $e_i$
on the space of link patterns $\mathbb{C}P_{2n}$
(vector space with canonical basis the $\ket{\pi}$ indexed by link patterns),
it is simpler
to view them graphically as $e_i=\vcenterbox{\hdrTLgenerator}$; then the relations
of the Temperley--Lieb algebra, as well as the representation on the space of link patterns, become
natural on the picture; for example, we find
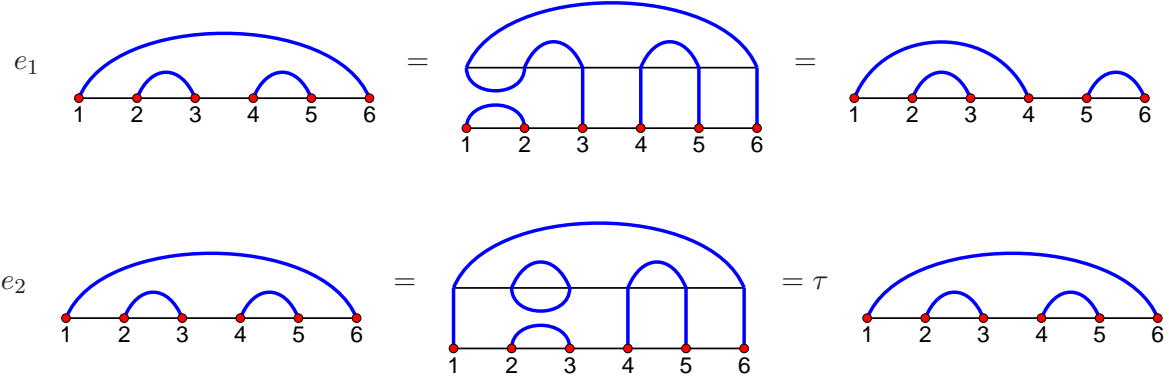
\[
e_1\
\vcenterbox{\hdrlinepthree{2}}
=
\vcenterbox{\hdrTLactionone}
=
\vcenterbox{\hdrlinepthree{4}}
\]
\[
e_2\
\vcenterbox{\hdrlinepthree{2}}
=
\vcenterbox{\hdrTLactiontwo}
=
\tau\ \vcenterbox{\hdrlinepthree{2}}
\]
As before, the role of the parameter $\tau$ is that each time a closed loop is formed, it can be erased at the price
of a multiplication by $\tau$.

The two representations we have just defined are of course related: the transformation of section \ref{secequiv} makes the representation
on link patterns a sub-representation of the one on spins.

Finally, we need to define {\em affine}\/ versions of Temperley--Lieb and Hecke algebras. 
It is convenient to do so by starting from their non-affine counterparts and adding
an extra generator $\rho$, as well as relations $\rho e_{i}=e_{i+1}\rho$, $i=1,\ldots,L-1$, and $\rho^L=1$.
Note that this allows us to define a new element $e_L=\rho e_{L-1}\rho^{-1}=\rho^{-1} e_1 \rho$ such that all defining relations
of the algebra become true modulo $L$. In fact, 
a more standard approach would be to introduce only $e_L$ and not $\rho$ itself. 
Adding $\rho$ leads to a slightly extended
affine Hecke/Temperley--Lieb algebra, which is more convenient for our purposes (see the discussion in \cite{Pas-scatt}).

In the link pattern representation, $\rho$ simply rotates link patterns counterclockwise:
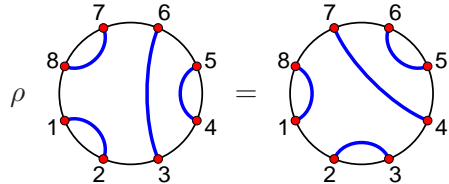
\[
\rho\ \vcenterbox{\hdrlpfour{12}}=\vcenterbox{\hdrlpfour{3}}
\]
In the spin representation, 
it rotates the factors of the tensor product forward one step and twists the last one (that moves to the first position)
by $\Omega$. Once again, for these two representations to be equivalent, one needs $\Omega$ to be of the form
$\Omega=\omega^{\sigma^z}$ where $\omega=-q$.

As an application, consider the Hamiltonian $H_L$, obtained as the logarithmic derivative
of the transfer matrix $T_L(x)$ of \eqref{tmparam} evaluated at $x=1$.
We find that with periodic boundary conditions, it is simply given up to additive and multiplicative constants by
\[
H_L=-\sum_{i=1}^L e_i
\]
In the spin representation, using \eqref{eipauli},
we recognize in $H_L$ the Hamiltonian of the {\em XXZ spin chain} (with the so-called ferromagnetic sign convention). So the Hamiltonian of the loop model, which has
the same form, is equivalent to the XXZ spin chain Hamiltonian, but with
twisted periodic boundary conditions, which in terms of Pauli matrices means that
$\sigma^\pm_{L+1}=\omega^{\pm 2} \sigma^\pm_1$.

\subsection{Some boundary observables for loop models}
Here we consider the CPL model at $\tau=1$ with some specific boundary conditions
which will play an important role since the observables we shall compute live at the boundary.
Several geometries are possible and lead to interesting combinatorial results \cite{dG-review}, 
but here we only consider the case of a cylinder.
\subsubsection{Loop model on the cylinder}\label{loopcyl}
\begin{figure}
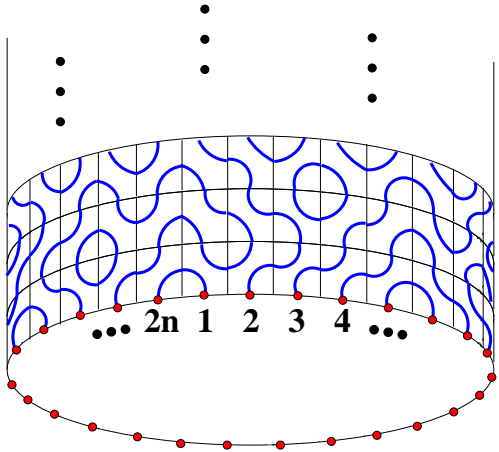

\hdrloopcylinder
\caption{The CPL model on a cylinder.}\label{figcplcyl}
\end{figure}

We consider the model of completely packed loops (CPL) on a semi-infinite cylinder with a finite even number of sites $L=2n$ around the cylinder,
see Fig.~\ref{figcplcyl}. It is convenient to draw the square lattice dual to the vertex lattice, so that the cylinder is divided into {\em plaquettes}.
Each plaquette can contain one of the two drawings $\vcenterbox{\hdrturnone}$ and $\vcenterbox{\hdrturntwo}$.

We furthermore set $\tau=1$ (or $q=\e{2\pi i/3}$), that is we do not put any weights on the loops. There are no more non-local weights, 
and in fact plaquettes are independent of
each other. So we can reformulate this model as a purely probabilistic model, in which one draws independently at random each plaquette,
with say probability $p$ for $\vcenterbox{\hdrturnone}$ and $1-p$ for $\vcenterbox{\hdrturntwo}$.\looseness=-1

Finally, we define the observables we are interested in. We consider the {\em connectivity}\/ of the boundary points, i.e.,\ the endpoints of loops (which are
in this case not loops but paths) lying on the bottom circle. We encode them into link patterns (see section \ref{lpdef}).
In the present context, they can be visualized as follows. Project the cylinder onto a disk in such a way that the boundaries coincide and the infinity is
somewhere inside the disk. Remove all loops except the boundary paths. Up to deformation of these resulting paths, what one obtains is a 
link pattern. The probabilities of occurrence of the various link patterns can be encoded as one vector with $c_n$ entries:
\[
\Psi_L=\sum_{\pi\in P_{2n}} \Psi_\pi \ket{\pi}
\]
where $P_{2n}$ is the set of link patterns of size $2n$ and $\Psi_\pi$ is the probability of link pattern $\pi$.

\subsubsection{Markov process on link patterns}
We now show that $\Psi_L$ can be reinterpreted as the steady state of a Markov process on link patterns.
This is easily understood by considering a transfer matrix formulation of the model. As in section \ref{secequiv}, let us introduce the transfer matrix:
it corresponds here to creating one extra row
to the semi-infinite cylinder, and encoding not the actual plaquettes but the effect of the new plaquettes on the connectivity
of the endpoints. That is, $T_{\pi,\pi'}(p)$ is
the probability that starting from a configuration of the cylinder whose endpoints are connected via the link pattern 
$\pi'$ and adding a row of plaquettes, one obtains a new configuration whose endpoints are connected via the link pattern $\pi$. This forms a $c_n\times c_n$ matrix $T_L(p)$.

This transfer matrix is actually stochastic in the sense that
\begin{equation}\label{consproba}
\sum_{\pi\in P_{2n}} T_{\pi,\pi'}(p)=1\qquad \forall \pi'
\end{equation}
which expresses the conservation of probability. This is of course a special feature of the transfer matrix at $\tau=1$.
Note that \eqref{consproba} says that $T_L(p)^T$ has eigenvector $(1,\ldots,1)$ with eigenvalue $1$.

The matrix $T_L(p)$ has non-negative entries; it is easy to show that it is primitive (the entries of $T_L(p)^n$ are positive). These are the hypotheses
of the Perron--Frobenius theorem. Therefore, $T_L(p)$ possesses a unique eigenvector $\Psi_L$ with positive entries; the corresponding eigenvalue is positive and
is larger in modulus than all other eigenvalues.
Now the theorem also applies to $T_L(p)^T$ and by uniqueness we conclude that the largest eigenvalue of $T_L(p)$ and of $T_L(p)^T$ is $1$. 
In conclusion, we find that the eigenvector with positive entries of $T_L(p)$, which with a bit of foresight we call $\Psi_L$ again, satisfies
\begin{equation}\label{sseq}
T_L(p)\Psi_L=\Psi_L
\end{equation}
(In fact the whole reasoning in the previous paragraph is completely
general and applies to any Markov process, $\Psi_L$ being up to normalization the steady state of the Markov process defined by $T_L(p)$.)

Two more observations are needed. Firstly, \eqref{sseq} is clearly satisfied by the vector of probabilities that we defined in the previous paragraph (the
semi-infinite cylinder being invariant by addition of one extra row);
it is in fact defined uniquely up to normalization by \eqref{sseq}. This explains that we have used the same notation.

Secondly, $\Psi_L$ is in fact independent of $p$. The easiest way to see this is to note that
$p$ now plays the role of spectral parameter (explicitly, with the conventions of the six-vertex $R$-matrix,
$p=\frac{q\,x-q^{-1}x^{-1}}{q\,x^{-1}-q^{-1}x}$). In particular we conclude
that, as in the six-vertex model, $[T_L(p),T_L(p')]=0$, so that these transfer matrices have a common Perron--Frobenius eigenvector.

\subsubsection{Properties of the steady state: some empirical observations}\label{secobs}
We begin with an example in size $L=2n=8$. By brute force diagonalization of the stochastic matrix $T_L$ one obtains the vector $\Psi_L$ of probabilities:
\begin{align*}
\Psi_8=
\frac{1}{42}&\left(
\vcenterbox{\hdrlpfour{1}}
+\vcenterbox{\hdrlpfour{4}}
+\vcenterbox{\hdrlpfour{8}}
+\vcenterbox{\hdrlpfour{10}}
\right)\\
+\frac{3}{42}&\left(
\vcenterbox{\hdrlpfour{2}}
+\vcenterbox{\hdrlpfour{7}}
+\vcenterbox{\hdrlpfour{13}}
+\vcenterbox{\hdrlpfour{5}}
\right.\\
&\left.+\vcenterbox{\hdrlpfour{12}}
+\vcenterbox{\hdrlpfour{3}}
+\vcenterbox{\hdrlpfour{9}}
+\vcenterbox{\hdrlpfour{11}}
\right)\\
+\frac{7}{42}&\left(
\vcenterbox{\hdrlpfour{6}}
+\vcenterbox{\hdrlpfour{14}}
\right)
\end{align*}
We recognize some of our favorite numbers $A_n$, namely 7 and 42.

In fact, Batchelor, de Gier and Nienhuis \cite{BdGN-XXZ-ASM-PP} 
conjectured the following properties for all system sizes $L=2n$:
\begin{enumerate}
\item The smallest probability is $1/A_n$, and corresponds to link patterns with all parallel pairings.
\item All probabilities are integer multiples of the smallest probability. 
\item The largest probability is $A_{n-1}/A_n$, and corresponds to link patterns which pair nearest neighbors.
\end{enumerate}
By now all these properties have been proven \cite{artic31,artic41}, as will be discussed in section \ref{secqkz}.

\subsubsection{The general conjecture}
A question however remains: according to property 2 above, if one multiplies the probabilities by $A_n$, we obtain a collection of integers.
The smallest one is $1$ and the largest one is $A_{n-1}$, but what can we say about the other ones?

Recall that $A_n$ also counts the number of six-vertex model configurations with DWBC.
Furthermore, we showed that there is a one-to-one correspondence between six-vertex model configurations 
and FPL configurations (cf.\ section \ref{FPLequiv}). 
In this correspondence, the DWBC become alternating occupied and empty external edges for FPL configurations (in short, FPLs).
Let us now draw explicitly the 42 FPLs of size $4\times 4$:
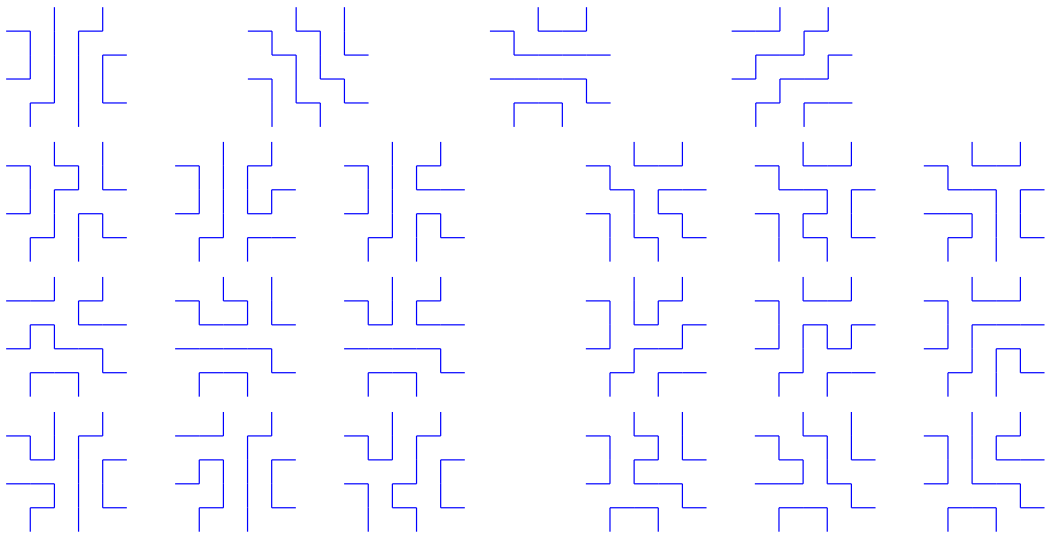
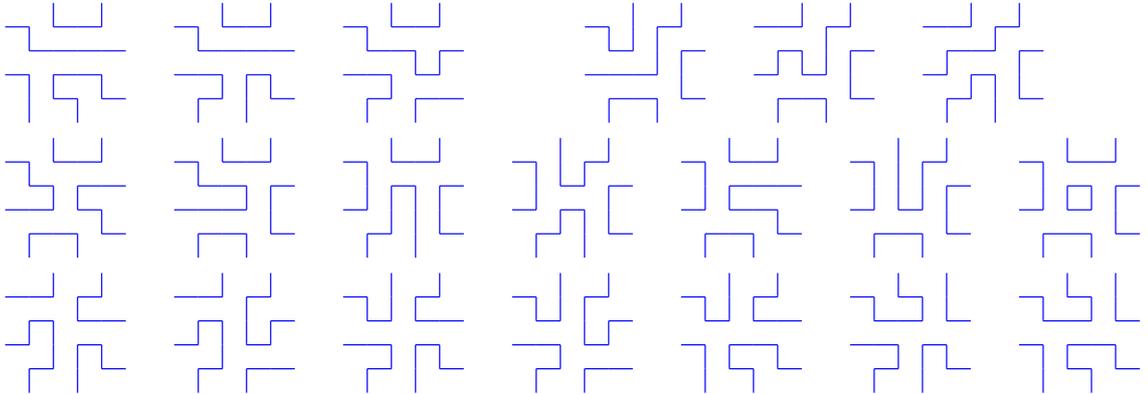
\begin{align*}
&\vcenterbox{\hdrfplfourgroup{1}}
\qquad\qquad
\vcenterbox{\hdrfplfourgroup{4}}
\qquad\qquad
\vcenterbox{\hdrfplfourgroup{8}}
\qquad\qquad
\vcenterbox{\hdrfplfourgroup{10}}
\\
&\vcenterbox{\hdrfplfourgroup{2}}
\qquad\qquad
\vcenterbox{\hdrfplfourgroup{7}}
\\
&\vcenterbox{\hdrfplfourgroup{13}}
\qquad\qquad
\vcenterbox{\hdrfplfourgroup{5}}
\\
&\vcenterbox{\hdrfplfourgroup{12}}
\qquad\qquad
\vcenterbox{\hdrfplfourgroup{3}}
\\
&\vcenterbox{\hdrfplfourgroup{9}}
\qquad\qquad
\vcenterbox{\hdrfplfourgroup{11}}
\\
&\vcenterbox{\hdrfplfourgroup{6}}
\\
&\vcenterbox{\hdrfplfourgroup{14}}
\end{align*}

Interestingly, we find that the reformulation in terms of FPLs allows us to introduce once again a notion of connectivity. Indeed, there are $2n$ occupied edges
on the exterior square and they are paired by the FPL. We can therefore count separately FPLs with a given link pattern $\pi$; let us denote this by $A_\pi$.
The Razumov--Stroganov conjecture \cite{RS-conj} then states that
\[
\Psi_\pi=\frac{A_\pi}{A_n}
\]
thus relating two different models of loops (CPL and FPL) with completely different boundary conditions. And even though both models are equivalent to the six-vertex 
model, the values of $\Delta$ are also different (they differ by a sign).

The Razumov--Stroganov conjecture remains open, though some special cases have been proved, see \ref{fewlittle} below.

The relation to the conjectured properties of the previous section is as follows.
It is easy to show that if $\pi$ is a link pattern with all pairings parallel, then there exists a unique
FPL configuration with connectivity $\pi$. Thus the RS conjecture implies property (1). Furthermore, since all $A_\pi$
are integer, it obviously implies property (2). Property (3) however remains non-trivial, since even assuming the
RS conjecture it amounts to saying that $A_\pi=A_{n-1}$ in the case of the two link patterns $\pi$ that pair 
nearest neighbors, which has not been proven.

\section{The quantum Knizhnik--Zamolodchikov equation}\label{secqkz}
We now introduce a new equation whose solution will roughly correspond to a double generalization
of the ground state eigenvector $\Psi_L$ of the loop model introduced above:
(i) it contains inhomogeneities and (ii) it is a continuation of the original eigenvector to an arbitrary
value of $q$, the original value being $q=\e{2\pi i/3}$.

\newcommand\cR{{\check R}}
\subsection{Basics}
\subsubsection{The \texorpdfstring{$q$}{q}KZ system}\label{qkzsys}
Let $L=2n$ be an even positive integer.
Introduce once again the $R$-matrix, but this time rotated by 45 degrees and acting a bit differently than before.
Namely, it acts on the vector space $\mathbb{C}P_{2n}$ spanned by link patterns, in the following way:
\begin{equation}\label{rmat}
\cR_i(z)= \phi(z)
\frac{(q^{-1}-q z) +(1-z)e_i}{q^{-1}z-q}
\qquad i=1,\ldots,L-1
\end{equation}
where we refer to section \ref{sectl} for the definition of $e_i$.
$\phi(z)$ is a scalar function that we do not need to specify explicitly here,
see section \ref{normR} for a discussion.
Redrawing slightly the latter as 
$e_i=\vcenterbox{\rotatebox{45}{\hdrturnone}}$,
and similarly the identity as 
$1=\vcenterbox{\rotatebox{45}{\hdrturntwo}}$,
we recognize the two (rotated) CPL plaquettes. Note that in this section, it is convenient to use
spectral parameters $z$ that are the {\em squares}\/ of our old spectral parameters $x$.
Indeed, using the equivalence to the six-vertex model described
in section \ref{secequiv}, which amounts to the representation \eqref{tlspin}
for the Temperley--Lieb generators (acting on the $i^{\rm th}$ and $(i+1)^{\rm st}$ spaces),
we essentially recover the $R$-matrix of the six-vertex model, cf.\ (\ref{6vR},\ref{6vwei}),
after the change of variables $z=x^2$:
\[
\check R(x)=\frac{\phi(x^2)}{q\,x^{-1}-q^{-1}x}
\begin{pmatrix}
q\,x-q^{-1}x^{-1}&0&0&0\\
0&(q-q^{-1})x&x-x^{-1}&0\\
0&x-x^{-1}&(q-q^{-1})x^{-1}&0\\
0&0&0&q\,x-q^{-1}x^{-1}
\end{pmatrix}
\]
provided one performs the following transformation:
$\check R(z)\propto\mathcal{P} x^{\kappa/2}R(x)x^{-\kappa/2}$ where $\mathcal{P}$ permutes the factors of the tensor product,
and $\kappa={\rm diag}(0,-1,1,0)$.

The Yang--Baxter equation can be rewritten as follows:
\[
\check R_i(z) \check R_{i+1}(z\,w) \check R_i(w)
=
\check R_{i+1}(w) \check R_{i}(z\,w) \check R_{i+1}(z)
\qquad
i=1,\ldots,L-2
\] 

We also require that $\phi(z)\phi(1/z)=1$, so that $\check R_i$ satisfies
the unitary equation:
\[
\check R_i(z)\check R_i(1/z)=1
\]

Consider now a vector-valued function $\Psi_L$, depending on the variables
$z_1,\ldots,z_L$ and the parameters $q,q^{-1}$. It satisfies the following
system of equations for $i=1,\ldots,L-1$:
\begin{align}
\cR_i(z_{i+1}/z_i)\Psi_L(z_1,\ldots,z_L)&=
\Psi_L(z_1,\ldots,z_{i+1},z_i,\ldots,z_L)\label{qkza}\\
\rho^{-1} \Psi_L(z_1,\ldots,z_L)&=\kappa\,\Psi_L(z_2,\ldots,z_L,s\,z_1)\label{qkzb}
\end{align}
where it is recalled that $\rho$ rotates link patterns counterclockwise,\footnote{We use $\rho^{-1}$ in the l.h.s.\ because
$(\rho^{-1}\Psi_L)_\pi=\Psi_{\rho(\pi)}$ !}
and $\kappa$ is a constant that is needed for homogeneity. $s$ is a parameter of the equation:
if one sets $s=q^{2(k+\ell)}$ with $k=2$ (technically, this is the dual Coxeter number of the underlying 
quantum group $U_q(\widehat{\mathfrak{sl}(2)})$),
then $\ell$ is called the {\em level} of the $q$KZ equation. \eqref{qkza} is the {\em exchange}\/ equation, which is ubiquitous
in integrable models. \eqref{qkzb} has something to do with our (periodic) boundary conditions, 
or equivalently with an appropriate affinization of the underlying algebra.

This system of equations first appeared in \cite{Smi} in the study of form factors in integrable models.
It is not what is usually called the quantum Knizhnik--Zamolodchikov ($q$KZ) equation; the latter was
introduced in the seminal paper \cite{FR-qKZ} as a $q$-deformation of the Knizhnik--Zamolodchikov (KZ) equation
($q$KZ is to quantum affine algebras what KZ is to affine algebras). 
The usual $q$KZ equation involves the operators $S_i$, which can be defined pictorially as
\[
S_i=\cdots
\vcenterbox{\hdrblankplaquette}
\vcenterbox{\hdrblankplaquette}
\vcenterboxlabel{\hdrturntwo}{$i$}
\vcenterbox{\hdrblankplaquette}
\vcenterbox{\hdrblankplaquette}
\cdots
\]
where the empty box is just the ``face'' 
graphical representation of the $R$-matrix (dual to the
``vertex'' representation we have used for the six-vertex model):
\[
\vcenterbox{\hdrblankplaquette}=\frac{q^{-1}-q\,z}{q^{-1}z-q}
\vcenterbox{\hdrturnone}
+\frac{1-z}{q^{-1}z-q}
\vcenterbox{\hdrturntwo}
\]
and the spectral parameters $z$ to be used in $S_i$ are as follows: for the box numbered $j$,
$z_j/z_i$ if $j>i$ or $z_j/(s\,z_i)$ if $j<i$. Loosely, $S_i$ is the ``scattering matrix for the $i^{\rm th}$ particle''.

Alternatively, $S_i$ can be expressed as a product of $\cR_i$ and $\rho$:
\begin{equation}\label{defS}
S_i=
\cR_{i-1}(z_{i-1}/(s\,z_i))
\cdots
\cR_2(z_2/(s\,z_i))
\cR_1(z_1/(s\,z_i))
\rho
\cR_{L-1}(z_L/z_i)
\cdots
\cR_{i+1}(z_{i+2}/z_i)
\cR_i(z_{i+1}/z_i)
\end{equation}

The quantum Knizhnik--Zamolodchikov equation can then be written
\begin{equation}\label{qkz-fr}
S_i(z_1,\ldots,z_L) \Psi_L(z_1,\ldots,z_i,\ldots,z_L)=\Psi_L(z_1,\ldots,s\,z_i,\ldots,z_L)
\end{equation}

It is a simple exercise to check using \eqref{defS} that the system (\ref{qkza}--\ref{qkzb}) 
implies the $q$KZ equation \eqref{qkz-fr}. Naively, the converse is untrue. However one can show that, 
up to some linear recombinations,
a complete set of (meromorphic) solutions of \eqref{qkz-fr} can always be reduced to a complete
set of solutions of (\ref{qkza}--\ref{qkzb}), see \cite{Che-qKZ}.

\subsubsection{Normalization of the \texorpdfstring{$R$}{R}-matrix}\label{normR}
Usually, the normalization $\phi(z)$ of the
$R$-matrix is chosen such that it satisfies an additional constraint,
the crossing symmetry -- which amounts
to the identity $\phi(q^{-2}z^{-1})=\phi(z)\frac{q^2z-q^{-2}}{q^{-1}z-q}$ --
as well as a requirement of analyticity.
Explicitly, for $|q|<1$ there is a unique choice which is analytic 
in $x=z^{1/2}$ in the region $|q|<|x|<|q|^{-1}$, namely
\begin{equation}\label{truenorm}
\phi(z)=z^{1/2}\frac{(q^4/z;q^4)_\infty(q^6z;q^4)_\infty}{(q^4 z;q^4)_\infty(q^6/z;q^4)_\infty}
\end{equation}
This normalization also corresponds to setting the partition function per site
to be $1$. There is also a different formula for $|q|=1$ \cite{JM-gapless} which gives $\phi=1$ at $q=e^{2\pi i/3}$.

In the context of the $q$KZ equation, there are correlated choices
of normalization of $\check R$ and $\Psi$.
Indeed, consider a solution $\Psi$ of the system
(\ref{qkza},\ref{qkzb}) and redefine
\[
\tilde\Psi=\prod_{1\le i<j\le L} f(z_j/z_i)\ \Psi
\]
where $f(z)$ is some scalar function. Then $\tilde\Psi$ satisfies the same system of equations:
the first equation remains formally identical,
but with the normalization of the $R$-matrix modified from $\phi(z)$ to 
$\tilde\phi(z)=\phi(z)f(z)/f(1/z)$. For the second equation to remain true
imposes a constraint on the function $f(z)$, so that we find
\begin{align*}
f(z)/f(1/z)&=\tilde\phi(z)/\phi(z)\\
f(s\,z)&=f(1/z)
\end{align*}
With reasonable analyticity conditions on $\tilde\phi/\phi$, this system
of equations can be solved (e.g.,\ if $|s|<1$, writing $\tilde\phi(z)/\phi(z)=f_0(z)/f_0(1/z)$, one finds $f(z)=(\prod_{i=0}^\infty f_0(s^i/z)\prod_{i=1}^\infty f_0(s^iz))^{-1}$). 

In what follows we shall only be interested in homogeneous {\em polynomial}\/ solutions of the system (\ref{qkza}--\ref{qkzb}).
This implies the choice of the normalization $\phi=1$, 
which we shall adopt from now on.

\subsubsection{Relation to affine Hecke algebra}
The role of the
affine Hecke algebra (and even double affine Hecke algebra) was emphasized in the work
of Cherednik \cite{Che} in relation to Macdonald polynomials and the quantum 
Knizhnik--Zamolodchikov equation, and in the work of Pasquier \cite{Pas-scatt} on integrable models.
In particular, its use in the context of the
Razumov--Stroganov conjecture was advocated by Pasquier \cite{Pas-RS}.

Start from the $q$KZ system (\ref{qkza}--\ref{qkzb}), and rewrite it in such a way that the action
on the finite-dimensional part (on the space of link patterns) is separated from the action on the variables
(on the space of polynomials of $L$ variables). Start with \eqref{qkza}; after simple rearrangements it becomes
\begin{equation}\label{qkzc}
(\tau-e_i) \Psi_L
=
-(q\,z_i-q^{-1}z_{i+1})
\partial_i \Psi_L
\end{equation}
where $\partial_i\equiv {1\over z_{i+1}-z_i}(s_i - 1)$ and $s_i$
is the operator that switches variables $z_i$ and $z_{i+1}$, so that the l.h.s.\ only acts on
link patterns, whereas the r.h.s.\ only acts on the polynomial part of $\Psi_L$.

The operators $\tau-\hat e_i:=-(q\,z_i-q^{-1}z_{i+1})\der_i$ acting on the space of polynomials 
form a representation of the Hecke algebra (with parameter $\tau$); 
i.e.,\ they satisfy the relations of \eqref{defhecke}. Equivalently, the $\hat e_i$ satisfy them and form a second set of generators.

As to \eqref{qkzb}, it is already in a separated form; 
and one can add an extra operator on the space of polynomials, the one that appears in the r.h.s.\ of \eqref{qkzb}, that is
the cyclic shift $\hat\rho$ of spectral parameters $z_1\mapsto z_2\mapsto\cdots\mapsto z_L\mapsto s\, z_1$ (with the multiplication by $\kappa$ included).
The $\hat e_i$ together with $\hat\rho$ generate a representation of the affine Hecke algebra. We can write formally:
\begin{align}\label{qkzd}
e_i \Psi_L &= \hat e_i \Psi_L
\\
\rho^{-1} \Psi_L &= \hat\rho \Psi_L\label{qkze}
\end{align}

We can now interpret (\ref{qkzd},\ref{qkze}) as follows: we have on the one hand a representation
of the affine Hecke algebra on $\mathbb{C}P_{2n}$, the space of link patterns (with generators $e_i$ and $\rho$);
and on the other hand a representation of the same algebra on $\mathbb{C}[z_1,\ldots,z_L]$, 
the polynomials of $L$ variables (the $\hat e_i$ and $\hat\rho$).
$\Psi_L$ provides a bridge between these two representations: it is essentially an invariant object in the
tensor product of the two, that is it provides a sub-representation of the space of polynomials (explicitly,
the span of the $\Psi_\pi$) which is isomorphic to the {\em dual}\/ of the
representation on the link patterns. By dual we mean, defined by composing
the anti-isomorphism that sends the $e_i$ to themselves and $\rho$ to
$\rho^{-1}$ (but reverses the order of the products) and transposition. 
The role of the anti-isomorphism is most evident when we try to compose equations (\ref{qkzd},\ref{qkze}) 
and find that $\hat e_i \hat e_j \Psi = e_j e_i \Psi$ (in fact, it is often advocated to make one of the sets of 
operators act on the right, though we find the notation too cumbersome to use here).

So the search for polynomial solutions of (\ref{qkza},\ref{qkzb}) is equivalent to finding certain irreducible
sub-representations of the action of affine Hecke on the space of polynomials.

{\em Remark}: the direct relation between $q$KZ and representations of an appropriate affine algebra
only works for the $A_n$ series of algebras, i.e.,\ affine Hecke. For more complicated situations such as 
the Birman--Wenzl--Murakami (BWM) 
algebra, it does not work so well because one cannot separate the two different actions \cite{Pas-BWM}.

\subsection{Construction of the solution}\label{preconstrsol}
On general grounds, we only expect polynomial solutions for integer values of the level.
Here we shall only need a solution at level $1$, that is $s=q^6$.

Note that iterating \eqref{qkzb} $L$ times implies that $\kappa^L s^d=1$ where $d$ is the degree of $\Psi$.

\subsubsection{\texorpdfstring{$q$}{q}KZ as a triangular system}\label{constrsol} 
\begin{figure}
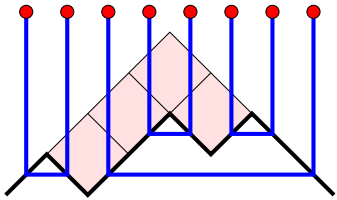

\hdrlinkpatterntoyoung
\caption{From a Young diagram to a link pattern: in this example, from the partition $(2,1,1)$ to the pairings
$(1,2)$, $(3,8)$, $(4,5)$, $(6,7)$.}\label{figlptoyoung}
\end{figure}

We shall build this solution in several steps. First,
we use a ``nice'' property of our basis of link patterns, that is, the fact
that \eqref{qkzc} can be written as a {\em triangular}\/ linear system in the components of $\Psi_L$.
This requires to define an order on link patterns, which is most conveniently described as follows.
Draw once again link patterns as pairings of points on a line and consider the operation
described on Fig.~\ref{figlptoyoung}. It gives a bijection between link patterns of size $2n$ and Young diagrams inside
the staircase diagram $1^n=(n-1,n-2,\ldots,1)$. Then the order is inclusion of Young diagrams.
The smallest element, corresponding to the empty Young diagram, is denoted by $0^n$; it connects
$i$ and $2n+1-i$ (note that it is one of the link patterns with all pairings parallel, which correspond
to the smallest probability $1/A_n$ in the loop model). 
Consider now the exchange equation \eqref{qkzc} and write it in components; we find
two possibilities:
\begin{itemize}
\item $i$ and $i+1$ are not paired. Then we find that \eqref{qkzc} only involves $\Psi_\pi$, and implies
that $q\,z_i-q^{-1}z_{i+1}$ divides $\Psi_\pi$, and furthermore $\Psi_\pi/(q\,z_i-q^{-1}z_{i+1})$ is symmetric
in the exchange of $z_i$ and $z_{i+1}$.
\item $i$ and $i+1$ are paired. Then
\[
(q\,z_i-q^{-1}z_{i+1})\der_i \Psi_\pi 
= \sum_{\pi'\ne \pi, e_i \cdot\pi'=\pi} \Psi_{\pi'}
\]
that is it involves the sum over preimages of $\pi$ by $e_i$ viewed as acting on the set of link patterns.
It turns out there are two types of preimages of a given $\pi$: in terms of Young diagrams, there is the Young
diagram obtained from $\pi$ by adding one box at $i,i+1$ (which is always possible
unless $\pi$ is the largest element); and there are other Young diagrams that are included
in $\pi$. So we can write the equation
\begin{equation}\label{trisys}
\Psi_{\pi+\text{one box at }(i,i+1)}=(q\,z_i-q^{-1}z_{i+1})\der_i \Psi_\pi -\sum_{\substack{\pi'\subset\pi\\ e_i\cdot\pi'=\pi}}\Psi_{\pi'}
\end{equation}
which has the desired triangular structure and allows us to build the $\Psi_\pi$ one by one by adding boxes
to the corresponding Young diagram. However there is no equation for $\Psi_{0^n}$.
In fact this triangular system can be explicitly solved \cite{KL-KL} (see also \cite{dGP-factor}) in the sense that
every $\Psi_\pi$ can be written as a series of operators acting on $\Psi_{0^n}$. We shall not use this explicit solution here.
\end{itemize}

From the discussion above, we find that all we need is to fix $\Psi_{0^n}$.
We use the following simple observation, which generalizes the first case in the dichotomy above:
\begin{quote}
If there are no pairings between points $i,i+1,\ldots,j$ in $\pi$, then
$\prod_{i\le p<q\le j} (q\, z_p - q^{-1} z_q)$ divides $\Psi_\pi$.
\end{quote}
(the proof is by induction on $j-i$).

We now make the ``minimality'' assumption that in the case of $0^n$, these factors
form all of $\Psi_{0^n}$, so that we find
\begin{equation}\label{basecase}
\Psi_{0^n}=\prod_{1\le i<j\le n} (q\, z_i-q^{-1}z_j)
\prod_{n+1\le i<j\le 2n} (q^{-1}z_j-q\, z_i)
\end{equation}
where we recall that the system size is $L=2n$.

If a solution of the $q$KZ system with such a $\Psi_{0^n}$ exists, then it will be the {\em only}\/ solution of degree $n(n-1)$
(up to multiplication by a constant).
It remains however a non-trivial fact that with such a choice of $\Psi_{0^n}$, the construction above is consistent
(i.e.,\ independent of the order in which one adds boxes), and that
\eqref{qkzb} is satisfied, with $s=q^6$. This is the subject of the next section. 

\goodbreak{\em Example:} we find in size $L=6$
\begin{align*}
\Psi_{\smash{\hdrminilpthree{1}}}={}&
-\frac{\left(q^2 z_1-z_2\right) \left(q^2 z_1-z_3\right) \left(q^2 z_2-z_3\right)}{q^6}\\[-0.1cm]
&\quad\times \left(q^2 z_4-z_5\right) \left(q^2 z_4-z_6\right) \left(q^2 z_5-z_6\right)\\[0.5cm]
\Psi_{\smash{\hdrminilpthree{2}}}=&
\frac{\left(q^2 z_1-z_2\right) \left(q^2 z_3-z_4\right) \left(q^2 z_5-z_6\right)}{q^7}
 (z_1 z_2 z_3 q^8+z_1 z_2 z_4 q^8\\
&\quad{}-z_1 z_3 z_4 q^6-z_2 z_3 z_4 q^6-z_1 z_2 z_5 q^6-z_1 z_2 z_6 q^6
+z_3 z_4 z_5 q^2+z_3 z_4 z_6 q^2\\
&\quad{}+z_1 z_5 z_6 q^2+z_2 z_5 z_6 q^2-z_3 z_5 z_6-z_4 z_5 z_6)\\[0.3cm]
\Psi_{\smash{\hdrminilpthree{3}}}={}&
-\frac{\left(q^2 z_2-z_3\right) \left(q^2 z_2-z_4\right) \left(q^2 z_3-z_4\right)}{q^8}\\[-0.1cm]
&\quad\times \left(q^4 z_1-z_5\right) \left(q^4 z_1-z_6\right) \left(q^2 z_5-z_6\right)\\[0.5cm]
\Psi_{\smash{\hdrminilpthree{4}}}={}&
-\frac{\left(q^2 z_1-z_2\right) \left(q^2 z_3-z_4\right) \left(q^2 z_3-z_5\right)}{q^8}\\[-0.1cm]
&\quad\times \left(q^2 z_4-z_5\right) \left(q^4 z_1-z_6\right) \left(q^4 z_2-z_6\right)\\[0.5cm]
\Psi_{\smash{\hdrminilpthree{5}}}=&
\frac{\left(q^2 z_2-z_3\right) \left(q^2 z_4-z_5\right) \left(q^4 z_1-z_6\right)} {q^9}
(z_1 z_2 z_3 q^{10}-z_2 z_3 z_4 q^6-z_2 z_3 z_5 q^6\\
&\quad{}-z_1 z_4 z_5 q^6-z_1 z_2 z_6 q^6-z_1 z_3 z_6 q^6
+z_2 z_4 z_5 q^4+z_3 z_4 z_5 q^4\\
&\quad{}+z_2 z_3 z_6 q^4+z_1 z_4 z_6 q^4+z_1 z_5 z_6 q^4-z_4 z_5 z_6)
\\
\end{align*}

\subsubsection{Consistency and Jucys--Murphy elements}
Let us go back to the $q$KZ system (\ref{qkzd},\ref{qkze}).
In order to show that this system of equations is consistent, one must simply check 
that one obtains the same representation of affine Hecke on both sides (up to duality). 
Thus, we shall now check that the center has the same value in both representations.
This will fix $\kappa$ and more importantly $s$. 

Let us first define a new set of elements of the affine Hecke algebra
(Jucys--Murphy elements), namely
\[
y_i=t_i\cdots t_{L-1}\rho^{-1} t_1^{-1}\cdots t_{i-1}^{-1}\qquad i=1,\ldots,L
\]
For this section we shall use ``knot-theoretic'' drawings to explain the equations we write.
It is easy to see that if the $t_i=q^{-1/2}e_i+q^{1/2}$ are crossings in the knot theoretic sense
(this definition being itself the skein relation for the Jones polynomial), 
then the $y_i$ have an equally simple description:
\[
t_i=\vcenterbox{\hdraffinecrossing}
\qquad
y_i=\vcenterbox{\hdraffineyi}
\]
where the second picture is embedded in a strip that is periodic in the horizontal direction.
Then it becomes graphically clear that the $y_i$ commute:
\[
y_i y_j = \vcenterbox{\hdraffinecommuting}=y_j y_i
\]

One can also show that the symmetric polynomials of the $Y_i=-q^{5/2-i} y_i$ form the center of the affine Hecke algebra.\footnote{This extra factor
of $q^i$ comes from our normalization of the $t_i$. In the Hecke algebra they have no preferred normalization, and one usually chooses it
to make this factor disappear, but in Temperley--Lieb there is one (which makes the skein relation rotationally invariant), 
that we have chosen here, and it is unfortunately different. The extra constant is so that $Y_1=1$.}
Thus, assuming that we have an irreducible representation 
such that we can diagonalize the action of the $Y_i$, the (unordered) set of their $L$ eigenvalues
is independent of the eigenvector and characterizes the representation.
\footnote{In fact, for an irrep given by a Young diagram $\mu$, 
the eigenvalues of the $Y_i$ are the $q^{-2c}$ where $c$ runs over
the contents of the boxes of $\mu$.
Here one can check that $\mu$ is the rectangle with 2 rows and
$n$ columns.}

Let us now apply this to our two representations. First, let us make the $y_i$ act on link patterns. They have an obvious common eigenvector:
the link pattern $1^n$ which pairs neighbors $2i-1$ and $2i$, $i=1,\ldots,n$. For example for $i=3$:
\[
y_i 1^n=\vcenterbox{\hdraffineyeigenvector}=-q^{-3/2}\vcenterbox{\hdrlpfour{14}}
\]
i.e.,\ the action of $y_i$ is equivalent to a Reidemeister move I (with negative orientation for $i$ even and positive orientation for $i$ odd).
A move I multiplies the Jones polynomial by $-q^{-3/2}$ (this amounts to the calculation $t_i e_i=-q^{-3/2}e_i$), so that we obtain
\begin{equation}\label{JMa}
y_i 1^n=-q^{(-1)^{i} 3/2} 1^n
\end{equation}

On the polynomial side, define similarly $\hat y_i=\hat t_{i-1}^{-1}\cdots \hat t_1^{-1}\hat\rho \hat t_{L-1}\cdots \hat t_i$,
where $\hat t_i=q^{-1/2}\hat e_i+q^{1/2}$.
One can show that there is an order on monomials (see \cite{Pas-RS}) such that the operators $\hat y_i$ are upper triangular.
Thus it suffices to evaluate their diagonal elements in the basis on monomials. The action of $\hat y_i$ is most easily computed
on monomials $\prod_{i=1}^L z_i^{\lambda_i}$ with non-increasing powers $\lambda_{i+1}\le\lambda_{i}$. In this case we find
\begin{equation}\label{JMb}
\hat y_i \prod_{i=1}^L z_i^{\lambda_i}=
-\kappa s^{\lambda_i} q^{3(n+\oh-i)} \prod_{i=1}^L z_i^{\lambda_i} + \text{lower terms}
\end{equation}
Comparing the two expressions \eqref{JMa} and \eqref{JMb}, we find that $s=q^6$ and $\lambda=(\lambda_i)$ is, up to a global shift, equal to
$\lambda^{(n)}=(n-1,n-1,\ldots,1,1,0,0)$. This is the minimal degree solution that we shall consider -- the degree being
$n(n-1)$, as assumed earlier. We also deduce $\kappa=q^{-3(n-1)}$.

The latter computation looks very similar to a computation in nonsymmetric Macdonald polynomials \cite{Macdo-nonsym}.
This is no coincidence. In fact the eigenvectors of $\hat y_i$ are exactly nonsymmetric Macdonald polynomials,
see \cite{KT-qKZ}.  If we restrict ourselves to leading terms $\prod_{i=1}^L z_i^{\lambda_i(\pi)}$ which are in bijection with link patterns by (separately) labelling closings and openings in decreasing order, e.g.,
\[
\pi=\hdrleadingmonomialpattern\quad\longmapsto\quad \lambda(\pi)=(3,2,3,1,2,1,0,0)
\]
then their span is exactly the span of the $\Psi_\pi$,
even though they are distinct from the $\Psi_\pi$ -- there is a triangular change of basis, so that only $\Psi_{0^n}$ is itself
a nonsymmetric Macdonald polynomial.

\subsubsection{Wheel condition}
In \cite{Pas-RS}, a particularly attractive characterization is obtained for a certain
space of polynomials: the span of the $\Psi_\pi(z_1,\ldots,z_L)$ with $\pi\in P_{2n}$.
It will give an independent check of some of the results of the last section, as we shall see.

Consider the following subspace of homogeneous polynomials:
\begin{equation}\label{wheelcond}
M_n=\{ f(z_1,\ldots,z_{2n}),\ \deg_{z_i} f\le n-1 : f(\ldots,z,\ldots, q^2 z,\ldots, q^4 z,\ldots)=0 \}
\end{equation}
that is polynomials that vanish as soon as an ordered triplet of variables forms the sequence $1,q^2,q^4$. This vanishing condition is a so-called
{\em wheel condition}, see \cite{Kas-wheel} for more general ones. We have also imposed a bound of $n-1$ on the degree in each variable,
which is a little bit more restrictive than what is really needed: imposing on the (total) degree $\deg f\le n(n-1)$
would suffice, but this condition is less convenient for our purposes.\footnote{In fact, 
one can show that having a partial degree greater than $n-1$ while
retaining a total degree of $n(n-1)$ would lead to a contradiction with our representation of affine Hecke using \eqref{JMb}.}

It is easy to check that $\Psi_{0^n}\in M_n$; more interestingly, the action of $\tau-\hat e_i=-(q\,z_i-q^{-1}z_{i+1})\der_i$ 
preserves $M_n$ (check the vanishing property by discussing
separately the three cases: (i) neither $i$ nor $i+1$ are in the triplet: trivial; (ii) both $i$ and $i+1$ are in the triplet:
then the prefactor vanishes; and (iii) one of them is: then use the property for the original triplet and the one with $i$ replaced with $i+1$).
Since the $\Psi_\pi$ can be built from $\Psi_{0^n}$ by action of the $\tau-\hat e_i$, one concludes that they are all in 
$M_n$.\footnote{A more visual proof of this result is given in \cite{artic37}, where it is noted that since 
$(i,i+1)$ and $(i+1,i+2)$ cannot be both pairings, $\Psi_L(z_i,z_{i+1}=q^2z_i,z_{i+2}=q^4z_i)=0$ and then use \eqref{qkza} to permute the
arguments and get the general wheel condition.}
In fact, note that the action of $\hat\rho$ also leaves $M_n$ invariant on condition that $s=q^6$, so that $M_n$ is a representation of the
affine Hecke algebra.

We have an inclusion of the span of the $\Psi_\pi$ inside $M_n$. In order to prove the equality,
we shall state certain useful properties.

\subsubsection{Recurrence relation and specializations}\label{secrec}
First, we need the following simple observation. 
According to the dichotomy of section \ref{constrsol}, if $\pi$ is a link pattern
such that the points $i$ and $i+1$ are unconnected, then $q\,z_i-q^{-1}z_{i+1}$ divides $\Psi_\pi$,
so that $\Psi_\pi(z_{i+1}=q^2 z_i)=0$. But if $i$ and $i+1$ are connected (we shall call such a pairing of neighbors a ``little arch''), 
then we can use the wheel condition to say that
$\Psi_\pi(z_{i+1}=q^2 z_i)$ vanishes when $z_j=q^4 z_i$, $j>i+1$, or $z_j=q^{-2}z_i$, $j<i$, so that it is of the form
$\prod_{j=1}^{i-1}(z_i-q^2 z_j)\prod_{j=i+2}^{2n}(z_i-q^{-4}z_j)\Phi$ where by degree consideration $\Phi$ can only depend
on the $z_j$ with $j\ne i,i+1$. It is already clear that $\Phi$ is in $M_{n-1}$ and so is a linear combination of the entries of $\Psi_{L-2}$;
but there is better.

Call $\varphi_i$ the mapping from $P_{L-2}$ to $P_L$ which inserts an extra little arch at $(i,i+1)$. Then we claim that
\begin{equation}\label{recrel}
\Psi_\pi(\ldots,z_{i+1}=q^2 z_i,\ldots)=
\begin{cases}
0&\pi\not\in\Im\varphi_i\\
\begin{aligned}
q^{-(n-1)}\prod_{j=1}^{i-1}&(z_i-q^2 z_j)\prod_{j=i+2}^{2n}(q^3 z_i-q^{-1}z_j)\\
&\Psi_{\pi'}(z_1,\ldots,z_{i-1},z_{i+2},\ldots,z_L)
\end{aligned}
&\pi=\varphi_i\pi'
\end{cases}
\end{equation}
A direct proof of this formula is particularly tedious. Let us instead use the following trick. 
Consider the following cyclic invariance property: the formulae \eqref{recrel} for different values of
$i=1,\ldots,L-1$ can be deduced from each other by application of \eqref{qkzb}.
In fact, one can even extend this way \eqref{recrel} to $i=L$; 
being a little careful of the factors of $s=q^6$ that appear here and there, we obtain:
%
\begin{equation}\label{recrelb}
\Psi_{\varphi_L(\pi')}(z_1=q^{-4}z_{L},\ldots,z_{L})=q^{-4(n-1)}
\prod_{j=2}^{2n-1}(z_{L}-q^2 z_j) \Psi_{\pi'}(z_2,\ldots,z_{L-1})
\end{equation}
where $\varphi_L$ adds an arch between $L$ and $1$ (on top of all other arches in the half-plane picture of link patterns).
So, to prove \eqref{recrel}, it is enough to prove \eqref{recrelb}.

We shall use the construction of section \ref{constrsol}. Note that the mapping $\varphi_L$ is particularly natural
via the bijection to Young diagrams, since it becomes the embedding of the set of diagrams
inside $1^{n-1}$ to that of diagrams inside $1^n$. In particular $0^{n-1}$ is sent to $0^n$; so the first check is compatibility
with \eqref{basecase}, i.e.,\ that
when $\pi=0^n$ in \eqref{recrelb} we obtain in the r.h.s.\ the correct prefactors times $\Psi_{0^{n-1}}$.
The second check is that the formula \eqref{trisys}, which allows us to compute the components of $\Psi_L$, when restricted
to diagrams inside $1^{n-1}$, produces the components of $\Psi_{L-2}$.
This is rather obvious graphically: the action of $e_i$ is compatible with the natural inclusion of diagrams
(it can be defined directly in terms of Young diagrams). Furthermore,
in order to remain inside $1^{n-1}$,
one must never add boxes at $(1,2)$ or $(L-1,L)$. So we never affect the variables $z_1$, $z_L$ and up to a shift of indices
$i\to i-1$, this is just the procedure in size $L-2$. This is enough to characterize entirely the r.h.s.\ of \eqref{recrelb}.

Next, we need two more closely related properties:
\begin{itemize}
\item 
If $f\in M_n$, then
\begin{equation}\label{speczero}
f(q^{-\epsilon_1},\ldots,q^{-\epsilon_{2n}})=0\quad \forall (\epsilon_i) \text{ Dyck path increments }
\quad\Rightarrow\quad f=0
\end{equation}
(a Dyck path increment is a sequence $(\epsilon_i)_{i=1,\ldots,2n}$ of $\pm 1$ such that $\sum_{i=1}^{2n} \epsilon_i=0 $ and $\sum_{i=1}^j\epsilon_i\ge 0$ for all $j\le 2n$).
So a polynomial in $M_n$ is entirely determined by its values at these $c_n$ specializations.

\item We have the specializations:
\begin{equation}\label{specpsi}
\Psi_\pi(q^{-\epsilon_1},\ldots,q^{-\epsilon_{2n}})=
\begin{cases}
(-1)^{n(n-1)/2}
(q-q^{-1})^{n(n-1)} \tau^{|\pi|}&\text{if $\epsilon_i={\rm sign}(\pi(i)-i)$, $i=1,\ldots,2n$}\\
0&\text{otherwise}
\end{cases}
\end{equation}
($|\pi|$ is the number of boxes in the correspondence with Young diagrams of Fig.~\ref{figlptoyoung}).
In other words, these values are the entries in the basis of the $\Psi_\pi$.
\end{itemize}
The first point is shown explicitly in appendix C of \cite{artic45} and we shall not reproduce the proof here; 
the second one is simply an application of the recurrence formula \eqref{recrel},
by removing little arches one by one (see also \cite{artic41}).

\subsubsection{Wheel condition continued}
Using the properties of the previous section, 
we can now conclude that $\dim M_n\le c_n$ and that the $\Psi_\pi$ are independent polynomials, so that $M_n$ is the span of the $\Psi_\pi$.

One could use more representation-theoretic arguments to prove the equality of $M_n$ and of the span
of the $\Psi_\pi$, for example based on the fact
that $M_n$, as a representation of the affine Hecke algebra, is irreducible for generic $q$ (in fact,
it corresponds to the rectangular Young diagram with 2 rows and $n$ columns), but this is not our philosophy here.

The wheel condition has the following physical interpretation (borrowed once again from \cite{Pas-RS}).
Consider the case $q=\pm 1$. Then the $q$KZ equation \eqref{qkz-fr} becomes the ordinary KZ equation at level $1$.
It is most conveniently expressed in the spin basis:
\[
3{\der\over\der z_i} \Psi= \sum_{\substack{j=1\\j\ne i}}^L \frac{\sigma_{i,j}+1}{z_i-z_j} \Psi
\]
where $\sigma_{i,j}$ exchanges spins $i$ and $j$. The KZ equation is now in its usual form
for correlation functions in
$\widehat{\mathfrak{sl}(2)}_1$, except for a trivial change of $\sigma_{i,j}-1/2$ to $\sigma_{i,j}+1$, which is the same as $\Psi \to
\Psi \prod_{i<j} (z_i-z_j)^{1/2}$ (which is needed since we want a polynomial solution and not the usual one involving square roots).
The solution is a product of two Vandermonde determinants:
\begin{multline}\label{kzsol}
\Psi^{spin}_{\alpha_1,\ldots,\alpha_L}(z_1,\ldots,z_L)=(-1)^{\#\{i<j: \alpha_i=-,\alpha_j=+\}} \Delta(z_i)_{\alpha_i=+}\Delta(z_i)_{\alpha_i=-}
\\(\alpha_1,\ldots,\alpha_L)\in \{+,-\}^L
\end{multline}
This is simply the ground state wave function of two species of free fermions $\psi_\pm$, i.e.,\ with notations 
similar to section 1
\[
\Psi^{spin}_{\alpha_1,\ldots,\alpha_{2n}}(z_1,\ldots,z_{2n})=\bra{0,0}\psi_{\alpha_1}(z_1)\cdots \psi_{\alpha_{2n}}(z_{2n})\ket{n,n}
\]
In this language the wheel condition is just the Pauli exclusion principle (if three fermions sit at the same spot, at least
two of them must be of the same kind!). Of course the $q=\pm 1$ case is trivial in the sense that starting from say $\Psi_{0^n}$,
one can apply the exchange equation \eqref{qkza} which simply says in this limit, using $\check R_i=-\sigma_{i,i+1}$, 
that exchanging variables is the same as exchanging
spins (up to a fermionic minus sign), which immediately results in \eqref{kzsol}.

For a general value of $q$, one has a $q$-deformed version of this system of free fermions in which the vanishing condition
of the wave function occurs not at coinciding points but when ratios of coordinates are ordered successive powers of $q^2$.
The usual fermionic statistic is turned into a non-trivial braid group statistic.

\subsection{Connection to the loop model}
In general, the two problems of diagonalizing the transfer matrix of the CPL loop model and finding solutions of the $q$KZ system
are unrelated. However there is exactly one value of $q$ where a solution of $q$KZ does in
fact provide an eigenvector of the transfer matrix (with periodic boundary conditions). 
This is when the parameter $s=1$, which here occurs
when $q=\e{2\pi i/3}$ (other sixth roots of unity are possible but they are either trivial or
give the same result as the one we picked). In this case note that \eqref{qkzb} becomes a simple
rotational invariance condition. Furthermore the real $q$KZ equation \eqref{qkz-fr} becomes
an eigenvector equation for the scattering matrices:
\[
S_i(z_1,\ldots,z_L) \Psi_L(z_1,\ldots,z_L) = \Psi_L(z_1,\ldots,z_L)
\]
These scattering matrices do not involve any extra shifts of the spectral parameters, 
and as is well-known in Bethe Ansatz, are just specializations of the
inhomogeneous transfer matrix. Indeed if we define $T_L(z;z_1,\ldots,z_L)$ to be simply
\[
T_L=
z\,
\vcenterboxlabel{\hdrblankplaquette}{$z_1$}
\vcenterboxlabel{\hdrblankplaquette}{$z_2$}
\cdots
\vcenterboxlabel{\hdrblankplaquette}{$z_L$}
\]
(with periodic boundary conditions), where the box represents the $R$-matrix evaluated at the ratio of vertical and horizontal spectral parameters,
then observe that
$S_i(z_1,\ldots,z_L)=T_L(z_i;z_1,\ldots,z_L)$. By a Lagrange interpolation argument, we conclude that
\[
T_L(z;z_1,\ldots,z_L)
\Psi_L(z_1,\ldots,z_L) = \Psi_L(z_1,\ldots,z_L)
\]
i.e.,\ $\Psi_L$ is up to normalization the steady state of the inhomogeneous Markov process defined by $T_L(z;z_1,\ldots,z_L)$.
In order to recover the original homogeneous Markov process, one simply sets all $z_i=1$.

Note that the reasoning above is valid for {\em any}\/ solution of $q$KZ, not just the one that was discussed in section
\ref{constrsol}. But there is no point in considering higher degree polynomial solutions since at $q=\e{2\pi i/3}$ they
will simply be the minimal-degree solution of section \ref{constrsol} times a symmetric polynomial.

We now show how to apply the $q$KZ technology to prove some of the statements formulated in section \ref{seclooprs}.

\subsubsection{Proof of the sum rule}\label{sumrule}
As a simple application of the above formalism, we explain how to recover the ``sum rule'' at $q=\e{2\pi i/3}$.
Let us explain what we mean by that.
We start by noting that the normalization of $\Psi$ is fixed once the value of $\Psi_{0^n}$ is specified; in particular,
if all $z_i=1$, using \eqref{basecase}
we find $\Psi_{0^n}=3^{n(n-1)/2}$, to be compared with the (conjectured) probability $1/A_n$ associated to
$0^n$, cf.\ section \ref{secobs}. In other words, with this normalization, one should have $\sum_{\pi\in P_{2n}}\Psi_\pi=3^{n(n-1)/2} A_n$.
This is what we are going to show now. In fact, following \cite{artic31}, we are going to show more generally
the inhomogeneous sum rule that $\sum_{\pi\in P_{2n}} \Psi_\pi(z_1,\ldots,z_{2n})$ is equal to the 
Schur function $s_{\lambda^{(n)}}(z_1,\ldots,z_{2n})$, where $\lambda^{(n)}$ is defined in \eqref{defstairs}, that is up to
a constant the partition
function of the six-vertex model with DWBC at $q=\e{i\pi/3}$. Instead of the inductive method of \cite{artic31}, i.e.,\ the comparison
of the recurrence relations of sections \ref{secrec} and \ref{koreprec}, we shall here proceed directly.

Define the covector $v$ with entries $v_\pi=1$, $\pi\in P_{2n}$. The stochasticity property of a matrix
is the fact that $v$ is a left eigenvector with eigenvalue $1$; and it is satisfied at $\tau=1$ 
by all the $e_i$ and hence by $\check R_i(z)$:
\[
v e_i = v \quad \Rightarrow\quad v \check R_i(z)=v\qquad i=1,\ldots,L-1,\quad q=\e{\scriptstyle 2\pi i/3}
\]
Applying this identity to \eqref{qkza}, we immediately conclude that $v\cdot\Psi=\sum_{\pi\in P_{2n}}\Psi_\pi$ is a symmetric
polynomial of its arguments $z_1,\ldots,z_L$.

$\sum_{\pi\in P_{2n}}\Psi_\pi(z_1,\ldots,z_{2n})$ is a symmetric polynomial of degree $n(n-1)$ which satisfies the wheel condition
of \eqref{wheelcond}. We claim that this defines it uniquely up to normalization (see \cite{artic37} for a similar
claim in a more general fused model). The simplest proof is to use once again property \eqref{speczero}, i.e.,\ that polynomials
in $M_n$ are characterized by the specializations $(q^{-\epsilon_1},\ldots,q^{-\epsilon_{2n}})$.
But for a symmetric polynomial, all these specializations reduce to only one. This proves the claim.

Next, one notes that $s_{\lambda^{(n)}}(z_1,\ldots,z_{2n})$ also satisfies
these properties. The degree bound is clear from the definition \eqref{defschurb} of Schur functions; as to the wheel condition,
it follows from formula \eqref{defweyl} by noting that if $z_k=q^2 z_j=q^4 z_i$, the sub-matrix of the matrix of the numerator with
columns $\{i,j,k\}$ is of rank 2 and thus the determinant of the whole matrix vanishes. 

Finally, to fix the normalization constant, note that thanks to \eqref{specpsi},
all components of $\Psi$ vanish at $(z_1,\ldots,z_{2n})=(q^{-1},\ldots,q^{-1},q,\ldots,q)$ except 
$\Psi_{0^n}(q^{-1},\ldots,q^{-1},q,\ldots,q)=3^{n(n-1)/2}$; on the other hand, using \eqref{recschur}, we find
the same value $s_{\lambda^{(n)}}(q^{-1},\ldots,q^{-1},q,\ldots,q)=3^{n(n-1)/2}$. We conclude that
\[
\sum_{\pi\in P_{2n}} \Psi_\pi(z_1,\ldots,z_{2n}) = s_{\lambda^{(n)}}(z_1,\ldots,z_{2n})
\]

{\it Remark:} from the representation-theoretic point of view, $v$ is an invariant element (under affine Hecke action) in $(\mathbb{C}P_{2n})^\star$;
and $\sum_{\pi\in P_{2n}} \Psi_\pi$ is its image in $\mathbb{C}[z_1,\ldots,z_L]$. This shows that at $q=\e{2\pi i/3}$,
the representation of affine Hecke on $\mathbb{C}P_{2n}$ (resp.\ the span of the $\Psi_\pi$) is {\em not}\/ irreducible (though it remains
indecomposable), since it has a codimension one (resp.\ dimension one) stable subspace.

\subsubsection{Case of few little arches}\label{fewlittle}
Note a remarkable property of recurrence relation \eqref{recrel}: it can only decrease the number of little arches!
So, if one considers the subset of link patterns with a given maximum number of little arches, that subset is closed under these
relations.

The link patterns that possess only two little arches (which is the minimum) are the link pattern $0^n$ and its rotations,
i.e.,\ the $n$ link patterns for which all the pairings are parallel and for which we know explicitly the components:
they are given by the action of $\hat\rho$ (cyclic rotation of variables plus shift of $s$) on $\Psi_{0^n}$ which is
\eqref{basecase}.

These are the smallest components in the homogeneous limit. They correspond via the Razumov--Stroganov conjecture to a single
FPL. As such they are clearly the ``easiest'' components.

\begin{figure}
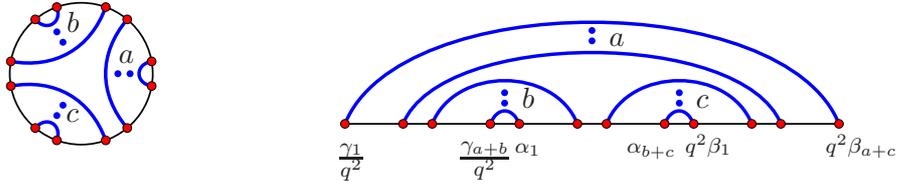

\hdrthreelittlearchescircle
\hskip2cm
\hdrthreelittlearchesline
\caption{Link patterns with three little arches.}\label{fig3l}
\end{figure}

It is natural to look at what happens when one considers more little arches. Link patterns with three little arches
are of the form of Fig.~\ref{fig3l}. In fact, the case of three little arches
was first investigated in \cite{artic27} on the other side of the Razumov--Stroganov conjecture: there, the problem of the
enumeration of FPLs with such a connectivity was solved, with the remarkable result that the number of FPLs with 
link pattern $(a,b,c)$ is simply equal to the number of plane partitions inside an $a\times b\times c$ box, that is given
by the MacMahon formula \eqref{maceq}. The proof is bijective and we shall not reproduce it here. Let us just comment on it.
The crux of the bijection is the observation, known as ``de Gier's lemma'', that in an FPL with given connectivity
many edges are fixed by the simple requirement of the connectivity of the endpoints. Removing all these fixed edges
we obtain a simpler enumeration problem, typically of plane partitions.
Now the little arches play a key role in the sense that they are the ones that limit application of de Gier's lemma.
In other words, the more little arches there are, the fewer
fixed edges in FPLs, and therefore the more complicated the enumeration is. Once again, 
the number of little arches is a measure of complexity.

Closing this philosophical parenthesis, let us go back to the $q$KZ equation and try to compute the entries $\Psi_{a,b,c}$
of $\Psi_L$ corresponding to link patterns with three series of $a$, $b$, $c$ nested arches.
This is performed in \cite{artic38} for $q=\e{2\pi i/3}$ and we generalize it here to arbitrary $q$.\looseness=-1

Up to rotation and use of \eqref{qkzb}, one can always assume that one of the little arches, say the one that is part of the series of $a$
nested arches, is between $L$ and $1$. Let us further rename the $L$ variables $z_i$ as follows: they become 
$(q^{-2}\gamma_1,\ldots,q^{-2}\gamma_{a+b},\alpha_1,\ldots,\alpha_{b+c},q^2\beta_1,\ldots,q^2\beta_{a+c})$, see Fig.~\ref{fig3l}.
Then according to the results of section \ref{constrsol}, 
\begin{multline*}
\Psi_{a,b,c}=
\prod_{1\le i<j\le b+c} (q\,\alpha_i-q^{-1}\alpha_j)
\prod_{1\le i<j\le a+c} (q\,\beta_i-q^{-1}\beta_j)
\prod_{1\le i<j\le a+b} (q\,\gamma_i-q^{-1}\gamma_j)
\\
q^{-b(a+b-1)+c(a+3b-1)+c^2}
(-1)^{(a+b)(a+b-1)/2+c(b+(c-1)/2)}
\Phi_{a,b,c}
\end{multline*}
where $\Phi_{a,b,c}$ is a polynomial that is symmetric in the $\{\alpha_i\}$, in the $\{\beta_i\}$ and in the $\{\gamma_i\}$.
The powers of $q$ and sign have been chosen carefully so that
when we rewrite recurrence relation \eqref{recrel} for $z_i=\alpha_{b+c}$, $z_{i+1}=q^2\beta_1$ in terms of $\Phi_{a,b,c}$,
everything cancels out and we are left with
\begin{equation}\label{recurkagome}
\Phi_{a,b,c}|_{\beta_1=\alpha_{b+c}}=\prod_{k=1}^{a+b}(\alpha_{b+c}-\gamma_k) \Phi_{a,b,c-1}
\end{equation}
where the arguments $\beta_1$, $\alpha_{b+c}$ are missing from $\Phi_{a,b,c-1}$. Now since $\Phi_{a,b,c}$ is symmetric in the $\alpha_i$,
this relation provides $\Phi_{a,b,c}$ at $b+c$ values of $\beta_1$; and it is not hard to check that $\Phi_{a,b,c}$ is of degree $b$ in $\beta_1$.
Thus it is entirely determined by these values (and can for example be obtained by Lagrange interpolation).

To complete the (implicit) calculation of $\Phi_{a,b,c}$ one must provide an initial condition. At $c=0$, the link pattern $(a,b,0)$
is nothing but the base link pattern $0^n$, and in this case we find
\begin{equation}\label{initkagome}
\Phi_{a,b,0}=\prod_{i=1}^a \prod_{j=1}^b (\alpha_j-\beta_i)
\end{equation}
Some explicit formulae for $\Phi_{a,b,c}$ are given in \cite{artic38}. In fact they are ``triple Schur functions''
in the sense of multi-Schur functions of \cite{Las}. \rem{or maybe cite Mac91?}

\begin{figure}
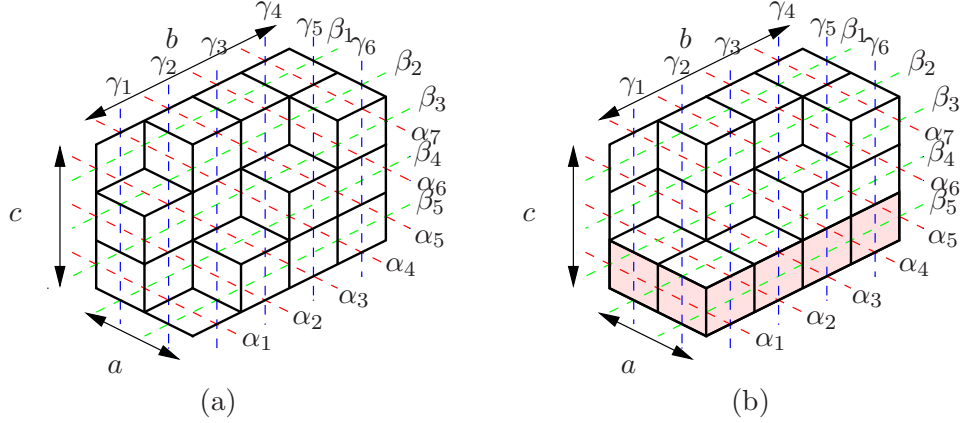

\hbox to \textwidth{\hfil\hdrkagomepanelone\qquad\hdrkagomepaneltwo\hfil}
\caption{The three sets of spectral parameters associated to lozenge tilings.}\label{figkagome}
\end{figure}

From the discussion above,
it is natural to try to put non-trivial Boltzmann weights on lozenge tilings of a hexagon of sides $a$, $b$, $c$ 
in such a way that their partition function
inside an $a\times b\times c$ box coincides with the inhomogeneous component $\Phi_{a,b,c}$.
Such a model is introduced in \cite{artic38}. We point it out here because we believe this model is of some interest
and might deserve further study. The spectral parameters live on the medial lattice, which is the {\em Kagome}\/
lattice, see Fig.~\ref{figkagome}(a). 
Each lozenge has a weight which is the difference of spectral parameters crossing at the center of the lozenge,
with the sign convention $\alpha-\beta$, $\beta-\gamma$, $\alpha-\gamma$.\footnote{This is only a convention
because the total number of lozenges of each orientation is conserved.
The choice that we made does not respect the $\mathbb{Z}_3$
symmetry of the model but turns out to be convenient.}

One can show that the integrability of the underlying model on the Kagome lattice implies that the partition function
is a symmetric function of the $\{\alpha_i\}$, of the $\{\beta_i\}$ and of the $\{\gamma_i\}$. We choose not to do so
here and refer instead to the appendix of \cite{artic38} for a ``manual'' proof of symmetry.
\rem{what is integrability of this model? could be a future project with dGN. is there a connection
with my LR tilings? not obviously because where is the Kagome lattice in LR?}

Finally, it is easy to see that the partition function of lozenge tilings with such weights satisfies the
recurrence relation \eqref{recurkagome}, as illustrated on Fig.~\ref{figkagome}(b); and that when $c=0$, there is a 
unique tiling of the $a\times b$ parallelogram, resulting in \eqref{initkagome}. Note the similarity
with the properties of the partition function of the six-vertex model with DWBC (symmetry
and recurrence) found by Korepin.

Thus, $\Phi_{a,b,c}$ and this partition function coincide.
Finally, if $\alpha_i=1$, $\beta_i=q^{-2}$, $\gamma_i=q^2$, using the fact that the number of lozenges
of each orientation is fixed and equal to $ab$, $bc$, $ca$, we find:
\begin{align*}
\Phi_{a,b,c}&=(1-q^{-2})^{ab}(q^{-2}-q^2)^{bc}(1-q^2)^{ca}\,N_{a,b,c}\\
            &=(q-q^{-1})^{ab+bc+ca}q^{a(c-b)}(-1)^{ca}N_{a,b,c}&&q=\e{2\pi i/3}\\
  \intertext{and therefore}
\Psi_{a,b,c}&=q^{-(a+c)(a+c-1)+(a+b)(a+b-1)-b(a+b-1)+c(a+3b-1)+c^2}(-1)^{(a+b)(a+b-1)/2+c(b+(c-1)/2)}\\
&\quad(q-q^{-1})^{\frac{(b+c)(b+c-1)}{2}+\frac{(c+a)(c+a-1)}{2}+\frac{(a+b)(a+b-1)}{2}}\,\Phi_{a,b,c}\\
&=(q-q^{-1})^{n(n-1)}(-1)^{\frac{n(n-1)}{2}}\,N_{a,b,c}&&q=\e{2\pi i/3}\\
\end{align*}
We conclude that $\Psi_{a,b,c}/\Psi_{0^n}=N_{a,b,c}$, as expected.

The case of four little arches is similar to that of three arches and is sketched in \cite{artic38}.
On the one hand, the recurrence relations are once again enough to determine all $\Psi_\pi$ uniquely.
On the other hand, the enumeration of FPL configurations with four little arches is doable (after appropriate
use of the so-called Wieland rotation \cite{Wieland}) and can be once
again reduced to a lozenge tiling enumeration \cite{artic30}. So the same strategy applies.
Starting with five little arches, however, it fails on both sides: the recurrence relations are not sufficient
to determine the corresponding components $\Psi_\pi$, and the enumeration of FPLs becomes non-trivial due to
an insufficient number of fixed edges.

\subsection{Integral formulae}\label{secintform}
We now want to show that
the formalism of the $q$KZ equation allows us to prove
the properties discussed in \ref{secobs}, as well as to reconnect
the three models that we have found in which the same numbers $A_n$ appear.
A particularly useful tool to exploit these solutions of $q$KZ has been 
introduced in \cite{artic41,artic42}: 
it consists of writing integral formulae for them.

\subsubsection{Integral formulae in the spin basis}
There are various types of known integral formulae for solutions of the quantum Knizhnik--Zamolodchikov equation.
Here we are only interested in a very specific one, which only exists at level $1$ and is obtained by ($q$-)bosonization,
see \cite{JM-book}.

The following formula is valid in even size $L=2n$:
\begin{multline}\label{intspin}
\Psi^{spin}_{a_0,\ldots,a_{n-1}}(z_1,\ldots,z_L)=(q-q^{-1})^n
\prod_{1\le i<j\le L}(q\,z_i-q^{-1}z_j)\\
\oint\cdots\oint \prod_{\ell=0}^{n-1} {w_\ell\,d w_\ell\over 2\pi i}
{
\prod_{0\le \ell<m\le n-1} (w_m-w_\ell)(q\,w_\ell-q^{-1}w_m)
\over
\prod_{\ell=0}^{n-1} \prod_{1\le i\le a_\ell} (w_\ell-z_i) 
\prod_{a_\ell\le i\le L} (q\,w_\ell-q^{-1}z_i)
}
\end{multline}
where the contours surround the $z_i$ counterclockwise, but not the $q^{-2}z_i$. The reader must be warned
that we have changed the labelling of the entries of $\Psi$: the index, instead of being a sequence of $L$ spins,
is the ordered sequence of locations of plus spins. In other words the correspondence is $(\alpha_1,\ldots,\alpha_L)
\mapsto \{a_0<\cdots<a_{n-1}\}=\{i:\alpha_i=+\}$.

For the case of odd size see \cite{artic42}.

\subsubsection{The partial change of basis}
Next we would like to write similar expressions for the entries of $\Psi$ in the basis of link patterns.
Unfortunately this would require to {\em invert}\/ the change of basis from link patterns to spins 
that was described in section \ref{secequiv}; 
and there is no simple explicit formula for the inverse 
(this is the famous problem of the computation of Kazhdan--Lusztig polynomials
\cite{KL}; in this case there is a combinatorial formula \cite{LS-KL} but it is not obvious how to
combine it with integral formulae).

Instead we shall do here something more modest: 
we introduce an intermediate basis between spins and link patterns. We skip here the details,
which can be found in \cite{artic41} (see in particular the appendices) and \cite{artic43}.
The bottom line is the introduction of a slight improvement of formula \eqref{intspin}:
\begin{multline}\label{intform}
\Psi_{a_0,\ldots,a_{n-1}}(z_1,\ldots,z_L)=
\prod_{1\le i<j\le L}(q\,z_i-q^{-1}z_j)\\
\oint\cdots\oint \prod_{\ell=0}^{n-1} {d w_\ell\over 2\pi i}
{
\prod_{0\le \ell<m\le n-1} (w_m-w_\ell)(q\,w_\ell-q^{-1}w_m)
\over
\prod_{\ell=0}^{n-1} \prod_{1\le i\le a_\ell} (w_\ell-z_i) 
\prod_{a_\ell< i\le L} (q\,w_\ell-q^{-1}z_i)
}
\end{multline}
(note the subtle modification: the factors of $w_\ell$ disappeared and an inequality became strict in the denominator).
Here $(a_0,\ldots,a_{n-1})$ is an arbitrary non-decreasing sequence of integers. The connection
with the entries of $\Psi$ in the bases of link patterns and spins is as follows:
\begin{align*}
\Psi^{spin}_{a_0,\ldots,a_{n-1}}&=
\sum_{\varepsilon_0,\ldots,\varepsilon_{n-1}\in\{0,1\}}
(-q)^{-\sum_i \varepsilon_i}
\,\Psi_{a_0-\varepsilon_0,\ldots,a_{n-1}-\varepsilon_{n-1}}\\
\Psi_{a_0,\ldots,a_{n-1}}&=\sum_{\pi\in P_{2n}}
\Bigg(\prod_{i,j\text{ paired in }\pi} U_{\# \{ \ell:\, i\le a_\ell<j\} - (j-i+1)/2}\Bigg)
\,\Psi_\pi
\end{align*}
where $U_k$ is the Chebyshev polynomial of the second kind evaluated at $-\tau$:
$U_k=\frac{q^{k+1}-q^{-(k+1)}}{q-q^{-1}}$.

Note an interesting property of the second change of basis, which is
that the coefficients are polynomials in $\tau$ with integer coefficients.

A second related property appears when one takes the homogeneous
limit $z_i=1$ in the integral formula \eqref{intform}. It is convenient to
send the two poles ($1$ and $q^{-2}$) to zero and infinity respectively by the homographic transformation
$u_\ell=(w_\ell-1)/(q\,w_\ell-q^{-1})$, so that:
\[
\frac{\Psi_{a_0,\ldots,a_{n-1}}}{\Psi_{0^n}}
(1,\ldots,1)=
\oint\cdots\oint \prod_{\ell=0}^{n-1} {du_\ell\over 2\pi i
\, u_\ell^{a_\ell}}
\prod_{0\le \ell<m\le n-1} (u_m-u_\ell)(1+\tau u_m+u_\ell u_m)
\]
where $\tau=-q-q^{-1}$, and the contours surround zero. This can be rewritten
\[
\frac{\Psi_{a_0,\ldots,a_{n-1}}}{\Psi_{0^n}}
(1,\ldots,1)=
\prod_{0\le \ell<m\le n-1} (u_m-u_\ell)(1+\tau u_m+u_\ell u_m)
\big|_{\prod_{i=0}^{n-1} u_i^{a_i-1}}
\]
As shown in \cite{artic43}, one can find a subset of sequences $(a_0,\ldots,a_{n-1})$ such that the matrix of change of basis
to link patterns is upper triangular with $1$'s on the diagonal. Combining the two properties above,
we reach the important conclusion that, normalized
so that $\Psi_{0^n}=1$, the entries $\Psi_\pi$ of the level $1$ polynomial solution of $q$KZ in the homogeneous limit $z_i=1$
are polynomials of $\tau$ with integer coefficients.
In fact, these coefficients are non-negative, but no combinatorial meaning is known for them. We shall however see
in the next section that we can interpret coefficients of certain partial sums of entries in terms of plane partitions.

\goodbreak{\em Example:} in size $L=6$, we find
the entries of $\Psi$ to be,
with the same ordering of link patterns as in sections \ref{lpdef} and \ref{constrsol},
the polynomials $1$, $2\tau$, $\tau^2$, $\tau^2$, $\tau+\tau^3$.

\subsubsection{Sum rule and largest component}
\begin{figure}
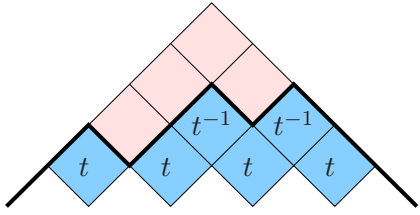

\hdryoungweight
\caption{Weight of a component expressed in terms of the associated Young diagram.}\label{figyoungweight}
\end{figure}

The integral formulae above allow us to compute various linear combinations of the $\Psi_\pi$. Here,
we concentrate on a single quantity of particular interest (which is considered in \cite{artic41,artic44}).
Consider the linear combination $F_n(t)=\sum_{\pi\in P_{2n}} c_\pi(t) \Psi_\pi(1,\ldots,1)$,
with the normalization $\Psi_{0^n}=1$, and
where the coefficient $c_\pi(t)$ is most simply expressed in terms
of the corresponding Young diagram by assigning a weight of $t$ or $t^{-1}$
to the boxes of even/odd rows in the {\em complement}\/ inside $1^n$, 
see Fig.~\ref{figyoungweight}. Three special cases are of particular
interest: $F(1)$ is just the sum $\sum_{\pi\in P_{2n}} \Psi_\pi$ of all link
patterns; $F(0)$ is just $\Psi_{1^n}$, one of the two largest components;
and $F(t)|_{t^{n-1}}$, the leading term of $F(t)$, is just $\Psi_{\rho(1^n)}$, the other largest component.

We then claim that $\sum_{\pi\in P_{2n}} c_\pi(t) \Psi_\pi$ can
be naturally expressed in the intermediate basis as
$\sum_{\epsilon_0,\ldots,\epsilon_{n-1}\in\{0,1\}} t^{\sum_{i=0}^{n-1} \epsilon_i}
\Psi_{1-\epsilon_0,3-\epsilon_1,\ldots,2n-1-\epsilon_{n-1}}$ (this is a simple calculation of the corresponding entries
of the matrix of change of basis), which results in
an integral formula. In the homogeneous limit we obtain
\[
F_n(t)=
\prod_{\ell=0}^{n-1} (1+t\,u_\ell)
\prod_{0\le \ell<m\le n-1} (u_m-u_\ell)(1+\tau u_m+u_\ell u_m)
\big|_{\prod_{i=0}^{n-1} u_i^{2i}}
\]

In fact one can simply set $u_0=0$ to get
\begin{equation}\label{qkzcounting}
F_n(t)=
\prod_{\ell=1}^{n-1}(1+t\,u_\ell)(1+\tau\, u_\ell)
\prod_{1\le \ell<m\le n-1} (u_m-u_\ell)(1+\tau u_m+u_\ell u_m)
\big|_{\prod_{i=1}^{n-1} u_i^{2i-1}}
\end{equation}
which looks strikingly similar to the formula for the weighted enumeration of TSSCPPs \eqref{tsscppcounting}.

There is however an important difference. Whereas \eqref{tsscppcounting} just contains a product of functions of one variable
times an antisymmetric function of the $u_i$ (which, ultimately, comes from the free fermionic nature of the model),
\eqref{qkzcounting} does not possess any particular symmetry w.r.t.\ exchange of its variables,
due to the factors $1+\tau u_m+u_\ell u_m$ (which come from the $q$-deformed Vandermonde product in the original 
integral formula). One can however antisymmetrize formula \eqref{qkzcounting}, as conjectured in \cite{artic41}
and then subsequently proved in \cite{Zeil-qKZ} and in a slightly stronger version in \cite{artic45}.
The identity, as formulated in \cite{artic44}, is:
\begin{quote}If
$AS(f(u_1,\ldots,u_{n-1}))=\sum_{\sigma\in \mathcal{S}_{n-1}} (-1)^\sigma f(u_{\sigma(1)},\ldots,u_{\sigma(n-1)})$, and $(\cdots)_{\le 0}$ means keeping only non-positive powers of a Laurent
polynomial in the variables
$u_\ell$, then the following equality holds:
\begin{multline}\label{zeilid}
\left\{\prod_{1\le\ell\le m\le n-1}(1-u_\ell u_m)\
{\rm AS}\left(\prod_{\ell=1}^{n-1} u_\ell^{-2\ell+1} \prod_{1\le\ell<m\le n-1}
(1+u_\ell u_m+\tau u_m)\right)\right\}_{\le 0}\\
={\rm AS}\left(\prod_{\ell=1}^{n-1}
\left(u_\ell^{-\ell}(\tau +u_\ell^{-1})^{\ell-1}\right)\right)=
\prod_{\ell=1}^{n-1} u_\ell^{-1}\prod_{1\le\ell<m\le n-1}(u_m^{-1}-u_\ell^{-1})
(\tau+u_\ell^{-1}+u_m^{-1})
\end{multline}
\end{quote}

Applying \eqref{zeilid} to \eqref{qkzcounting}, we find
\begin{equation}\label{fingen}
F_n(t)=\prod_{1\le i<j\le n-1} \frac{u_j-u_i}{1-u_i u_j} \prod_{i=1}^{n-1}(1+t\,u_i)(1+\tau u_i)^i
\big|_{\prod_{i=1}^{n-1} u_i^{2i-1}}
\end{equation}
If $t=1$, this coincides exactly with \eqref{tsscppcounting}. What we have found is
a very non-trivial combinatorial interpretation of this polynomial solution of $q$KZ, 
which goes beyond the special value $q=\e{2\pi i/3}$:
for generic $q$, the sum of components of this solution reproduces the weighted enumeration of TSSCPPs
with weight $\tau=-q-q^{-1}$, proving a conjecture formulated in \cite{DF-qKZ-TSSCPP}.

Next, we deduce from \eqref{fingen} that the leading term of $F_n(t)$ is
\[
F_n(t)\big|_{t^{n-1}}
=\prod_{1\le i<j\le n-1} \frac{u_j-u_i}{1-u_i u_j} \prod_{i=1}^{n-1}(1+\tau u_i)^i
\big|_{\prod_{i=1}^{n-1} u_i^{2i-2}}
=
F_{n-1}(\tau)
\]
where the last equality is obtained by setting $u_1=0$ and relabelling $i\to i-1$.
With a bit more work, one can also prove that $F_n(0)=\tau^{n-1}F_{n-1}(1/\tau)$.

As a result, at $\tau=1$, $F_n(1)$, the sum of components, is $A_n$, and $
F_n(t)|_{t^{n-1}}=F_n(0)$, the largest
components, are $A_{n-1}$.

\subsubsection{Refined enumeration}
For general $t$, one can show that $F_n(t)$ corresponds to a {\em refined}\/ weighted enumeration of TSSCPPs,
that is to adding extra boundary weights on the TSSCPPs. The details can be found in \cite{artic41}.

Remarkably enough, if one sets $\tau=1$, then $F_n(t)$ also coincides with the refined enumeration of ASMs.
For ASMs, the refinement is simple to explain and consists of giving a weight of $t^{i-1}$ to an ASM whose
1 in the first row is at column $i$ (which is essentially equivalent to
keeping one spectral parameter free in the partition function of the
six-vertex model with DWBC and sending the others to $1$).
This strange coincidence was observed in \cite{artic42} and proved via
rather indirect arguments (making use of the Bethe Ansatz calculations of \cite{RS-bethe}).

One can push this idea further: in 1986,
Mills, Robbins and Rumsey conjectured that the {\em double}\/ refinements of TSSCPPs and ASMs coincide
\cite{MRR-TSSCPP}. Again, the double refinement of ASMs is easy to explain, consisting in weights
$t^{i-1} u^{n-j}$ for an ASM with a 1 at $(1,i)$ and at $(n,j)$
(which is equivalent to keeping two, say horizontal, spectral parameters free). The double refinement of TSSCPPs can be formulated in
multiple ways and we refer the reader to appendix A of \cite{artic45} for details.
And in fact,
using slight generalizations of the integrals above, one can give a direct proof of
this conjecture and turn it into a theorem, as is done in
\cite{artic45}.

%
%


%
%
%

\newcommand\A{\textsc{a}}
\newcommand\B{\textsc{b}}

\section{Integrability and geometry}\label{secgeom}
The goal in this section is to reinterpret the families of polynomials that give rise to a solution of the $q$KZ
equation in terms of geometric data. For simplicity we shall stick to cohomology (as opposed to $K$-theory),
which means that we only consider {\em rational}\/ solutions of the Yang--Baxter equation.
In terms of the parameter $q$, it means taking an appropriate $q\to \pm 1$ limit (it is actually more convenient
to consider $q\to -1$ with our sign conventions).

We shall consider three examples in what follows. 
The first one (matrix Schubert varieties)
is there as a warm-up, though it already contains many of the key ingredients.
The second example (orbital varieties) will correspond to
the $q\to -1$ limit of the solution of $q$KZ of section \ref{secqkz}. 
The third one (Brauer loop scheme) in fact contains the first two as special cases and is the most
interesting one.
But first we need to introduce some technology, i.e.,\ an algebraic analogue of equivariant cohomology 
for affine schemes with a linear group action: multidegrees \cite{MS-mdeg}.
Note that in this section we give almost no proofs and refer to the papers for details.

\subsection{Multidegrees}
Our group will always be a torus $T=(\mathbb{C}^\times)^r$. $T$ acts linearly on a complex vector space $W$.
To a closed $T$-invariant sub-scheme $X\subseteq W$ we will
assign a polynomial $\mdeg_W X \in \Sym(T^*) \iso \integers[z_1,\ldots,z_r]$ (here $T^*$ is viewed as a lattice
inside the dual of the Lie algebra of $T$)
called the {\dfn multidegree} of $X$.

\subsubsection{Definition by induction}
This assignment can be computed inductively using the following properties 
(as in \cite{Jos-mdeg}):
\begin{enumerate}
\item[1.] If $X=W=\{0\}$, then $\mdeg_W X = 1$.
\item[2.] If the scheme $X$ has top-dimensional components $X_i$, 
  where $m_i>0$ denotes the multiplicity of $X_i$ in $X$, 
  then $\mdeg_W X = \sum_i m_i\ \mdeg_W X_i$. This lets one reduce from
  the case of schemes to the case of varieties (reduced irreducible schemes).
\item[3.] Assume $X$ is a variety, 
  and $H$ is a $T$-invariant hyperplane in $W$.
  \begin{enumerate}
  \item If $X\not\subset H$, then $\mdeg_W X = \mdeg_H (X\cap H)$.
  \item If $X\subset H$, then 
    $ \mdeg_W X = (\mdeg_H X) \cdot ( \hbox{the weight of $T$ on $W/H$}). $
  \end{enumerate}
\end{enumerate}
One can readily see from these properties that $\mdeg_W X$ is 
homogeneous of degree $\codim_W X$, and is a positive sum of products
of the weights of $T$ on $W$. 

\subsubsection{Integral formula}\label{secintmdeg}
We can also reformulate the multidegree as an integral
(generalizing the idea that the degree of a projective variety is essentially its volume).
To the torus $T$ acting linearly on the complex vector space $W$ is naturally associated
a moment map $\mu$, which is quadratic. With reasonable assumptions
(i.e.,\ that the multigrading associated to the torus action is {\em positive}),
one can map generators of $T^*$ to 
complex numbers in such a way that this quadratic form is positive.
Then one can formally write
\[
\mdeg_W X = 
\frac{
\int_X dx\,\e{\textstyle -\pi\mu(x)}
}
{
\int_W dx\,\e{\textstyle -\pi\mu(x)}
}
\]
where it is implied that on both sides, this evaluation map has been applied.
More explicitly, suppose that $(x_i)_{i=1,\ldots,n}$ is a set of coordinates
on $W$ which are eigenvectors of the torus action,
with weights $\alpha_i$; then
\begin{align*}
\mdeg_W X 
&= 
\frac{
\int_X dx\,\e{\textstyle -\pi\sum_{i=1}^n \alpha_i |x_i|^2}
}
{
\int_W dx\,\e{\textstyle -\pi\sum_{i=1}^n \alpha_i |x_i|^2}
}
\\
&=\prod_{i=1}^n \alpha_i\
\int_X dx\,\e{\textstyle -\pi\sum_{i=1}^n \alpha_i |x_i|^2}
\end{align*}
where the $\alpha_i$ must be evaluated to positive real numbers for the integral to make sense.
So, the multidegree is just a Gaussian integral. The subtlety comes from the fact that
in general $X$ is a non-trivial variety to integrate on (in particular, it will be singular
at the torus fixed point $0$, the critical point of the function in the exponential!). In the
case where $X$ is a linear subspace of $W$, 
given by equations $x_{i_1}=\cdots=x_{i_k}=0$,
which is the only case where the integral is really an ordinary Gaussian integral,
we compute immediately $\mdeg_W X=\alpha_{i_1}\ldots\alpha_{i_k}$, as expected.

\newcommand\C{{\mathbb C}}
\newcommand\gothg{{\mathfrak g}}
\newcommand\gothb{{\hat{\mathfrak b}_+}}
\newcommand\gothbp{{{\mathfrak b}_+}}
\newcommand\gothbm{{{\mathfrak b}_-}}
\newcommand\gothn{{\hat{\mathfrak n}_+}}
\newcommand\gotht{{\mathfrak t}}
\newcommand\gothnp{{{\mathfrak n}_+}}
\newcommand\gothB{{\hat{B}_+}}
\subsection{Matrix Schubert varieties}\label{secschub}
Matrix Schubert varieties are a slight variation of the classical Schubert varieties, whose study initiated Schubert calculus.
They are simpler objects to define, being affine schemes (i.e.,\ given by a set of polynomial equations
in a vector space).
\subsubsection{Geometric description}
Let $\gothg=\mathfrak{gl}(L,\C)$, $\gothbp=\{\text{upper triangular matrices}\}\subset\gothg$, $\gothbm=\{\text{lower triangular matrices}\}\subset\gothg$, 
and similarly we can define the groups $G=GL(L,\C)$ and
$B_\pm =\{\text{invertible upper/lower triangular matrices}\}\subset G$.

The matrix Schubert variety $S_\sigma$ associated to 
a permutation $\sigma\in \mathcal{S}_L$ is a closed sub-variety of $\gothg$ defined by the equations:
\newcommand\upperleft{{\hdrupperleftcorner}}
\begin{equation}\label{eqschub}
S_\sigma=
\{ M\in\gothg: \rank M_{\upperleft}\le \rank \sigma^T{}_{\!\upperleft},\ 
i,j=1,\ldots,L\}
\end{equation}
where $M_{\upperleft}$ is the sub-matrix above and to the left of $(i,j)$,
and $\sigma^T$ is the transpose of the permutation matrix of $\sigma$. 

Explicitly, $S_\sigma$ is defined by a set of polynomial equations of the form ``a determinant of a sub-matrix of $M$ is zero'',
which express the rank conditions.

One can also describe $S_\sigma$ as a group orbit closure:
\begin{equation}\label{schuborb}
S_\sigma=\overline{B_-\sigma^T B_+}
\end{equation}
Evidently, we have $S_{\sigma^{-1}}=(S_\sigma)^T$.
Less evidently, the codimension of $S_\sigma$ is the inversion number $|\sigma|$ of $\sigma$.

\goodbreak{\em Examples:}
\begin{itemize}
\item
$\sigma=1$: then $\rank \sigma^T{}_{\!\upperleft}=\min(i,j)$ which is the maximum possible rank of an $i\times j$ matrix.
So there is no rank condition and $S_\sigma=\gothg$.

\item
$\sigma=\sigma_0$, the longest permutation, $\sigma_0(i)=L+1-i$: this time
$\rank \sigma^T{}_{\!\upperleft}=\max(0,i+j-L)$ and we find that $S_{\sigma_0}$ is a linear subspace
\def\addots{\mathinner{\mkern1mu\raise7pt\vbox{\kern7pt\hbox{.}}\mkern-12mu\raise4pt\hbox{.}\mkern-12mu\raise1pt\hbox{.}\mkern1mu}}
\[
S_{\sigma_0}=\left\{ M=(M_{ij})_{i,j=1,\ldots,L} : M_{ij}=0\ \forall i,j,\ i+j\le L \right\}
=\left\{\begin{pmatrix}&0&\star\\0&\addots&\vdots\\\star&\cdots&\star\end{pmatrix}\right\}
\]

\item All the other matrix
Schubert varieties are somewhere in between. For example,
if $\sigma=(4132)$, then
\[
\sigma^T=\begin{pmatrix}0&0&0&1\\ 1&0&0&0\\ 0&0&1&0\\ 0&1&0&0\end{pmatrix}
\qquad
\Rightarrow
\qquad
S_{(4132)}=\left\{ M=(M_{ij})_{i,j=1,\ldots,4} : \begin{gathered}M_{11}=M_{12}=M_{13}=0\\M_{21}M_{32}-M_{22}M_{31}=0 \end{gathered}\right\}
\]
Note that among the multiple determinant vanishing conditions, one can extract in this case four that imply all the others.
In general, there are simple graphical rules to determine which conditions to keep.
\end{itemize}

There is a torus $T=(\C^\times)^{2L}$ acting on $\gothg$, by multiplication on the left and on the right
by diagonal matrices. The corresponding generators of its dual are denoted by
$x_1,\ldots,x_L$ (left) and $y_1,\ldots,y_L$ (right).
In fact, since multiplication by a scalar does not see the distinction between
left and right, the torus acting is really of dimension $2L-1$, and this amounts to saying that
all multidegrees we shall consider only depend on differences $x_i-y_j$.

The $S_\sigma$ are $T$-invariant; let us define the {\em double Schubert polynomials} $\Xi_\sigma$ to
be their multidegrees:
\[
\Xi_\sigma = \mdeg_\gothg S_\sigma
\]
Note that
\[
\Xi_{\sigma^{-1}}(x_1,\ldots,x_L|y_1,\ldots,y_L)=(-1)^{|\sigma|}\Xi_\sigma(y_1,\ldots,y_L|x_1,\ldots,x_L),
\]
so the two sets of variables play symmetric roles, up to this sign.

\subsubsection{Pipe dreams}
In \cite{FK-Schubert} (see also \cite{KM-Schubert}), a combinatorial formula for the calculation of
double Schubert polynomials was provided. It can be described as configurations
of a simple statistical lattice model coined in \cite{KM-Schubert} (reduced) ``pipe dreams''.

The pipe dreams are made of plaquettes, similarly to loop models of section \ref{seclooprs}. The two allowed plaquettes
are $\vcenterbox{\hdrpipedreamturn}$ and $\vcenterbox{\hdrpipedreamcrossing}$ (where the second one
should be considered as two lines actually crossing). Furthermore, pipe dreams have a specific geometry: they live
in a right-angled triangle, as shown on the examples below. On the hypotenuse, the plaquettes are forced to be of the first
type (and only one half of them is represented).

A pipe dream is {\em reduced}\/ iff no two lines cross more than once. Alternatively, one can think of putting
a non-local weight of zero to ``bubbles'' formed by two lines crossing twice. We assign weights to reduced pipe dreams
as follows: a crossing at row/column $(i,j)$ has weight $x_i-y_j$. We can now state the formula
for double Schubert polynomials:
\[
\Xi_\sigma(x_1,\ldots,x_L|y_1,\ldots,y_L)=
\sum_{\substack{\text{reduced pipe dreams of size $L$}\\\text{such that point $i$ on vertical axis}\\\text{is sent to point $\sigma(i)$ on horizontal axis}\\i=1,\ldots,L}}
\text{(weight of pipe dream)}
\]

A sketch of proof of this formula will be given in the next two sections.

\goodbreak{\em Examples:}
\begin{itemize}
\item $\sigma=1$: then only one pipe dream contributes, the one with no crossings, of the
type 
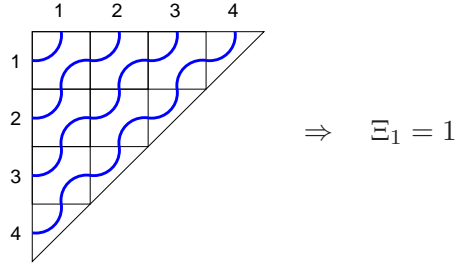
\[ 
\vcenterbox{\hdrpipedream{1}}
\quad\Rightarrow\quad
\Xi_1=1
\]

\item $\sigma=\sigma_0$: again, only one pipe dream contributes, this time with only crossings:
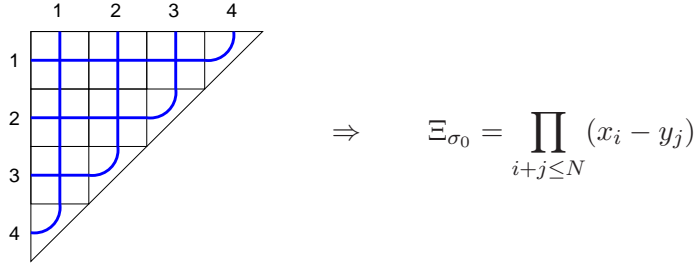
\[
\vcenterbox{\hdrpipedream{2}}
\qquad\Rightarrow\qquad
\Xi_{\sigma_0}=\prod_{i+j\le L} (x_i-y_j)
\]
Each factor of $x_i-y_j$ has the meaning of the weight of the equation $M_{ij}=0$.

\item If $\sigma=(4132)$, there are two (reduced) pipe dreams:
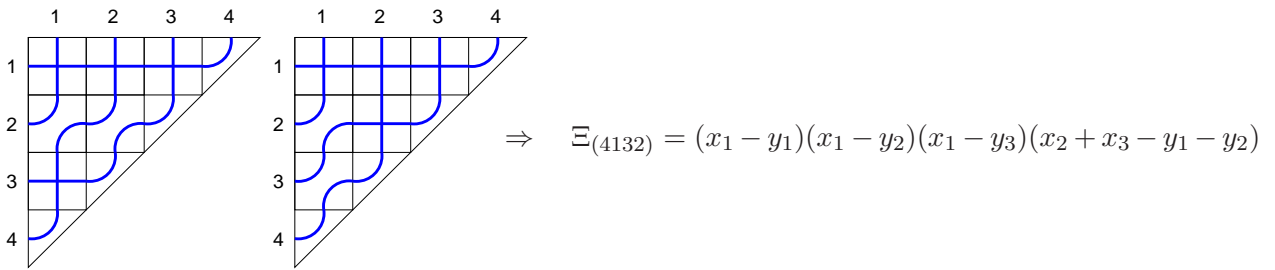
\[
\vcenterbox{\hdrpipedream{3}}
\ 
\vcenterbox{\hdrpipedream{4}}
\mskip-20mu\Rightarrow\quad
\Xi_{(4132)}=(x_1-y_1)(x_1-y_2)(x_1-y_3)(x_2+x_3-y_1-y_2)
\]
Again, we recognize in each factor the weight of one equation of $S_{(4132)}$.
In general, as long as $S_\sigma$ is a complete intersection (as many defining equations as the codimension), 
$\Xi_\sigma$ is just the product of weights of its equations
(i.e.,\ linear factors).
\end{itemize}

\subsubsection{The nil-Hecke algebra}
In \cite{FS-Schubert,FK-Schubert}, double Schubert polynomials are related to a solution of the Yang--Baxter
equation (YBE) based on the nil-Hecke algebra. We shall reformulate this connection here in terms of our exchange relation.

Define the nil-Hecke algebra by generators $t_i$, $i=1,\ldots,L-1$, and relations
\[
t_i^2=0\qquad t_i t_{i+1}t_i = t_{i+1}t_i t_{i+1}\qquad t_i t_j = t_j t_i \quad |i-j|>1
\]
As a vector space, the nil-Hecke algebra is isomorphic to $\C[\mathcal{S}_L]$
(the $t_i$ being identified with elementary transpositions). It has an obvious
graphical depiction in which basis elements $\sigma\in \mathcal{S}_L$
permute lines; the product is the usual product in $\mathcal{S}_L$
except that when lines cross twice, the result is zero:
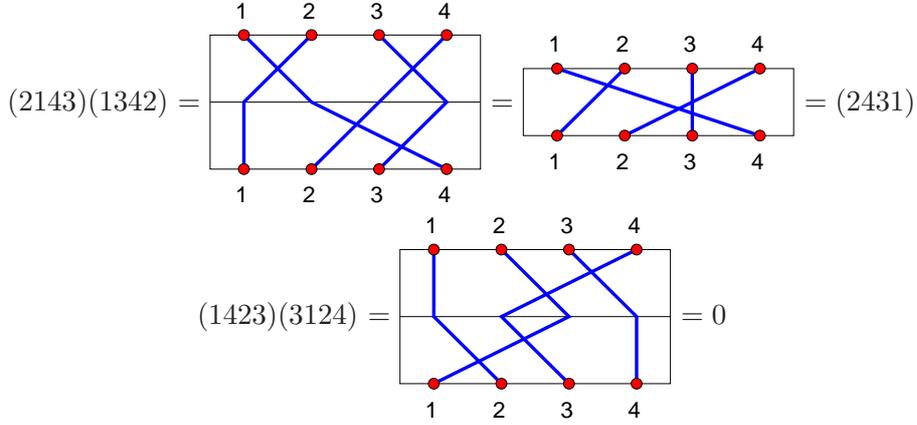
\begin{gather*}
(2143)(1342)
=\vcenterbox{\hdrnilheckeone}
=\vcenterbox{\hdrnilhecketwo}
=(2431)
\\
(1423)(3124)
=\vcenterbox{\hdrnilheckethree}=0
\end{gather*}
(the expressions from right to left correspond to the pictures from bottom to top). The inversion number of $\sigma$
is graphically its number of crossings.

Define
\[
\check R_i(u)=1+u\, t_i
\]
which is the associated solution to the {\em rational}\/ Yang--Baxter equation, that is with additive spectral
parameters:
\[
\check R_i(u) \check R_{i+1}(u+v) \check R_i(v)
=
\check R_{i+1}(v) \check R_{i}(u+v) \check R_{i+1}(u)
\]
Note that 
$t_i=\vcenterbox{\rotatebox{45}{\hdrcrossingplaquette}}$,
and the identity is $1=\vcenterbox{\rotatebox{45}{\hdrturntwo}}$,
that is 45 degrees rotated versions of our plaquettes, cf.\ a similar remark in section \ref{qkzsys}.
As usual, our solution of YBE also satisfies the unitarity equation:
\[
\check R_i(u) \check R_i(-u)=1
\]

Now consider the formal object $\Xi=\sum_{\sigma\in \mathcal{S}_L} \Xi_\sigma \sigma$ viewed as an element of nil-Hecke.
We then claim that the following formulae hold:
\begin{align}
\Xi(x_1,\ldots,x_L|y_1,\ldots,y_L)\cR_i(x_{i+1}-x_{i}) 
&=\Xi(x_1,\ldots,x_{i+1},x_i,\ldots,x_L|y_1,\ldots,y_L)\label{schubex}\\
\cR_i(y_{i}-y_{i+1}) \Xi(x_1,\ldots,x_L|y_1,\ldots,y_L)
&=\Xi(x_1,\ldots,x_L|y_1,\ldots,y_{i+1},y_i,\ldots,y_L)\label{schubexb}
\end{align}
which is nothing but exchange relations. Note that if we consider
$\Xi$ as a function of the first set of variables $x_1,\ldots,x_L$, the nil-Hecke algebra must act {\em on the right}\/ on itself (ultimately, this is because of the transposition
in \eqref{eqschub} and \eqref{schuborb}). 

These formulae are obvious if one defines $\Xi$ in terms of pipe dreams,
that is if one writes $\Xi$ as \cite{FK-Schubert}
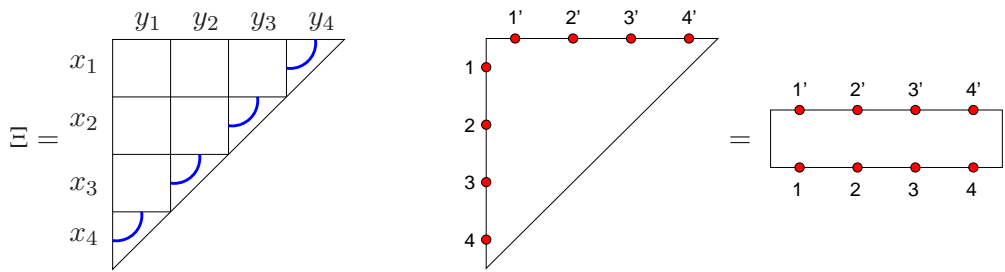
\[
\Xi=\vcenterbox{\hdrpipedreamvariables}
\qquad\qquad
\vcenterbox{\hdrpipeendpoints}=\ \vcenterbox{\hdrnilheckeidentity}
\]
where each plaquette is an $R$-matrix evaluated at $x_i-y_j$ (row $i$, column $j$), and
each pipe dream is interpreted as an element of nil-Hecke according to the convention of endpoints
shown next to the equality.
The proof of \eqref{schubex} is then the usual argument
(cf.\ for example section \ref{koreprec}) of applying the $\check R$ matrix between rows $i$ and $i+1$,
moving it across using YBE and checking that it disappears once it reaches the hypotenuse;
and similarly for \eqref{schubexb}.

To conform with the literature, we shall now focus on the first
form \eqref{schubex}.
Writing it explicitly in components, we have to distinguish as usual cases. Let us denote as before
$s_i$ the permutation of variables $x_i$ and $x_{i+1}$.
\begin{itemize}
\item If $\sigma(i)<\sigma(i+1)$ (the lines starting at $(i,i+1)$ do not cross):
then $\sigma$ is not the image of $t_i$ acting by right multiplication, so that we find
\begin{equation}\label{nilheckea}
\Xi_\sigma = s_i \Xi_\sigma
\end{equation}
i.e.,\ $\Xi_\sigma$ is symmetric under exchange of $x_i$ and $x_{i+1}$.

\item If $\sigma(i)>\sigma(i+1)$ (the lines starting at $(i,i+1)$ cross):
this time we find that $\Xi_\sigma+(x_{i+1}-x_i)\Xi_{\sigma'}=s_i \Xi_\sigma$, where $\sigma'$ is the unique
preimage of $\sigma$ under right multiplication by $t_i$. Thus,
\begin{equation}\label{nilheckeb}
\Xi_{\sigma'}= \der_i \Xi_\sigma
\end{equation}
where $\der_i=\frac{1}{x_{i+1}-x_i}(s_i-1)$, as before. Note that $\sigma'$ satisfies
$\sigma'(i)<\sigma'(i+1)$, so that \eqref{nilheckea} can be deduced from \eqref{nilheckeb} (the image of
$\der_i$ is symmetric in $x_i$, $x_{i+1}$).
\end{itemize}
In \eqref{nilheckeb}, $\sigma'$ has one fewer crossing than $\sigma$.
It is easy to show this way that, starting from $\Xi_{\sigma_0}=\prod_{i+j\le L}(x_i-y_j)$, where $\sigma_0$ is the permutation
with the most crossings (the highest inversion number), one can compute any $\Xi_\sigma$ by
repeated use of \eqref{nilheckeb}.

We now explain what the geometric meaning of \eqref{nilheckeb} is. 
This will prove that the pipe dreams do compute the multidegrees of matrix Schubert varieties.

\subsubsection{The Bott--Samelson construction}\label{secbs}
Let $L_i$ be the subgroup of $G$ 
consisting of matrices which are the identity everywhere
except in the entries $M_{ab}$ with $a,b\in\{i,i+1\}$, and $B_{\pm,i}=B_\pm \cap L_i$.
We use the following lemma, which is similar to lemma 1 of \cite{artic33} (see also lemma 8 of \cite{artic39}):
(throughout the lemma the sign $\pm$ is fixed)
\begin{quote}
Let $V$ be a left $L_i$-module, and let $X$ be a variety in $V$ that is invariant under scaling and under 
the action of $B_{\pm,i}\subset L_i$.
Define the map
$\mu: (L_i \times X)/B_{\pm,i} \to V$ that sends classes of pairs $(g,x)$ (where $B_{\pm,i}$
acts on the right on $L_i$ and on the left on $X$) to $g\cdot x$. 
If $\mu$ is generically one-to-one, then
\[
\mdeg_V \Im \mu = \mp \der_i \mdeg_V X
\]
If on the other hand $X$ is $L_i$-invariant, then
\[
\der_i \mdeg_V X=0
\]
\end{quote}

Here we apply the lemma with $V=\gothg$, $L_i$ acting on it by left multiplication, and $X=S_\sigma$.
According to \eqref{schuborb}, $X$ is $B_{-,i}$-invariant. Furthermore,
if $\sigma(i)>\sigma(i+1)$, one can check that the map $\mu$ is generically one-to-one,
and the image is precisely $S_{\sigma'}$ where $\sigma=\sigma't_i$. Hence \eqref{nilheckeb} holds.
On the contrary, if $\sigma(i)<\sigma(i+1)$, $X$ is simply $L_i$-invariant,
and we conclude that \eqref{nilheckea} holds.

\subsubsection{Factorial Schur functions}\label{secdoubleschur}
Pipe dreams contain non-local information carried by the connectivity of the lines, just like the loop
models of section \ref{seclooprs}. It is natural to wonder if there is a ``vertex'' representation (similar to the
six-vertex model) for pipe dreams. As this model is based on a Hecke algebra (which we have considered so far
in the regular representation),
there are in fact plenty of vertex representations corresponding to various quotients of the Hecke algebra. The most general one corresponds to giving a different color to every pipe:
\[
\vcenterbox{\hdrmulticolouredpipedream}
\]
The connectivity is now locally encoded in the colors,
and the permutation can be read off the colors at endpoints.

Inversely, the simplest nontrivial vertex representation, related to the Temperley--Lieb quotient, is as follows.
Suppose that $\sigma$ is a {\em Grassmannian permutation}. By definition, this means that $\sigma$
has at most one descent, i.e.,\ there is a $k$ such that
$i\ne k$ implies $\sigma(i)<\sigma(i+1)$. Then one can group lines into two subsets depending on whether
they start on the vertical axis at position $i\le k$ or $i>k$. If we use different colors for them, we get a picture
such as
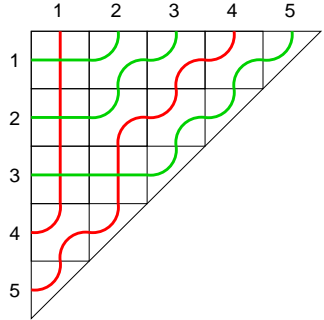
\[
\vcenterbox{\hdrcolouredpipedream}
\]
Several observations are in order. First, there is indeed no more non-local information in the sense
that the connectivity of the endpoints can be entirely determined from the sequences of colors
on the horizontal axis. Indeed, if the green endpoints are $a_1<\cdots<a_k$ and the red endpoints
are $b_1<\cdots<b_{L-k}$, there is only one Grassmannian permutation which is compatible with these colors,
namely $(a_1,\ldots,a_k,b_1,\ldots,b_{L-k})$.
Secondly, there is never any crossing below row $k$, so $\Xi_\sigma$ does not depend on $x_{k+1},\ldots,x_L$.
According to \eqref{nilheckea}, we also know that it is a symmetric polynomial of $x_1,\ldots,x_k$.
Thirdly, there are now five types of plaquettes:
the four colorings of the non-crossing plaquette, but only one crossing plaquette (lines of the
same color cannot cross each other!). In fact, by identifying green with occupied and red with empty,
we recognize the six-vertex model configurations under the form of north-east going paths,
in which the weight $b_1=0$ (green paths cannot go straight up). Taking into account the weights,
this is exactly the free fermionic five-vertex
model considered at the end of section \ref{sec5v} (Fig.~\ref{fig5v2}). Furthermore,
if we continue the green lines to the left so they all end on the same horizontal line, we find
exactly the configurations that contribute to the Schur function $s_\lambda$, where $\lambda$
is encoded by the sequence of red and green endpoints at the top (which one can extend into an infinite
sequence by filling with green dots at the left and red dots at the right). Finally, by comparing
the weights, we find the equality:
\[
\Xi_\sigma(x_1,\ldots,x_k|0,\ldots,0)=
s_\lambda(x_1,\ldots,x_k)
\]
In this case, the $\Xi_\sigma(x_1,\ldots,x_k|y_1,\ldots,y_L)$ are usually called factorial Schur functions. They are essentially
the same as double Schur functions, see \cite{MS-doubleschur,Molev-coproduct} for a detailed discussion.


\subsection{Orbital varieties}\label{secorbvar}
We shall move on to more sophisticated objects.
In general, orbital varieties are the irreducible components of the intersection of a nilpotent
orbit (by conjugation) with a Borel sub-algebra inside a Lie algebra. We cannot possibly
reproduce the general theory of orbital varieties, and here, we shall restrict
ourselves to a very special type of orbital varieties which corresponds to the Temperley--Lieb
algebra (for more general orbital varieties from an ``integrable'' point of view, see \cite{artic34,artic35}).

\subsubsection{Geometric description}
We use the same notations as in section \ref{secschub}; but here
all matrices are of even size $L=2n$. 
Furthermore, call $\gothnp=\{\text{strict upper triangular matrices}\}\subset\gothbp$.

Consider the affine scheme
\[
\mathcal{O}_L=\{ M\in \gothnp: M^2=0 \}
\]
that is upper triangular matrices that square to zero. We have the following description of its
irreducible components: they are indexed by link patterns $\pi\in P_{2n}$. To each $\pi$ we associate
the upper triangular matrix $\pi_<$ with entries $(\pi_<)_{ij}$ equal to $1$ if $i<j$ and $(i,j)$ paired by $\pi$, 
$0$ otherwise.
For example,
\def\smallddots{\mathinner{\mkern0.1mu\raise2.33pt\vbox{\hbox{.}}\mkern-4mu%
    \raise1.333pt\hbox{.}\mkern-4mu\raise0.33pt\hbox{.}\mkern0.1mu}}%
\[
\pi=\vcenterbox{\hdrlinepfour[0.44cm]{3}}
\qquad
\pi_<=\left(\begin{smallmatrix}%
0&0&0&0&0&0&0&1\\
&\smallddots&1&0&0&0&0&0\\
&&&0&0&0&0&0\\
&&&&0&0&1&0\\
&&&&&1&0&0\\
&&&&&&0&0\\
&&&&&&\smallddots&0\\
&&&&&&&0\\
\end{smallmatrix}\right)
\]

\newcommand\lowerleft{{\hdrlowerleftcorner}}
Then the corresponding irreducible component $\mathcal{O}_\pi$ is given by the equations
\begin{equation}\label{orbeq}
\mathcal{O}_\pi = \{ M\in \gothg: M^2=0\text{ and }\rank M^{\lowerleft}\le \rank \pi_<{}^{\!\lowerleft},
i,j=1,\ldots,L\}
\end{equation}
where $M^\lowerleft$ is the sub-matrix of $M$ below and to the left of $(i,j)$.
Alternatively, 
$\mathcal{O}_\pi$ can be defined as an orbit closure:
\begin{equation}\label{orborb}
\mathcal{O}_\pi
=\overline{B_+\cdot \pi_<}\qquad
\text{$B_+$ acts by conjugation}
\end{equation}
Like all orbital varieties, $\mathcal{O}_L$ is equidimensional, and the codimension of $\mathcal{O}_\pi$ in $\gothnp$ is
$n(n-1)$.

There is a torus $T=(\C^\times)^{L+1}$ acting on $\gothnp$, by conjugation
by diagonal matrices and by scaling. The corresponding generators of its dual are denoted by
$x_1,\ldots,x_L$ and $\A$.
In fact, since conjugation by a scalar is trivial,
the torus acting is really of dimension $L$, and this amounts to saying that
all multidegrees we shall consider only depend on differences $x_i-x_j$ (and on $\A$).

Finally, the $\mathcal{O}_\pi$ are $T$-invariant; we define $\Omega_\pi$, the {\em Joseph--Melnikov} polynomial, to be
\[
\Omega_\pi=\mdeg_\gothnp \mathcal{O}_\pi
\]
These are extended Joseph polynomials, in the sense that 
the original Joseph polynomials correspond to $\A=0$ (no scaling action). 
Melnikov is the one that initiated the study of these particular orbital varieties \cite{Meln}.

The specialization $\A=1$, $x_i=0$, corresponding to considering the action by scaling only,
gives the {\em degree}\/ of $\mathcal{O}_\pi$ (i.e.,\ the number of intersections
with a generic linear subspace of complementary dimension;
in the case of a complete intersection, it is just
the product of degrees of defining equations).

\goodbreak{\em Examples}:
\begin{itemize}
\item If $\pi=0^n$, the base link pattern, then $\mathcal{O}_{0^n}$ is a linear subspace where the
upper-right $n\times n$ block is free while all other entries are zero. Thus,
\begin{equation}\label{ratbasecase}
\mathcal{O}_{0^n}=\left\{ \begin{pmatrix}0&\star\\ 0 &0\end{pmatrix} \right\}
\qquad
\Omega_{0^n}=\prod_{1\le i<j\le n} (\A+x_i-x_j)\prod_{n+1\le i<j\le 2n} (\A+x_i-x_j)
\end{equation}
We note the similarity with \eqref{basecase}.

\item A random example in size $L=6$:
\begin{equation}\label{exjo}
\pi=\vcenterbox{\hdrlinepthree[0.72cm]{4}}
\qquad
\mathcal{O}_\pi=
\left\{ M
\in\gothnp : \begin{gathered}M_{12}=M_{34}=M_{35}=M_{45}=0\\
(M^2)_{16}=(M^2)_{26}=0 \end{gathered}\right\}
\end{equation}
Again, there are simple rules to determine which equations are actually needed.
Here we have six equations left which is the codimension, so that
\[
\Omega_\pi=(\A+x_1-x_2)(\A+x_3-x_4)(\A+x_3-x_5)(\A+x_4-x_5)(2\A+x_1-x_6)(2\A+x_2-x_6)
\]
Compare with the example at the end of section \ref{constrsol}.
\end{itemize}

\subsubsection{The Temperley--Lieb algebra revisited}
In general, Joseph polynomials are known to be related to representations of the corresponding
Weyl group, in our case the symmetric group. Note that the latter is nothing but the limit $q\to\pm1$
of the Hecke algebra. For the particular nilpotent orbit we have considered, we find unsurprisingly
a representation of the Temperley--Lieb quotient of the symmetric group. However our extra scaling
action makes the story much more interesting and produces a connection to the rational $q$KZ equation.

Consider the Temperley--Lieb algebra with parameter $\tau=2$ (or $q=-1$) acting as in section \ref{sectl}
on the space of link patterns. The rational $R$-matrix is of the form
\[
\cR_i(u)=\frac{(\A-u)+u \, e_i}{\A+u}
\]
where we recall that graphically, 
$e_i=\vcenterbox{\rotatebox{45}{\hdrturnone}}$,
and the identity is $1=\vcenterbox{\rotatebox{45}{\hdrturntwo}}$.
It can be deduced from the trigonometric $R$-matrix \eqref{rmat} by sending $q\to -1$ in the following way
\[
q=-\e{-\hbar \A/2},\quad z=\e{\hbar u},\quad \hbar\to 0
\]

The {\em rational $q$KZ system} is
\begin{align}
\check R_i(x_i-x_{i+1}) \Omega_L(x_1,\ldots,x_L)&=\Omega_L(x_1,\ldots,x_{i+1},x_i,\ldots,x_L)\label{ratqkza}\\
\rho^{-1} \Omega_L(x_1,\ldots,x_L)&=(-1)^{n-1} \Omega_L(x_2,\ldots,x_L,x_1+3\A)\label{ratqkzb}
\end{align}
It can be obtained from the (trigonometric) $q$KZ system (\ref{qkza},\ref{qkzb})
by once again sending $q$ to $-1$:
\[
q=-\e{-\hbar \A/2},\quad
z_i=\e{-\hbar x_i},\quad \hbar\to 0
\]
The claim is that $\Omega_L$, the vector of Joseph--Melnikov polynomials defined in the previous section, solves (\ref{ratqkza},\ref{ratqkzb}),
and in fact coincides with the $\hbar\to0$ limit of $(-1)^{n(n-1)/2} \hbar^{-n(n-1)}\Psi_L$, where $\Psi_L$ is
the solution of the trigonometric $q$KZ system that was discussed in section \ref{preconstrsol}.

We can proceed analogously to section \ref{constrsol} and rewrite \eqref{ratqkza} in components by separating it
into two cases:
\begin{align}
(i,i+1)&\text{ not paired in }\pi: & \der_i \left(\frac{\Omega_\pi}{\A+x_i-x_{i+1}}\right)&=0\label{ratqkzc}\\
(i,i+1)&\text{ paired in }\pi:& -(\A+x_i-x_{i+1})\der_i \Omega_\pi&=\sum_{\pi'\ne\pi,e_i\cdot\pi'=\pi} \Omega_{\pi'}\label{ratqkzd}
\end{align}

We shall now discuss the geometric meaning of (\ref{ratqkzc},\ref{ratqkzd}).
Since we know the base case \eqref{ratbasecase}, this will suffice to prove that $\Omega_L$,
the vector of multidegrees, satisfies the whole $q$KZ system. Still, it would be satisfactory to have a geometric
interpretation of \eqref{ratqkzb} too. Unfortunately, it is currently unknown.
Note that the effect of the r.h.s.\ of \eqref{ratqkzb} on multidegrees can be quite drastic: for example, starting from
the example \eqref{exjo}, one goes back to the base case $0^3$, cf.\ \eqref{ratbasecase}; 
but so doing, two quadratic equations have turned into linear equations!

{\em Remark:} one can write a rational $q$KZ equation 
which is analogous to \eqref{qkz-fr} but with additive spectral parameters.
It should not be confused with the Knizhnik--Zamolodchikov
(KZ) equation: the latter is recovered by the further limit $\A\to 0$, turning
the difference equation into a differential equation.

\subsubsection{The Hotta construction}
The Hotta construction \cite{Hotta} is intended to explain the Joseph representation
(Weyl group representation on Joseph polynomials) for an arbitrary orbital variety. When extended to our scaling action,
it will produce the exchange relation \eqref{ratqkza}. For more details see \cite{artic35,artic39}.

The idea, as in the Schubert case, is to try to ``sweep'' with a subgroup $L_i$, here acting by
conjugation, and to apply the lemma of section \ref{secbs}. We need to be a little careful that our embedding space,
$\gothnp$, is not $L_i$-invariant, so we must translate multidegrees in $\gothnp$ to multidegrees in $\gothg$.
It is easy to see that
\[
\mdeg_\gothg W =\prod_{i\ge j}(\A+x_i-x_j)\, \mdeg_\gothnp W \qquad W\subset \gothnp
\]
so that the effect of sweeping with $L_i$ amounts to a ``gauged'' divided difference operator $\tilde\der_i$:
\[
\tilde\der_i f 
=\frac{1}{\A+x_{i+1}-x_i}\partial_i ((\A+x_{i+1}-x_i)f)
=(\A+x_{i}-x_{i+1})\der_i \frac{f}{\A+x_{i}-x_{i+1}}
\]

Now, letting $L_i$ and $B_{+,i}$ act by conjugation,
start with an orbital variety $\mathcal{O}_\pi$ which according to \eqref{orborb}, is $B_{+,i}$-invariant.
If one tries to sweep it with $L_i$,
one can in general produce non-upper triangular matrices. In fact, a small calculation
shows that this occurs exactly if $M_{i\,i+1}\ne 0$.
So, we must distinguish two cases:
\begin{itemize}
\item If $(i,i+1)$ are not paired in $\pi$, this means that the rank condition at $(i,i+1)$ in \eqref{orbeq} says
that $M_{i\,i+1}=0$; which in turn implies that $\A+x_i-x_{i+1}\,|\,\Omega_\pi$. Furthermore, in this case $\mathcal{O}_\pi$
is $L_i$-invariant, so that $\tilde\der_i \Omega_\pi=0$. This is exactly \eqref{ratqkzc}.

\item If $(i,i+1)$ is a pair in $\pi$, then generically $M_{i\,i+1}\ne 0$ in $\mathcal{O}_\pi$. We then proceed in two steps.
{\em Cutting}: first we intersect $\mathcal{O}_\pi$ with the hyperplane $M_{i\,i+1}=0$. Since the intersection
is transverse, the effect on the multidegree is to multiply by the weight of the hyperplane, which is
$\A+x_i-x_{i+1}$.
{\em Sweeping}: now we can sweep with $L_i$ and we stay inside $\mathcal{O}_L$.
One can check that the map $\mu$ is generically one-to-one, and by dimension argument the image
must be a union of orbital varieties (plus possibly some lower-dimensional pieces).
The claim, which we shall not attempt to justify here, is that the orbital varieties thus obtained
are exactly the proper preimages of $\pi$ under $e_i$. So we find $-\tilde\der_i ((\A+x_i-x_{i+1})\Omega_\pi)=
\sum_{\pi'\ne\pi,e_i\cdot\pi'=\pi} \Omega_{\pi'}$, which is equivalent to \eqref{ratqkzd}.
\end{itemize}

\subsubsection{Recurrence relations and wheel condition}\label{recgeom}
Several other constructions have simple geometric meaning. We mention
in passing here the meaning of recurrence relations and of the wheel condition.

Set $x_{i+1}=x_i+\A$. This gives the weight of $0$ to $M_{i\,i+1}$. Roughly,
this corresponds to looking at what happens when $M_{i\,i+1}\to\infty$ (this is more or less
clear from the integral formula of section \ref{secintmdeg};
for a more precise statement, see \cite{KMY}). To remain inside $\mathcal{O}_L$, 
writing the equations $(M^2)_{j\,i+1}=0$ and $(M^2)_{ij}=0$, one concludes that one must
have $M_{ji}=0$ and $M_{i+1\,j}=0$ for all $j$. At this stage one notes that
the entries $M_{j\,i+1}$ and $M_{ij}$ are unconstrained by the equations. Removing the $i^{\rm th}$
and $(i+1)^{\rm st}$ rows and columns reduces $\mathcal{O}_L$ to $\mathcal{O}_{L-2}$.

What we have just shown is that when $M_{i\,i+1}\to\infty$, being in $\mathcal{O}_L$ amounts
to setting a certain number of entries to zero and then the result is some irrelevant
linear space times $\mathcal{O}_{L-2}$. One can be a bit more careful and do this reasoning
at the level of each irreducible component: it is easy to see that only the components
that have a pair $(i,i+1)$ survive (the others satisfy $M_{i\,i+1}=0$), and that 
$\mathcal{O}_{\varphi_i\pi}$ gets sent to $\mathcal{O}_\pi$. Finally, we get the recurrence
relations:
\begin{equation}\label{ratrecrel}
\Omega_\pi(\ldots,x_{i+1}=x_i+\A,\ldots)=
\begin{cases}
0&\pi\not\in\Im\varphi_i\\
\begin{aligned}
\prod_{j=1}^{i-1}&(\A+x_j-x_i)\prod_{j=i+2}^{2n}(2\A+x_i-x_j)\\
&\Omega_{\pi'}(x_1,\ldots,x_{i-1},x_{i+2},\ldots,x_L)
\end{aligned}
&\pi=\varphi_i\pi'
\end{cases}
\end{equation}
to be compared with \eqref{recrel}.

Similarly, if one sets $x_k=x_j+\A=x_i+2\A$, $i<j<k$, 
then the equation $(M^2)_{ik}=0$ cannot possibly be satisfied since
it contains the infinite term $M_{ij}M_{jk}$, so all multidegrees must vanish,
which is nothing but the wheel condition.

\subsection{Brauer loop scheme}
In this section we introduce a new affine scheme whose existence was suggested by the underlying integrable model.
The subject has an interesting history, which we summarize now. In 2003, Knutson, in his study of the commuting variety,
introduced the upper-upper scheme \cite{Kn-uu}; one of its components is closely related to the commuting variety,
and in particular has the same degree. 
In an a priori unrelated development, de Gier and Nienhuis \cite{dGN-brauer} studied the Brauer loop model, a model of crossing
loops which is completely similar to the one that we described in section \ref{loopcyl}, except crossing plaquettes
($\vcenterbox{\hdrcrossingplaquette}$) are allowed. They found that the entries of the ground state
(or steady state of the corresponding Markov process) are again integers, and observed
empirically that certain entries coincide with the degrees of the irreducible components of the upper-upper scheme.
This mysterious connection required an explanation. A partial one was given in \cite{artic32}, where
the corresponding inhomogeneous model was introduced and it was suggested that this generalization corresponds
to going over from degrees to multidegrees. Also, an idea of the geometric action of the Brauer algebra was given.
But some entries remained unidentified. In \cite{artic33},
the Brauer loop scheme was introduced, and it was proved that its top-dimensional components produce all the entries
of the ground state. Finally, in the recent preprint \cite{artic39}, it is shown that more generally,
an appropriate polynomial solution of the $q$KZ system associated to the Brauer algebra produces multidegrees
of these components. 

\subsubsection{Geometric description}
There are several equivalent descriptions of the Brauer loop scheme, see \cite{artic33,artic39}.
Here we try to present it in a way which best respects the underlying symmetries.

Let $L$ be an integer.
Consider complex upper triangular matrices that are infinite in both directions and that are periodic by shift
by $(L,L)$:
\[
R_{\integers\bmod L}
=\{ M=(M_{ij})_{i,j\in\integers}: M_{ij}=M_{i+L\,j+L}\ \forall i,j\in\integers\}
\]
$R_{\integers\bmod L}$ is an algebra.
Let $S=(\delta_{i,j-1})\in R_{\integers\bmod L}$ 
be the shift operator. Then we define the algebra $\gothb$ to be the quotient
\[
\gothb=
R_{\integers\bmod L}/\left< S^L\right>
\]
$\gothb$ is finite-dimensional: $\dim \gothb=L^2$.
Inside $\gothb$ we have its radical $\gothn$,
which consists of (classes of) matrices with zero diagonal entries,
and the group $\gothB$ of its invertible elements, i.e.,\ with non-zero diagonal entries.

The Brauer loop scheme is then defined as
\[
E_L=\{ M\in\gothn: M^2=0 \}
\]
For a different point of view on the origin of $E_L$, see section 1.3 of \cite{artic39}.

In all that follows, we take $L$ to be even, $L=2n$. We define a {\em crossing link pattern}\/ to be an involution
of $\integers/L\integers$ without fixed points. Their set is denoted by $P^{cr}_{2n}$ and is of cardinality $(2n-1)!!$.
They have a similar graphical representation as link patterns, but in which crossings are allowed:
\[
P^{cr}_4=\left\{
\vcenterbox{\hdrlptwo{1}},
\vcenterbox{\hdrlptwo{2}},
\vcenterbox{\hdrlptwo{3}}
\right\}
\]

The irreducible components of $E_L$ are known to be indexed by
crossing link patterns \cite{artic33,rothbach}. If $\pi\in P^{cr}_{2n}$, then
\[
E_\pi=\overline{\{ M\in E_L: (M^2)_{i,i+L}=(M^2)_{j,j+L}\ \Leftrightarrow\ i=j\text{ or }i=\pi(j)\bmod L\}}
\]
Note that this definition (i) makes sense because $(M^2)_{i,i+L}$, despite being undefined for a generic
element of $\gothb$ (it is killed by the quotient by $S^L$), is actually well-defined for elements of $\gothn$
and (ii) implies that the $L$ numbers $(M^2)_{i,i+L}$ always come in pairs for $M\in E_L$, which is somewhat surprising
(an elementary proof would be nice, as opposed to the proof of \cite{artic33}).

We can also describe $E_\pi$ as the closure of a union of orbits:
\begin{equation}\label{brauerorb}
E_\pi=\overline{\gothB\cdot \gotht\pi_<},\quad \text{$\gothB$ acts by conjugation, }
\gotht\pi_<=\{ M\in \gothb: M_{ij}\ne 0\ \Rightarrow\ i=\pi(j)\bmod L\}
\end{equation}

Finally, every matrix $M\in E_\pi$ satisfies the following equations:
\begin{equation}\label{eqbrauer}
\begin{gathered}
M^2=0,\\
(M^2)_{i,i+L}=(M^2)_{\pi(i),\pi(i)+L}
\qquad (i\in\integers),\\
\rank M^{\lowerleft}\le \rank \pi_<{}^{\!\lowerleft}
\qquad (i,j\in\integers,\ j<i+L).
\end{gathered}
\end{equation}
where $\pi_<$ is any generic matrix in $\gotht\pi_<$, for example the one made of zeroes and ones.
These equations do not, in general, suffice to define $E_\pi$.

Let us now turn to the torus action on $\gothb$. First there is the usual scaling action, with generator $\A$.
But the remaining $L$-dimensional torus action is more difficult to explain.
Let $(x_i)_{i\in\integers}$ be a set of formal variables satisfying the relations $x_{i+L}=x_i+\epsilon$, where $\epsilon$ is another
formal variable. Then the action of the full torus $T=(\C^\times)^{L+1}$ is defined by writing that
\[
\wt(M_{ij})=\A+x_i-x_j\qquad i,j\in\integers, i\le j
\]

It is slightly non-trivial but true that $E_L$ and the $E_\pi$ are $T$-invariant. 
We define the {\em Brauer loop polynomials}\/ $\Upsilon_\pi$ to be their multidegrees:
\[
\Upsilon_\pi=\mdeg_{\gothn} E_\pi
\]
$\Upsilon_\pi$ is a homogeneous polynomial in $L+1$ variables, which we can choose to be
$\A$, $x_i-x_1$ ($i=2,\ldots,L$) and $\epsilon$. Its degree is the codimension of $E_\pi$, which is $2n(n-1)$.

\goodbreak{\em Examples}:
\begin{itemize}
\item The maximally crossed link pattern $\chi^n$ corresponds to $\chi^n(i)=i+n$.
In this case, $E_{\chi^n}$ is a linear subspace defined by the equations $M_{ij}=0$ if $j\le i+n-1$.
Thus,
\begin{equation}\label{brauerbasecase}
\chi^n=\vcenterbox{\hdrmaximallycrossed}
\qquad
\Upsilon_{\chi^n}=\prod_{i=1}^L \prod_{j=i+1}^{i+n-1}(\A+x_i-x_j)  
\end{equation}

\item At the opposite end, we find the non-crossing link patterns of $P_{2n}$.
Among them, we have $0^n$ (or any of its rotations), which has all pairings parallel.
In the Brauer loop scheme, they play a special role: Knutson proved, in the context
of the upper-upper scheme \cite{Kn-uu}, that there is a Gr\"obner degeneration
of $C_n\times V$ to $E_{0^n}$, where $V$ is some irrelevant vector space, and
$C_n$ is the commuting variety:
\[
C_n=\{ (X,Y)\in \mathfrak{gl}(n,\C)^2: XY=YX \}
\]
 
\item Explicit examples can become quite complicated; even in size $L=4$, we find, with the notation 
$\B=\A-\epsilon$:
\begin{align*}\label{exbrauer}
\chi^2=&\vcenterbox{\hdrlptwo[0.92cm]{3}}
\qquad
E_{\chi^2}=
\left\{ M\in\gothn : M_{12}=M_{23}=M_{34}=M_{45}=0\right\}\\
\Upsilon_{\chi^2}=&(\A+x_1-x_2)(\A+x_2-x_3)(\A+x_3-x_4)(\B+x_4-x_1)\\[0.3cm]
0^2=&\vcenterbox{\hdrlptwo[0.92cm]{1}}
\qquad
E_{0^2}=\left\{
M\in\gothn : \begin{gathered}M_{12}=M_{34}=0\\ M_{23}M_{35}+M_{24}M_{45}=0\\ M_{45}M_{57}+M_{46}M_{67}=0\\ M_{13}M_{35}-M_{24}M_{46}=0\end{gathered}
\right\}\\
\Upsilon_{0^2}=&(\A+x_1-x_2)(\A+x_3-x_4)\\
&(\A^2+\A\B+\B^2-\B x_1+\A x_2+x_1x_2-\A x_3-x_2x_3+\B x_4-x_1x_4+x_3x_4)\\[0.3cm]
1^2=&\rho(0^2)=\vcenterbox{\hdrlptwo[0.92cm]{2}}
\qquad
E_{1^2}=\left\{
M\in\gothn : \begin{gathered}M_{23}=M_{45}=0\\ M_{12}M_{24}+M_{13}M_{34}=0\\ M_{34}M_{46}+M_{35}M_{56}=0\\ M_{13}M_{35}-M_{24}M_{46}=0\end{gathered}
\right\}\\
\Upsilon_{1^2}=&(\A+x_2-x_3)(\B+x_4-x_1)\\
&(\A^2+2\A\B+\B x_1-\A x_2-x_1x_2+\A x_3+x_2x_3-\B x_4+x_1x_4-x_3x_4)
\end{align*}
The last two varieties are not complete intersections.
\end{itemize}

\subsubsection{The Brauer algebra}
The 
{\dfn Brauer algebra}\/ is 
defined by generators $f_i$, $e_i$, $i=1,\ldots,L-1$, and relations
\begin{equation}\label{eqn:defbrauer}
\begin{split}
&e_i^2=\tau e_i\qquad e_ie_{i\pm 1}e_i=e_i\\
&f_i^2=1\qquad (f_if_{i+1})^3=1\\
&f_ie_i=e_if_i=e_i\quad e_i f_i f_{i+1}=e_i e_{i+1}=f_{i+1}f_ie_{i+1}\quad e_{i+1}f_i f_{i+1}=e_{i+1}e_i=f_if_{i+1}e_i\\
&e_i e_j=e_j e_i\qquad f_i f_j=f_j f_i\qquad e_i f_j=f_je_i\qquad |i-j|>1
\end{split}
\end{equation}
where indices take all values for which the identities make sense. 
Note that the $f_i$ are generators of the symmetric group $\mathcal{S}_L$, whereas the $e_i$ generate a Temperley--Lieb
algebra.

There is an associated solution of the Yang--Baxter equation, namely
\begin{equation}\label{brauerRmat}
  \check R_i(u)
  ={\A(\A-u)+\A\, u\, e_i + (1-\tau/2)u(\A-u) f_i\over (\A+u)(\A-(1-\tau/2)u)}
\end{equation}

The Brauer algebra acts naturally on crossing link patterns.
The graphical rules are the same as in the previous sections, and we shall not illustrate them again.
If $\pi$ is viewed as an involution, then $f_i$ acts by conjugation by the transposition $(i,i+1)$, whereas
$e_i$ creates new cycles $(i,i+1)$ and $(\pi(i),\pi(i+1))$ -- unless $\pi(i)=i+1$, in which case it multiplies
the state by $\tau$.

We now claim that the vector $\Upsilon_L$ of multidegrees $\Upsilon_\pi$ of the irreducible components of
the Brauer loop scheme solves the $q$KZ system associated to the Brauer loop model \cite{artic39},
which we can write
\begin{equation} \label{brauerqkz}
\check R_i(x_{i}-x_{i+1})\Upsilon_L=s_i \Upsilon_L\qquad\forall i\in\integers
\end{equation}
where $s_i$ permutes $x_{i+kL}$ and $x_{i+1+kL}$ for all $k$, on condition that the following identification
of the parameters is made:
\begin{equation}\label{taueps}
\tau=\frac{2(\A-\epsilon)}{2\A-\epsilon}
\end{equation}
We could also add the cyclicity condition (which is obviously satisfied by $\Upsilon_L$ by rotational invariance),
but note that it could at most be marginally stronger than \eqref{brauerqkz} (and in fact, it is not), because we
have imposed \eqref{brauerqkz} for all integer values of $i$. In other words, this is already an ``affinized'' version of
the exchange relation because of our shifted periodicity property $x_{i+L}=x_i+\epsilon$.

For future use, let us write in components \eqref{brauerqkz}. We find the usual dichotomy:
\begin{align}
\pi(i)&\ne i+1:
\quad
-(\A+x_i-x_{i+1})((\A+\B)\der_i+s_i)\left({\Upsilon_\pi\over \A+x_i-x_{i+1}}\right)=\Upsilon_{f_i\cdot\pi}
\label{brauerqkzc}\\
\pi(i)&=i+1:
\quad
-(\A+x_i-x_{i+1})(\A+\B+x_{i+1}-x_i)\der_i \Upsilon_\pi=(\A+\B)\sum_{\pi'\ne\pi,e_i\cdot\pi'=\pi} \Upsilon_{\pi'}
\label{brauerqkzd}
\end{align}
with the convenient notation $\B=\A-\epsilon$.

{\em Remark:} The Brauer algebra is the rational limit of the BWM algebra considered in \cite{Pas-BWM}.

\subsubsection{Geometric action of the Brauer algebra}
This is the most technical part of \cite{artic39}, which we shall only sketch here. Once again,
it follows the same general idea of trying to ``sweep'' with a subgroup $\hat L_i$ 
analogous to the $L_i$ used so far. In our setting of infinite periodic matrices,
this $\hat L_i$ consists of invertible matrices which are the identity everywhere
except in the entries $M_{ab}$ with $a,b\in\{i+kL,i+1+kL\}$ for some $k$; and $\hat{B}_{+,i} = \hat L_i \cap \gothB$.
The additional subtlety comes from the fact that contrary to the case of orbital varieties, the condition
$M^2=0$ is {\em not}\/ stable by conjugation by $\hat L_i$ (i.e.,\ $\left<S^L\right>$ is not stable by conjugation by
$\hat L_i$, since the latter is outside $\gothb$). We shall have to reimpose one equation, namely
$(M^2)_{i+1\,i+L}=0$, after sweeping.

So, letting $\hat L_i$ and $\hat B_{+,i}$ act by conjugation,
start with a component $E_\pi$ which according to \eqref{brauerorb}, is $\hat B_{+,i}$-invariant.
As usual we have to distinguish the two cases:

\begin{itemize}
\item $\pi(i)\ne i+1$: in this case we can directly {\em sweep}\/ with $\hat L_i$, and then {\em cut}\/
with $(M^2)_{i+1,i+L}$. As shown in \cite{artic32,artic33,artic39} with varying levels of rigor and clarity,
the result is precisely $E_\pi \cup E_{f_i\cdot\pi}$. So we find, at the level of multidegrees,
\[
-(\A+\B+x_{i+1}-x_i)\tilde\der_i \Upsilon_\pi=\Upsilon_\pi + \Upsilon_{f_i\cdot\pi}
\]
which reduces after a few manipulations to \eqref{brauerqkzc}.

\item $\pi(i)=i+1$: this time we first {\em cut}\/ with $M_{i\,i+1}=0$, 
{\em sweep}\/ with $\hat L_i$ (throwing away the $\hat L_i$-invariant piece which cannot
contribute to the multidegree calculation), 
and then {\em cut}\/ again with $(M^2)_{i+1\,i+L}$.
We lost one dimension in the process, so the result cannot simply be a union of components.
In fact, one can show that it is $\bigcup_{\pi'\ne\pi, e_i\cdot\pi'=\pi} E_{\pi'}
\cap \{ (M^2)_{i\,i+L}=(M^2)_{i+1\,i+L+1}\}$. This results directly in the multidegree identity \eqref{brauerqkzd}.
\end{itemize}

Note that the second case can be reduced to the first case after the first step, at least at the level of multidegrees, since the sweeping/cutting procedure is the same and we have an explicit description of the intermediate object
$E_\pi \cap \{ M_{i,i+1}=0 \}$ (up to $L_i$-invariant pieces), see
\cite{artic39} for details.

Recurrence relations can also be written and interpreted geometrically as in section \ref{recgeom}, but we shall
omit them for brevity. Let us simply mention the wheel condition: the $\Upsilon_\pi$ satisfy
\[
\Upsilon_\pi(x_i=x,x_j=x+\A,x_k=x+2\A)=0
\qquad i<j<k<i+L
\]
(cf.\ the appendix B of \cite{Pas-BWM}). This stems geometrically from the equation $(M^2)_{ik}=0$ (which is not
killed by the quotient by $S^L$ since $k<i+L$), which cannot be satisfied when $M_{ij},M_{jk}\to\infty$.

\subsubsection{The degenerate limit}
There are two special values of the parameter $\tau$. One, on which we shall not dwell,
is $\tau=1$, which corresponds to $\epsilon=0$ in \eqref{taueps}. This is the value for which $\Upsilon_L$
becomes the ground state eigenvector of an integrable transfer matrix, that of \cite{dGN-brauer}, and is
the subject of \cite{artic32,artic33}. Various interesting results can be obtained, including sum rules.
In particular, in the homogeneous limit $x_i\to0$ multidegrees become degrees and we recover
the integer numbers observed in \cite{dGN-brauer}.

However, there is another special point: $\tau=2$. Indeed, we see that the coefficient of $f_i$ in the
$R$-matrix \eqref{brauerRmat} vanishes. In fact we recover this way the Temperley--Lieb solution
of the rational Yang--Baxter equation, which was used in connection with orbital varieties. What is the geometric
meaning of this reduction?

Plugging $\tau=2$ into \eqref{taueps}, we note that there is no solution for $\epsilon$, since
$\epsilon\to\infty$ when $\tau\to 2$. Let us carefully take the limit $\epsilon\to\infty$ in the polynomials
$\Upsilon_\pi$. We consider them as polynomials in $x_1,\ldots,x_L$, $\A$ and $\epsilon$ (this choice breaks
the rotational invariance $x_i\to x_{i+1}$) and keep the former fixed while sending the latter to infinity.
Geometrically, the situation is as follows. Let us compute the weights of the entries $M_{ij}$ of a matrix
$M\in\gothb$ in terms of $x_1,\ldots,x_L$, $\A$ and $\B=\A-\epsilon$. Due to the periodicity, we can always assume
$i$ to be between $1$ and $L$, and due to the quotient we should then consider $i\le j<i+L$. The result is that
one finds two categories of entries:
\[
\wt(M_{ij})=
\begin{cases}
\A+x_i-x_j&1\le i \le j \le L \\
\B+x_i-x_{j-L}& 1\le j-L< i\le L 
\end{cases}
\]
This amounts to subdividing $\gothb$ (a vector space of dimension $L^2$) as
\[
\gothb=\left\{ \left(\vcenterbox{\hdrupperlowerblock}\right) \right\}
\]
where $U$ is an upper triangular matrix, and $L$ a strict lower triangular matrix. These two pieces
have quite different destinies as $\tau\to2$: the weights of $U$ remain unchanged and are identical to those
that we have used in section \ref{secorbvar} for orbital varieties, whereas the weights of $L$
diverge. It turns out that this limit corresponds to killing off this lower triangular part (as is clear from the definition
of the multidegree as an integral, see section \ref{secintmdeg}); and after taking the
quotient, it is easy to see that we are left with $\gothbp$, the algebra of upper triangular matrices. Let us call
$p$ this projection $(U,L)\mapsto U$ from $\gothb$ to $\gothbp$.

At the level of components, here is what happens: if $\pi$ is a non-crossing link pattern, then $p(E_\pi)
=\mathcal{O}_\pi$, the corresponding
orbital variety. Translated into multidegrees, this means:
\[
\Upsilon_\pi(x_1,\ldots,x_L,\A,\B) 
\buildrel B\to\infty\over \sim
\B^{n(n-1)} \Omega_\pi(x_1,\ldots,x_L,\A)
\]
In general, crossing link patterns will project to lower dimensional $B_+$-orbit closures \cite{artic39}. 

Interestingly, the matrix Schubert varieties and double Schubert polynomials of section \ref{secschub}
also appear as a special case in the $\tau=2$ limit. Indeed, consider the {\em permutation sector},
that is the subset of $P^{cr}_{2n}$ consisting of involutions such that $\pi(\{1,\ldots,n\})=\{n+1,\ldots,2n\}$.
Such involutions are in bijection with permutations $\sigma\in\mathcal{S}_n$, according to
$\sigma(i)=\pi(n+1-i)-n$, $i=1,\ldots,n$:
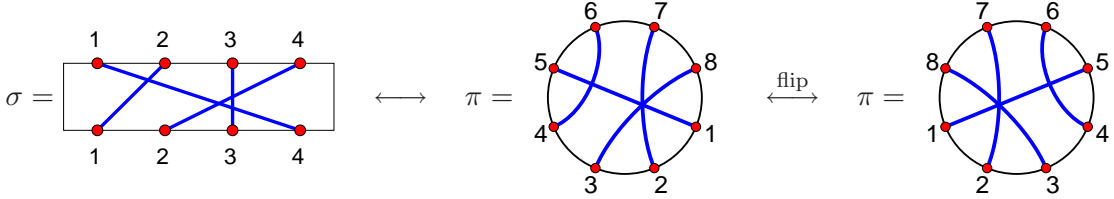
\[
\sigma=\vcenterbox{\hdrnilhecketwo}
\quad\longleftrightarrow\quad
\pi=\vcenterbox{\hdrlpfourexampleone}
\quad\buildrel\text{flip}\over\longleftrightarrow\quad
\pi=\vcenterbox{\hdrlpfourexampletwo}
\]
(the last flip is purely cosmetic and due to the fact that we have written labels of link patterns
counterclockwise up to now).

In the permutation sector, the structure of $E_\pi$ is as follows:
\[
E_\pi=\left\{ \left(\vcenterbox{\hdrbrauerblock}\right) \right\}
\]
If one discards the unconstrained entries (represented by $\star$), one recognizes the pairs of $n\times n$
matrices $(X,Y)$ in terms
of which the upper-upper scheme of Knutson is defined \cite{Kn-uu}. 
In fact the union of $E_\pi$ where $\pi$ runs over the permutation sector
is exactly up to these irrelevant entries the upper-upper scheme.

Now the projection of such components $E_\pi$, that is here $(X,Y)\mapsto X$,
is essentially the corresponding matrix Schubert variety $S_\sigma$.
The description in terms of orbits gives a proof. Also compare the defining equations \eqref{eqschub} of matrix Schubert varieties
with the equations \eqref{eqbrauer} satisfied on $E_\pi$
(in the permutation sector, these are the equations in the first conjecture of section 3 of \cite{Kn-uu}). More precisely, there is a vertical flip, so that
$p(E_\pi)\simeq S_\sigma \sigma_0$. In terms of multidegrees, this means that
\begin{multline*}
\Upsilon_\pi(x_1,\ldots,x_{2n},\A,\B) 
\buildrel B\to\infty\over \sim
\B^{n(n-1)-|\sigma|} 
\prod_{1\le i<j\le n}(\A+x_i-x_j)\prod_{n+1\le i<j\le 2n}(\A+x_i-x_j)
\\
\Xi_\sigma(\A+x_n,\ldots,\A+x_1|x_{n+1},\ldots,x_{2n})
\end{multline*}
(note that the prefactor is nothing but $\Omega_{0^n}$).
So the exchange relation \eqref{schubex} satisfied by double Schubert polynomials (as well as the symmetric
one \eqref{schubexb}) should be somehow contained in the exchange relation \eqref{brauerqkz}
of Brauer loop polynomials. In order to see this, one must redefine the generator $f_i$ to $t_i=(1-\tau/2)f_i$ when
taking the limit $\tau\to 2$. The operators $t_i$ then satisfy the nil-Hecke algebra, and one can check that
in the permutation sector, where the $e_i$ part of the $R$-matrix never contributes when $i\ne n$, the exchange relation
\eqref{brauerqkz} reduces to \eqref{schubex} if $i<n$ or to \eqref{schubexb} if $i>n$.


\bibliography{biblio}

\def\cprime{$'$}
\providecommand{\bysame}{\leavevmode\hbox to3em{\hrulefill}\thinspace}
\begin{thebibliography}{100}

\bibitem{AR}
D.~Allison and Nicolai Reshetikhin, \emph{Numerical study of the 6-vertex model
  with domain wall boundary conditions}, Ann. Inst. Fourier (Grenoble)
  \textbf{55} (2005), no.~6, 1847--1869,
  \href{http://arxiv.org/abs/cond-mat/0502314}{\path{arXiv:cond-mat/0502314}}.
  \MR{MR2187938 (2006i:82007)}.

\bibitem{And-TSSCPP}
George Andrews, \emph{Plane partitions. {V}. {T}he {TSSCPP} conjecture}, J.
  Combin. Theory Ser. A \textbf{66} (1994), no.~1, 28--39. \MR{MR1273289
  (95g:05010)}.

\bibitem{BdGN-XXZ-ASM-PP}
Murray Batchelor, Jan de~Gier, and Bernard Nienhuis, \emph{The quantum
  symmetric {XXZ} chain at {$\Delta=-1/2$}, alternating-sign matrices and plane
  partitions}, J. Phys. A \textbf{34} (2001), no.~19, L265--L270,
  \href{http://arxiv.org/abs/cond-mat/0101385}{\path{arXiv:cond-mat/0101385}}.
  \MR{MR1836155 (2002e:82014b)}.

\bibitem{Baxter}
R.~Baxter, \emph{Exactly solved models in statistical mechanics}, Academic
  Press, 1982.

\bibitem{BPZ}
A.~Belavin, A.~Polyakov, and A.~Zamolodchikov, \emph{Infinite conformal
  symmetry in two-dimensional quantum field theory}, Nucl. Phys. B \textbf{241}
  (1984), no.~2, 333,
  {\scriptsize\href{http://dx.doi.org/10.1016/0550-3213\%2884\%2990052-X}{\path{doi:10.1016/0550-3213\%2884\%2990052-X}}}.

\bibitem{BF}
P.~Bleher and V.~Fokin, \emph{Exact solution of the six-vertex model with
  domain wall boundary conditions. {D}isordered phase}, Comm. Math. Phys.
  \textbf{268} (2006), no.~1, 223--284,
  \href{http://arxiv.org/abs/math-ph/0510033}{\path{arXiv:math-ph/0510033}}.
  \MR{MR2249800 (2007j:82022)}.

\bibitem{BL2}
P.~Bleher and K.~Liechty, \emph{Exact solution of the six-vertex model with
  domain wall boundary conditions. {C}ritical line between ferroelectric and
  disordered phases}, J. Stat. Phys. \textbf{134} (2009), no.~3, 463--485,
  \href{http://arxiv.org/abs/0802.0690}{\path{arXiv:0802.0690}}.
  \MR{MR2485725}.

\bibitem{BL}
\bysame, \emph{Exact solution of the six-vertex model with domain wall boundary
  conditions. {F}erroelectric phase}, Comm. Math. Phys. \textbf{286} (2009),
  no.~2, 777--801,
  \href{http://arxiv.org/abs/0712.4091}{\path{arXiv:0712.4091}}.
  \MR{MR2472044}.

\bibitem{Bressoud}
David Bressoud, \emph{Proofs and confirmations: The story of the alternating
  sign matrix conjecture}, MAA Spectrum, Mathematical Association of America,
  Washington, DC, 1999. \MR{MR1718370 (2000i:15002)}.

\bibitem{HC}
Harish Chandra, \emph{Differential operators on a semi-simple {L}ie algebra},
  Amer. J. Math. \textbf{79} (1957), 87--120.

\bibitem{Che-qKZ}
Ivan Cherednik, \emph{Quantum {K}nizhnik--{Z}amolodchikov equations and affine
  root systems}, Comm. Math. Phys. \textbf{150} (1992), no.~1, 109--136,
  \url{http://projecteuclid.org/getRecord?id=euclid.cmp/1104251785}.
  \MR{MR1188499 (94a:17019)}.

\bibitem{Che}
\bysame, \emph{Double affine {H}ecke algebras}, London Mathematical Society
  Lecture Note Series, vol. 319, Cambridge University Press, Cambridge, 2005.
  \MR{MR2133033 (2007e:32012)}.

\bibitem{CLP}
Henry Cohn, Michael Larsen, and Jim Propp, \emph{The shape of a typical boxed
  plane partition}, New York J. Math. \textbf{4} (1998), 137--165,
  \href{http://arxiv.org/abs/math/9801059}{\path{arXiv:math/9801059}}.
  \MR{MR1641839 (99j:60011)}.

\bibitem{CP}
Filippo Colomo and Andrei Pronko, \emph{The {A}rctic {C}ircle revisited},
  Integrable systems and random matrices, Contemp. Math., vol. 458, Amer. Math.
  Soc., Providence, RI, 2008, pp.~361--376,
  \href{http://arxiv.org/abs/0704.0362}{\path{arXiv:0704.0362}}. \MR{2411918
  (2010d:82018)}.

\bibitem{CP2}
\bysame, \emph{Emptiness formation probability in the domain-wall six-vertex
  model}, Nuclear Phys. B \textbf{798} (2008), no.~3, 340--362,
  \href{http://arxiv.org/abs/0712.1524}{\path{arXiv:0712.1524}}.
  \MR{MR2411855}.

\bibitem{CP3}
\bysame, \emph{The limit shape of large alternating sign matrices}, 2008,
  \href{http://arxiv.org/abs/0803.2697}{\path{arXiv:0803.2697}}.

\bibitem{DCN}
D.~Cont and Bernard Nienhuis, \emph{The packing of two species of polygons on
  the square lattice}, J. Phys. A \textbf{37} (2004), no.~9, 3085--3100,
  {\scriptsize\href{http://dx.doi.org/10.1088/0305-4470/37/9/002}{\path{doi:10.1088/0305-4470/37/9/002}}}.
  \MR{2042607 (2004m:82014)}.

\bibitem{dG-review}
Jan de~Gier, \emph{Loops, matchings and alternating-sign matrices}, Discrete
  Math. \textbf{298} (2005), no.~1-3, 365--388,
  \href{http://arxiv.org/abs/math/0211285}{\path{arXiv:math/0211285}}.
  \MR{MR2163456 (2006i:05007)}.

\bibitem{dGN-brauer}
Jan de~Gier and Bernard Nienhuis, \emph{Brauer loops and the commuting
  variety}, J. Stat. Mech. Theory Exp. (2005), no.~1, 006, 10 pp,
  \href{http://arxiv.org/abs/math.AG/0410392}{\path{arXiv:math.AG/0410392}}.
  \MR{MR2114232 (2005m:82038)}.

\bibitem{dGP-factor}
Jan de~Gier and Pavel Pyatov, \emph{Factorised solutions of {T}emperley--{L}ieb
  $q${KZ} equations on a segment}, 2007,
  \href{http://arxiv.org/abs/0710.5362}{\path{arXiv:0710.5362}}.

\bibitem{artic44}
Jan de~Gier, Pavel Pyatov, and Paul Zinn-Justin, \emph{Punctured plane
  partitions and the {$q$}-deformed {K}nizhnik--{Z}amolodchikov and {H}irota
  equations}, J. Combin. Theory Ser. A \textbf{116} (2009), no.~4, 772--794,
  \href{http://arxiv.org/abs/0712.3584}{\path{arXiv:0712.3584}},
  {\scriptsize\href{http://dx.doi.org/10.1016/j.jcta.2008.11.008}{\path{doi:10.1016/j.jcta.2008.11.008}}}.
  \MR{2513634 (2010k:05028)}.

\bibitem{DF-qKZ-TSSCPP}
Philippe Di~Francesco, \emph{Totally symmetric self-complementary plane
  partitions and the quantum {K}nizhnik--{Z}amolodchikov equation: a
  conjecture}, J. Stat. Mech. Theory Exp. (2006), no.~9, P09008, 14 pp,
  \href{http://arxiv.org/abs/cond-mat/0607499}{\path{arXiv:cond-mat/0607499}}.
  \MR{MR2278472 (2007j:82023)}.

\bibitem{artic31}
Philippe Di~Francesco and Paul Zinn-Justin, \emph{Around the
  {R}azumov--{S}troganov conjecture: proof of a multi-parameter sum rule},
  Electron. J. Combin. \textbf{12} (2005), Research Paper 6, 27 pp,
  \href{http://arxiv.org/abs/math-ph/0410061}{\path{arXiv:math-ph/0410061}}.
  \MR{MR2134169 (2005m:82017)}.

\bibitem{artic34}
\bysame, \emph{Quantum {K}nizhnik--{Z}amolodchikov equation, generalized
  {R}azumov--{S}troganov sum rules and extended {J}oseph polynomials}, J. Phys.
  A \textbf{38} (2005), no.~48, L815--L822,
  \href{http://arxiv.org/abs/math-ph/0508059}{\path{arXiv:math-ph/0508059}},
  {\scriptsize\href{http://dx.doi.org/10.1088/0305-4470/38/48/L02}{\path{doi:10.1088/0305-4470/38/48/L02}}}.
  \MR{MR2185933 (2006m:82019)}.

\bibitem{artic35}
\bysame, \emph{From orbital varieties to alternating sign matrices},
  Proceedings of the 18th {C}onference on {F}ormal {P}ower {S}eries and
  {A}lgebraic {C}ombinatorics ({S}an {D}iego, 2006), 2006,
  \href{http://arxiv.org/abs/math-ph/0512047}{\path{arXiv:math-ph/0512047}}.

\bibitem{artic32}
\bysame, \emph{Inhomogeneous model of crossing loops and multidegrees of some
  algebraic varieties}, Comm. Math. Phys. \textbf{262} (2006), no.~2, 459--487,
  \href{http://arxiv.org/abs/math-ph/0412031}{\path{arXiv:math-ph/0412031}}.
  \MR{MR2200268 (2006k:82068)}.

\bibitem{artic43}
\bysame, \emph{Quantum {K}nizhnik--{Z}amolodchikov equation: reflecting
  boundary conditions and combinatorics}, J. Stat. Mech. Theory Exp. (2007),
  no.~12, P12009, 30 pp,
  \href{http://arxiv.org/abs/0709.3410}{\path{arXiv:0709.3410}},
  {\scriptsize\href{http://dx.doi.org/10.1088/1742-5468/2007/12/P12009}{\path{doi:10.1088/1742-5468/2007/12/P12009}}}.
  \MR{MR2367185}.

\bibitem{artic41}
\bysame, \emph{Quantum {K}nizhnik--{Z}amolodchikov equation, {T}otally
  {S}ymmetric {S}elf-{C}omplementary {P}lane {P}artitions and {A}lternating
  {S}ign {M}atrices}, Theor. Math. Phys. \textbf{154} (2008), no.~3, 331--348,
  \href{http://arxiv.org/abs/math-ph/0703015}{\path{arXiv:math-ph/0703015}},
  {\scriptsize\href{http://dx.doi.org/10.1007/s11232-008-0031-x}{\path{doi:10.1007/s11232-008-0031-x}}}.

\bibitem{artic27}
Philippe Di~Francesco, Paul Zinn-Justin, and Jean-Bernard Zuber, \emph{A
  bijection between classes of fully packed loops and plane partitions},
  Electron. J. Combin. \textbf{11} (2004), no.~1, Research Paper 64, 11 pp,
  \href{http://arxiv.org/abs/math/0311220}{\path{arXiv:math/0311220}}.
  \MR{MR2097330 (2006d:05021)}.

\bibitem{artic30}
\bysame, \emph{Determinant formulae for some tiling problems and application to
  fully packed loops}, Ann. Inst. Fourier (Grenoble) \textbf{55} (2005), no.~6,
  2025--2050,
  \href{http://arxiv.org/abs/math-ph/0410002}{\path{arXiv:math-ph/0410002}}.
  \MR{MR2187944 (2007f:05037)}.

\bibitem{faddeev}
L.~D. Faddeev, \emph{How {A}lgebraic {B}ethe {A}nsatz works for integrable
  model}, 1996,
  \href{http://arxiv.org/abs/hep-th/9605187}{\path{arXiv:hep-th/9605187}}.

\bibitem{FK-Schubert}
Sergey Fomin and Anatol Kirillov, \emph{The {Y}ang--{B}axter equation,
  symmetric functions, and {S}chubert polynomials}, Discrete Math. \textbf{153}
  (1996), no.~1-3, 123--143, Proceedings of the 5th {C}onference on {F}ormal
  {P}ower {S}eries and {A}lgebraic {C}ombinatorics ({F}lorence, 1993),
  {\scriptsize\href{http://dx.doi.org/10.1016/0012-365X(95)00132-G}{\path{doi:10.1016/0012-365X(95)00132-G}}}.
  \MR{MR1394950 (98b:05101)}.

\bibitem{FS-Schubert}
Sergey Fomin and Richard Stanley, \emph{Schubert polynomials and the
  nil-{C}oxeter algebra}, Adv. Math. \textbf{103} (1994), no.~2, 196--207.
  \MR{MR1265793 (95f:05115)}.

\bibitem{artic45}
Tiago Fonseca and Paul Zinn-Justin, \emph{On the doubly refined enumeration of
  {A}lternating {S}ign {M}atrices and {T}otally {S}ymmetric
  {S}elf-{C}omplementary {P}lane {P}artitions}, Electron. J. Combin.
  \textbf{15} (2008), Research Paper 81, 35 pp,
  \href{http://arxiv.org/abs/0803.1595}{\path{arXiv:0803.1595}}.
  \MR{MR2411458}.

\bibitem{FR-qKZ}
Igor Frenkel and Nicolai Reshetikhin, \emph{Quantum affine algebras and
  holonomic difference equations}, Comm. Math. Phys. \textbf{146} (1992),
  no.~1, 1--60,
  \url{http://projecteuclid.org/getRecord?id=euclid.cmp/1104249974}.
  \MR{1163666 (94c:17024)}.

\bibitem{GV}
Ira Gessel and Xavier Viennot, \emph{Binomial determinants, paths, and hook
  length formulae}, Adv. in Math. \textbf{58} (1985), no.~3, 300--321.
  \MR{MR815360 (87e:05008)}.

\bibitem{Hotta}
Ryoshi Hotta, \emph{On {J}oseph's construction of {W}eyl group
  representations}, Tohoku Math. J. (2) \textbf{36} (1984), no.~1, 49--74.
  \MR{MR733619 (86h:20061)}.

\bibitem{HWKK-5v}
H.Y. Huang, F.Y. Wu, H.~Kunz, and D.~Kim, \emph{Interacting dimers on the
  honeycomb lattice: an exact solution of the five-vertex model}, Physica A
  \textbf{228} (1996), 1--32,
  \href{http://arxiv.org/abs/cond-mat/9510161}{\path{arXiv:cond-mat/9510161}}.

\bibitem{IZ}
Claude Itzykson and Jean-Bernard Zuber, \emph{The planar approximation. {II}},
  J. Math. Phys. \textbf{21} (1980), no.~3, 411--421. \MR{MR562985
  (81a:81068)}.

\bibitem{Iz-6v}
Anatoli Izergin, \emph{Partition function of a six-vertex model in a finite
  volume}, Dokl. Akad. Nauk SSSR \textbf{297} (1987), no.~2, 331--333.
  \MR{MR919260 (88m:82040)}.

\bibitem{ICK}
Anatoli Izergin, David Coker, and Vladimir Korepin, \emph{Determinant formula
  for the six-vertex model}, J. Phys. A \textbf{25} (1992), no.~16, 4315--4334.
  \MR{MR1181591 (94d:82016)}.

\bibitem{JK-FPL2}
Jesper Jacobsen and Jan{\'e} Kondev, \emph{Field theory of compact polymers on
  the square lattice}, Nucl. Phys. B \textbf{532} (1998), 635--688,
  \href{http://arxiv.org/abs/cond-mat/9804048}{\path{arXiv:cond-mat/9804048}}.

\bibitem{JK-flory}
\bysame, \emph{Conformal field theory of the {F}lory model of polymer melting},
  Phys. Rev. E \textbf{69} (2004), no.~6, 066108,
  \href{http://arxiv.org/abs/cond-mat/0209247}{\path{arXiv:cond-mat/0209247}},
  {\scriptsize\href{http://dx.doi.org/10.1103/PhysRevE.69.066108}{\path{doi:10.1103/PhysRevE.69.066108}}}.

\bibitem{artic28}
Jesper Jacobsen and Paul Zinn-Justin, \emph{Algebraic {B}ethe {A}nsatz for the
  {FPL$^2$} model}, J. Phys. A \textbf{37} (2004), no.~29, 7213--7225,
  \href{http://arxiv.org/abs/math-ph/0402008}{\path{arXiv:math-ph/0402008}},
  {\scriptsize\href{http://dx.doi.org/10.1088/0305-4470/37/29/004}{\path{doi:10.1088/0305-4470/37/29/004}}}.
  \MR{MR2078954 (2005e:82014)}.

\bibitem{JM-ff}
Michio Jimbo and Tetsuji Miwa, \emph{Solitons and infinite-dimensional {L}ie
  algebras}, Publ. Res. Inst. Math. Sci. \textbf{19} (1983), no.~3, 943--1001.
  \MR{MR723457 (85i:58060)}.

\bibitem{JM-book}
\bysame, \emph{Algebraic analysis of solvable lattice models}, CBMS Regional
  Conference Series in Mathematics, vol.~85, Published for the Conference Board
  of the Mathematical Sciences, Washington, DC, 1995. \MR{MR1308712
  (96e:82037)}.

\bibitem{JM-gapless}
\bysame, \emph{Quantum {KZ} equation with {$\vert q\vert =1$} and correlation
  functions of the {$XXZ$} model in the gapless regime}, J. Phys. A \textbf{29}
  (1996), no.~12, 2923--2958,
  {\scriptsize\href{http://dx.doi.org/10.1088/0305-4470/29/12/005}{\path{doi:10.1088/0305-4470/29/12/005}}}.
  \MR{1398600 (97i:82021)}.

\bibitem{JPS}
William Jockusch, Jim Propp, and Peter Shor, \emph{Random domino tilings and
  the arctic circle theorem}, 1998,
  \href{http://arxiv.org/abs/math.CO/9801068}{\path{arXiv:math.CO/9801068}}.

\bibitem{Joh-arctic}
Kurt Johansson, \emph{The arctic circle boundary and the {A}iry process}, Ann.
  Probab. \textbf{33} (2005), no.~1, 1--30,
  \href{http://arxiv.org/abs/math/0306216}{\path{arXiv:math/0306216}}.
  \MR{MR2118857 (2005k:60304)}.

\bibitem{Jos-mdeg}
Anthony Joseph, \emph{Orbital varieties, {G}oldie rank polynomials and unitary
  highest weight modules}, Algebraic and analytic methods in representation
  theory ({S}\o nderborg, 1994), Perspect. Math., vol.~17, Academic Press, San
  Diego, CA, 1997, pp.~53--98. \MR{MR1415842 (97k:17017)}.

\bibitem{Kas-wheel}
Masahiro Kasatani, \emph{Subrepresentations in the polynomial representation of
  the double affine {H}ecke algebra of type {$GL\sb n$} at {$t\sp {k+1}q\sp
  {r-1}=1$}}, Int. Math. Res. Not. (2005), no.~28, 1717--1742,
  \href{http://arxiv.org/abs/math/0501272}{\path{arXiv:math/0501272}}.
  \MR{MR2172339 (2007c:20012)}.

\bibitem{KT-qKZ}
Masahiro Kasatani and Yoshihiro Takeyama, \emph{The quantum
  {K}nizhnik--{Z}amolodchikov equation and non-symmetric {M}acdonald
  polynomials}, Funkcial. Ekvac. \textbf{50} (2007), no.~3, 491--509,
  \href{http://arxiv.org/abs/math/0608773}{\path{arXiv:math/0608773}}.
  \MR{MR2381328}.

\bibitem{Kast}
Pieter Kasteleyn, \emph{Graph theory and crystal physics}, Graph {T}heory and
  {T}heoretical {P}hysics, Academic Press, London, 1967, pp.~43--110.
  \MR{MR0253689 (40 \#6903)}.

\bibitem{KL}
David Kazhdan and George Lusztig, \emph{Representations of {C}oxeter groups and
  {H}ecke algebras}, Invent. Math. \textbf{53} (1979), no.~2, 165--184,
  {\scriptsize\href{http://dx.doi.org/10.1007/BF01390031}{\path{doi:10.1007/BF01390031}}}.
  \MR{MR560412 (81j:20066)}.

\bibitem{Kenyon-notes}
Richard Kenyon, \emph{Lectures on dimers}, Statistical mechanics, IAS/Park City
  Math. Ser., vol.~16, Amer. Math. Soc., Providence, RI, 2009, pp.~191--230,
  \url{http://www.math.brown.edu/~rkenyon/papers/dimerlecturenotes.pdf}.
  \MR{2523460 (2010j:82023)}.

\bibitem{KL-KL}
Alexander Kirillov, Jr. and Alain Lascoux, \emph{Factorization of
  {K}azhdan--{L}usztig elements for {G}rassmannians}, Combinatorial methods in
  representation theory ({K}yoto, 1998), Adv. Stud. Pure Math., vol.~28,
  Kinokuniya, Tokyo, 2000, pp.~143--154,
  \href{http://arxiv.org/abs/math.CO/9902072}{\path{arXiv:math.CO/9902072}}.
  \MR{MR1864480 (2002i:20007)}.

\bibitem{Kn-uu}
Allen Knutson, \emph{Some schemes related to the commuting variety}, J.
  Algebraic Geom. \textbf{14} (2005), no.~2, 283--294,
  \href{http://arxiv.org/abs/math.AG/0306275}{\path{arXiv:math.AG/0306275}}.
  \MR{MR2123231 (2005j:14076)}.

\bibitem{KM-Schubert}
Allen Knutson and Ezra Miller, \emph{Gr\"obner geometry of {S}chubert
  polynomials}, Ann. of Math. (2) \textbf{161} (2005), no.~3, 1245--1318.
  \MR{MR2180402 (2006i:05177)}.

\bibitem{KMY}
Allen Knutson, Ezra Miller, and A.~Yong, \emph{Gr\"obner geometry of vertex
  decompositions and of flagged tableaux}, J. Reine Angew. Math. \textbf{630}
  (2009), 1--31,
  \href{http://arxiv.org/abs/math.CO/0502144}{\path{arXiv:math.CO/0502144}},
  {\scriptsize\href{http://dx.doi.org/10.1515/CRELLE.2009.033}{\path{doi:10.1515/CRELLE.2009.033}}}.
  \MR{2526784}.

\bibitem{artic33}
Allen Knutson and Paul Zinn-Justin, \emph{A scheme related to the {B}rauer loop
  model}, Adv. Math. \textbf{214} (2007), no.~1, 40--77,
  \href{http://arxiv.org/abs/math.AG/0503224}{\path{arXiv:math.AG/0503224}}.
  \MR{MR2348022 (2008j:14097)}.

\bibitem{artic39}
\bysame, \emph{The {B}rauer loop scheme and orbital varieties}, J. Geom. Phys.
  \textbf{78} (2014), 80--110,
  \href{http://arxiv.org/abs/1001.3335}{\path{arXiv:1001.3335}},
  {\scriptsize\href{http://dx.doi.org/10.1016/j.geomphys.2014.01.006}{\path{doi:10.1016/j.geomphys.2014.01.006}}}.

\bibitem{Kor}
Vladimir Korepin, \emph{Calculation of norms of {B}ethe wave functions}, Comm.
  Math. Phys. \textbf{86} (1982), no.~3, 391--418. \MR{MR677006 (83m:81078)}.

\bibitem{artic12}
Vladimir Korepin and Paul Zinn-Justin, \emph{Thermodynamic limit of the
  six-vertex model with domain wall boundary conditions}, J. Phys. A
  \textbf{33} (2000), no.~40, 7053--7066,
  \href{http://arxiv.org/abs/cond-mat/0004250}{\path{arXiv:cond-mat/0004250}},
  {\scriptsize\href{http://dx.doi.org/10.1088/0305-4470/33/40/304}{\path{doi:10.1088/0305-4470/33/40/304}}}.
  \MR{MR1792450 (2001h:82018)}.

\bibitem{Kup-ASM}
Greg Kuperberg, \emph{Another proof of the alternating-sign matrix conjecture},
  Internat. Math. Res. Notices (1996), no.~3, 139--150,
  \href{http://arxiv.org/abs/math/9712207}{\path{arXiv:math/9712207}}.
  \MR{MR1383754 (97c:05009)}.

\bibitem{Las}
Alain Lascoux, \emph{Symmetric functions and combinatorial operators on
  polynomials}, CBMS Regional Conference Series in Mathematics, vol.~99,
  Published for the Conference Board of the Mathematical Sciences, Washington,
  DC, 2003. \MR{MR2017492 (2005b:05217)}.

\bibitem{LS-KL}
Alain Lascoux and Marcel-Paul Sch{\"u}tzenberger, \emph{Polyn\^omes de
  {K}azhdan \& {L}usztig pour les grassmanniennes}, Young tableaux and {S}chur
  functors in algebra and geometry ({T}oru\'n, 1980), Ast\'erisque, vol.~87,
  Soc. Math. France, Paris, 1981, pp.~249--266. \MR{MR646823 (83i:14045)}.

\bibitem{LSW1}
G.~Lawler, O.~Schramm, and W.~Werner, \emph{Values of {B}rownian intersection
  exponents. {I}. {H}alf-plane exponents}, Acta Math. \textbf{187} (2001),
  no.~2, 237--273. \MR{MR1879850 (2002m:60159a)}.

\bibitem{LSW2}
\bysame, \emph{Values of {B}rownian intersection exponents. {II}. {P}lane
  exponents}, Acta Math. \textbf{187} (2001), no.~2, 275--308. \MR{MR1879851
  (2002m:60159b)}.

\bibitem{LSW3}
\bysame, \emph{Values of {B}rownian intersection exponents. {III}. {T}wo-sided
  exponents}, Ann. Inst. H. Poincar\'e Probab. Statist. \textbf{38} (2002),
  no.~1, 109--123. \MR{MR1899232 (2003d:60163)}.

\bibitem{Lieb2}
Elliott Lieb, \emph{Exact solution of the {$F$} model of an antiferroelectric},
  Phys. Rev. Lett. \textbf{18} (1967), no.~24, 1046--1048,
  {\scriptsize\href{http://dx.doi.org/10.1103/PhysRevLett.18.1046}{\path{doi:10.1103/PhysRevLett.18.1046}}}.

\bibitem{Lieb1}
\bysame, \emph{Exact solution of the problem of the entropy of two-dimensional
  ice}, Phys. Rev. Lett. \textbf{18} (1967), no.~17, 692--694,
  {\scriptsize\href{http://dx.doi.org/10.1103/PhysRevLett.18.692}{\path{doi:10.1103/PhysRevLett.18.692}}}.

\bibitem{Lieb3}
\bysame, \emph{Exact solution of the two-dimensional {S}later {KDP} model of a
  ferroelectric}, Phys. Rev. Lett. \textbf{19} (1967), no.~3, 108--110,
  {\scriptsize\href{http://dx.doi.org/10.1103/PhysRevLett.19.108}{\path{doi:10.1103/PhysRevLett.19.108}}}.

\bibitem{Lieb4}
\bysame, \emph{Residual entropy of square ice}, Phys. Rev. \textbf{162} (1967),
  no.~1, 162--172,
  {\scriptsize\href{http://dx.doi.org/10.1103/PhysRev.162.162}{\path{doi:10.1103/PhysRev.162.162}}}.

\bibitem{Lind}
Bernt Lindstr{\"o}m, \emph{On the vector representations of induced matroids},
  Bull. London Math. Soc. \textbf{5} (1973), 85--90. \MR{MR0335313 (49 \#95)}.

\bibitem{Macdo-nonsym}
Ian~G. Macdonald, \emph{Affine {H}ecke algebras and orthogonal polynomials},
  Ast\'erisque (1996), no.~237, Exp.\ No.\ 797, 4, 189--207, S{\'e}minaire
  Bourbaki, Vol. 1994/95. \MR{MR1423624 (99f:33024)}.

\bibitem{MacM}
Percy MacMahon, \emph{Combinatory analysis}, Cambridge University Press, 1915.

\bibitem{Meln}
A.~Melnikov, \emph{{$B$}-orbits in solutions to the equation {$X^2=0$} in
  triangular matrices}, J. Algebra \textbf{223} (2000), no.~1, 101--108.
  \MR{MR1738254 (2001b:20083)}.

\bibitem{MS-mdeg}
Ezra Miller and Bernd Sturmfels, \emph{Combinatorial commutative algebra},
  Graduate Texts in Mathematics, vol. 227, Springer-Verlag, New York, 2005.
  \MR{MR2110098 (2006d:13001)}.

\bibitem{MRR-ASM}
W.~H. Mills, David Robbins, and Howard Rumsey, Jr., \emph{Alternating sign
  matrices and descending plane partitions}, J. Combin. Theory Ser. A
  \textbf{34} (1983), no.~3, 340--359. \MR{MR700040 (85b:05013)}.

\bibitem{MRR-TSSCPP}
\bysame, \emph{Self-complementary totally symmetric plane partitions}, J.
  Combin. Theory Ser. A \textbf{42} (1986), no.~2, 277--292. \MR{MR847558
  (88b:05008)}.

\bibitem{Miwa-transf}
Tetsuji Miwa, \emph{On {H}irota's difference equations}, Proc. Japan Acad. Ser.
  A Math. Sci. \textbf{58} (1982), no.~1, 9--12,
  \url{http://projecteuclid.org/euclid.pja/1195516178}. \MR{MR649054
  (83f:58042)}.

\bibitem{Molev-coproduct}
Alexander Molev, \emph{Comultiplication rules for the double {S}chur functions
  and {C}auchy identities}, 2008,
  \href{http://arxiv.org/abs/0807.2127}{\path{arXiv:0807.2127}}.

\bibitem{MS-doubleschur}
Alexander Molev and Bruce Sagan, \emph{A {L}ittlewood--{R}ichardson rule for
  factorial {S}chur functions}, Transactions of the American Mathematical
  Society \textbf{351} (1999), no.~11, 4429--4443,
  \href{http://arxiv.org/abs/q-alg/9707028}{\path{arXiv:q-alg/9707028}}.

\bibitem{Nien-CG}
Bernard Nienhuis, \emph{{C}oulomb gas formulation of two-dimensional phase
  transitions}, Phase Transitions and Critical Phenomena (C.~Domb and J.~L.
  Lebowitz, eds.), vol.~11, Academic Press, London, 1987, pp.~1--53.

\bibitem{NK-5v}
J.~D. Noh and D.~Kim, \emph{Interacting domain walls and the five-vertex
  model}, Phys. Rev. E \textbf{49} (1994), no.~3, 1943--1961,
  \href{http://arxiv.org/abs/cond-mat/9312001}{\path{arXiv:cond-mat/9312001}},
  {\scriptsize\href{http://dx.doi.org/10.1103/PhysRevE.49.1943}{\path{doi:10.1103/PhysRevE.49.1943}}}.

\bibitem{Nolden-6v}
I.~Nolden, \emph{The asymmetric six-vertex model}, J. Stat. Phys. \textbf{67}
  (1992), 155--201,
  {\scriptsize\href{http://dx.doi.org/10.1007/BF01049030}{\path{doi:10.1007/BF01049030}}}.

\bibitem{Oka}
Soichi Okada, \emph{Enumeration of symmetry classes of alternating sign
  matrices and characters of classical groups}, J. Algebraic Combin.
  \textbf{23} (2006), no.~1, 43--69,
  \href{http://arxiv.org/abs/math/0408234}{\path{arXiv:math/0408234}},
  {\scriptsize\href{http://dx.doi.org/10.1007/s10801-006-6028-3}{\path{doi:10.1007/s10801-006-6028-3}}}.
  \MR{MR2218849 (2007d:05155)}.

\bibitem{OR}
Andrei Okounkov and Nicolai Reshetikhin, \emph{Correlation function of {S}chur
  process with application to local geometry of a random 3-dimensional {Y}oung
  diagram}, J. Amer. Math. Soc. \textbf{16} (2003), no.~3, 581--603,
  \href{http://arxiv.org/abs/math.CO/0107056}{\path{arXiv:math.CO/0107056}}.
  \MR{MR1969205 (2004b:60033)}.

\bibitem{Pas-scatt}
Vincent Pasquier, \emph{Scattering matrices and affine {H}ecke algebras},
  Low-dimensional models in statistical physics and quantum field theory
  ({S}chladming, 1995), Lecture Notes in Phys., vol. 469, Springer, Berlin,
  1996, pp.~145--163,
  \href{http://arxiv.org/abs/q-alg/9508002}{\path{arXiv:q-alg/9508002}}.
  \MR{MR1477945}.

\bibitem{Pas-BWM}
\bysame, \emph{Incompressible representations of the
  {B}irman--{W}enzl--{M}urakami algebra}, Ann. Henri Poincar\'e \textbf{7}
  (2006), no.~4, 603--619,
  \href{http://arxiv.org/abs/math/0507364}{\path{arXiv:math/0507364}}.
  \MR{MR2232366 (2007d:16026)}.

\bibitem{Pas-RS}
\bysame, \emph{Quantum incompressibility and {R}azumov {S}troganov type
  conjectures}, Ann. Henri Poincar\'e \textbf{7} (2006), no.~3, 397--421,
  \href{http://arxiv.org/abs/cond-mat/0506075}{\path{arXiv:cond-mat/0506075}}.
  \MR{MR2226742 (2007d:82012)}.

\bibitem{RS-conj}
Alexander Razumov and Yuri Stroganov, \emph{Combinatorial nature of the
  ground-state vector of the {$O(1)$} loop model}, Teoret. Mat. Fiz.
  \textbf{138} (2004), no.~3, 395--400,
  \href{http://arxiv.org/abs/math/0104216}{\path{arXiv:math/0104216}},
  {\scriptsize\href{http://dx.doi.org/10.1023/B:TAMP.0000018450.36514.d7}{\path{doi:10.1023/B:TAMP.0000018450.36514.d7}}}.
  \MR{MR2077318 (2005f:82022)}.

\bibitem{RS-bethe}
\bysame, \emph{Bethe roots and refined enumeration of alternating-sign
  matrices}, J. Stat. Mech. Theory Exp. (2006), no.~7, P07004, 12 pp,
  \href{http://arxiv.org/abs/math-ph/0605004}{\path{arXiv:math-ph/0605004}}.
  \MR{MR2244324 (2007h:82014)}.

\bibitem{artic42}
Alexander Razumov, Yuri Stroganov, and Paul Zinn-Justin, \emph{Polynomial
  solutions of {$q$KZ} equation and ground state of {$XXZ$} spin chain at
  {$\Delta=-1/2$}}, J. Phys. A \textbf{40} (2007), no.~39, 11827--11847,
  \href{http://arxiv.org/abs/0704.3542}{\path{arXiv:0704.3542}},
  {\scriptsize\href{http://dx.doi.org/10.1088/1751-8113/40/39/009}{\path{doi:10.1088/1751-8113/40/39/009}}}.
  \MR{MR2374053}.

\bibitem{Resh-lectures}
Nicolai Reshetikhin, \emph{Lectures on the six-vertex model}, Les Houches
  lecture notes.

\bibitem{rothbach}
B.~Rothbach, \emph{Equidimensionality of the {B}rauer loop scheme}, Electron.
  J. Combin. \textbf{17} (2010), no.~1, Research Paper 75, 11,
  \url{http://www.combinatorics.org/Volume_17/Abstracts/v17i1r75.html}.
  \MR{2651728}.

\bibitem{SB-6v}
J.~Shore and D.~J. Bukman, \emph{Coexistence point in the six-vertex model and
  the crystal shape of fcc materials}, Phys. Rev. Lett. \textbf{72} (1994),
  no.~5, 604--607,
  {\scriptsize\href{http://dx.doi.org/10.1103/PhysRevLett.72.604}{\path{doi:10.1103/PhysRevLett.72.604}}}.

\bibitem{Smi}
Fedor Smirnov, \emph{A general formula for soliton form factors in the quantum
  sine-{G}ordon model}, J. Phys. A \textbf{19} (1986), no.~10, L575--L578,
  {\scriptsize\href{http://dx.doi.org/10.1088/0305-4470/19/10/003}{\path{doi:10.1088/0305-4470/19/10/003}}}.
  \MR{851469 (87j:81115)}.

\bibitem{Stan}
Richard Stanley, \emph{A baker's dozen of conjectures concerning plane
  partitions}, Combinatoire \'enum\'erative ({M}ontreal, {Q}ue., 1985/{Q}uebec,
  {Q}ue., 1985), Lecture Notes in Math., vol. 1234, Springer, Berlin, 1986,
  pp.~285--293. \MR{MR927770 (89h:05008)}.

\bibitem{Strog-IK}
Yuri Stroganov, \emph{{I}zergin--{K}orepin determinant at a third root of
  unity}, Teoret. Mat. Fiz. \textbf{146} (2006), no.~1, 65--76,
  \href{http://arxiv.org/abs/math-ph/0204042}{\path{arXiv:math-ph/0204042}}.
  \MR{MR2243403 (2007f:82026)}.

\bibitem{Suth}
Bill Sutherland, \emph{Exact solution of a two-dimensional model for
  hydrogen-bonded crystals}, Phys. Rev. Lett. \textbf{19} (1967), no.~3,
  103--104,
  {\scriptsize\href{http://dx.doi.org/10.1103/PhysRevLett.19.103}{\path{doi:10.1103/PhysRevLett.19.103}}}.

\bibitem{Wieland}
Benjamin Wieland, \emph{A large dihedral symmetry of the set of alternating
  sign matrices}, Electron. J. Combin. \textbf{7} (2000), Research Paper 37, 13
  pp, \href{http://arxiv.org/abs/math/0006234}{\path{arXiv:math/0006234}}.
  \MR{MR1773294 (2001g:05016)}.

\bibitem{Wu-5v}
F.~Y. Wu, \emph{Remarks on the modified potassium dihydrogen phosphate model of
  a ferroelectric}, Phys. Rev. \textbf{168} (1968), no.~2, 539--543,
  {\scriptsize\href{http://dx.doi.org/10.1103/PhysRev.168.539}{\path{doi:10.1103/PhysRev.168.539}}}.

\bibitem{Zeil-ASM}
Doron Zeilberger, \emph{Proof of the alternating sign matrix conjecture},
  Electron. J. Combin. \textbf{3} (1996), no.~2, Research Paper 13, 84 pp, The
  Foata Festschrift,
  \href{http://arxiv.org/abs/math/9407211}{\path{arXiv:math/9407211}}.
  \MR{MR1392498 (97d:05012)}.

\bibitem{Zeil-qKZ}
\bysame, \emph{Proof of a conjecture of {P}hilippe {D}i {F}rancesco and {P}aul
  {Z}inn-{J}ustin related to the {$q$KZ} equations and to {D}ave {R}obbins' two
  favorite combinatorial objects}, 2007,
  \url{http://www.math.rutgers.edu/~zeilberg/mamarim/mamarimhtml/diFrancesco.html}.

\bibitem{artic13}
Paul Zinn-Justin, \emph{Six-vertex model with domain wall boundary conditions
  and one-matrix model}, Phys. Rev. E \textbf{62} (2000), no.~3, part A,
  3411--3418,
  \href{http://arxiv.org/abs/math-ph/0005008}{\path{arXiv:math-ph/0005008}}.
  \MR{MR1788950 (2001g:82037)}.

\bibitem{artic21}
\bysame, \emph{H{CIZ} integral and 2{D} {T}oda lattice hierarchy}, Nuclear
  Phys. B \textbf{634} (2002), no.~3, 417--432,
  \href{http://arxiv.org/abs/math-ph/0202045}{\path{arXiv:math-ph/0202045}}.
  \MR{MR1912027 (2003f:81113)}.

\bibitem{artic20}
\bysame, \emph{The influence of boundary conditions in the six-vertex model},
  unpublished preprint, 2002,
  \href{http://arxiv.org/abs/cond-mat/0205192}{\path{arXiv:cond-mat/0205192}}.

\bibitem{artic38}
\bysame, \emph{Proof of the {R}azumov--{S}troganov conjecture for some infinite
  families of link patterns}, Electron. J. Combin. \textbf{13} (2006), no.~1,
  Research Paper 110, 15 pp,
  \href{http://arxiv.org/abs/math/0607183}{\path{arXiv:math/0607183}}.
  \MR{MR2274325 (2007k:82034)}.

\bibitem{artic37}
\bysame, \emph{Combinatorial point for fused loop models}, Comm. Math. Phys.
  \textbf{272} (2007), no.~3, 661--682,
  \href{http://arxiv.org/abs/math-ph/0603018}{\path{arXiv:math-ph/0603018}},
  {\scriptsize\href{http://dx.doi.org/10.1007/s00220-007-0225-3}{\path{doi:10.1007/s00220-007-0225-3}}}.
  \MR{MR2304471}.

\bibitem{artic46}
\bysame, \emph{{L}ittlewood--{R}ichardson coefficients and integrable tilings},
  Electron. J. Combin. \textbf{16} (2009), Research Paper 12,
  \href{http://arxiv.org/abs/0809.2392}{\path{arXiv:0809.2392}}.

\end{thebibliography}
\bibliographystyle{amsplainhyper}

\end{document}